\documentstyle[12pt,a4,cite,axodraw,psfig]{article}
\makeatletter
 
\renewenvironment{thebibliography}[1]
{\small
 \begin{list}{[\arabic{enumi}]}
 {\usecounter{enumi} \setlength{\parsep}{6pt}
  \setlength{\itemsep}{6pt} \settowidth{\labelwidth}{[#1]}
  \settowidth{\leftmargin}{[100]}
  \sloppy}}
 {\end{list}}
  
\renewcommand{\theequation}{\thesection.\arabic{equation}}
\newcommand{\cleqn}{\setcounter{equation}{0}}

\newtoks\@stequation

\def\subequations{\refstepcounter{equation}%
  \edef\@savedequation{\the\c@equation}%
  \@stequation=\expandafter{\theequation}
  \edef\@savedtheequation{\the\@stequation}
  \edef\oldtheequation{\theequation}%
  \setcounter{equation}{0}%
  \def\theequation{\oldtheequation\alph{equation}}}

\def\endsubequations{%
  \ifnum\c@equation < 2 \@warning{Only \the\c@equation\space subequation
    used in equation \@savedequation}\fi
  \setcounter{equation}{\@savedequation}%
  \@stequation=\expandafter{\@savedtheequation}%
  \edef\theequation{\the\@stequation}%
  \global\@ignoretrue}

\def\eqnarray{\stepcounter{equation}\let\@currentlabel\theequation
\global\@eqnswtrue\m@th
\global\@eqcnt\z@\tabskip\@centering\let\\\@eqncr
$$\halign to\displaywidth\bgroup\@eqnsel\hskip\@centering
     $\displaystyle\tabskip\z@{##}$&\global\@eqcnt\@ne
      \hfil$\;{##}\;$\hfil
     &\global\@eqcnt\tw@ $\displaystyle\tabskip\z@{##}$\hfil
   \tabskip\@centering&\llap{##}\tabskip\z@\cr}

\def\dlinepattern#1#2{%
\ifdim#2<#1
   \errmessage{the 1st argument is less than the 2nd argument.}%
\else
   \gdef\dline@solid{#1}\gdef\dline@period{#2}%
\fi}

\def\dline#1{\@dline[#1]}
\def\@dline[#1-#2]{\noalign{\global\@dla#1\relax
\global\advance\@dla\m@ne
\ifnum\@dla>\z@\global\let\@gtempa\@dlinea\else
  \global\let\@gtempa\@dlineb\fi
\global\@dlb#2\relax
\global\advance\@dlb-\@dla}\@gtempa
\noalign{\vskip-\arrayrulewidth}}

\def\@dlinea{\multispan\@dla&\multispan\@dlb
\unskip\cleaders\hbox to \dline@period
{\hss\rule{\dline@solid}{\arrayrulewidth}\hss}\hfill\cr}
 
\newcount\@dla
\newcount\@dlb

\def\@dlineb{\multispan\@dlb
\unskip\cleaders\hbox to \dline@period
{\hss\rule{\dline@solid}{\arrayrulewidth}\hss}\hfill\cr}

\dlinepattern{2pt}{5pt}

\makeatother

\newenvironment{oneitem}[1]{\begin{quote} {\bf #1\/:\pr} }{\end{quote}}

\newcommand\commentout[1]{}
\newcommand\COMMENTOUT[1]{}
\newcommand{\pr}{\hspace{\parindent}}

\def\bea{\begin{eqnarray}}
\def\eea{\end{eqnarray}}
\def\bsub{\begin{subequations}}
\def\esub{\end{subequations}}

\def\ol{\overline}

\def\simgt{\rlap{\lower 3.5 pt \hbox{$\mathchar \sim$}} \raise 1pt \hbox {$>$}}
\def\simlt{\rlap{\lower 3.5 pt \hbox{$\mathchar \sim$}} \raise 1pt \hbox {$<$}}

\def\mev{{\,\rm MeV}}

\def\gev{{\,\rm GeV}}
\def\GeV{{\,\rm GeV}}

\def\msbar{\overline{\rm MS }}	
\def\to{\rightarrow}
\def\epem{\ifmmode{ e^{+}e^-} \else{$ e^{+}e^- $ } \fi}
\def\ttbar{\ifmmode{t\bar{t}} \else{$t\bar{t}$} \fi}

\def\qqquad{\qquad \quad}

\def\sp3{\ifmmode{\:\:\:}\else{$\:\:\:$} \fi}

\def\pthree {\:\:\:}

\def\pfive  {\:\:\:\:\:}
\def\psix   {\:\:\:\:\:\:}

\def\mm    {\!\!}

\def\mfive {\!\!\!\!\!}

\def\mv{m_V^{}}
\def\mz{m_Z^{}}
\def\mw{m_W^{}}
\def\mh{m_H^{}}
\def\mmv{m_V^2}
\def\mmz{m_Z^2}
\def\mmw{m_W^2}
\def\mmh{m_H^2}

\def\ebar{\bar{e}}
\def\sbar{\bar{s}}
\def\cbar{\bar{c}}
\def\gzbar{\bar{g}_Z}
\def\gwbar{\bar{g}_W}

\def\ehat{\hat{e}}
\def\shat{\hat{s}}
\def\chat{\hat{c}}
\def\gzhat{\hat{g}_Z}
\def\ghat{\hat{g}}

\def\msbar{\overline{\rm MS}}

\def\pibar{\overline{\Pi}}
\def\gambar{\overline{\Gamma}}
\def\delb{\bar{\delta}_{b}}
\def\delg{\bar{\delta}_{G}^{}}
\def\zbb{Zb_L^{}b_L^{}}

\def\matrixelement#1#2#3#4{{M}^{#1#2\vphantom{f}}_{\!#3#4}}
\def\boxfactor#1#2#3#4{{B}^{#1#2\vphantom{f}}_{\!#3#4}}

\def\MefLL{\matrixelement efLL}
\def\MefLR{\matrixelement efLR}
\def\MefRL{\matrixelement efRL}
\def\MefRR{\matrixelement efRR}

\def\MeqLL{\matrixelement eqLL}
\def\MeqLR{\matrixelement eqLR}
\def\MeqRL{\matrixelement eqRL}
\def\MeqRR{\matrixelement eqRR}

\def\MelLL{\matrixelement e{\ell}LL}
\def\MelLR{\matrixelement e{\ell}LR}
\def\MelRL{\matrixelement e{\ell}RL}
\def\MelRR{\matrixelement e{\ell}RR}

\def\MnqLa{\matrixelement {\nu\!_\mu}qL{\alpha}}

\def\MneLL{\matrixelement {\nu\!_\mu}eLL}
\def\MneLR{\matrixelement {\nu\!_\mu}eLR}

\def\BeqLL{\boxfactor eqLL}
\def\BeqLR{\boxfactor eqLR}
\def\BeqRL{\boxfactor eqRL}
\def\BeqRR{\boxfactor eqRR}

\def\BnqLa{\boxfactor {\nu\!_\mu}qL\alpha}
\def\BnuLL{\boxfactor {\nu\!_\mu}uLL}
\def\BnuLR{\boxfactor {\nu\!_\mu}uLR}
\def\BndLL{\boxfactor {\nu\!_\mu}dLL}
\def\BndLR{\boxfactor {\nu\!_\mu}dLR}
\def\BneLL{\boxfactor {\nu\!_\mu}eLL}
\def\BneLR{\boxfactor {\nu\!_\mu}eLR}

\def\squared#1{\Bigl| #1 \Bigr|^2}
\def\alphasoverpi{\biggl(\frac{\alpha_s}{\pi}\biggr)}

\def\uL{u_L^{}}
\def\dL{d_L^{}}
\def\uR{u_R^{}}
\def\dR{d_R^{}}

\begin{document}
\thispagestyle{empty}
\vglue 10mm
\vspace*{-30mm}
\baselineskip10pt
\begin{flushright}
\begin{tabular}{l}
{\bf\small KEK-TH-375 }\\[-1mm]
{\bf\small KEK preprint 93--159 }\\[-1mm]
{\bf\small KANAZAWA-94-19}\\[-1mm]
{\bf\small DESY 94--002 }\\[-1mm]
{\bf\small YUMS 94--22 }\\[-1mm]
{\bf\small SNUTP 94--82 }\\[-1mm]
September 1994
\end{tabular}
\end{flushright}
\baselineskip18pt 
\vglue 4mm 
\begin{center}
{\large \bf A Novel Approach to Confront Electroweak Data and Theory  }\\
\vglue 10mm
\def\thefootnote{\fnsymbol{footnote}}
\setcounter{footnote}{0}
{\bf K.~Hagiwara,~~~S.~Matsumoto%
\footnote{%
  On leave from Department of Physics, Kanazawa University, 
  Kanazawa, 920-11, Japan}, } \\
\vglue 1mm
{\it Theory Group, KEK, Tsukuba, Ibaraki 305, Japan }\\ 
\vglue 3mm
{\bf D.~Haidt } \\
\vglue 1mm
{\it DESY, Notkestra{\ss}e 85, D-22603 Hamburg, Germany }\\
\vglue 3mm
{and} \\
\vglue 3mm
{\bf C.S.~Kim }\\
\vglue 1mm
{\it Department of Physics, Yonsei University, Seoul 120-749, Korea }\\
\vglue 10mm
{\bf ABSTRACT} \\
\vglue 4mm
\begin{minipage}{14cm}
{
A novel approach to study electroweak physics at one-loop level in
generic ${\rm SU(2)_L \times U(1)_Y}$ theories is introduced. It
separates the 1-loop corrections into two pieces: process specific ones
from vertex and box contributions, and universal ones from
contributions to the gauge boson propagators. The latter are
parametrized in terms of four effective form factors $\bar{e}^2(q^2)$,
$\bar{s}^2(q^2)$, $\bar{g}_Z^2(q^2)$ and $\bar{g}_W^2(q^2)$
corresponding to the $\gamma\gamma$, $\gamma Z$, $ZZ$ and $WW$
propagators. Under the assumption that only the Standard Model
contributes to the process specific corrections, the magnitudes of the
four form factors are determined at $q^2=0$ and at $q^2=\mmz$ by
fitting to all available precision experiments. These values are then
compared systematically with predictions of ${\rm SU(2)_L \times
U(1)_Y}$ theories.
In all fits $\alpha_s(\mz)$ and $\bar{\alpha}(\mmz)$ are treated as
external parameters in order to keep the interpretation as flexible
as possible. The treatment of the electroweak data is presented in
detail together with the relevant theoretical formulae used to interpret
the data. No deviation from the Standard Model has been identified. 
Ranges of the top quark and Higgs boson masses are derived as functions
of $\alpha_s(\mz)$ and $\bar{\alpha}(\mmz)$.
Also discussed are consequences of the recent precision measurement
of the left-right asymmetry at SLC as well as the impact of a top quark
mass and an improved $W$ mass measurement.
}
\end{minipage}
\\[8mm]
{\large {\it To be published in Zeitschrift f\"ur Physik C} }
\end{center}
\vglue 15mm
\baselineskip18pt  

\def\dataofzlineshape{%
  \begin{eqnarray}
     \label{dataofzlineshape}
     \left.
     \begin{array}{lll}
         m_Z({\rm GeV})      &\!=\!& 91.187 \pm 0.007  \\[1mm]
         \Gamma_Z({\rm GeV}) &\!=\!&  2.489 \pm 0.007  \\[1mm]
         \sigma_h^0({\rm nb})&\!=\!& 41.56  \pm 0.14   \\[1mm]
         R_\ell = \sigma_h^0/\sigma_\ell^0
                             &\!=\!& 20.763\pm  0.049  \\[1mm]
         A_{\rm FB}^{0,\ell} &\!=\!& 0.0158 \pm 0.0018
        \end{array}
        \right. 
        \rho_{\rm corr} =
        \left(
        \begin{array}{rrrrr}
          1 & -0.157 &  0.017 &  0.012 &  0.075 \\[1mm]
            &  1     & -0.070 &  0.003 &  0.006 \\[1mm]
            &        &  1     &  0.137 &  0.003 \\[1mm]
            &        &        &  1     &  0.008 \\[1mm]
            &        &        &        &  1    
        \end{array}
        \right). \hspace{4mm}
  \end{eqnarray}
}%
\def\dataofzothers{%
  \begin{subequations}
    \label{dataofzothers}
  \begin{eqnarray}
      \label{dataofptau}
  	P_\tau           &=& -0.139  \pm 0.014 ,\\
      \label{dataofoldalr}
  	A_{\rm LR}^0     &=&  0.10   \pm 0.044
                                 \psix \qquad (\mbox{SLD\cite{alr92}}),\\
      \label{dataofafbb}
  	A_{\rm FB}^{0,b} &=&  0.099  \pm 0.006 ,\\
      \label{dataofafbc}
  	A_{\rm FB}^{0,c} &=&  0.075  \pm 0.015 ,\\
      \label{dataofrb}
  	R_b =\sigma_b^0/\sigma_h^0 
	     &=&  0.2203 \pm 0.0027 \pthree \quad ({\rm LEP + SLD}).
  \end{eqnarray}
  \end{subequations}
}%
\def\fitofzparameter{%
  \begin{subequations}
  \label{fitofzparameter}
  \begin{eqnarray}
     \left.
     \begin{array}{lll}
        \gzbar^2(m_Z^2)&\mm =&
            0.5542 -0.00030\,\frac{\alpha_s-0.12}{0.01} \pm 0.0017 \\[1mm]
        \sbar^2(m_Z^2) &\mm =&
            0.2313 +0.00008\,\frac{\alpha_s-0.12}{0.01} \pm 0.0007 \\[1mm]
        \delb(m_Z^2)   &\mm =&\mfive
           -0.0061 -0.00430\,\frac{\alpha_s-0.12}{0.01} \pm 0.0034
     \end{array}
     \right. 
        \rho_{\rm corr} =
        \left(
        \begin{array}{rrr}
          1 &  0.14 & -0.36 \\[1mm]
            &  1    &  0.20 \\[1mm]
            &       &  1
       	\end{array}
        \right) \,,\quad
   \\[2mm]
     \label{chisqofzparameter}
	\chi^2_{\rm min} =
            1.53 +\biggl(\frac{\alpha_s-0.1029}{0.0128}\biggr)^2,
         \hspace{65mm}
     \end{eqnarray}
  \end{subequations}
}%
\def\fitofzparametersummary{%
  \begin{subequations}
  \label{fitofzparametersummary}
  \begin{eqnarray}
     \left.
     \begin{array}{lll}
        \gzbar^2(m_Z^2)&\mm =&
           0.5542 -0.00030\,\frac{\alpha_s-0.12}{0.01} \pm 0.0017 \\[1mm]
        \sbar^2(m_Z^2) &\mm =&
           0.2313 +0.00008\,\frac{\alpha_s-0.12}{0.01} \pm 0.0007 \\[1mm]
        \delb(m_Z^2)   &\mm =&\mfive
          -0.0061 -0.00430\,\frac{\alpha_s-0.12}{0.01} \pm 0.0034
     \end{array}
     \right. 
        \rho_{\rm corr} =
        \left(
        \begin{array}{rrr}
          1 &  0.14 & -0.36 \\[1mm]
            &  1    &  0.20 \\[1mm]
            &       &  1
       	\end{array}
        \right) \,,\quad
   \\[2mm]
     \label{chisqofzparametersummary}
	\chi^2_{\rm min}/{\rm (d.o.f.)} =
          \biggl[ 
              2.48 +\biggl(\frac{\alpha_s-0.1017}{0.0127}\biggr)^2
          \biggr] \bigg/6.
         \hspace{65mm}
     \end{eqnarray}
  \end{subequations}
}%
\def\fitofzasymmetries{%
  \begin{subequations}
     \label{fitofzasymmetries}
  \begin{eqnarray}
     \label{fitofsbarfromafbl}
	\sbar^2(m_Z^2) &=& 0.2309 \hspace{30.5mm} \pm 0.0010
		\quad\:\:  ({\rm from}\: A_{\rm FB}^{0,\ell}) ,\\
     \label{fitofsbarfromptau}
	\sbar^2(m_Z^2) &=& 0.2316 \hspace{30.5mm} \pm 0.0018
		\quad\:\:  ({\rm from}\: P_{\tau}) ,\\
     \label{fitofsbarfromalr}
  	\sbar^2(m_Z^2) &=& 0.2365 \hspace{30.5mm} \pm 0.0055
		\quad\:\:  ({\rm from}\: A_{\rm LR}^0) ,\\ 
     \label{fitofsbarfromafbb}
  	\sbar^2(m_Z^2) &=& 0.2313 
                          +\mbox{$0.00004\frac{\alpha_s-0.12}{0.01}$}
                           \pm 0.0011
	       	\quad\:\:  ({\rm from}\: A_{\rm FB}^{0,b}) ,\\
     \label{fitofsbarfromafbc}
  	\sbar^2(m_Z^2) &=& 0.2302
                         +\mbox{$0.00004\frac{\alpha_s-0.12}{0.01}$}
                           \pm 0.0035
	       	\quad\:\:  ({\rm from}\: A_{\rm FB}^{0,c}),
  \end{eqnarray}
  \end{subequations}
}%
\def\fitofzasymmetriesall{%
  \begin{subequations}
     \label{fitofzasymmetriesall}
  \begin{eqnarray}
     \label{fitofsbarfromlepton}
	\sbar^2(m_Z^2) &=& 0.2312 \hspace{30.5mm} \pm 0.0009\quad\: 
        ({\rm from}\: A_{\rm FB}^{0,\ell}, P_{\tau}, A_{\rm LR}^0) ,\\
     \label{fitofsbarfromasymmetries}
  	\sbar^2(m_Z^2) &=& 0.2312
                          +\mbox{$0.00002\frac{\alpha_s-0.12}{0.01}$}
                          \pm 0.0007
	\quad\: ({\rm from}\: 
	A_{\rm FB}^{0,\ell}, P_{\tau}, A_{\rm LR}^0, 
        A_{\rm FB}^{0,b}, A_{\rm FB}^{0,c}) . \quad
  \end{eqnarray}
  \end{subequations}
}%
\def\fitofdelbalps{%
  \begin{eqnarray}
    \label{fitofdelbalps}
    \left.
    \begin{array}{lll}
      \delb(m_Z^2)  &\!\!=&\! 0.0014 \pm 0.0070 \\[1mm]
      \alpha_s(\mz) &\!\!=&\, 0.103  \pm 0.013
    \end{array}
    \right\}\;
    \rho_{\rm corr}=-0.85. 
  \end{eqnarray}
}
\def\fitofdelb{%
  \begin{subequations}
  \label{fit_delb}
  \begin{eqnarray}
	\delb(\mmz) &=& 
		-0.0068 
		-0.0084\,\mbox{$\frac{\gzbar^2(\mmz)-0.5542}{0.0017}$}
		+0.0020\,\mbox{$\frac{\sbar^2(\mmz)-0.2313}{0.0007}$}
		-0.0061\,\mbox{$\frac{\alpha_s-0.12}{0.01}$}
			\:\pm 0.0077
	\nonumber \\ & &
	\mbox{\hspace{100mm}(from $\Gamma_Z$)}\,,
  \label{delb_gamz}\\
	\delb(\mmz) &=&
		-0.0210
		+0.0000\,\mbox{$\frac{\gzbar^2(\mmz)-0.5542}{0.0017}$}
		+0.0004\,\mbox{$\frac{\sbar^2(\mmz)-0.2313}{0.0007}$}
		-0.0063\,\mbox{$\frac{\alpha_s-0.12}{0.01}$}
			\:\pm 0.0168
	\nonumber\\ &&
	\mbox{\hspace{100mm}(from $\sigma_h^0$ )}\,,
  \label{delb_sigh}\\
	\delb(\mmz) &=&
		-0.0078 
		+0.0000\,\mbox{$\frac{\gzbar^2(\mmz)-0.5542}{0.0017}$}
		+0.0011\,\mbox{$\frac{\sbar^2(\mmz)-0.2313}{0.0007}$}
		-0.0061\,\mbox{$\frac{\alpha_s-0.12}{0.01}$}
			\:\pm 0.0044
	\nonumber\\ &&
	\mbox{\hspace{100mm}(from $R_\ell$ )}\,,
  \label{delb_rl}\\
	\delb(\mmz) &=&
		 \hphantom{-}0.0012 	
		+0.0000\,\mbox{$\frac{\gzbar^2(\mmz)-0.5542}{0.0017}$}
		-0.0001\,\mbox{$\frac{\sbar^2(\mmz)-0.2313}{0.0007}$}
		-0.0001\,\mbox{$\frac{\alpha_s-0.12}{0.01}$}
			\:\pm 0.0068
	\nonumber\\ &&
	\mbox{\hspace{100mm}(from $R_b$ )}\,.
  \label{delb_rb}
  \end{eqnarray}
  \end{subequations}
}
\def\fitofdelball{%
  \begin{eqnarray} 
	\delb(\mmz) &=&
		-0.0062	
		-0.0014\,\mbox{$\frac{\gzbar^2(\mmz)-0.5542}{0.0017}$}
		+0.0009\,\mbox{$\frac{\sbar^2(\mmz)-0.2313}{0.0007}$}
		-0.0046\,\mbox{$\frac{\alpha_s-0.12}{0.01}$}
			\,\pm 0.0031,
  \nonumber\\ && 
  \label{delb_all}
  \end{eqnarray}
}
\def\dataofnuqorig{%
  \begin{equation}
     \label{dataofnuqorig}
     \begin{array}{llr}
        g^2_L    = 
                [\quad 0.2982 -0.0058(m_c-1.5)\,]\pm 0.0028\pm 0.0029,\\
        g^2_R    = 
                [\quad 0.0309 -0.0053(m_c-1.5)\,]\pm 0.0034\pm 0.0028,\\
        \delta^2_L = 
                [\,   -0.0588 -0.0025(m_c-1.5)\,]\pm 0.0233\pm 0.0042,\\
        \delta^2_R =
                [\quad 0.0206 +0.0010(m_c-1.5)\,]\pm 0.0155\pm 0.0039,
     \end{array}
  \end{equation}
}
\def\dataofnuqorigcorr{%
  \begin{equation}
        \rho_{\rm corr}^{\rm (exp)} =
        \left(
        \begin{array}{rrrr}
          1 &  -0.751 & -0.100 &  0.118 \\
            &   1     &  0.064 &  0.097 \\
            &         &  1     & -0.436 \\
            &         &        &  1    
        \end{array}
        \right), \;\;
        \rho_{\rm corr}^{\rm (par)} =
        \left(
        \begin{array}{rrrr}
          1 &  -0.914 & -0.975 &  0.606 \\
            &   1     &  0.945 & -0.677 \\
            &         &  1     & -0.712 \\
            &         &        &  1    
        \end{array}
        \right).
  \end{equation}
}%
\def\dataofnuq{%
  \begin{equation}
     \label{dataofnuq}
	\left.
	\begin{array}{llr}
	g^2_L 	   &=&  0.2980 \pm 0.0044 \\
	g^2_R 	   &=&  0.0307 \pm 0.0047 \\
	\delta^2_L &=& -0.0589 \pm 0.0237 \\
	\delta^2_R &=&  0.0206 \pm 0.0160 
      	\end{array}
   	\right. \quad
   	\rho_{\rm corr} =
   	\left(
       	\begin{array}{rrrr}
          1 &  -0.559 & -0.163 &  0.162 \\
            &   1     &  0.156 & -0.037 \\
	    &         &  1     & -0.447 \\
	    &         &        &  1    
       	\end{array}
   	\right),
  \end{equation}
}%
\def\fitofnuq{
  \begin{subequations}
     \label{fitofnuq}
  \begin{eqnarray}
	&&\left.
	\begin{array}{lll}
	\gzbar^2(0) &=& 0.5497 \pm 0.0080 \\
	\sbar^2(0)  &=& 0.2413 \pm 0.0136
	\end{array}
	\right\} \quad
	\rho_{\rm corr} =0.916 ,
  \\
     \label{chisqofnuq}
     &&\hspace{1cm}\chi^2_{\rm min} =0.13,
  \end{eqnarray}
  \end{subequations}
}%
\def\fitofnuqnotgauss{%
  \begin{subequations}
  \begin{eqnarray}
     &&\left.
     \begin{array}{lll}
       \gzbar^2(0) &=&  0.5500^{+0.0077}_{-0.0083}\\[2mm]
       \sbar^2(0)  &=&  0.2420^{+0.0130}_{-0.0142}
     \end{array}
     \right\} \quad \rho_{\rm corr} = 0.916, 
   \label{fitofnuqnotgauss}
     \\
     &&\hspace{1cm}\chi^2_{\rm min} = 0.13,
  \end{eqnarray}
  \end{subequations}
}%
\def\dataofnue{%
  \begin{equation}
     \label{dataofnue}
	\left.
	\begin{array}{llr}
	\rho_{\nu e}\equiv
	(\rho)^{\nu_\mu e}_{\rm eff}           &=& 1.007\pm 0.028 \\[1mm]
	s^2_{\nu e}\equiv 
	(\sin^2\theta_W)^{\nu_\mu e}_{\rm eff} &=& 0.233\pm 0.008 
      	\end{array}
	\right\} \quad
	\rho_{\rm corr} = 0.09.
  \end{equation}
}%
\def\fitofnue{%
  \begin{equation}
     \label{fitofnue}
	\left.
	\begin{array}{lll}
	\gzbar^2(0)	&=&     0.5459 \pm 0.0154 \\
	\sbar^2(0)	&=&     0.2416 \pm 0.0079
	\end{array}
	\right\} \quad
	\rho_{\rm corr} =0.09.
  \end{equation}
}%
\def\dataofapv{%
 \begin{equation}
    \label{dataofapv}
	Q_W(135,55) = -71.04 \pm 1.81 \,.
 \end{equation}
}%
\def\fitofapv{%
 \begin{equation}
    \label{fitofapv}
      \sbar^2(0) = 0.2294 -0.6178\,[\,\gzbar^2(0)-0.5486\,]\pm 0.0082.
 \end{equation}
}%
\def\dataofed{%
  \begin{equation}
     \label{dataofed}
	\left.
	\begin{array}{llr}
	2\,C_{1u}-C_{1d} &=& +0.94 \pm 0.26 \\
	2\,C_{2u}-C_{2d} &=& -0.66 \pm 1.23 
	\end{array}
	\right\} \quad
	\rho_{\rm corr} =-0.975
  \end{equation}
}%
\def\fitofed{%
  \begin{subequations}
     \label{fitofed}
  \begin{eqnarray}
     \sbar^2(0) &=& 0.2273+0.3067\,[\,\gzbar^2(0)-0.5486\,]\pm 0.0092,
     \\
     \label{chisqofed}
      \chi^2_{\rm min} &=& 0.46 -1.77\,[\,\gzbar^2(0)-0.5486\,]\,,
  \end{eqnarray}
  \end{subequations}
}%
\def\fitoflenc{%
  \begin{subequations}
     \label{fitoflenc}
  \begin{eqnarray}
	&&\left.
	\begin{array}{lll}
	\gzbar^2(0)  \!\! &=&  \!\!  0.5462 \pm 0.0036 \\[1mm]
	\sbar^2(0)   \!\! &=&  \!\!  0.2353 \pm 0.0044
	\end{array}
	\right\} \quad
	\,\rho_{\rm corr} =0.53,
     \label{fitoflenc_a}
  \\
     \label{chisqoflenc}
	&&\hspace{1cm}\chi^2_{\rm min} = 2.22. 
  \end{eqnarray}
  \end{subequations}
}%
\def\fitoflencsummary{%
  \begin{subequations}
     \label{fitoflencsummary}
  \begin{eqnarray}
      &&\left.
	\begin{array}{lll}
	\gzbar^2(0)  \!\! &=&  \!\!  0.5462 \pm 0.0036 \\[1mm]
	\sbar^2(0)   \!\! &=&  \!\!  0.2353 \pm 0.0044
	\end{array}
	\right\} \quad
	\,\rho_{\rm corr}=0.53,
     \label{fitoflenc_a_summary}
  \\
     \label{chisqoflencsummary}
     &&\hspace{1cm}\chi^2_{\rm min}/{\rm (d.o.f.)}= 2.22/7.
  \end{eqnarray}
  \end{subequations}
}%
\def\fitofnuonly{%
  \begin{subequations}
     \label{fitofnuonly}
  \begin{eqnarray}
	&&\left.
	\begin{array}{lll}
	\gzbar^2(0)  \!\! &=&  \!\! 0.5496 \pm 0.0068 \\[1mm]
	\sbar^2(0)   \!\! &=&  \!\! 0.2414 \pm 0.0047
	\end{array}
	\right\} \,\quad
	\rho_{\rm corr} =0.75,
     \label{fitofnuonly_a}
  \\
     \label{chisqofnuonly}
	&&\hspace{1cm}\chi^2_{\rm min} = 0.19.
  \end{eqnarray}
  \end{subequations}
}%
\def\fitofeqonly{%
  \begin{subequations}
     \label{fitofeqonly}
  \begin{eqnarray}
	&&\left.
	\begin{array}{lll}
	\gzbar^2(0)  \!\! &=&  \!\!  0.5510 \pm 0.0165 \\[1mm] 
	\sbar^2(0)   \!\! &=&  \!\!  0.2280 \pm 0.0088 
	\end{array}
	\right\} \,\quad
	\rho_{\rm corr} = -0.62,
     \label{fitofeqonly_a}
  \\
     \label{chisqofeqonly}
	&&\hspace{1cm}\chi^2_{\rm min} = 0.46.
  \end{eqnarray}
  \end{subequations}
}%
\def\dataofmw{%
  \begin{equation}
     \label{dataofmw}
        \mw = 80.24 \pm 0.16 \gev .
  \end{equation}
}%
\def\fitofmw{%
  \begin{equation}
     \label{fitofmw}
	\gwbar^2(0) = 0.4225 -0.0031\,\frac{\delg-0.0055}{\alpha}
       \pm 0.0017\,,
  \end{equation}
}%
\def\fitofmwsummary{%
  \begin{subequations}
     \label{fitofmwsummary}
  \begin{eqnarray}
	&&\gwbar^2(0) = 0.4225 \pm 0.0017,
         \\
        &&\chi^2_{\rm min}/{\rm (d.o.f.)} = 0/0.
  \end{eqnarray}
  \end{subequations}
}%
\def\fitofrunning{%
  \begin{eqnarray}
     \label{fitofrunning}
     \left.
     \begin{array}{ccc}
       \displaystyle{
             \frac{4\pi}{\gzbar^2(m_Z^2)} -\frac{4\pi}{\gzbar^2(0)}
       }
          &\!\!=\!\!&  -0.33 + 1.2 (\alpha_s -0.12)  \pm 0.17
       \\[5mm]
       \displaystyle{
           \frac{\sbar^2(m_Z^2)}{\bar{\alpha}(m_Z^2)_{\rm SM}}
          -\frac{\sbar^2(0)}{\alpha}
       }
          &\!\!=\!\!&  -2.47 + 1.1 (\alpha_s -0.12)  \pm 0.62
     \end{array}
     \right\} \, \quad
     \rho_{\rm corr} =  -0.49.
  \end{eqnarray}
}%
\def\smrunning{%
  \begin{subequations}
    \label{smrunning}
  \begin{eqnarray}
       \frac{4\pi}{\gzbar^2(m_Z^2)} -\frac{4\pi}{\gzbar^2(0)}
          &=&  -0.2998 + 0.0013 \biggl(\frac{150}{m_t}\biggr)^2
                       + 0.0021 \biggl(\frac{100}{\mh}\biggr)^2 \,,
       \\[1mm]
       \frac{\sbar^2(m_Z^2)}{\bar{\alpha}(m_Z^2)_{\rm SM}}
       -\frac{\sbar^2(0)}{\alpha}
          &=&  -3.0760 + 0.0058 \biggl(\frac{150}{m_t}\biggr)^2 \,,
  \end{eqnarray}
  \end{subequations}
}%
\def\fitoflencatmz{%
  \begin{eqnarray}
	\left.
	\begin{array}{lll}
	\gzbar^2(\mmz)  \!\! &=&  \!\!  0.5533 \pm 0.0037 \\[1mm]
	\sbar^2(\mmz)   \!\! &=&  \!\!  0.2266 \pm 0.0047
	\end{array}
	\right\} \quad
	\,\rho_{\rm corr} =0.53.
     \label{fitoflencatmz}
  \end{eqnarray}
}%
\def\fitofnuonlyatmz{%
  \begin{eqnarray}
	\left.
	\begin{array}{lll}
	\gzbar^2(\mmz)  \!\! &=&  \!\! 0.5568 \pm 0.0048 \\[1mm]
	\sbar^2(\mmz)   \!\! &=&  \!\! 0.2331 \pm 0.0072
	\end{array}
	\right\} \,\quad
	\rho_{\rm corr} =0.75,
     \label{fitofnuonlyatmz}
  \end{eqnarray}
}%
\def\fitofeqonlyatmz{%
  \begin{eqnarray}
	\left.
	\begin{array}{lll}
	\gzbar^2(\mmz)  \!\! &=&  \!\!  0.5583 \pm 0.0170 \\[1mm] 
	\sbar^2(\mmz)   \!\! &=&  \!\!  0.2188 \pm 0.0093
	\end{array}
	\right\} \,\quad
	\rho_{\rm corr} = -0.62.
     \label{fitofeqonlyatmz}
  \end{eqnarray}
}%
\def\fitofallnc{%
  \begin{subequations}
     \label{fitofallnc}
  \begin{eqnarray}
	\left.
	\begin{array}{llr}
	\gzbar^2(m_Z^2)\mm &=& \mm 
		0.5544 -0.00023\frac{\alpha_s-0.12}{0.01} \pm 0.0015 \\
	\sbar^2(m_Z^2) \mm &=& \mm
		0.2312 +0.00008\frac{\alpha_s-0.12}{0.01} \pm 0.0007 \\
	\delb(m_Z^2)   \mm &=& \mm
		-0.0064 -0.00437\frac{\alpha_s-0.12}{0.01} \pm 0.0034
      	\end{array}
   	\right. \:\:
   	\rho_{\rm corr} =
   	\left(
       	\begin{array}{rrr}
          1 &   0.16 & -0.32 \\
            &   1    &  0.20 \\
	    &        &  1     
       	\end{array}
   	\right), \hspace{6mm} \\
	\label{chisqofallnc}
	\chi^2_{\rm min} =4.67
         +\biggl(\frac{\alpha_s-0.1024}{0.0127}\biggr)^2. \hspace{70mm}
  \end{eqnarray}
  \end{subequations}
}%
\def\fitofstu{%
  \begin{subequations}
     \label{fitofstu}
  \begin{eqnarray}
  &&\!\!
    \begin{array}{clllll}
         S  =&  \hspace{-3mm} -0.35 &
                \hspace{-3mm} -0.016\,\frac{\alpha_s-0.12}{0.01} & 
                \hspace{-3mm} +0.067\,\frac{\delta_\alpha}{0.10} &
                \hspace{-3mm} \pm 0.33 \\[1mm]
         T  =&  \hspace{-3mm} \hphantom{-} 0.39 &
                \hspace{-3mm} -0.058\,\frac{\alpha_s-0.12}{0.01} &
                \hspace{-3mm} -0.004\,\frac{\delta_\alpha}{0.10} &
                \hspace{-3mm} \pm 0.36  \\[1mm]
         U  =&  \hspace{-3mm} \hphantom{-}  0.41 &
                \hspace{-3mm} +0.058\,\frac{\alpha_s-0.12}{0.01} &
                \hspace{-3mm} +0.024\,\frac{\delta_\alpha}{0.10} &
                \hspace{-3mm} \pm 0.54 \\[1mm]
         \ol{\delta}_b =&
                \hspace{-3mm} -0.0064 &
                \hspace{-3mm} -0.0043\,\frac{\alpha_s-0.12}{0.01} &
              & \hspace{-3mm} \pm 0.0034
    \end{array}
   \rho_{\rm corr} = \left(
      \begin{array}{rrrr}
           1    &  0.83 & -0.18 & -0.12 \\[1mm]
                &  1    & -0.40 & -0.32 \\[1mm]
                &       &  1    &  0.20 \\[1mm]
                &       &       &  1    \\[1mm]
      \end{array}
      \right),
   \nonumber \\ && \hspace{-5mm}
   \label{fitofstu_a}
  \end{eqnarray}
}
\def\fitofstuchisqmin{%
  \begin{equation}
     \chi^2_{\rm min} =
       4.67 + \biggl(\frac{\alpha_s-0.1024}{0.0127}\biggr)^2
       +\biggl(\frac{\delta_\alpha}{0.10}\biggr)^2\,.
     \label{chisqofstu}
  \end{equation}
  \end{subequations}
}%
%
%
\def\chisqsm{
  \begin{subequations}
  \label{total_chisqsm}
  \begin{eqnarray}
     \chi^2_{\rm SM}(m_t,\mh,\alpha_s,\delta_\alpha)
     &=& \biggl(\frac{m_t -\langle m_t\rangle}{\Delta m_t} \biggr)^2
         +\chi^2_{H}(\mh,\alpha_s,\delta_\alpha)\,,
     \label{chisqsm}
  \end{eqnarray}
}
%
%
\def\fitofmt{%
  \begin{eqnarray}
     \label{fitofmt}
    \langle m_t \rangle &=& 145.2 +12.5\, \ln\frac{\mh}{100}
                            +0.9 \, \ln^2\frac{\mh}{100}
           -1.9\,\biggl(\frac{\alpha_s-0.12}{0.01}\biggr)
           -4.6\,\biggl(\frac{\delta_\alpha}{0.10}\biggr)\,,
     \label{fitofmtbest}
     \\
     \Delta m_t   &=&14.6 - 0.23\, \ln \frac{\mh}{100}
          - \Bigl(0.38 -0.05\,\ln \frac{\mh}{100}\Bigr)\,
            \frac{m_t -150}{10}\,,
     \label{fitofmterror}
  \end{eqnarray}
}%
%
%
\def\chisqminsm{%
  \begin{eqnarray}
     \chi^2_H(\mh,\alpha_s,\delta_\alpha) &=&
      6.11 + \biggl(\frac{\delta_\alpha -0.31}{0.43} \biggr)^2
          + \biggl(\frac{\alpha_s-0.1173+0.005\,\delta_\alpha}{0.0060}
            \biggr)^2
     \nonumber \\
     && \!\!\!\!\!
          - \biggl(\frac{\alpha_s-0.1244+0.025\,\delta_\alpha}{0.0136} 
                   \biggr) \ln \frac{\mh}{100}
          - \biggl(\frac{\alpha_s-0.1322}{0.0700}\biggr)
                   \ln^2 \frac{m_H^{}}{100}
          +\biggl(\frac{\delta_\alpha}{0.10}\biggr)^2.
     \nonumber \\ &&   \label{chisqminsm}
  \end{eqnarray}
  \end{subequations}
}%
 
 
\def\dataofnewalr{%
  \begin{eqnarray}
      \label{dataofnewalr}
      A_{\rm LR}^0   &=&  0.1656   \pm 0.0076,
  \end{eqnarray}
}
 
 
\def\fitofzparameterwithnewalr{%
  \begin{subequations}
  \label{fitofzparameterwithnewalr}
  \begin{eqnarray}
     \left.
     \begin{array}{lll}
        \gzbar^2(m_Z^2)&\mm =&
          0.5538 -0.00031\,\frac{\alpha_s-0.12}{0.01} \pm 0.0017 \\[1mm]
        \sbar^2(m_Z^2) &\mm =&
          0.2303 +0.00006\,\frac{\alpha_s-0.12}{0.01} \pm 0.0005 \\[1mm]
        \delb(m_Z^2)   &\mm =&\mfive
         -0.0071 -0.00432\,\frac{\alpha_s-0.12}{0.01} \pm 0.0035
     \end{array}
     \right. 
        \rho_{\rm corr} =
        \left(
        \begin{array}{rrr}
          1 &  0.11 & -0.37 \\[1mm]
            &  1    &  0.16 \\[1mm]
            &       &  1
       	\end{array}
        \right) \,,\quad
   \\[2mm]
     \label{chisqofzparameterwithnewalr}
	\chi^2_{\rm min} =
            5.78 +\biggl(\frac{\alpha_s-0.1000}{0.0127}\biggr)^2.
         \hspace{65mm}
     \end{eqnarray}
  \end{subequations}
}%
%
 
 
\def\fitofsbarfromnewalr{%
  \begin{eqnarray}
     \label{fitofsbarfromnewalr}
      \sbar^2(m_Z^2) = 0.2282 \pm  0.0010.
  \end{eqnarray}
}
 
 
\def\fitofstuwithnewalr{%
  \begin{subequations}
     \label{fitofstuwithnewalr}
  \begin{eqnarray}
  &&\!\!
    \begin{array}{clllll}
         S  =&  \hspace{-3mm} -0.67 &
                \hspace{-3mm} -0.024\,\frac{\alpha_s-0.12}{0.01} & 
                \hspace{-3mm} +0.066\,\frac{\delta_\alpha}{0.10} &
                \hspace{-3mm} \pm 0.30 \\[1mm]
         T  =&  \hspace{-3mm} \hphantom{-} 0.30 &
                \hspace{-3mm} -0.060\,\frac{\alpha_s-0.12}{0.01} &
                \hspace{-3mm} -0.004\,\frac{\delta_\alpha}{0.10} &
                \hspace{-3mm} \pm 0.36  \\[1mm]
         U  =&  \hspace{-3mm} \hphantom{-}  0.24 &
                \hspace{-3mm} +0.053\,\frac{\alpha_s-0.12}{0.01} &
                \hspace{-3mm} +0.024\,\frac{\delta_\alpha}{0.10} &
                \hspace{-3mm} \pm 0.54 \\[1mm]
         \ol{\delta}_b =&
                \hspace{-3mm} -0.0074 &
                \hspace{-3mm} -0.0044\,\frac{\alpha_s-0.12}{0.01} &
              & \hspace{-3mm} \pm 0.0034
    \end{array}
   \rho_{\rm corr} = \left(
      \begin{array}{rrrr}
           1    &  0.87 & -0.25 & -0.19 \\[1mm]
                &  1    & -0.42 & -0.33 \\[1mm]
                &       &  1    &  0.19 \\[1mm]
                &       &       &  1    \\[1mm]
      \end{array}
      \right),
   \nonumber \\ && \hspace{-5mm}
  \end{eqnarray}
\vspace{-6mm}
  \begin{equation}
     \chi^2_{\rm min} =
       8.60 + \biggl(\frac{\alpha_s -0.0998}{0.0126}\biggr)^2
       +\biggl(\frac{\delta_\alpha}{0.10}\biggr)^2\,,
     \label{chisqofstuwithnewalr}
  \end{equation}
  \end{subequations}
}%
 
%
%
\def\chisqsmwithnewalr{
  \begin{subequations}
  \label{total_chisqsmwithnewalr}
  \begin{eqnarray}
     \chi^2_{\rm SM}(m_t,\mh,\alpha_s,\delta_\alpha)
     &=& \biggl(\frac{m_t -\langle m_t\rangle}{\Delta m_t} \biggr)^2
         +\chi^2_H(\mh,\alpha_s,\delta_\alpha)\,,
     \label{chisqsmwithnewalr}
  \end{eqnarray}
}
%
%
\def\fitofmtwithnewalr{%
  \begin{eqnarray}
     \label{fitofmtwithnewalr}
    \langle m_t\rangle  &=& 162.2 +12.6\, \ln\frac{\mh}{100}
                            +0.8 \, \ln^2\frac{\mh}{100}
           -1.2\,\biggl(\frac{\alpha_s-0.12}{0.01}\biggr)
           -4.8\,\biggl(\frac{\delta_\alpha}{0.10}\biggr)\,,
     \label{fitofmtbestwithnewalr}
     \\
     \Delta m_t   &=&12.0 - 0.09\, \ln \frac{\mh}{100}
          - \Bigl(0.31 -0.05\, \ln \frac{\mh}{100}\Bigr)\,
            \frac{m_t -175}{10} \,,
     \label{fitofmterrorwithnewalr}
  \end{eqnarray}
}%
%
%
\def\chisqminsmwithnewalr{%
  \begin{eqnarray}
     \chi^2_H(\mh,\alpha_s,\delta_\alpha) &=&
      9.56 + \biggl(\frac{\delta_\alpha -0.75}{0.39} \biggr)^2
          + \biggl(\frac{\alpha_s-0.1164+0.005\,\delta_\alpha}{0.0060}
            \biggr)^2
     \nonumber \\
     && \!\!\!\!\!
          - \biggl(\frac{\alpha_s-0.1365+0.030\,\delta_\alpha}{0.0144} 
                   \biggr) \ln \frac{\mh}{100}
          - \biggl(\frac{\alpha_s-0.1255}{0.0639}\biggr)
                   \ln^2 \frac{m_H^{}}{100}
          +\biggl(\frac{\delta_\alpha}{0.10}\biggr)^2.
     \nonumber \\ &&   \label{chisqminsmwithnewalr}
  \end{eqnarray}
  \end{subequations}
}%

 
\def\fitofzparameterwithmtfixed{%
  \begin{subequations}
     \label{fitofzparameterwithmtfixed}
  \begin{eqnarray}
	&&\left.
	\begin{array}{lll}
	\gzbar^2(m_Z^2) \!\!&=&\!\! 0.55430
           -0.00109\,\frac{\alpha_s-0.12-0.0023\frac{m_t -174}{16}}{0.01}
           \pm 0.00156 \\[1mm]
	\sbar^2(m_Z^2)  \!\!&=&\!\! 0.23023
           +0.00016\,\frac{\alpha_s-0.12-0.0021\frac{m_t -174}{16}}{0.01}
           \pm 0.00054
	\end{array}
	\right\} \;
	\rho_{\rm corr} = 0.19,\quad
  \\
     \label{chisqofzparameterwithmtfixed}
	&&\chi^2_{\rm min}
          = 6.86 +\bigg(\frac{m_t-90}{60}\biggl)^2
           +\biggl(
            \frac{\alpha_s-0.1187-0.0022\frac{m_t-174}{16} }{0.0068}
            \biggr)^2\,,
  \end{eqnarray}
  \end{subequations}
}%
 
 
\def\fitofstuwithmtfixed{%
  \begin{subequations}
     \label{fitofstuwithmtfixed}
  \begin{eqnarray}
  &&\!\! \left.
    \begin{array}{clllll}
         S  =&  \hspace{-3mm} -0.62 &
                \hspace{-3mm} -0.097\,\frac{\alpha_s-0.12
                                     -0.0022\frac{m_t-174}{16}}{0.01} & 
                \hspace{-3mm} +0.066\,\frac{\delta_\alpha}{0.10} &
                \hspace{-3mm} \pm 0.30
         \\[1mm]
         T  =&  \hspace{-3mm} \hphantom{-} 0.39 &
                \hspace{-3mm} -0.214\,\frac{\alpha_s-0.12
                                     -0.0022\frac{m_t-174}{16}}{0.01} &
                \hspace{-3mm} -0.004\,\frac{\delta_\alpha}{0.10} &
                \hspace{-3mm} \pm 0.34
         \\[1mm]
         U  =&  \hspace{-3mm} \hphantom{-}  0.17 &
                \hspace{-3mm} +0.182\,\frac{\alpha_s-0.12
                                     -0.0022\frac{m_t-174}{16}}{0.01} & 
                \hspace{-3mm} +0.023\,\frac{\delta_\alpha}{0.10} &
                \hspace{-3mm} \pm 0.53 \\[1mm]
    \end{array}
    \right\}
   \rho_{\rm corr} = \left(
      \begin{array}{rrr}
           1    &  0.87 & -0.22  \\[1mm]
                &  1    & -0.38  \\[1mm]
                &       &  1     \\[1mm]
      \end{array}
      \right),
   \nonumber \\ && \hspace{-5mm}
   \\
	&&\chi^2_{\rm min}
          = 9.58 +\bigg(\frac{m_t-84}{63}\biggl)^2
           +\biggl(
            \frac{\alpha_s-0.1185-0.0022\frac{m_t-174}{16} }{0.0067}
            \biggr)^2
       +\biggl(\frac{\delta_\alpha}{0.10}\biggr)^2\,.
     \label{chisqofstuwithmtfixed}
  \end{eqnarray}
  \end{subequations}
}%
%

\def\thefootnote{\arabic{footnote}}
\setcounter{footnote}{0}
\setcounter{page}{1}

\section*{\large \bf 1  Introduction }
\renewcommand{\theequation}{1.\arabic{equation}}
\pr
The Standard Model (SM) of the electroweak interactions has been with
us for nearly two decades. Despite the general belief that it should be
an effective theory valid at energies below the Fermi scale, so far no
unambiguous sign of physics beyond the SM has been found nor any clue
to the origin of the underlying gauge symmetry breaking mechanism.
On the other hand, the accuracy of the experiments testing the
electroweak theory has improved significantly in the past decade both
in low energy neutral current experiments and in high energy collider
experiments on the $W$ and $Z$ boson properties. The precision of these
experiments has reached the level, where meaningful searches for new physics
through the investigation of quantum effects can be carried out.
 
The effects may be significant, if there are new particles with masses
as light as weak bosons, or if many new particles contribute
constructively, or if there exist new strong interactions among them.
Even in the absence of such a signal, constraints on certain new
physics possibilities can be derived and tightened in future precision
experiments. With this motivation to study electroweak radiative
corrections several groups have made efforts towards comprehensive and
systematic analyses\cite{lps,kl,lep1,stu,stu_mod,kms,ab,peccei,
marciano,gw,hisz,susy_thr,susy_ellis,abc,amaldi,costa,llm,stuvw}.
 
In this report a novel approach to confront electroweak data and theory
is presented with the aim of a systematic look for new physics effects.
In the following, the conditions imposed on the electroweak analysis
scheme are outlined.

Since it is the aim to search for new physics effects in the electroweak
precision data, a model-independent framework to analyse the data is
required. As both the experimental accuracy and the new physics effects
looked for are of similar size as the SM radiative effects, it is
essential to take account of the SM radiative effects as
accurately as possible. 
For testing grand unification of the three gauge 
couplings\cite{sgut91,sgut92,sgut92_hy,sgut_ellis,sgut93} 
the fits should be studied quantitatively as a function of $\alpha_s$. 
Furthermore, the
level of precision accessible in the near future is such that the
present uncertainty in the hadronic vacuum polarization contribution to
the running of the effective QED coupling constant $\alpha(q^2)$
severely limits the ability to study new physics through quantum
effects. In order to assess the effects of possible future improvements
in the $\epem$ hadroproduction experiments at low and intermediate
energies, the consequences of varying $\alpha(\mmz)$ should be
examined quantitatively. During the course of this study, 
sometimes the published results of earlier theoretical analysis
could not be reproduced easily. This happened in most cases
because not all the details of the assumptions and approximations
underlying the analysis have been clearly stated in the literature. The
quantum effects studied are so sensitive to details of the exact
treatment of higher order effects and 
to uncertainties in the analysis
that equally sensible looking assumptions often lead to a significant
numerical difference. We therefore make every effort to render the
report self-contained so that all our results can be reproduced
unambiguously.
 
In order to comply with all the above requirements, our comprehensive
analysis of electroweak precision experiments is performed according
to the following steps, by systematically strengthening the underlying
theoretical assumptions.
\begin{oneitem}{1}
All electroweak data are expressed in terms of model-independent
parameters.
\end{oneitem}
\noindent
For the choice of model-independent parameters, we basically follow the
strategy of ref.~\cite{old_ncfit} for low energy neutral current
experiments, and that of the LEP electroweak working group\cite{lep1}
for the $Z$ parameters. In addition, the $W$ boson mass, the fine structure
constant $\alpha$ and the Fermi coupling constant $G_F$ are used 
as inputs of the analysis.
Some of these parameters are directly related to experimental
observables up to corrections due to known physics, such as the
external QED bremsstrahlung effects and the quark-parton model, and
uncertainties in these correction factors are included as part of the
errors of the experimentally measured parameters.
\begin{oneitem}{2}
 The model-independent parameters are then expressed in terms of the
 pole positions of the $W$ and $Z$ propagators, and the $S$-matrix
 elements of four external fermions, quarks or leptons, which are
 approximated as products of two external standard $V\pm A$ currents
 and the scalar transition form factors.
\end{oneitem}
\noindent
All electroweak precision measurements that have been performed
so far can be expressed in terms of the $S$-matrix elements of quark
and lepton processes whose masses are negligible compared to the
weak boson masses.  To an excellent approximation, chirality-flip terms
in the loop amplitudes can be neglected and the relevant $S$-matrix
elements can be expressed in terms of the scalar product of the
standard $V\pm A$ currents multiplied by transition form factors
depending on the flavors and chiralities of the currents as well as the
momentum transfer of the process under consideration. External QCD and
QED corrections can hence be applied exactly as in the SM, and
electroweak models can be confronted with experiment, once the
transition form factors are determined in a particular model.
The dependence of the fit on the QCD parameter $\alpha_s$ and quark
masses is taken into account by introducing appropriate external
parameters. Up to this stage, our analysis is quite general, as the
formulae are valid for any electroweak model respecting the flavor and
chirality conservation laws of the SM, that is, for all new physics
contributions which can interfere with the leading SM amplitudes.
 
Although one may attempt to constrain these model-independent transition
form factors directly by experiment, we find it impractical, since the
number of independent transition form factors exceeds by far the
effective number of degrees of freedom 
provided by precision measurements. Hence,
we perform the quantitative comparison of data with theory in a more
restricted class of models 
which are minimal extensions from the SM, 
i.e. 
those models which respect the ${\rm SU(2)_L\times U(1)_Y}$ gauge
symmetry broken spontaneously down to ${\rm U(1)_{EM}}$.
\begin{oneitem}{3}
 The transition form factors are expanded perturbatively
 in ${\rm SU(2)_L\!\times U(1)_Y}$ gauge couplings,
 and radiative effects are classified either as the universal
 gauge boson propagator corrections or as the process specific
 vertex and box corrections.  The universal propagator
 correction factors are then parameterized by four charge
 form factors, $\ebar^2(q^2)$, $\sbar^2(q^2)$, $\gzbar^2(q^2)$
 and $\gwbar^2(q^2)$, corresponding to the $\gamma\gamma$,
 $\gamma Z$, $ZZ$ and $WW$ propagator degrees of freedom.
\end{oneitem}
\noindent
The restriction to the electroweak gauge group 
${\rm SU(2)_L\times U(1)_Y}$ implies at the tree
level that all fermions, quarks and leptons, couple to the electroweak
gauge bosons universally with the same coupling constant as long as
they have common ${\rm SU(2)_L\times U(1)_Y}$ quantum numbers. This
universality of the gauge boson coupling to quarks and leptons can in
general be violated at the quantum level, because the gauge symmetry
breaks spontaneously down to ${\rm U(1)_{EM}}$. It has been widely
recognized, however, that this universality of the couplings holds true
even at one-loop level in a wider class of models where new
particles affect the precision experiments only via their effects on
the electroweak gauge boson  propagators
\cite{lps,kl,lep1,kms,stu,stu_mod,ab,peccei,marciano,gw}. 
This class of new
physics effects is often called oblique\cite{lps,stu} or propagator
\cite{ab} corrections, or those satisfying generalized universality
\cite{gw}. This concept of universality can be generalized to certain
vertex corrections with 
non-standard weak boson interactions\cite{hisz}. It is also often
useful in theories with non-standard vertex and box corrections, such as
the supersymmetric SM (SUSY-SM), since the propagator corrections can
be larger than the vertex/box ones:  propagator corrections can be
significant either because of a large multiplicity of contributing
particles or by the presence of a relatively light new particle.
 
When confronting the electroweak theory with experiment, we adopt this
distinction between new physics contributions to the gauge boson
propagators and those to the rest, 
where we allow the most general contributions
in the former, whereas we consider only the SM contributions to the
latter (vertex and box corrections).
\begin{oneitem}{4}
 By assuming that the well-known SM contributions
 dominate the process specific vertex and box corrections, apart from
 the $Zb_Lb_L$ vertex for which new physics contributions are allowed,
 we determine from precision experiments the four universal charge form
 factors at the typical momentum transfer scales,
 $q^2=0$ and $\mmz$.
\end{oneitem}
\noindent
The new physics contributions may either prevent our ability to fit
the experimental data within our approach, or lead to non-standard
values of the fitted four charge form factors and the $\zbb$ vertex form
factor, $\delb(q^2)$. 
At this stage, the whole body of electroweak precision data 
can be expressed in terms of 
the two weak boson masses $\mw$ and $\mz$, 
and these five form factors, that is, 
the four universal charge form factors and $\delb(q^2)$. 
Although the form factors could be determined at any point on
the momentum scale $q^2$, they are actually  measured with adequate
precision only at two 
specific $q^2$ ranges, namely all four charge form
factors at $q^2=0$ or $q^2 \ll \mmz$, while $\sbar^2(q^2)$, $\gzbar^2(q^2)$
and $\delb(q^2)$ at $q^2 = \mmz$.
Hence, there are just 9 electroweak
parameters measured accurately enough to be used for testing theories:
$\mw$ and $\mz$, $\ebar^2(0)$, $\sbar^2(0)$, $\gzbar^2(0)$ and
$\gwbar^2(0)$, $\sbar^2(\mmz)$, $\gzbar^2(\mmz)$ and $\delb(\mmz)$.
 
Apart from the vertex form factor $\delb(\mmz)$ the remaining 8
parameters characterise the universal propagator corrections. On the
experimental side, the three quantities $\alpha$, $G_F$ and $\mz$ are
measured so accurately that it is justifiable to treat them as constants:
$\alpha=1/137.0359895$ and $G_F=1.16639\times 10^{-5}{\rm GeV}^{-2}$
from the PDG listing\cite{pdg92}, and $\mz=91.187~{\rm GeV}$ from
the LEP results\cite{lep93}. Among the 8 universal parameters above,
$\ebar^2(0)=4\pi\alpha$ and $\mz$ are fixed immediately, while $G_F$
fixes the ratio $\gwbar^2(0)/\mmw$,
once we assume the SM dominance of the vertex and box corrections
($\bar{\delta}_G$) to the muon decay lifetime.
Since the gauge boson properties are fixed  at tree level by only
three parameters in general models with the ${\rm SU(2) \times U(1)}$
symmetry broken by a vacuum expectation value, the remaining 5
universal parameters serve to test the theory at the quantum level 
(see Table~1).
We therefore first determine from precision experiments the 5 parameters,
$\sbar^2(0)$, $\gzbar^2(0)$, $\gwbar^2(0)$, $\sbar^2(\mmz)$ and
$\gzbar^2(\mmz)$, together with $\delb(\mmz)$, and then confront
their values with various theoretical predictions.
 
In the fit to the $Z$ boson parameters the strong coupling
constant $\alpha_s(\mz)$ is treated as external parameter 
which can be varied 
within certain limits. In this way the analysis
remains transparent and easy to update. The fitted electroweak
parameters $\sbar^2(\mmz)$, $\gzbar^2(\mmz)$ and $\delb^2(\mmz)$
are thus presented as parametrizations in $\alpha_s$ (see Table~2).
 
When the new physics scale is significantly higher than the scale
($\simlt \mmz$) of precision measurements, new physics contributions
to the running of the charge form factors can be neglected.
\begin{oneitem}{5}
By assuming further that the running of the charge form factors
between $q^2=0$ and $q^2=\mmz$
are governed only by SM physics, three universal
 parameters sensitive to radiative effects can be determined. We adopt
 a modified version of the $S$, $T$, $U$ parameters of ref.~\cite{stu}
 by including the SM radiative effects as well as new physics
 contributions.
\end{oneitem}
\noindent
Among the 5 universal parameters, the values of
$\sbar^2(0)/\alpha$ and $\gzbar^2(0)$ can then be calculated from
$\sbar^2(\mmz)/\bar{\alpha}(\mmz)$ and $\gzbar^2(\mmz)$, respectively,
using SM physics only.
There are then 3 remaining
universal free parameters which correspond 
to the parameters $S$, $T$ and $U$ of ref.~\cite{stu}, 
$\epsilon_1$, $\epsilon_2$ and $\epsilon_3$ of ref.~\cite{ab}, 
or the other related triplets of parameters in
refs.\cite{stu_mod}. When the scale of new physics that couples to gauge
boson propagators is near to the weak boson masses, its signal can be
identified as an anomalous running of the charge form factors. This
point has been stressed in refs.\cite{susy_thr} in connection with
the possible existence of light SUSY particles. It has also been
pointed out that when new physics effects to the electroweak gauge
boson sector are parametrized by the four dimension-six
operators of ref.~\cite{gw}, there occurs anomalous running of the charge form
factors\cite{hisz}. The triplet parametrizations are then no longer
sufficient to account for new physics degrees of freedom, and all 5
parameters in Table~1 should be regarded as free. Several alternative
approaches to the same problem have been proposed in
refs.\cite{susy_thr,abc,stuvw}.
 
Note that in order to obtain the charge form factors from the three
known parameters $\alpha$, $G_F$, $\mz$ and the radiative
parameters $S$, $T$ and $U$ that are calculable in a given model, 
the effective QED coupling at the $Z$ mass
scale, $\bar{\alpha}(\mmz)$ is needed. Its value is calculable from
$\alpha$ in the SM but suffers from uncertainty in the hadronic
vacuum polarization contribution\cite{piqq_h,piqq_h_latest,pi3q_h}. 
The effect of
this uncertainty on the final results turns out to be non-negligible.
In order to gauge the effects due to this uncertainty quantitatively,
we introduce $\delta_\alpha = 1/\bar{\alpha}(\mmz) -128.72$ as
external parameter and allow it to vary in the fit (see Table~2).
It is then straightforward to examine the effects of shifts in
the $\bar{\alpha}(\mmz)$ value and the impact of future improvements
in its measurement.
 
In the minimal SM, the three universal parameters $S$, $T$, $U$ and 
the $\zbb$ vertex form factor $\delb(\mmz)$
depend on just two parameters: $m_t$ and $\mh$.
\begin{oneitem}{6}
 Finally, by assuming that no new physics contributes significantly
 to electroweak precision experiments, we can express all the
 radiative effects in terms of the two parameters of the minimal
 SM, $m_t$ and $\mh$. The $\chi^2$ curves of the global fit are
 shown as a function of these two parameters, for several values
 of $\alpha_s(\mz)$ and $\bar{\alpha}(\mmz)$.
\end{oneitem}
\noindent
The preferred range of $m_t$ is presented as a function of $\mh$,
$\alpha_s$, $\delta_\alpha$, that of $\mh$ as a function of $m_t$,
$\alpha_s$, $\delta_\alpha$, that of $\alpha_s$ as a function of
$m_t$, $\mh$, $\delta_\alpha$, and that of $\delta_\alpha$ 
as a function of $m_t$, $\mh$, $\alpha_s$.
The chosen value for the parameter $\delta_\alpha$ is
essential, since it is not well constrained by the present precision
measurements alone.
 
A clear advantage of this approach is that we can test the electroweak
theory at three qualitatively distinct levels. 
If we cannot 
fit all the data at a given $q^2$ with common form factor values, we
should either look for new physics affecting the relevant vertex/box
corrections significantly or else we should introduce new tree level
interactions such as those induced by an exchange of a new heavy boson,
or from new strong interactions that bind common constituents of quarks
and leptons. If the 'universality' in terms of the above four charge
form factors holds, but their $q^2$-dependence does not agree with the
expectations of the Standard Model, we may anticipate a new physics
scale very near to the present experimental limit\cite{susy_thr}, or
effective higher dimensional interactions among the gauge bosons
\cite{gw,hisz}. 
New physics contributions which decouple at low energies
can thus be identified by their anomalous running of
the charge form factors. If the running of the form factors is
found to be consistent with the SM, then our approach reduces to the
standard three parameter analyses\cite{stu,stu_mod,ab}, or those with
three plus one parameter\cite{susy_thr,abc} when including the
$\zbb$ vertex parameter $\delb(\mmz)$ as well. Deviation from the
SM is still possible, since the SM has only two free parameters, $m_t$
and $\mh$. In this case sensitivity to those new physics
contributions which do not decouple at low energies remains.
 
As emphasized at the beginning of this section, we present at all
stages of our quantitative analysis the best-fit values of the model
parameters, including a parametrization of the $\chi^2$ goodness of the
fit around its minimum as functions of 
the external parameters 
$\alpha_s = \alpha_s(\mz)_{\msbar}$ and
$\delta_\alpha \equiv 1/\bar{\alpha}(\mmz)-128.72
\simeq (1/\bar{\alpha}(\mmz)-1/\alpha)_{\rm hadrons} +3.88 $. One can
examine consequences 
of possible future improvements in the measurement of 
$\alpha_s$\cite{alpha_s} and those 
of hadronic contribution to $\delta_\alpha$ by adding to the
quoted $\chi^2$ function terms of the form
$[(\alpha_s-\overline{\alpha_s})/(\Delta\alpha_s)]^2$ and
$[(\delta_\alpha-\overline{\delta_\alpha})/(\Delta \delta_\alpha)]^2$.
 
The paper is organized as follows.
In section~2,
we present our formalism in detail. The helicity amplitudes are stated
for general four-fermion processes in terms of the universal charge
form factors and process-dependent vertex and box corrections.
Definitions of the form factors and the $S$, $T$, $U$ parameters
are given and their SM values are shown.
Section~3
contains theoretical formulae for the electroweak observables 
expressed in terms of the helicity amplitudes of section~2, with QCD/QED
corrections. Numerical predictions are also given for wide ranges of
the form factor values, and also in the minimal SM.
In section~4,
we present our model-independent parametrizations of all
experimental data and, confront them with our theoretical predictions.
The universal charge form factors and $\delb(\mmz)$ are determined by
assuming SM dominance in the remaining vertex and box corrections.
Section~5
presents  a systematic analysis of the electroweak data by gradually
tightening the theoretical assumptions. First the running of the charge
form factors  $\gzbar^2(q^2)$ and $\sbar^2(q^2)$ is tested, then
the 4-parameter ($S,T,U$ and $\delb(\mmz)$) fit to all electroweak
data is performed by assuming SM running of the charge form factors.
Finally, constraints on $m_t$ and $\mh$ are discussed in the SM fit.
The total $\chi^2$ of the SM is parametrized in terms of $m_t$, $\mh$,
$\alpha_s$ and $\delta_\alpha$.
In section~6, 
consequences of the new precision measurement of the
left-right asymmetry\cite{alr94} and the impact of 
a top quark mass measurement
are considered.
Section~7 summarizes our observations.
Details of the theoretical formulae used are collected in the
appendices.
In appendix A,
we give all the SM radiative correction terms completely at 
one-loop level, and partly at two-loop level for $O(\alpha\alpha_s)$
terms. They are classified into three parts, the propagator corrections,
the vertex corrections and the box corrections.
In appendix B,
we discuss the renormalization group improvement of the charge form factors
and hadronic contributions to the gauge boson propagators.
Appendix C
gives the complete analytical formulae for the $S,T,U$ parameters and
the $Zb_Lb_L$ vertex form factor $\delb(\mmz)$ in the SM. Here all the
known two-loop level corrections are included. We also give convenient
approximations to the exact formulae.
Appendix D
provides explicit expressions for the $A,B,C,D$ functions\cite{pv}
that are used to express all the one-loop correction factors.
\vspace{-5mm}


\vspace*{1cm}
\section*{\large \bf 2  Basic formalism}
\renewcommand{\theequation}{2.\arabic{equation}}
\cleqn
\subsection*{\normalsize \bf 2.1
S-matrix elements, weak boson masses, and charge form factors }
 
\pr
All the precision experiments sensitive to electroweak physics
at one-loop level so far are concerned with processes involving
external fermions, that is, leptons or quarks (excluding top quarks),
whose masses can safely be neglected in the correction terms as
compared to the weak boson masses. There are the $Z$ boson properties
as measured at LEP and SLC, the neutral current ($NC$) processes at low
energies ($\ll \mz$), the measurements of 
charged current ($CC$) processes at low
energies and those of the $W$ mass at $p\bar{p}$ colliders. The
relevant observables in these processes are then expressed in terms of
the $S$-matrix elements of four external fermions which form a scalar
product of two chirality conserving currents. All the information on
electroweak physics is contained in the scalar amplitudes which
multiply these current-current products.
 
For example, consider the $S$-matrix element responsible for the generic
4-fermion $NC$ process $ij \to ij$ (or any one of its crossed channels). 
This includes $e^+e^- \rightarrow f\overline{f}$ as
well as $\nu_{\mu}q \rightarrow \nu_{\mu}q$. The matrix element has the
form
\begin{equation}
\label{t_ij}
 T_{ij} = M_{ij} J_i \cdot J_j,
\end{equation}
where $J^\mu_i$ and $J^\mu_j$ denote currents without 
coupling factors, that is,
$J^\mu_i = \bar{\psi}_f\gamma^\mu P_\alpha \psi_f$ for $i=f_\alpha$,
where $P_\alpha = (1+\alpha \gamma_5)/2$ with $\alpha=\pm 1$ are 
the chiral projectors.\footnote{\normalsize\baselineskip 18pt
We use the chirality index $\alpha=+1$ for right-handedness and
$\alpha=-1$ for left-handedness throughout the paper; e.g.,
$P_{+1}=P_{+}=P_R$ and $P_{-1}=P_{-}=P_L$ for the chiral projectors,
$f_{+}=f_R$ and $f_{-}=f_L$ for chiral fermions.}
All radiative effects interfering with the tree-level SM amplitudes can
be cast into the above form as long as terms of order $m_f^2/\mmz$ in
the one-loop amplitudes are neglected ($m_f$ denoting the external
fermion mass).
The one-loop corrections then appear in the scalar amplitudes $M_{ij}$
which depend on flavor and chirality of the currents and on the
invariant momentum transfers $s$ and $t$ of the process.
 
In neutral current amplitudes, the photonic corrections attached
only to the external fermion lines are ${\rm U(1)}$ gauge invariant by
themselves\cite{lep1}. Therefore, finite and gauge invariant amplitudes
can be obtained by excluding all the external photonic corrections. We
find the following closed form for the generic neutral current amplitude
$M_{ij}$ of (\ref{t_ij}) at one-loop order (see details in
Appendix~A)~:
\begin{eqnarray}
 M^{NC}_{ij} &=& \frac{Q_iQ_j}{s}
 \Bigl[\,\ebar^2(s) +\ehat^2(\Gamma^i_1 +\Gamma^j_1)(s)
 -i\ehat^2 \Delta_{\gamma\gamma}(s) \Bigr]
 +\ehat^2\Bigl[(Q_iI_{3j}) \frac{\gambar_2^j(s)}{s}
        +(I_{3i}Q_j) \frac{\gambar_2^i(s)}{s} \Bigr]\hspace{5mm}
\nonumber\\
 &&+\frac{1}{s-\mmz +is\frac{\Gamma_Z}{\mz}\theta(s)}
\nonumber \\
    && \quad \times \{
 (I_{3i}-Q_i\shat^2)(I_{3j}-Q_j\shat^2)
 \Bigl[\,\gzbar^2(s) +\gzhat^2(\Gamma^i_1 +\Gamma^j_1)(s)
 -i\gzhat^2\Delta_{ZZ}(s) \Bigr]
\nonumber\\
 &&\hspace{10mm} +(I_{3i}-Q_i\shat^2)\gzhat^2
 \Bigl[\,I_{3j}(\chat^2\gambar_2^j +\Gamma_3^j)(s) +\Gamma_4^j(s)
 -Q_j(\sbar^2(s)-\shat^2 +i\Delta_{\gamma Z}(s))\Bigr]
\nonumber\\
 &&\hspace{10mm} +(I_{3j}-Q_j\shat^2)\gzhat^2
 \Bigl[\,I_{3i}(\chat^2\gambar_2^i +\Gamma_3^i)(s) +\Gamma_4^i(s)
 -Q_i(\sbar^2(s)-\shat^2 +i\Delta_{\gamma Z}(s))\Bigr] \}
\nonumber
\end{eqnarray}
\begin{equation}
\label{m_nc}
 +B^{NC}_{ij}(s,t). \mbox{\hspace{105mm}}
\end{equation}
Here $s$ is the momentum transfer of the current $J_i^\mu$, and
$t$ is the momentum transfer between the fermions $i$ and $j$.
The hatted couplings $\ehat \equiv \ghat\shat \equiv \gzhat^{}\shat\chat$
and all the ultraviolet singular loop functions are renormalized in
the $\msbar$ scheme, and hence they depend either implicitly or
explicitly on the unit-of-mass $\mu$. Three of the four charge form
factors of Table~1, $\ebar^2(s)$, $\sbar^2(s)$ and $\gzbar^2(s)$,
appear in the $NC$ amplitudes~:
\begin{subequations}
\label{ncbarcharges}
\begin{eqnarray}
 \ebar^2(q^2) &=& \ehat^2
 \Bigl[\,1 -{\rm Re}\,\overline{\Pi}_{T,\gamma}^{\gamma\gamma}(q^2)\Bigr],
\\
 \sbar^2(q^2) &=& \shat^2
 \Bigl[\,1 +{\chat \over \hat{s}}
 {\rm Re}\, \overline{\Pi}_{T,\gamma}^{\gamma Z}(q^2)\Bigr],
\\
 \gzbar^2(q^2) &=& \gzhat^2
 \Bigl[\,1 -{\rm Re}\, \overline{\Pi}_{T,Z}^{ZZ}(q^2)\Bigr].
\end{eqnarray}
\end{subequations}
Imaginary parts of the propagator correction factors denoted by
$\Delta_{\gamma\gamma}(s)$, $\Delta_{\gamma Z}(s)$ and $\Delta_{ZZ}(s)$
are defined as follows~:
\begin{subequations}
\label{delta}
\begin{eqnarray}
\label{delta_aa}
 \Delta_{\gamma\gamma}(q^2) &=&
 {\rm Im}\, \overline{\Pi}_{T,\gamma}^{\gamma\gamma}(q^2),
\\
\label{delta_az}
 \Delta_{\gamma Z}(q^2) &=& \shat\chat\,
 {\rm Im}\, \overline{\Pi}_{T,\gamma}^{\gamma Z}(q^2),
\\
\label{delta_zz}
 \Delta_{ZZ}(q^2) &=&
 {\rm Im}\, \overline{\Pi}_{T,Z}^{ZZ}(q^2)
 -\frac{{\rm Im}\, \overline{\Pi}_{T}^{ZZ}(\mmz)}{\mmz}.
\end{eqnarray}
\end{subequations}
The vertex functions $\Gamma_n^{f_\alpha}(s)$ and the box functions
$B_{f_\alpha f'_\beta}(s,t)$ are process specific.
The SM contributions to all the two-, three-, and four-point 
functions in eq.~(\ref{m_nc}) are calculated in the 't~Hooft-Feynman
gauge. 
Their explicit forms are found in appendix~A.
 
The residues of the $\gamma$- and $Z$-poles are separately
$\mu$-independent and gauge invariant, and therefore physical
observables. For $q^2 = 0$, the vanishing of the vertex functions
\begin{equation}
\label{ward}
 \Gamma_1^{f_\alpha}(0)  =0, \pfive
 \gambar_2^{f_\alpha}(0) =0,
\end{equation}
is ensured for all $f_\alpha$ by the Abelian and non-Abelian parts of
the Ward identities, respectively. The universal residue of the photon
pole gives the square of the unit electric charge
$\bar{e}^2(0) = 4\pi\alpha$.
 
Likewise, the charged current ($CC$) process $ij \to i'j'$ can be
expressed by
\begin{equation}
\label{m_cc}
 M^{CC}_{ij} =\frac{1}{s -\mmw +is\frac{\Gamma_W}{\mw}\theta(s)} \Bigl\{
 \gwbar^2(s) +\ghat^2\,[\ol{\Gamma}^{ii'}\!(s) +\ol{\Gamma}^{jj'}\!(s)
                        -i\Delta_{WW}(s)] \Bigr\}
 +B^{CC}_{ij}(s,t),
\end{equation}
with an appropriate CKM factor $V_{ii'}V^*_{jj'}$ accounting for
quark family mixing. The $W$ propagator corrections appear in the
charge form factor $\gwbar^2(s)$ and in the imaginary part
$\Delta_{WW}(s)$:
\begin{eqnarray}
\label{ccbarcharge}
 \gwbar^2(q^2) &=& \ghat^2\,\Bigl[\,1 -{\rm Re}\pibar^{WW}_{T,W}(q^2)
 \Bigr]\,,
\\
\label{delta_ww}
 \Delta_{WW}(q^2) &=& {\rm Im}\pibar^{WW}_{T,W}(q^2)
 -\frac{{\rm Im}\pibar^{WW}_{T}(\mmw)}{\mmw}\,.
\end{eqnarray}
Factorization of the external photonic corrections does not hold
for the charged current processes, and hence all the one-loop correction
terms are included in eq.~(\ref{m_cc}). Explicit forms of the SM
contributions to the propagator function  $\pibar^{WW}_T(q^2)$ are found
in Appendix~A.
 
The gauge boson two-point functions appearing in
eqs.~(\ref{ncbarcharges}), (\ref{delta}), (\ref{ccbarcharge}) and
(\ref{delta_ww}) are defined as follows~:
\begin{equation}
\label{pitv}
 \pibar_{T,V}^{AB}(q^2)
 = { \pibar_T^{AB}(q^2) -\pibar_T^{AB}(\mmv) \over q^2 -\mmv }\,,
\end{equation}
where $\mv$ is the physical mass of the gauge boson $V$ (that is, 
$\mw$, $\mz$ and $m_{\gamma}$ with $m_\gamma =0$)
and the subscript $T$ stands for the transverse part of the vacuum
polarization tensor $\Pi_{\mu \nu}(q)$,
\begin{equation}
\label{pimunu}
 \Pi^{AB}_{\mu\nu}(q) =
 \Bigl(-g_{\mu\nu} +\frac{q^\mu q^\nu}{q^2}\Bigr) \Pi^{AB}_T(q^2)
       +\frac{q^\mu q^\nu}{q^2}  \Pi^{AB}_L(q^2)\,.
\end{equation}
Contributions from the longitudinal part of the gauge boson propagators
are consistently neglected in the one-loop corrections, because they
give terms of order $m_f^2/m_V^2$ ($V=Z$ or $W$) in the weak
amplitudes.
 
The gauge boson propagators are calculated in the 't~Hooft-Feynman gauge
and the so-called pinch terms\cite{kl,pt1,pt2} of the vertex functions
arising from diagrams with the weak boson self-couplings are included
in the overlined functions $\pibar_T^{AB}(q^2)$:
\begin{subequations}
\label{pibar}
\begin{eqnarray}
 \pibar_T^{\gamma\gamma}(q^2) &=& \Pi_T^{\gamma\gamma}(q^2)
 -\frac{\ehat^2}{4\pi^2} q^2 B_0(q^2;\mw,\mw),
\\
 \pibar_T^{\gamma Z}(q^2) &=& \Pi_T^{\gamma Z}(q^2)
 -\frac{\ehat\gzhat^{}\,\chat^2}{4\pi^2}
 (q^2-\frac{\mmz}{2}) B_0(q^2;\mw,\mw),
\\
 \pibar_T^{ZZ}(q^2) &=& \Pi_T^{ZZ}(q^2)
 -\frac{\gzhat^2\,\chat^4}{4\pi^2} (q^2-\mmz) B_0(q^2;\mw,\mw),
\\
 \pibar_T^{WW}(q^2) &=& \Pi_T^{WW}(q^2)
 -\frac{\ghat^2}{4\pi^2} (q^2-\mmw)
 [\chat^2 B_0(q^2;\mw,\mz) +\shat^2 B_0(q^2;\mw,m_\gamma)].
 \mbox{\hspace{8mm}}
\end{eqnarray}
\end{subequations}
Here $B_0$ is a Passarino-Veltman function\cite{pv} in the notation of
appendix~D. The overlines on the vertex functions $\gambar_2^{f_L}(s)$
in eqs.~(\ref{m_nc}) and (\ref{m_cc}) 
and $\gambar^{ii'}\!(s)$ in eq.~(\ref{m_cc}) 
indicate the {\em subtraction} of the
pinch term associated with this prescription
(note, the pinch terms in eqs.~(\ref{pibar}) have a negative sign 
in our convention). 

The absorption of the above $q^2$ dependent propagator-like parts of
the vertex functions into the effective charges\cite{kl} 
improves over the usual method of absorbing the relevant vertex term
at zero momentum transfer\cite{lep1} in two ways. One is that
the remaining vertex parts do no longer give rise to large logarithms
of the type $\ln(-q^2/\mmw)$ at $|q^2| \gg \mmw$, and hence the
effective charges are useful in making the improved Born approximation
\cite{kl} even at very high energies. The second is that the effective
charges are now {\em gauge invariant}\cite{kl,pt2}, and hence their
properties can be discussed independently of the other process specific
corrections of the same order. Most importantly, we can obtain
explicitly renormalization group invariant relations between the
$\msbar$ couplings and the effective charges
\begin{subequations}
\label{rg}
\begin{eqnarray}
 \frac{1}{\ebar^2(q^2)} &=& \frac{1}{\ehat^2(\mu)}
 \Bigl[\,1 +{\rm Re}\pibar_{T,\gamma}^{\gamma\gamma}(q^2)\,\Bigr]\,,
\\
 \sbar^2(q^2) &=& \shat^2(\mu)
 +\frac{\ebar^2(q^2)}{\ehat(\mu)\gzhat^{}(\mu)}
 \,{\rm Re}\,\pibar_{T,\gamma}^{\gamma Z}(q^2),
\end{eqnarray}
\end{subequations}
within the 't~Hooft-Feynman gauge of the electroweak theory.
This enables us to discuss the renormalization group improvement of
the above two effective charges as a whole, that is, 
without separating the
contributions from the SM fermions and the rest. The trajectories of
all the $\msbar$ couplings ($\ehat=\ghat\shat=\gzhat^{}\shat\chat$) are
completely fixed by the above two equations
at one-loop level, which can be used to study
quantitatively the heavy particle threshold corrections 
in GUT theories\cite{sgut92_hy}.
 
In the analysis presented here the $\msbar$ couplings act as the
expansion parameters of the perturbation series, since we find them
the most convenient when studying consequences of various theoretical
models beyond the SM. Their usefulness in the SM analysis has been
emphasized in ref.~\cite{msb_ew}, and they are often used in the
analysis of new physics contributions to the precision
experiments\cite{stu_mod}. However, it is not convenient to use the
$\msbar$ couplings at a specific unit-of-mass ($\mu$) scale, such as
$\mu=\mz$, when dealing with a theory with particles much heavier than
the weak bosons because of the appearance of large logarithms of their
masses.  Hence, we adopt the following renormalization conditions
\begin{equation}
\label{rc}
 \ehat^2 =\ebar^2(\mmz), \pfive \shat^2 =\sbar^2(\mmz),
\end{equation}
consistently for all processes studied.  The above conditions
renormalize all the logarithms of large masses with the help of the
renormalization group identities (\ref{rg}) at $|q^2|<O(\mmz)$. Note
that the running of $\ebar^2(q^2)$ and $\sbar^2(q^2)$ at low energies
arises from the QED$\,\times\,$QCD interactions\cite{4f_rc}, 
and hence the ratio
$\ebar^2(q^2)/\sbar^2(q^2)$ is not an appropriate expansion parameter
of the weak corrections at $|q^2|\ll \mmz$. Note further that,
apart from details concerning the higher order terms, the effective
charges $\ebar^2(q^2)$ and $\sbar^2(q^2)$ (\ref{rg}) are the same as
the real parts of the corresponding star-scheme\cite{kl} charges,
$e_*^2(q^2)$ and $s_*^2(q^2)$, respectively. More details on the
treatment of the renormalization group improvement and the hadronic
contributions to the charge form factors are given in appendix~B.
 
Since we adopt the LEP convention\cite{lep1} regarding mass and width
($\mv$ and $\Gamma_V$) for both $Z$ and $W$,  the Breit-Wigner
propagator factors in eqs.~(\ref{m_nc}) and (\ref{m_cc}) have the
running width factor, and the imaginary parts (\ref{delta_zz}) and
(\ref{delta_ww}) have the associated subtraction terms. These masses and
widths can also be defined in terms of the more
conventional pole masses and widths\cite{mz_pole}, 
denoted by $m_{V,p}$ and
$\Gamma_{V,p}$, as follows\cite{mz_lep}:
\begin{subequations}
\label{mv_lep}
\begin{eqnarray}
 \mmv &=& m_{V,p}^2 +\Gamma_{V,p}^2 \,,
\\
 \Gamma_V &=& \Gamma_{V,p} \, \sqrt{1+(\Gamma_{V,p}/m_{V,p})^2}  \,.
\end{eqnarray}
\end{subequations}
The Breit-Wigner propagator function with the fixed width and that
with the running width are then related by the exact
relation\cite{mz_lep}
\begin{equation}
\label{bw_relation}
 \frac{1}{s-m_{V,p}^2+im_{V,p}\Gamma_{V,p}} =
 \frac{1+i\Gamma_V/\mv}{s-\mmv+is\Gamma_V/\mv} \,.
\end{equation}
The imaginary part of the numerator $\Delta_{ZZ}(q^2)$ (\ref{delta_zz})
and $\Delta_{WW}(q^2)$ (\ref{delta_ww}) are arranged such that the
imaginary parts of the full amplitudes vanish exactly at zero momentum
transfer: $\Delta_{VV}(0)=0$. 
The theta function $\theta(s)$ 
($\theta(s)=1$ for $s>0$ and $\theta(s)=0$ for $s\le 0$)
in the running width factor of eqs.~(\ref{m_nc}) and (\ref{m_cc}) 
then ensures the reality of the amplitudes at $s<0$. 
It should be noted that the imaginary
part $\Delta_{VV}(q^2)$ vanishes at $q^2=\mmv$
at one-loop level, 
if all the contributing
particle masses can be neglected. As long as the relations
(\ref{mv_lep}) and (\ref{bw_relation}) are respected, physical
consequences for observables near the $W$- or $Z$-poles remain unchanged.
When constraining the electroweak parameters, however, we often refer
to the weak currents at zero momentum transfer. The masses in the LEP
convention are more appropriate to use in this case\cite{mz_lep}, since they
absorb reducible higher order contributions from the $W$ and $Z$ widths.
 
\vspace*{5mm}
\subsection*{\normalsize \bf 2.2  Vertex and box corrections}
\pr
In this subsection, the vertex and box corrections are numerically
estimated in the SM, while their explicit forms are given in appendix~A.
First the neutral current ($NC$) amplitudes near the $Z$-pole and at
low energies ($|q^2| \ll \mmz$) are discussed, then the charged
current ($CC$) amplitudes in the zero momentum transfer limit. Except
for the $\zbb$ vertex, all the vertex and box corrections are
assumed to be dominated by these SM contributions in the following
analysis.

Four types of vertex form factors appear in the $NC$ amplitudes (\ref{m_nc}).
$\Gamma_1^f$ and $\gambar_2^f$ appear both in the $\gamma ff$ and
$Zff$ vertices, while $\Gamma_3^f$ and $\Gamma_4^f$ appear only in the
$Zff$ vertices:
\bsub
\label{gamma_vff}
\bea
\Gamma^{\gamma ff}(q^2) &=& 
    -\ehat\,\Bigl\{ Q_f [\,1+\Gamma_1^f(q^2)\,]
                  +I_{3f}\gambar_2^f(q^2) \Bigr\}\,,
\label{gamma_aff}
\\
\Gamma^{Zff}(q^2) &=& -\gzhat^{}\,\Bigl\{ 
          (I_{3f}-Q_f\shat^2)\, [\,1+\Gamma_1^f(q^2)\,]
             +I_{3f} [\,\chat^2\gambar_2^f(q^2) +\Gamma_3^f(q^2)\,] 
                      +\Gamma_4^f(q^2) \Bigr\}\,.\hspace{1.2cm}
\label{gamma_zff}
\eea
\esub
The SM contribution to the vertex form factors that are non-vanishing
at one-loop order are $\Gamma_1^{f_\alpha}(q^2)$, 
$\gambar_2^{f_L}(q^2)$ and $\Gamma_3^{f_L}(q^2)$.
They can be expressed by
\bsub
\label{gamma_f}
\bea
     \Gamma_1^{f_R}(q^2) &=&
        \biggl( \frac{g_R^{Zff}}{4 \pi} \biggr)^2
         \Gamma_{1Z}^f(q^2)\,,
\label{gamma1_fr}
\\
     \Gamma_1^{f_L}(q^2) &=&
        \biggl( \frac{g_L^{Zff}}{4 \pi} \biggr)^2
         \Gamma_{1Z}^f(q^2)
      + \sum_{f'} \biggl| \frac{g_L^{Wff'}}{4 \pi} \biggr|^2
         \Gamma_{1W}^{f'}(q^2)\,,
\label{gamma1_fl}
\\
     \ol{\Gamma}_2^{f_L}(q^2) &=&
          -2\,\sum_{f'} \biggl| \frac{g_L^{Wff'}}{4 \pi} \biggr|^2
          \gambar_{2W}^{f'}(q^2)\,,
\label{gamma2_fl}
\\[1mm]
     \Gamma_3^{f_L}(q^2) &=&
          \sum_{f'} \biggl| \frac{g_L^{Wff'}}{4 \pi} \biggr|^2
          \Gamma_{mW}^{f'}(q^2)\,,
\label{gamma3_fl}
\eea
\esub
with the gauge-boson--fermion coupling convention
\bea
\label{g_vff}
    \begin{array}{cc}
     g_L^{\gamma ff} = g_R^{\gamma ff} = \ehat\,Q_f\,, \quad &
     g_R^{Zff}  = \gzhat^{}\,(-Q_f\,\shat^2)\,,
     \\[3mm]
     g_L^{Zff}  = \gzhat^{}\,(I_{3f} -Q_f\,\shat^2)\,, \quad &
     g_L^{Wff'} = \displaystyle{\frac{\ghat}{\sqrt{2}}}\,V_{ff'}\,.
    \end{array}
\eea
Explicit forms of the functions $\Gamma_{1V}^f(q^2)$,
$\gambar_{2W}^{f'}(q^2)$ and $\Gamma_{mW}^{f'}(q^2)$ are given in
eqs.~(\ref{eq:gamma_fv})--(\ref{eq:gamma's}) in 
appendix~A.2. 
External
fermion self-energy corrections are included in the functions
$\Gamma_{1Z}^f(q^2)$ and $\Gamma_{1W}^{f'}(q^2)$. For right-handed
fermions  $\overline{\Gamma}_2^{f_R}(q^2)=0$ holds, since only those
diagrams with  $W$ exchange contribute to the vertex function
$\overline{\Gamma}_2$ at one-loop order. The vertex functions
$\Gamma_3^{f_\alpha}(s)$ are found to be proportional to the square of
the fermion mass inside the loop, and are non-vanishing only for
$f_\alpha=b_L$ in the SM, within our approximation of using diagonal
KM matrix elements and neglecting terms of order $(m_b/\mz)^2\alpha$.
For large $m_t$ ($m_t^2 \gg |s|$), the SM contribution to
$\Gamma_3^{b_L}(s)$ is proportional to $m_t^2/\mmw$\cite{zbb,lynnstuart}.
The functions $\Gamma_4^{f_\alpha}(s)$ can, in general, be present, but
happen to vanish for all $f_\alpha$ in the SM; they are, however,
found to be non-vanishing in some extended models such as the minimal
SUSY-SM.
These analytic expressions agree with the known results of 
refs.~\cite{lynnstuart,hollik90,fujimoto90,kniehlstuart,hollik93}%
\footnote{
\normalsize \baselineskip 18pt%
We note the following misprints in ref.~\cite{lynnstuart}.
In the last line of eq.~(2.7), 
the factor $1/(m^2-M^2)^2$ should read $1/(m^2-M^2)$.
In the first line of eq.~(2.8), 
the term $4q^4M^2$ should read $4q^2M^4$, and 
in the last line of eq.~(2.9), 
the term $m/2M^2$ should read $m^2/2M^2$. 
Our vertex functions $\Gamma_{1W}^f$,
$\gambar_{2W}^{f'}$ and $\Gamma_{mW}^{f'}$ 
are then related to their functions $\rho$, $\Lambda$ and $\Xi$ by 
the identities~: 
$  \Gamma_{1W}^{f'}(q^2) = \rho\,(-q^2,\mmw,m_{f'}^2)$, 
$  \ol{\Gamma}_{2W}^{f'}(q^2) = [\rho+\Lambda]\,(-q^2,\mmw,m_{f'}^2)
          +2\,[\,B_0(q^2;\mw,\mw)-B_0(0;\mw,\mw)]$,
and
$  \Gamma_{mW}^{f'}(q^2) = -2\,\Xi(-q^2,\mmw,m_{f'}^2)$.
}. 

The numerical values of the vertex form factors $\Gamma_1^f(q^2)$,
$\overline{\Gamma}_2^f(q^2)$ and $\Gamma_3^f(q^2)$ at $q^2=\mmz$ are
given in Table~\ref{tableofvertexformfactors}. All the numerical
results presented in this section and in the following sections are
obtained by setting
\bsub
\label{hatcouplings}
\bea
 4\pi/\ehat^2 &=& 128.72 \,,
\label{e2hat}\\
 \shat^2 &=& 0.2312 \,,
\label{s2hat}
\eea
\esub
with $\ehat=\ghat\shat=\gzhat^{}\shat\chat$ in the one-loop correction
terms. They are fixed by using the renormalization conditions
(\ref{rc}) and the SM predictions for $\bar{\alpha}(\mmz)$ and
$\sbar^2(\mmz)$ at $m_t=150$~GeV, $\mh=100$~GeV, $\alpha_s(\mz)=0.120$
and $\delta_\alpha=0$. We emphasize that we do not change the
numerical values of eq.~(\ref{hatcouplings}) when discussing
experimental constraints on the charge form factors $\ebar^2(\mmz)$
and $\sbar^2(\mmz)$. All our predictions for the $Z$ parameters can be
reproduced simply by using the numerical values listed in
Table~\ref{tableofvertexformfactors} and eq.~(\ref{hatcouplings}),
together with the imaginary parts of the gauge boson propagator
corrections
\bea
\begin{array}{|l|cccc|}
 \hline
 \alpha_s(\mz)_{\msbar}      & 0       & 0.11    & 0.12    & 0.13   \\
 \hline
 \Delta_{\gamma\gamma}(\mmz)& 0.01726 & 0.01760 & 0.01763 & 0.01766\\
 \Delta_{\gamma Z}(\mmz)    & 0.00248 & 0.00257 & 0.00257 & 0.00258\\
 \Delta_{ZZ}(\mmz)          & 0.00005 & 0.00003 & 0.00003 & 0.00003\\
 \hline
\end{array}
\label{delta_vv}
\eea
which are obtained by using the perturbative order $\alpha\alpha_s$
approximations of appendix A with the effective quark masses of
eqs.~(\ref{eq:mcmb}) and (\ref{eq:tableofqmass}). It is worth noting
that the real part of the vertex corrections
(Table~\ref{tableofvertexformfactors}) and the imaginary part
$\Delta_{\gamma\gamma}(\mmz)$ interfere with the leading $Z$-pole
amplitude:  the latter contribution has been subtracted in the $Z$
parameters\cite{lep93}, whereas the former contributions modify
the scattering amplitudes by as much as 0.5\%, and hence they can
contribute to the cross sections at the 1\% level.
 
Note further that the vertex correction without the pinch term
subtraction\cite{lep1,hollik90} $\Gamma_2^{f_L}(q^2)$ is related to
the $\gambar_2^{f_L}(q^2)$ function by
\bea
  \gambar_2^{f_L}(q^2) =  \Gamma_2^{f_L}(q^2)
 -\frac{\ghat^2}{8\pi^2}{\rm Re}[B_0(q^2;W,W)-B_0(0;W,W)]
\label{gamvsgambar}
\eea
in the t'~Hooft-Feynman gauge. The difference is universal
($f$-independent) and we find
$\gambar_2^{f_L}(\mmz) = \Gamma_2^{f_L}(\mmz) - 0.00134$. The vertex
corrections are slightly larger in magnitude after subtraction of the
pinch term.
 
It is
convenient to introduce the following special form factor
\bea
        \delb(s)
        =\Gamma_1^{b_L}(s) +\chat^2 \gambar_2^{b_L}(s) +\Gamma_3^{b_L}(s)
 +\mbox{\rm higher order terms},
\label{delb}
\eea
which is treated also as a free parameter in our fit at $s=\mmz$ 
to deal with the strong $m_t$ dependence of the $\zbb$ vertex
(see also ref.~\cite{susy_thr,abc}). 
In this way, the importance of the $\zbb$ vertex
correction\cite{zbb_lep} can be assessed independently of the specific
SM mechanism and also the data analysis is kept separate from the
evaluation of $\delb$ in a specific model.
In the SM, the parameter $\delb$ can be evaluated by including
$O(\alpha_s m_t^2)$\cite{zbb_qcd} 
and $O(m_t^4)$\cite{rho_zbb_ew,zbb_ew}
two-loop corrections of the SM, which are given explicitly in
appendix~C.4: see eq.~(\ref{eq:delta_b}).
 
At low energies, light fermion masses may not be neglected compared
to the momentum transfer $q^2$. In the limit of $|q^2|/\mmz\!\ll\! 1$ and
$m_f^2/\mmz\! \ll\! 1$ but at fixed $m_f^2/q^2$, the vertex functions reduce to
\bsub
\label{gamma_f_le}
\bea
     \Gamma_{1Z}^{f}(q^2) &=&
 \frac{-q^2}{\mmz}
        \Bigl[J_Z(q^2;f) +O\Bigl(\frac{q^2}{\mmz}\Bigr)\Bigr]\,,
\label{gamma1z_le}
\\
     \Gamma_{1W}^{f'}(q^2) &=&
      \frac{-q^2}{\mmw}
        \Bigl[J_W(q^2;f') +O\Bigl(\frac{q^2}{\mmz}\Bigr)\Bigr]\,,
\label{gamma1w_le}
\\
     \ol{\Gamma}_{2W}^{f'}(q^2) &=&
      \frac{-q^2}{\mmw}
        \Bigl[\ol{J}_W(q^2;f') +O\Bigl(\frac{q^2}{\mmz}\Bigr)\Bigr]\,.
\label{gamma2q_le}
\eea
\esub
The functions $J_Z(q^2;f)$, $J_W(q^2;f)$ and $\ol{J}_W(q^2;f)$ have
the same form as the fermionic contribution to the neutral gauge boson
vacuum polarization functions: see eq.~(\ref{eq:j}). The form factor
$\gambar_2^{\nu_L}(q^2)$ is often called the neutrino charge radius
term\cite{nu_chr}. The subtraction of the pinch term makes it
gauge invariant\cite{pt2}.
 
For the $NC$ process
$f_\alpha(p_1)f'_\beta(p_2) \to f_\alpha(p_3)f'_\beta(p_4)$,
as well as for its crossed channels, the box correction terms in
eq.~(\ref{m_nc}) can be expressed as
\bsub
\label{box}
\bea
     &B_{\alpha,\alpha}^{\,f\,f'}(s,t) &=
         \frac{1}{16 \pi^2}
         \left| g_\alpha^{Zff} g_\alpha^{Zf'f'} \right|^2
         \left[ I_1(u,s;\mz,0) - I_2(t,s;\mz,0) \right]
     \\
     &&+\frac{\delta_{\alpha L}}{16 \pi^2}
        \left| g_L^{Wff''} g_L^{Wf'f'''}  \right|^2
        \times
         \left\{
         \begin{array}{ll}
            \! +I_1(u,s;\mw,m_{f'''})
            & \:\:\mbox{ for }\; {I_3}_f{I_3}_{f'}<0
            \\[1mm]
            \! -I_2(t,s;\mw,m_{f'''})
            & \:\:\mbox{ for }\; {I_3}_f{I_3}_{f'}>0
  \end{array}
         \right. ,
\nonumber \\[-1mm]
\label{box_ll}\\
     &B_{\alpha,-\alpha}^{\,e\,f}(s,t) &=
        \frac{1}{16 \pi^2}
        \left| g_\alpha^{Zff} g_{-\alpha}^{Zf'f'} \right|^2
        \left[ I_2(u,s;\mz,0) - I_1(t,s;\mz,0) \right]\,,
\label{box_lr}
\eea
\esub
where $s=(p_1-p_3)^2$, $t=(p_1-p_4)^2$ and $u=(p_1+p_2)^2$ are the
Mandelstam variables 
satisfying $s+t+u=0$. In the second term of
(\ref{box_ll}), $f''$ and $f'''$ are the weak isospin partners of $f$
and $f'$, respectively, where all external and internal fermion masses
except for $m_{f'''}$
are neglected~: the upper term (${I_3}_f{I_3}_{f'}<0$) should be taken
for $(f,f')=(\ell,u)$, $(\nu,\ell)$, and $(\nu,d)$, whereas the lower term
(${I_3}_f{I_3}_{f'}>0$) for $(f,f')=(\ell,d)$, and $(\nu,u)$. The explicit
form of the box functions $I_1$ and $I_2$ are given in
eq.~(\ref{eq:i_box}) of appendix~A.3. 
These analytic expressions agree with the known results of 
refs.~\cite{lynnstuart,hollik90,fujimoto90,kniehlstuart,hollik93}.
It is worth noting here that the
box contributions to the helicity  amplitudes have the above simple
current product form only when the external fermion masses can be
neglected.
 
The numerical values of the box functions $B_{ij}(s,t)$ for the process
$e^+ e^- \to f \bar{f}$ are given in Table~\ref{tableofboxformfactors}
for $s=-2\,t=\mmz$. They contribute negligibly to the $Z$ parameters,
because they do not interfere with the dominant $Z$-pole amplitudes
being almost purely imaginary near the pole. The imaginary parts appear
in the box functions only above the $W$-pair production threshold.
 
The box contributions are found to be non-negligible in some low energy
$NC$ processes. In the $s=t=u=0$ limit, one finds
\bsub
\label{i_box_le}
\bea
 I_1(0,0;m_V,0) = -\frac{4}{m_V^2}\,,
\label{i1_le}\\
 I_2(0,0;m_V,0) = -\frac{1}{m_V^2}\,.
\label{i2_le}
\eea
\esub
The $WW$ box contributions to the processes with the $I_1$ function,
that is, 
the low energy $\nu$--$\ell$, $\nu$--$d$ and $e$--$u$ scattering processes
are found to be significant.
 
Precise values of the charged current matrix elements are needed only
at low energies. The muon decay constant is given by
\begin{equation}
\label{gf}
 G_F = \frac{\gwbar^2(0) +\ghat^2\delg} {4\sqrt{2}\mmw},
\end{equation}
where the factor $\delg$
\bea
     \delg = \frac{\ghat^2}{8\,\pi^2}
             \biggl[\,1 +\biggl(\frac{1}{4\,\shat^2}-1\biggr)
                     \ln\frac{1}{\chat^2} \biggr]
 \approx 0.0055
\label{delg}
\eea
denotes the sum of the vertex and the box contributions in the SM.
Its numerical value above is obtained for the couplings
of eq.~(\ref{hatcouplings}). 
The identity (\ref{gf}) gives the physical $W$ mass
in terms of $\gwbar^2(0)$, 
once the $\delg$ value is known for a given model.
The overline here again indicates the removal of the pinch terms with
the consequence that the numerical value is significantly (about 20\%)
smaller than the standard factor\cite{del_gf}
\bea
     \delta_G = \frac{\ghat^2}{8\,\pi^2}
             \biggl[\,3 -\biggl(\frac{7}{4\,\shat^2}-1\biggr)
                     \ln\frac{1}{\chat^2} \biggr]
          \approx 0.0068\,,
\label{delg_old}
\eea
which was obtained simply by subtracting the singular vertex function at
zero momentum transfer. 
The difference 
\bea
  \delg -\delta_G^{}=
   \frac{\ghat^2}{4\,\pi^2}
  \Bigl[B_0(0;W,W)-\chat^2\,B_0(0;W,Z)-\shat^2\,B_0(0;W,\gamma)\Bigr]
\eea
is the pinch term contribution\cite{pt2}.
Note that the sum of the propagator and the
vertex/box corrections is scheme-independent and that the correction
term $\delg$ of eq.~(\ref{delg}) should be used together with the
charge form factor $\gwbar^2(0)$ which contains the associated
pinch term.
\vspace*{5mm}
\subsection*{\normalsize \bf 2.3
  Constraints due to $\alpha$, $G_F$ and $\mz$ }
\pr
Among the electroweak observables the three quantities $\alpha$, $G_F$
and $\mz$ have been measured with outstanding precision, namely
$\Delta\alpha/\alpha \approx 5\times 10^{-8}$,
$\Delta G_F / G_F   \approx 2\times 10^{-5}$~\cite{pdg92}, and
$\Delta \mz / \mz   \approx 8\times 10^{-5}$~\cite{lep93}.
For this reason $\alpha$, $G_F$ and $\mz$ are chosen as our basic
electroweak parameters and treated as constants in the analysis 
(see Tables~1 and 2).
 
On the other hand, the tree-level properties of the gauge boson
propagators are fixed completely by three parameters, the two gauge
couplings $g$ and $g'$ for the ${\rm SU(2)_L}$ and ${\rm U(1)_Y}$
gauge groups, respectively, and one vacuum expectation value
$v\equiv (\sqrt{2}\,G_F)^{-1/2} \approx 246$~GeV, in models where the
electroweak symmetry breaking sector has the custodial ${\rm SU(2)}$
symmetry\cite{rho}. Consequently, the four charge form factors
$\ebar^2(q^2)$, $\sbar^2(q^2)$, $\gzbar^2(q^2)$, and $\gwbar^2(q^2)$
are completely determined by finite quantum corrections in this class
of models when expressed in terms of the three constants $\alpha$,
$G_F$ and $\mz$.
 
In this subsection, the prescription for calculating all charge form
factors in terms of ($\alpha$, $G_F$, $\mz$) is given explicitly
in an arbitrary model with the broken ${\rm SU(2)_L} \times {\rm U(1)_Y}$
gauge symmetry. Their numerical predictions are given in the SM.
 
The form factor $\ebar^2(q^2) \equiv 4\pi \bar{\alpha}(q^2)$ is fixed by
the following identity
\bea
 \frac{1}{\bar{\alpha}(q^2)} -\frac{1}{\alpha}
 = 4\pi {\rm Re} \biggl[ \pibar_{T,\gamma}^{QQ}(q^2)
                        -\pibar_{T,\gamma}^{QQ}(0)  \biggr]\,,
\label{runningalpha}
\eea
which gives the renormalization group improved running
$\bar{\alpha}(q^2)$ as explained in appendix~B. 
Here $\pibar_T^{QQ}(q^2)$ is the $\gamma\gamma$ propagator 
function without the overall coupling factor $\ehat^2$\cite{kl}: 
see eq.~(\ref{eq:pi_t}). 

In principle, the
effective coupling $\bar{\alpha}(\mmz)$ can be calculated from the
observed $\alpha$ value by using the above identity. In practice,
however, the right-hand side suffers from non-perturbative QCD
corrections to the light quark contributions. We make use of the
dispersion analyses\cite{piqq_h,piqq_h_latest,pi3q_h} 
to estimate the hadronic contributions
to the running of $\bar{\alpha}(q^2)$ and $\sbar^2(q^2)$ form factors
at $0<|q^2|<\mmz$. Details can be found in appendix~B.
 
In order to take account of uncertainty in the hadronic contribution
and also possible new physics contributions, the parameter $\delta_\alpha$
is introduced 
as an external parameter 
in the analysis~:
\bea
 \frac{1}{\bar{\alpha}(\mmz)} \equiv 128.72 +\delta_\alpha \,,
\label{delta_alpha}
\eea
which can be expressed by
\bea
 \delta_\alpha \approx \delta_{\rm had} +4\pi {\rm Re}
 \biggl[ \pibar_{T,\gamma}^{QQ}(\mmz)
        -\pibar_{T,\gamma}^{QQ}(0)  \biggr]_{\rm New Physics}\,,
\label{d_alpha}
\eea
for $m_t=150$--200~GeV as stated in eqs.~(\ref{eq:delta_alpha}) and
(\ref{eq:delta_alpha_sm}) of Appendix B.
Here $\delta_{\rm had}=0\pm 0.10$ (\ref{eq:delta_had}) is the present
estimate\cite{piqq_h_latest} for the hadronic contribution. 
The parameter
$\delta_\alpha$ being treated as an 
external parameter serves also to assess
future improvements in low energy $e^+e^-$ hadroproduction
experiments as well as possible new physics contributions. 
 
The remaining three charge form factors can be fixed by introducing the
three radiative parameters $S$, $T$ and $U$ 
that are defined by the following identities~:
\bsub
\label{stu_def}
\bea
&& \frac{\gwbar^2(0)}{\mmw}
 \frac{\mmz}{\gzbar^2(0)}
 \equiv 1 -\alpha T \,,
\label{t_def}\\[1mm]
&& \frac{4\pi}{\gzbar^2(0)}
 -\frac{\sbar^2(\mmz)\cbar^2(\mmz)}{\bar{\alpha}(\mmz)}
 \equiv -\frac{S}{4} \,,
\label{s_def}\\[1mm]
&& \frac{4\pi}{\gwbar^2(0)}
 -\frac{\sbar^2(\mmz)}{\bar{\alpha}(\mmz)}
 \equiv -\frac{S+U}{4} \,.
\label{u_def}
\eea
\esub
The parameters $S$, $T$ and $U$ can be calculated perturbatively
in any models from the gauge boson propagator functions of 
eq.~(\ref{eq:pi_t}) by
\bsub
\label{stu}
\bea
 S &=& 16 \,\pi\, {\rm Re}\,
 \Bigl[ \ol{\Pi}^{3Q}_{T,\gamma}(\mmz)
              -\ol{\Pi}^{33}_{T,Z}(0)  \Bigr] \,,
\label{s}\\
 T &=& \frac{4\sqrt{2}\,G_F}{\alpha}\,{\rm Re}\,
 \Bigl[\ol{\Pi}^{33}_T(0) -\ol{\Pi}^{11}_T(0) \Bigr] \,,
\label{t}\\
 U &=& 16 \,\pi\, {\rm Re}\,
          \Bigl[ \ol{\Pi}^{33}_{T,Z}(0)
                -\ol{\Pi}^{11}_{T,W}(0)  \Bigr] \,.
\label{u}
\eea
\esub
For models without custodial ${\rm SU(2)}$ symmetry, the $T$ parameter
is sensitive to the ultraviolet cut-off, and hence is un-calculable from
($\alpha$, $G_F$, $\mz$) alone. 
In this case it should be regarded as the
fourth basic parameter of the theory.
 
Our definitions (\ref{stu_def}) of the three parameters $S$, $T$, $U$
are inspired by the pioneering work of Peskin and Takeuchi\cite{stu}.
Our definition, in contrast to theirs, includes all radiative
effects from both SM and new physics contributions.
The original parameters, denoted below by the index $\rm PT$,
are approximately related to ours by subtracting 
the SM contributions evaluated at 
$m_t=150$~GeV and $\mh=1000$~GeV~:
\bsub
\label{stu_orig}
\bea
 S_{\rm PT} &\approx&
 S -S_{\rm SM}(m_t=150~{\rm GeV}, \mh=1000~{\rm GeV}) \,,
\label{s_orig}\\
 T_{\rm PT} &\approx&
 T -T_{\rm SM}(m_t=150~{\rm GeV}, \mh=1000~{\rm GeV}) \,,
\label{t_orig}\\
 U_{\rm PT} &\approx&
 U -U_{\rm SM}(m_t=150~{\rm GeV}, \mh=1000~{\rm GeV}) \,,
\label{u_orig}
\eea
\esub
provided the scale of new physics is much larger than $\mz$.
The expressions (\ref{stu}) agree with the modified 
$S$, $T$, $U$ parameters of ref.~\cite{pt2}.
The same form of the definitions without the pinch terms 
(in the 't~Hooft Feynman gauge) 
have been used in some earlier works\cite{hisz,kniehl,kniehlkohrs}. 
 
Explicit forms of the SM contributions to the $S$, $T$, $U$ parameters
are given in appendix~C, together with the SM contribution to the
$\zbb$ form factor $\delb(\mmz)$. All the known two-loop
corrections of order 
$\alpha\alpha_s$\cite{rho_pqcd,pivv_pqcd,rho_pqcd_mh2,zbb_qcd}
and order $m_t^4$\cite{rho_ew2,rho_ew3,rho_zbb_ew,zbb_ew} are included.
The recently found\cite{rho_pqcd_mh2,2loop_mh2} 
small two-loop corrections of order
$\mmh$ are neglected.
For practical reasons we adopt the perturbative order $\alpha\alpha_s$
\cite{rho_pqcd,pivv_pqcd,rho_pqcd_mh2,zbb_qcd} corrections at
$\alpha_s =\alpha_s(\mz)_{\msbar}$ in calculating all the parameters
$S$, $T$, $U$ and $\delb(\mmz)$. The reader can therefore
unambiguously reproduce our results. The
effects due to non-perturbative threshold corrections
\cite{rho_thr,rho_thr2,rho_thr3} should be evaluated separately,
and one can obtain more precise predictions of the SM from our formulae
by adjusting the effective top-quark mass to produce the same $S$, $T$, $U$,
and $\delb(\mmz)$ values. 
It should be noted that at present the uncertainty in the SM 
contribution to the $T$ parameter is such that $m_t$ can be
predicted with a few GeV uncertainty for a given $T$ 
value\cite{rho_thr3}. 
Fig.~\ref{figureofstusm} shows the SM
contributions to the $S$, $T$, $U$ and $\delb(\mmz)$ parameters as
functions of $m_t$  for $\mh=$1--1000~GeV at $\alpha_s(\mz)$=0.12.
It is worth noting that the $T$ and $\delb(\mmz)$ parameters are
proportional to $m_t^2$ for large $m_t$ ($m_t^2 \gg \mmz$),
the parameters $U$ and $\delb(\mmz)$ are almost independent of $\mh$,
the $T$ parameter decreases with increasing $\mh$,
and the $S$ parameter becomes negative for small $\mh$.
 
Once the $S$, $T$, $U$ parameters are calculated in a given model,
the three charge form factors can be predicted as follows~:
\bsub
\label{gbarfromstu}
\bea
 \frac{1}{\gzbar^2(0)}
        &=& \frac{1+\delg -\alpha \,T}{4\,\sqrt{2}\,G_F\,\mmz} \,,
\label{gzbarfromt}\\[2mm]
 \sbar^2(\mmz)
        &=& \frac{1}{2}
        -\sqrt{\frac{1}{4} -\ebar^2(\mmz)
               \biggl(\frac{1}{\gzbar^2(0)} +\frac{S}{16\,\pi}
               \biggr)  }\,,
\label{sbarfroms}\\[2mm]
 \frac{1}{\gwbar^2(0)}
        &=& \frac{\sbar^2(\mmz)}{\ebar^2(\mmz)}
            -\frac{1}{16\,\pi}\,(S+U) \,.
\label{gwbarfromu}
\eea
\esub
The expression (\ref{gzbarfromt}) follows from eqs.~(\ref{t_def}) and (\ref{gf})
up to terms of order $\alpha^2$. Its explicit form takes account of the
reducible order $m_t^4$ corrections\cite{rho_ew1}, and it makes clear
that the combination
\bea
 \delg -\alpha T
\label{delg-at}
\eea
determines the neutral current charge form factor $\gzbar^2(q^2)$
in terms of $G_F \mmz$. In fact, the pinch term contribution to $T$ in
eq.~(\ref{t}) and the one removed from the vertex contribution in
$\delg$ (\ref{delg}) cancel in the combination $\delg -\alpha T$.
 
It is clear from eqs.~(\ref{gbarfromstu}) that $\gzbar^2(0)$ is fixed by
$\delg -\alpha T$, $\sbar^2(\mmz)$ by $\gzbar^2(0)$,
$\bar{\alpha}(\mmz)$ and $S$, and $\gwbar^2(0)$ by  $\sbar^2(\mmz)$,
$\bar{\alpha}(\mmz)$ and $S+U$. It is instructive to express
these form factors approximately as linear combinations of the
parameters $S$, $T$, $U$ and $\delta_\alpha$~:
\bsub
\label{gbar_approx}
\bea
        \gzbar^2(0)    &=& 0.5456 \hphantom{ +0.0036S }\;\, +0.0040T  \,,
\label{gzbar_approx}\\
        \sbar^2(\mmz) &=& 0.2334            +0.0036S  -0.0024T
           \hphantom{ +0.0035U }\;\, -0.0026\delta_\alpha        \,,
\label{sbar_approx}\\
        \gwbar^2(0)    &=& 0.4183            -0.0030S  +0.0044T +0.0035U
                                      +0.0014\delta_\alpha      \,.
\label{gwbar_approx}
\eea
\esub
Expressed in this form, 
it becomes obvious that essentially $\gzbar^2(0)$ measures $T$,
$\sbar^2(\mmz)$ measures $S-0.7T$, and $\gwbar^2(0)$ measures
$T+0.8U-0.7S$, if the SM values of $\delg$ and $\delta_\alpha$ are
assumed. Here the coefficients are obtained by setting $\delg = 0.0055$.
Results for arbitrary $\delg$ are obtained by the replacement~:
\bea
 T \to T +\frac{0.0055 -\delg}{\alpha} \,.
\label{t_subst}
\eea
Note that the combination $\delg -\alpha T$ vanishes in the SM
($\delg \approx 0.0055$) for $T \approx 0.75$.
Fig.~\ref{figureofstusm} shows that this cancellation occurs at around
$m_t \approx 175$~GeV. The SM predictions for the neutral current
experiments can then be reproduced rather accurately by using the
'tree-level' predictions with $\delg-\alpha T = S = 0$ in
eqs.~(\ref{gbarfromstu}), since the SM contribution to $S$ is rather
small. This should not, however, be interpreted as absence of any
quantum corrections\cite{okun} (that is, $\delg=T=0$), but rather
as evidence for the large quantum correction $\alpha T \approx 0.0055$
within the SM
(see also section~5.3).
 
Finally, the running of the remaining three charge form factors
are calculated by
\bsub
\label{runninggbar}
\bea
      \frac{\sbar^2(q^2)}{\ebar^2(q^2)}
     -\frac{\sbar^2(\mmz)}{\ebar^2(\mmz)}
     &=& {\rm Re} \Bigl[ \ol{\Pi}^{3Q}_{T,\gamma}(q^2)
                        -\ol{\Pi}^{3Q}_{T,\gamma}(\mmz) \Bigr]\,,
\label{runningsb2}\\[1mm]
     \frac{1}{\gzbar^2(q^2)} -\frac{1}{\gzbar^2(0)}
     &=&         {\rm Re} \Bigl[ \ol{\Pi}^{33}_{T,Z}(q^2)
                                -\ol{\Pi}^{33}_{T,Z}(0) \Bigr]
     -2\,\shat^2\,{\rm Re}\Bigl[ \ol{\Pi}^{3Q}_{T,Z}(q^2)
                                -\ol{\Pi}^{3Q}_{T,Z}(0) \Bigr]
\nonumber \\
     && +\shat^4\,{\rm Re}\Bigl[ \ol{\Pi}^{QQ}_{T,Z}(q^2)
                                -\ol{\Pi}^{QQ}_{T,Z}(0) \Bigr]\,,
\label{runninggzb}\\[1mm]
     \frac{1}{\gwbar^2(q^2)} -\frac{1}{\gwbar^2(0)}
     &=& {\rm Re} \Bigl[ \ol{\Pi}^{11}_{T,W}(q^2)
                        -\ol{\Pi}^{11}_{T,W}(0) \Bigr]\,.
\label{runninggwb}
\eea
\esub
Equation (\ref{runningsb2}) is the solution of the RG equation (see
appendix~B), and hence is valid at arbitrary $q^2$.  At $|q^2|<\mmz$,
the parametrizations of the dispersive 
analysis\cite{piqq_h,piqq_h_latest,pi3q_h} are
used for the light quark contribution. Equations (\ref{runninggzb}) and
(\ref{runninggwb}) are valid perturbative expressions provided
$|q^2|\simlt O(\mmz)$. At very high energies ($|q^2| \gg \mmz$), the more
elaborate expressions  (\ref{eq:Q2})--(\ref{eq:szsw_def}) should be
used to estimate accurately the charge form factors $\gzbar^2(q^2)$ and
$\gwbar^2(q^2)$.
 
Fig.~\ref{figureofchargeformfactors} displays the four charge
form factors $1/\bar{\alpha}(q^2)$, $\sbar^2(q^2)$,
$\gzbar^2(q^2)$ and $\gwbar^2(q^2)$ as functions of $\sqrt{|q^2|}$
for both time-like ($q^2>0$) and space-like ($q^2<0$) momenta.
They are obtained in the SM for several $m_t$ and $\mh$ values, namely
$m_t=$~100, 150, 200~GeV, and $\mh=$~100, 1000~GeV. The trajectories
are fixed such that the known values of the three basic parameters
($\alpha$, $G_F$, and $\mz$) are reproduced for $\delg = 0.0055$,
$\delta_\alpha =0$ and $\alpha_s(\mz) = 0.12$. The running of the form
factors $\bar{\alpha}(q^2)$ and $\sbar^2(q^2)$ at $|q^2| \ll \mmz$ is
due to the QED$\times$QCD 
quantum effects\cite{4f_rc}, and its detailed treatment
is given in appendix~B. The threshold effects are clearly seen in the
time-like trajectories. Light hadron threshold effects do not show up
since we adopt the dispersion integral fit of the hadronic
contributions to the vacuum polarizations in the space-like
region\cite{piqq_h,piqq_h_latest,pi3q_h} 
also for their contribution in the time-like
region. The running of the $\gzbar^2(q^2)$ and $\gwbar^2(q^2)$
form factors freezes at $|q^2| \ll \mmz$.
It is clearly seen that the weak boson threshold effects are significant
for all the charge form factors in the time-like region\footnote{
\normalsize \baselineskip 18pt%
Note that the charge form factor $\gwbar^2(q^2)$ suffers
from an infrared singularity at $q^2 =\mmw$ due to the opening
of the $W+\gamma$ threshold on the pole\cite{mz_thr}.
The charged current cross section near the $W$-pole may be expressed
more conveniently in terms of $\gwbar^2(0)$, or $G_F \mmw$.}.
 
In section~4, the charge form factors (\ref{ncbarcharges}) and
(\ref{ccbarcharge}) are determined from the three sectors of the
electroweak precision experiments under the assumption that there
are no new physics contributions to the vertex and box corrections,
except for allowing the $\zbb$ vertex to take on arbitrary values.
 

\vspace*{1cm}
\section*{\large \bf 3  Predictions of electroweak observables}
\renewcommand{\theequation}{3.\arabic{equation}}
\cleqn
 
\pr
In this section, all electroweak observables are expressed in terms of
the helicity amplitudes of eqs.~(\ref{m_nc}) and (\ref{m_cc}), together
with the external QED and QCD correction factors. The predictions are
restricted to the models respecting ${\rm SU(2)_L \times U(1)_Y }$
gauge symmetry with spontaneous breakdown to ${\rm U(1)_{EM} }$
and presented as functions of the charge form factors $\sbar^2(0)$,
$\gzbar^2(0)$, $\gwbar^2(0)$, $\sbar^2(\mmz)$, $\gzbar^2(\mmz)$, and
the vertex form factor $\delb(\mmz)$. It is assumed that the remaining
vertex and box correction are dominated by the SM contributions.
 
\vspace*{5mm}
\subsection*{\normalsize \bf 3.1  $Z$ boson parameters}
 
\pr
The following observables on the $Z$-pole ($s=\mmz$) are used in the
fit~:
\bea
    \Gamma_Z,\; \sigma_h^0,\; R_\ell,\;  A_{\rm FB}^{0,\ell},\;
     P_\tau,\; A_{\rm LR},\;A_{\rm FB}^{0,b},\;
        A_{\rm FB}^{0,c},\;R_b \,.
\label{zobservables}
\eea
Since the $Z$ mass $\mz$ is measured very accurately, the value
$\mz=91.187$GeV is treated as a constant in the fits.
The contributions from the SM box corrections are very small on the
$Z$-pole (see Table~\ref{tableofboxformfactors}), thus the
$\cos\theta$-dependence of the box correction factors is neglected.
 
The total cross section for the process $e^+e^- \to f\overline{f}$ is
given by
\bea
     \sigma_f
     &\equiv& \sigma(e^+e^- \to f\ol{f})
     \nonumber \\
     &=& \frac{s}{48 \pi}
         \Biggl\{\;\,
         \left( \,\squared{\MefLL +\MefLR} +\squared{\MefRL +\MefRR}\,
         \right) \frac{C_{f V}}{2}
     \nonumber \\
     && \qquad
        +\left( \,\squared{\MefLL -\MefLR} +\squared{\MefRL -\MefRR}\,
         \right) \frac{C_{f A}}{2} \;
         \Biggr\}
     \biggl(1 +\frac{3}{4}\,Q_f^2\,\frac{\bar{\alpha}(s)}{\pi} \biggr) \,,
\label{sig_f}
\eea
for unpolarized beams, where the last term proportional to
$\bar{\alpha}(s)/\pi$ accounts for the final state QED correction.
Here and in the following
\bea
 M_{\alpha \beta}^{ef} \equiv M_{e_\alpha f_\beta}^{NC}
 ( s=(p_{e^-} +p_{e^+})^2, t=(p_{e^-} -p_f)^2 )
\label{m_ef}
\eea
denote the $NC$ amplitudes of eq.~(\ref{m_nc}). The factors $C_{qV}$ and
$C_{qA}$ for quarks contain the final state QCD corrections for the
vector \cite{qcd_4l} and axial vector current\cite{qcd_zgg_1,qcd_zgg_2}
contributions, respectively, together with the finite mass corrections
of the final state fermions\cite{qcd_mq}:~
\bsub
\label{cqv_cqa}
\bea
      C_{qV} &=& 3\; \biggl\{\,
        \frac{ \hat{\beta}_q \,(3 -\hat{\beta}_q^2 )}{2}
              +a +1.409\, a^2  -12.767\, a^3
      \nonumber \\
      && \qquad
        +12\, \frac{\hat{m}_q^2(\sqrt{s})}{s}
           \Bigl( a +8.736\, a^2 +45.146\, a^3 \,\Bigr)
      \biggl\} , \qquad
\label{cqv} \\
      C_{qA} &=& 3\; \biggl\{\, \hat{\beta}_q^3
           +a +1.409\, a^2  -12.767\, a^3
           -6\, \frac{\hat{m}_q^2(\sqrt{s})}{s}
               \Bigl( \frac{11}{3}a +14.286\, a^2 \Bigr)
      \nonumber \\
     && \qquad
      \mp a^2\,\biggl[\,f(m_t)+6\,\frac{\hat{m}_q^2(\sqrt{s})}{s}
                             \Bigl( 3+\ln \frac{m_t^2}{\mmz} \Bigr)
               \biggr]
      \biggr\} , \qquad
\label{cqa}
\eea
\esub
with
\bea
      a       &\equiv& a^{(5)}(\sqrt{s})\,
               \equiv \frac{\alpha_s(\sqrt{s})_{\msbar}}{\pi}\,,
\label{alphas_pi}\\
    \hat{\beta}_q &\equiv& \sqrt{1-\frac{4 \hat{m}_q^2(\sqrt{s})}{s}} \,,
\label{beta_q}
\eea
where $\hat{m}_q(\sqrt{s})$ denotes the $\msbar$ running quark mass
at $\mu=\sqrt{s}$. The masses of the three lightest quarks ($u$, $d$,
$s$) are neglected, while the bottom and charm quark running masses,
$\hat{m}_b(\sqrt{s})$ and $\hat{m}_c(\sqrt{s})$, are obtained from
$\hat{m}_b(m_b)$ and $\hat{m}_c(m_c)$ by the two-loop renormalization
group equations:
\bsub
\label{running_mq}
\bea
      \frac{\hat{m}_b(\sqrt{s})}{\hat{m}_b(m_b)}
     =\frac{\hat{m}_c(\sqrt{s})}{\hat{m}_c(m_b)}
      &=&  \biggl[\frac{a^{(5)}_{}(\sqrt{s})}{a^{(5)}_{}(m_b)}\biggl]
           ^{\frac{\gamma_0^{(5)}}{b_0^{(5)}}}
           \biggl[ \frac{b_0^{(5)} +b_1^{(5)}a^{(5)}_{}(\sqrt{s})}
                        {b_0^{(5)} +b_1^{(5)}a^{(5)}_{}(m_b)} \biggr]
           ^{\left( \frac{\gamma_1^{(5)}}{b_1^{(5)}}
                   -\frac{\gamma_0^{(5)}}{b_0^{(5)}} \right)}\,,
\label{running_mb} \\
      \frac{\hat{m}_c(m_b)}{\hat{m}_c(m_c)}
      &=&  \biggl[\frac{a^{(4)}_{}(m_b)}{a^{(4)}_{}(m_c)}\biggl]
           ^{\frac{\gamma_0^{(4)}}{b_0^{(4)}}}
           \biggl[ \frac{b_0^{(4)} +b_1^{(4)}a^{(4)}_{}(m_b)}
                        {b_0^{(4)} +b_1^{(4)}a^{(4)}_{}(m_c)} \biggr]
           ^{\left( \frac{\gamma_1^{(4)}}{b_1^{(4)}}
                   -\frac{\gamma_0^{(4)}}{b_0^{(4)}} \right)}\,,
\label{running_mc}
\eea
\esub
where\cite{beta_qcd,gammam_qcd}
\bsub
\label{rgfunction}
\bea
     & &  b_0^{(n_f)} = \frac{33  - 2\,n_f}{6}\,, \quad
          b_1^{(n_f)} = \frac{153 -19\,n_f}{12}\,,
\label{b0b1} \\
     & &  \gamma_0^{(n_f)} = 2\,, \quad
          \gamma_1^{(n_f)} = \frac{303 -10\,n_f}{36}\,,
\label{gamma0gamma1}
\eea
\esub
are the coefficients of the $\beta$-function and the anomalous mass
dimension in the effective $n_f$-flavor QCD. The running coupling
$a^{(4)}(\mu)$ of the effective $n_f=4$ theory is calculated from a
given $a^{(5)}_{}(\mz)\equiv \alpha_s(\mz)_{\msbar}/\pi$ by solving the
three-loop QCD renormalization group equation with the two-loop
matching condition\cite{qcd_matching}:
\bea
      a^{(4)}_{}(m_b)
     =a^{(5)}_{}(m_b)
      +\frac{5}{36} {\Bigl[a^{(5)}_{}(m_b)\Bigr]}^3\,,
\label{matching}
\eea
at $\mu=m_b$. The relation between the $\msbar$ quark mass
$\hat{m}_q(m_q)$ and the physical mass $m_q$ is given in
ref.~\cite{mq_pole} as
\bea
     \hat{m}_q(m_q) &=& m_q \,
      \biggl[1+\frac{4}{3}a^{(n_f)}_{}(m_q)
            +K_q [a^{(n_f)}_{}(m_q)]^2\biggr]^{-1}\,,
\label{mq_pole}
\eea
with $K_b \approx 12.4$ ($n_f=5$) and  $K_c \approx 13.3$ ($n_f=4$),
for bottom and charm quarks. The following table summarizes the
running quark masses, $\hat{m}_b(\mu)$ and $\hat{m}_c(\mu)$,
for $\alpha_s(\mz)_{\msbar}=0.11$, 0.12, 0.13, $m_b=4.7\pm 0.2 \gev$,
$m_c=1.4\pm 0.2\gev$ and $\mz = 91.187\gev$ (the difference $m_b-m_c$
is fixed to $3.3\gev$\cite{qcdpot} in evaluating $\hat{m}_c(\mu)$):
\bea
\begin{array}{|l|ccc|}
  \hline
  \alpha_s(\mz)_{\msbar} & 0.11        & 0.12         & 0.13  \\
  \hline
  m_c\mbox{ (GeV)} & 1.40\pm 0.20 & 1.40\pm 0.20 & 1.40\pm 0.20 \\
  \hline
  \hat{m}_c(m_c)     & 1.13\pm 0.18 & 1.03\pm 0.19 & 0.86\pm 0.20 \\
  \hat{m}_c(m_b)     & 0.90\pm 0.17 & 0.76\pm 0.17 & 0.56\pm 0.17 \\
  \hat{m}_c(\mz)     & 0.65\pm 0.13 & 0.53\pm 0.12 & 0.37\pm 0.12 \\
  \hline
  m_b\mbox{ (GeV)} & 4.70\pm 0.20 & 4.70\pm 0.20 & 4.70\pm 0.20 \\
  \hline
  \hat{m}_b(m_b)     & 4.17\pm 0.19 & 4.06\pm 0.18 & 3.92\pm 0.18 \\
  \hat{m}_b(\mz)     & 3.05\pm 0.16 & 2.83\pm 0.15 & 2.59\pm 0.15 \\
  \hline
\end{array}
\label{tab:mq_hat}
\eea
The function $f(m_t)$ in the $O(\alpha_s^2)$ axial part of 
eq.~(\ref{cqa}) is given by\cite{rho_pqcd_mh2,qcd_zgg_2}~:
\bea
    f(m_t) &=& 2 \,\ln\frac{\mz}{m_t} -\frac{37}{12}
         +\frac{28}{81} \left( \frac{\mz}{2\,m_t} \right)^2
           -0.5767
           \left( \frac{\mz}{2\,m_t} \right)^4
           +0.7873
           \left( \frac{\mz}{2\,m_t} \right)^6 ,
\label{f_mt}
\eea
The minus sign should be taken in front of $f(m_t)$ in eq.~(\ref{cqa})
for $u$, $c$ quarks, and the plus sign for $d$, $s$, $b$ quarks.
These formulae are sufficient to calculate the factors $C_{qV}$ and
$C_{qA}$ as functions of $\alpha_s(\mz)$, $m_b$ and $m_c$. For charged
leptons, the corresponding factors are
\bsub
\label{clv_cla}
\bea
      C_{\ell V} &=& \frac{\beta_\ell (3 -\beta_\ell^2)}{2} \,,
\label{clv} \\
      C_{\ell A} &=& \beta_\ell^3 \,.
\label{cla}
\eea
\esub
with
\bea
      \beta_\ell = \sqrt{1 -\frac{4 m_\ell^2 }{s}} \,.
\label{beta_l}
\eea
The effect of the charged lepton masses is negligible except for the
$\tau$ lepton.
 
Near the $Z$-pole, $s \sim \mmz$, the cross sections are sensitive to
the total $Z$ width, $\Gamma_Z$, and hence it should be evaluated at 
two-loop level\cite{gammaz,hollik90,hollik93}. The $Z$ width is
calculated in a similar way as the total cross section case
(\ref{sig_f}):
\bea
 \Gamma_Z &=& \sum_{f} \Gamma_f \,,
\label{gamma_z}\\
 \Gamma_f &=& \frac{\mz}{24\pi} \biggl\{
 \Big| M^f_L +M^f_R \Big|^2 \frac{C_{fV}}{2}+
 \Big| M^f_L -M^f_R \Big|^2 \frac{C_{fA}}{2} \biggr\}
 \biggl(1 +\frac{3}{4}\,Q_f^2\,\frac{\bar{\alpha}(s)}{\pi} \biggr) \,,
\label{gammaz_f}
\eea
by using the $Z \to f_\alpha \ol{f_\alpha}$ decay amplitudes
\bea
 T(Z \to f_\alpha \ol{f_\alpha}) &=&
 M^f_\alpha\; \epsilon_Z \cdot J_{f_\alpha} \,.
\label{t_zff}
\eea
Here $\epsilon_Z^\mu$ is the normalized $Z$ wave function,
$J_{f_\alpha}^\mu$ are the currents of eq.~(\ref{t_ij}),
and the scalar amplitudes $M^f_\alpha$ can be expressed by
\bea
 M^f_\alpha &=& (I_{3f_\alpha}-Q_f\shat^2)
 \Bigl[\,\gzbar^{}(\mmz) +\gzhat^{} {\rm Re} \Gamma_1^{f_\alpha}(\mmz)\Bigr]
\nonumber\\&&
 +\gzhat^{} {\rm Re} \Bigl[\,I_{3f_\alpha}(\chat^2\gambar_2^{f_\alpha}(\mmz)
 +\Gamma_3^{f_\alpha}(\mmz)) -Q_f(\sbar^2(\mmz)-\shat^2)\Bigr]\,.
\label{m_zff}
\eea
It is straightforward to evaluate the partial and total widths from the
above formulae, once the three form factors $\gzbar^2(\mmz)$,
$\sbar^2(\mmz)$, $\delb(\mmz)$ and $\alpha_s(\mz)$ are given.
Fig.~\ref{figureofgammaz} shows the predicted $\Gamma_Z$(GeV) in the
plane of $\sbar^2(\mmz)$ and $\gzbar^2(\mmz)$ for $\alpha_s(\mz)=0.11$,
0.12, 0.13 and $\delb(\mmz)$=0~(a), $-0.01$~(b) and $-0.02$~(c). 
In the SM,
$\delb(\mmz) \simlt -0.003$ holds for all $m_t$ (see
Fig.~\ref{figureofstusm}), $\delb(\mmz)=-0.01\,(-0.02)$ for
$m_t\approx$ 175 (270)~GeV.
                           \footnote{\normalsize \baselineskip 18pt
The $m_t$-dependences of the electroweak $Z$ boson observables are
not completely absorbed into the three form factors, $\gzbar^2(\mmz)$,
$\sbar^2(\mmz)$ and $\delb(\mmz)$.  Mild $m_t$-dependences remain
in the two-loop QCD correction factor $f(m_t)$ of eq.~(\protect\ref{f_mt})
and in the $\zbb$ vertex function $\Gamma_1^{b_L}(\mmz)$
(Table~\ref{tableofvertexformfactors}). When $\delb(\mmz)$ is allowed
to vary in the fit, these residual $m_t$-dependent terms are determined
by using the SM $m_t$-dependence of the $\delb(\mmz)$ form factor
(see Fig.~\ref{figureofstusm}), which can be inverted approximately as
$m_t({\rm GeV}) = 21.77 \sqrt{-10^4\delb(\mmz)-9.9} - 31.2$ valid in
the region 75~GeV$<\!\!m_t\!\!<$400~GeV.  We set $m_t=$75~GeV for
$\delb(\mmz)>-0.0036$ and $m_t=$400~GeV for $\delb(\mmz)<-0.0405$.
With this prescription the parameter $\delb(\mmz)$
covers the full $m_t$-dependences of the vertex corrections
within the SM, while it still allows $\delb(\mmz)$ to measure large new
physics contributions to the $\zbb$ vertex because of the relatively
mild $m_t$-dependences of the $f(m_t)$ and $\Gamma_1^{b_L}(\mmz)$
factors.}
It is clearly seen from the figure that $\Gamma_Z$ increases with
growing $\alpha_s$ and $\delb$, and that it remains roughly constant
when $\alpha_s$ increases by 0.01 and, simultaneously, $\delb$
decreases by about 0.006. The net effect is a strong anti-correlation
between the fitted $\alpha_s$ and $\delb$ values (see section~4.1).
 
In the SM, all the form factors are calculable in terms of $m_t$ and
$\mh$. In table~\ref{tableofzwidths}  the SM predictions are shown
for the partial and the total $Z$ widths for several $m_t$ and $\mh$
values, for $\alpha_s(\mz)=0.12$, $\delta_\alpha=0$, $\delg=0.0055$ and
($m_b$, $m_c$)~=~(4.7, 1.4)~GeV. The numerical values turn out to be
larger by about $1/5000$ than those quoted in ref.~\cite{hollik93}.
Uncertainties in our predictions are estimated as follows~:
(i)~Change of $m_b$ and $m_c$ by 0.2~GeV affects $\Gamma_b$ by
less than 0.2~MeV ($\simlt 1/2000$ of $\Gamma_b$)
and $\Gamma_c$ by about 0.03~MeV;
(ii)~Setting $\gzhat^{}=\gzbar^{}(\mmz)$ 
and $\shat^2=\sbar^2(\mmz)$ in the
amplitudes (\ref{m_zff}) affects the total width by about 0.2~MeV
for the $m_t$ and $\mh$ values of Table~\ref{tableofzwidths};
(iii)~If the imaginary parts in the amplitudes (\ref{m_zff})
are also included, the total width increases by about 0.01~MeV;
(iv)~QCD higher order effects may affect the hadronic widths at the
level of $\alpha_s^4 \sim 1/4000$;
(v)~The present uncertainty in $1/\bar{\alpha}(\mmz)$,
$\delta_\alpha=\pm 0.1$ affects $\sbar^2(\mmz)$ by $\mp 0.00026$
(\ref{sbar_approx}), and hence the $Z$ width by about $\pm 0.65$~MeV
($\sim 1/3000$ of $\Gamma_Z$).
These uncertainties are still an order of magnitude smaller than the
actual experimental error of $\Delta(\Gamma_Z)=7$~MeV\cite{lep93}
($\Delta(\Gamma_Z)/\Gamma_Z \sim 0.003$).
 
Note that we adopt the perturbative order $\alpha\alpha_s$
\cite{rho_pqcd,pivv_pqcd,rho_pqcd_mh2,zbb_qcd} corrections at
$\alpha_s\!=\!\alpha_s(\mz)_{\msbar}$ in calculating all the SM
predictions, since it allows the reader to reproduce our results
straightforwardly. The effects of non-perturbative threshold
corrections\cite{rho_thr,rho_thr2,rho_thr3} may be accounted for
by adjusting the effective top-quark mass to produce the same
$T$ parameter value. 
 
Once the $Z$ width, $\Gamma_Z$, is determined the formula (\ref{sig_f})
gives the total cross section for the process $e^+ e^- \to f \bar{f}$
at all energies, up to the $\cos\theta$-dependence of the box form
factors which can be safely neglected near the $Z$-pole.
At LEP, the on-pole cross sections $\sigma_f^0$ are obtained after
subtracting the $\gamma$-exchange contribution to the amplitudes.
Because of this subtraction, we cannot simply compare $\sigma_f(\mmz)$
of eq.~(\ref{sig_f}) with the corresponding published measurement.
In fact, the subtraction procedure is {\em not} completely
model-independent and the following two cases are examined~:
(i)~In the amplitudes (\ref{m_nc}) only those terms multiplying the $Z$
    propagator factor are retained;
(ii)~From the full amplitude (\ref{m_nc}) the $\gamma$-exchange
     amplitude
     $Q_iQ_j[\ebar^2(\mmz)-i\ehat^2\Delta_{\gamma\gamma}(\mmz)]/s$
     is subtracted.
The above two prescriptions differ by contributions from the $\gamma$
vertex corrections and the box corrections, but the numerical
predictions for $\sigma_h^0$ are found to differ by at most 0.0003~nb
and are thus 
negligibly small compared to the actual experimental error of
$\Delta(\sigma_h^0)=\,$0.14~nb. The pole amplitudes (i), the
term with the $Z$ propagator factor in eq.~(\ref{m_nc}), 
are used below when
confronting the theoretical predictions with the LEP/SLC experiments.
 
It must be pointed out here that the quantities quoted as $\sigma_f^0$ by
the LEP electroweak working group\cite{lep93} are {\em not} the peak
cross sections as obtained above, but that they are rather defined by
the following identities%
\footnote{\normalsize \baselineskip 18pt
We thank T.~Mori for pointing out our misunderstanding of $\sigma_h^0$
affecting the earlier version of the present work. The notation of the
LEP electroweak working group is misleading, since ref.\cite{lep93}
does not explicitly state that their $\sigma_h^0$ value is {\rm not}
the peak cross section. In order to avoid any ambiguity it would be
better to call this quantity
$(12\,\pi/\mmz)\Gamma_e\Gamma_h/\Gamma_Z^2$ and explain precisely
from which experimental quantities it is calculated. It is also
desirable to publish the total hadronic 
cross sections at $\sqrt{s}=\mz$ without subtracting the
$\gamma$-exchange contributions, since the full total cross sections
can be calculated unambiguously.
}:
\bea
 \sigma_f^0({\rm LEP}) \equiv
 \frac{12\pi}{\mmz}\frac{\Gamma_e \Gamma_f}{\Gamma_Z^2} \,.
\label{sig_f0_lep}
\eea
This quantity does not agree with the pole cross section $\sigma_f^0$
as calculated above,
but agrees rather accurately with the modified expression:
\bea
 \sigma_f^0({\rm LEP}) \approx \sigma_f^0 \,\cdot\,
 \biggl(1 +\frac{3}{4}\,\frac{\bar{\alpha}(s)}{\pi} \biggr) \,.
\label{sig_f0_approx}
\eea
For example, the SM predictions for $m_t=$175~GeV, $\mh=$100~GeV,
$\alpha_s(\mz)=0.12$ and $\delta_\alpha=0$ give~:
\bea
  \begin{array}{|c|c|c|c|}
     \hline
     f & \sigma_f^0 
     & \displaystyle{ \vphantom{\Bigg(}
         \sigma_f^0 \cdot
         \biggl(1+\frac{3}{4}\frac{\bar{\alpha}(\mmz)}{\pi}\biggr)}
     & \displaystyle{\vphantom{\Bigg(}\sigma_f^0\,({\rm LEP})}
    \\
     \hline
    \ell = e,\mu&\hphantom{4}1.995\;{\rm nb}&
                 \hphantom{4}1.998\;{\rm nb}&
                 \hphantom{4}1.997\;{\rm nb}\\
     \hline
     h          &           41.399\;{\rm nb}&
                            41.476\;{\rm nb}&
                            41.463\;{\rm nb}\\
     \hline
     b          &\hphantom{4}8.928\;{\rm nb}&
                 \hphantom{4}8.945\;{\rm nb}&
                 \hphantom{4}8.942\;{\rm nb}\\
     \hline
  \end{array} \,.
\label{tab:sig_f0}
\eea
The right-hand side of (\ref{sig_f0_approx}) reproduces the LEP
definition (\ref{sig_f0_lep}) with an accuracy of $1/3000$, while the
peak cross sections $\sigma_f^0$ as obtained from eq.~(\ref{sig_f}) 
with the $Z$-pole part of the amplitudes 
are
off by about $1/1000$ to $1/600$. The former uncertainty of about
$1/3000$ is typically on the order of the higher order corrections,
while the latter difference, especially the difference between
$\sigma_h^0$ and $\sigma_h^0({\rm LEP})$, shows up clearly in the fit as
a significant shift in the fitted $\delb(\mmz)$ and $\alpha_s$ values.
 
Figs.~\ref{figureofsigmah}, \ref{figureofrl}, \ref{figureofrb} show
\bea
 &&\sigma_h^0({\rm LEP}) = \sum_{f=u,d,s,c,b} \sigma_f^0({\rm LEP}) \,,
\label{sig_h0}\\
 &&R_\ell = \sigma_h^0({\rm LEP})\,/\,\sigma_\ell^0({\rm LEP})
 = \Gamma_h \,/\, \Gamma_\ell \,,
\label{rl}\\
 &&R_b    = \sigma_b^0({\rm LEP})\,/\,\sigma_h^0({\rm LEP})
 = \Gamma_b \,/\, \Gamma_h \,,
\label{rb}
\eea
respectively, in the plane of $\sbar^2(\mmz)$ and $\delb(\mmz)$
for $\alpha_s(\mz)=0.11$, 0.12 and 0.13. All the three quantities turn
out to be almost completely independent of $\gzbar^2(\mmz)$, as the
predictions at $\gzbar^2(\mmz)=0.55$  (solid lines) and
$\gzbar^2(\mmz)=0.57$ (dashed lines) are almost degenerate.
Fig.~\ref{figureofsigmah} shows that $\sigma_h^0$ is sensitive to both
$\alpha_s$ and $\delb$, but an increase of $\alpha_s(\mz)$ by 0.01 can
be compensated by a simultaneous decrease of $\delb$ by about 0.006,
just as for $\Gamma_Z$. Fig.~\ref{figureofrl} shows that the ratio
$R_\ell$ is only sensitive to a linear combination of $\sbar^2(\mmz)$
and $\delb(\mmz)$. At fixed $\sbar^2(\mmz)$, the correlated change of
$\alpha_s$ and $\delb$ leaving $\Gamma_Z$ and $\sigma_h^0$ unchanged,
keeps also the $R_\ell$ value roughly unchanged. The reason for this
behaviour is in the fact that the $\alpha_s$-dependences of the three
$Z$-resonance observables, $\Gamma_Z$, $\sigma_h^0$ and $R_l$, are
solely contained in just the quantity $\Gamma_h$ which depends on
$\alpha_s$ and
$\delta_b$ approximately in the combination
\bea
  \delb(\mmz) +0.6\,\alpha_s(\mz) \,.
\label{delb_alps}
\eea
Hence, in order to get $\alpha_s$ independently of $\delb$, the
measurement sensitive to another combination is required. For instance,
Fig.~\ref{figureofrb} shows that the ratio $R_b$ does measure
$\delb(\mmz)$ rather independently of $\alpha_s$ and $\sbar^2(\mmz)$.
An accurate measurement of $R_b$ offers the key to disentangle
$\alpha_s$ and $\delb$ (see also section~4.1).
 
The asymmetries on the $Z$-pole provide the measurement of the universal
parameter $\sbar^2(\mmz)$ almost independently of $\gzbar^2(\mmz)$ and
$\delb(\mmz)$ and with little or no dependence on the QCD coupling
$\alpha_s$.
 
The forward-backward (FB) asymmetry is given by
\bea
     A_{\rm FB}^{0,\ell} &=& \frac{3}{4}\,
     \frac{ \squared{\MelLL}+\squared{\MelRR}
           -\squared{\MelLR}-\squared{\MelRL}
          }{
            \squared{\MelLL}+\squared{\MelRR}
           +\squared{\MelLR}+\squared{\MelRL}
          } \,,
\label{afb_l}
\eea
for leptons, and
\bea
     A_{\rm FB}^{0,q} &=& \frac{3}{4}\,
     \frac{ 2 \beta_q
           \left\{     \squared{\MeqLL} \!+\!\squared{\MeqRR}
                  \!-\!\squared{\MeqLR} \!-\!\squared{\MeqRL} \right\}
          }{
           \displaystyle{\frac{3 -\beta_q^2}{2}}
           \left\{   \squared{\MeqLL \!+\!\MeqLR} \!\!
                  +\!\squared{\MeqRL \!+\!\MeqRR} \right\}
           +\beta_q^2
           \left\{   \squared{\MeqLL \!-\!\MeqLR} \!\!
                  +\!\squared{\MeqRL \!-\!\MeqRR} \right\}
          },
\nonumber\\
\label{afb_q}
\eea
for quarks ($q=b,c$). Here, the physical heavy quark masses $m_q$ are
used in the factor $\beta_q  = \sqrt{1 -4 m_q^2/\mmz}$. The QCD
corrections for the FB asymmetries\cite{qcd_afb} have not been included
in eq.~(\ref{afb_q}). The reported asymmetries from  LEP $A_{\rm
FB}^{0,b}(\rm{LEP})$ and $A_{\rm FB}^{0,c}({\rm LEP})$ have been
corrected for these effects assuming a linear $\alpha_s$-dependence
and $\alpha_s=0.12$. Therefore, we estimate the LEP asymmetries for a
given value of $\alpha_s(\mz)$ by using the following simple formula~:
\bea
      A_{\rm FB}^{0,q}({\rm LEP}) = A_{\rm FB}^{0,q}
      \frac{1 +k_A \displaystyle{\alphasoverpi}}
           {1 +k_A \displaystyle{\frac{0.12}{\pi}} }
\label{afb_q0}
\eea
with $k_A = 0.75$\cite{lep93}. The uncertainty in the coefficient
$\Delta k_A =\pm 0.25$ affects the above $\alpha_s$ dependence by less
than $1/1000$ in the range $0.11<\alpha_s(\mz)<0.13$. The QCD
correction depends on details of the final charm and bottom quark
tagging procedure, and hence it is desirable to have the
$\alpha_s$-dependence of the corrected asymmetry value from each
experiment.
 
The $\tau$ polarization asymmetry is defined by the ratio of the left-
and right-handed $\tau$ pair cross sections~:
\bea
     P_\tau &=& \frac{\sigma_{\tau_R} -\sigma_{\tau_L}}
                     {\sigma_{\tau_R} +\sigma_{\tau_L}}\,.
\label{p_tau_def}
\eea
By neglecting the $\tau$ mass one finds
\bea
     P_{\tau} &=&
     \frac{  \squared{\MelLR} +\squared{\MelRR}
            -\squared{\MelLL} -\squared{\MelRL}
          }{
             \squared{\MelLR} +\squared{\MelRR}
            +\squared{\MelLL} +\squared{\MelRL}
          } \,.
\label{p_tau}
\eea
 
Likewise, the left-right beam polarization asymmetry is defined by
\bea
     A_{\rm LR} &=&
     \frac{ \displaystyle{\sum_f}
              \left( \sigma^L_f -\sigma^R_f \right)
          }{\displaystyle{\sum_f}
              \left( \sigma^L_f +\sigma^R_f \right)
          } \,.
\label{a_lr}
\eea
where the cross sections for completely polarized beam are
expressed in terms of the helicity amplitudes by
\bsub
\label{sig_lr_f}
\bea
     \sigma^L_f
     &\equiv& \sigma(e^-_L e^+_R \to f\ol{f})
     \nonumber \\
     &=& \frac{s}{24 \pi}
         \Biggl\{\;\,\squared{\MefLL +\MefLR} \frac{C_{f V}}{2}
                 + \,\squared{\MefLL -\MefLR} \frac{C_{f A}}{2} \;
         \Biggr\}
     \biggl(1+\frac{3}{4}\,Q_f^2\,\frac{\bar{\alpha}(s)}{\pi}\biggr) \,,
\label{sig_l_f} \\
     \sigma^R_f
     &\equiv& \sigma(e^-_R e^+_L \to f\ol{f})
     \nonumber \\
     &=& \frac{s}{24 \pi}
         \Biggl\{\;\,\squared{\MefRL +\MefRR} \frac{C_{f V}}{2}
                 + \,\squared{\MefRL -\MefRR} \frac{C_{f A}}{2} \;
         \Biggr\}
     \biggl(1+\frac{3}{4}\,Q_f^2\,\frac{\bar{\alpha}(s)}{\pi}\biggr) \,.
\label{sig_r_f}
\eea
\esub
The cross section for the electron beam polarization $P_e$ is then
\bea
 \sigma_f(P_e) = \frac{1-P_e}{2}\,\sigma_f^L +\frac{1+P_e}{2}\,\sigma_f^R.
\label{sig_f_p}
\eea
We comment here that the factorization identities
\bea
 A_{\rm LR} &=& -P_\tau \,,
\label{alr_ptau}\\
 A_{\rm FB}^{0,\ell} &=& \frac{3}{4}(P_\tau)^2 \,,
\label{afb_ptau}
\eea
do not hold exactly even in our $Z$-pole approximation to the amplitudes
(\ref{m_nc}), since they do not factorize into $Z$ production and
$Z$ decay amplitudes at $s=\mmz$.  We find for instance for the SM predictions
at $m_t=175$~GeV, $\mh=100$~GeV, $\alpha_s(\mz)=0.12$ and
$\delta_\alpha=0$~:
\bea
  \begin{array}{|c|c|}
    \hline
    \bar{s}^2(\mmz)      & 0.23040 \\ \hline
    A_{\rm LR}            & 0.14801 \\
    -P_\tau               & 0.14802 \\ \hline
    A_{\rm FB}^{0,\ell}   & 0.01667 \\
    \frac{3}{4}(P_\tau)^2 & 0.01643 \\ \hline
  \end{array} 
\label{tab:alr_afb_ptau}
\eea
The identity (\ref{alr_ptau}) holds rather accurately, but
the identity (\ref{afb_ptau}) is violated by a factor of 1.4\%.
This is mainly because of the subtle cancellation among the squared
amplitudes of eq.~(\ref{afb_l}) rendering the asymmetry $A_{\rm FB}^\ell$
sensitive to our detailed treatment of the order $\alpha^2$ effects
such as the treatment of the imaginary part and the choice of
the couplings $\gzhat^2$ and $\shat^2$.
 
In Fig.~\ref{figureofasymmetry}, all asymmetry parameters on the 
$Z$-pole 
are plotted as functions of $\sbar^2(\mmz)$. For each asymmetry,
the contributions from both the $\gamma$-pole and $Z$-pole terms
are examined using the following helicity amplitudes~:
(i)
The full helicity amplitudes (\ref{m_nc}) including the $\gamma$
and $Z$ exchange as well as the box contributions.
(ii)
The helicity amplitudes obtained from the full amplitudes (\ref{m_nc})
by subtracting the real and imaginary parts of the $\gamma$ exchange
contribution $Q_iQ_j[\ebar^2(\mmz)-i\ehat^2\Delta_{\gamma\gamma}(\mmz)]/s$.
(iii)
The helicity amplitude retaining only the $Z$-pole term,
the term multiplying the $Z$ propagator factor in eq.~(\ref{m_nc}).
(iv)
The helicity amplitude in the improved Born approximation (IBA)
of the $Z$-exchange amplitudes~:
\bea
     \Bigl({\matrixelement ef{\alpha}{\beta}}\Bigr)_{\rm IBA}
     = \frac{\gzbar^2(\mmz)\,
             \Bigl[I_{3e_\alpha} -Q_e\,\sbar^2(\mmz)\Bigr]\,
             \Bigl[I_{3f_\beta } -Q_f\,\sbar^2(\mmz)\Bigr]}
            {s-\mmz +is\frac{\Gamma_Z}{\mz}}\,,
\label{m_iba}
\eea
on the $Z$-pole $s=\mmz$. In Fig.~\ref{figureofasymmetry}, the
predictions of
(i) are denoted by `Full',
(ii) by `$\mbox{Full}-\gamma$',
(iii) by `$Z$ only', and
(iv) by `IBA'.
The prescriptions (ii) and (iii) give almost identical predictions,
and we adopt (iii) in the fit.
It is worth noting that the subtraction of the $\gamma$-exchange
amplitudes affects the asymmetry $A_{\rm FB}^\ell$ significantly,
but not the other asymmetries. 
Note particularly that
the IBA gives
consistently larger asymmetries by as much as 10\% for
$A_{\rm FB}^\ell$, and by about 5\% for the rest. Hence, the
`process-dependent' effective $\sin^2\theta_W$ factor determined from
each asymmetry by making use of the IBA-like formula (\ref{m_iba})
differs significantly from the process-independent universal form
factor $\sbar^2(\mmz)$. We find approximately,
\bsub
\label{s2_eff}
\bea
 \sin^2\theta_W^{\rm eff}(A_{\rm FB}^{0,\ell}) &\approx&
 \sbar^2(\mmz)+0.0009\,,
\label{s2_afb_l}\\
 \sin^2\theta_W^{\rm eff}(A_{\rm LR}^0)&\approx&
 \sbar^2(\mmz)+0.0010\,,
\label{s2_alr}\\
 \sin^2\theta_W^{\rm eff}(A_{\rm FB}^{0,b}) &\approx&
 \sbar^2(\mmz)+0.0010\,,
\label{s2_afb_b}\\
 \sin^2\theta_W^{\rm eff}(A_{\rm FB}^{0,c}) &\approx&
 \sbar^2(\mmz)+0.0009\,.
\label{s2_afb_c}
\eea
\esub
A related study is found in ref.\cite{martinez}.
 
In the SM, all the form factors $\gzbar^2(\mmz)$, $\sbar^2(\mmz)$ and
$\delb(\mmz)$ are calculable as functions of $m_t$ and $\mh$ (see
appendix~C for details). The main uncertainty in these calculations
appears in the parameter $\delta_\alpha$ (\ref{delta_alpha}) which
parametrizes the uncertainty in the hadronic vacuum polarization
contribution to $1/\bar{\alpha}(\mmz)$. Hence, all $Z$ parameters can
be predicted accurately in the SM as functions of four parameters:
$m_t$, $\mh$, $\alpha_s$ and $\delta_\alpha$.
 
Figs.~\ref{figureofzparametersvsmt} shows the $m_t$-dependence of all
$Z$ parameters for three $\mh$ values 60~GeV (dashed lines), 300~GeV
(solid lines) and 1000~GeV (dash-dotted lines), at $\alpha_s(\mz)=0.11$,
0.12, 0.13 and $\delta_\alpha=0$ ($1/\bar{\alpha}(\mmz)=128.72$).
Shown by horizontal lines are the experimental data from LEP\cite{lep93}
and SLC\cite{alr94} (see sections~4 and 6).
The $m_t$-dependence is sizeable for all the observables.
In $R_b$ and $\sigma_h^0$, the $m_t$-dependence comes mainly from
the $\zbb$ form factor $\delb(\mmz)$, and hence these parameters have little
sensitivity to $\mh$ (see Fig.~\ref{figureofstusm}). The
$m_t$-dependences of all asymmetry parameters including $P_\tau$ come
from the form factor $\sbar^2(\mmz)$. $R_\ell$ receives $m_t$-dependences
from both $\delb(\mmz)$ and $\sbar^2(\mmz)$. Finally, the total $Z$
width is the only quantity sensitive to the form factor
$\gzbar^2(\mmz)$. In conclusion, the $m_t$-dependence of $\Gamma_Z$ is
a combined effect of all three form factors $\gzbar^2(\mmz)$,
$\sbar^2(\mmz)$ and $\delb(\mmz)$.
 
Likewise, Fig.~\ref{figureofzparametersvsalphas} shows the $\alpha_s$
dependences of the hadronic $Z$ parameters for the three $m_t$ values
100~GeV (dashed lines), 150~GeV (solid lines) and 200~GeV (dash-dotted
lines), all at $\mh=100$~GeV and $\delta_\alpha=0$. It can be seen that
the ratio $R_b$ and the asymmetries $A_{\rm FB}^{0,b}$ and
$A_{\rm FB}^{0,c}$ are almost independent of $\alpha_s$. $\Gamma_Z$ and
$R_\ell$ grow linearly with $\alpha_s$ because of the final state QCD
correction factor (\ref{cqv_cqa}). $\sigma_h^0$ decreases with
increasing $\alpha_s$, since it is proportional to the factor
$\Gamma_h/\Gamma_Z^2$. The ratio $R_\ell$ exhibits the strongest
dependence to $\alpha_s$. As emphasized above, however, the
$\alpha_s$-dependences of all $Z$ observables are approximately
proportional to a common factor $\delb(\mmz)+0.6\alpha_s(\mz)$, and
hence either an accurate determination of $\delb(\mmz)$ (via $R_b$) or
else the assumption of SM dominance to the form factor
$\delb(\mmz)$ is crucial for the extraction of $\alpha_s(\mz)$ from
these experiments.
 
\vspace*{5mm}
\subsection*{\normalsize \bf 3.2  Low energy neutral current
                                  experiments}
\pr
The data of four types of low energy neutral current experiments
are analysed~: neutrino-nucleon scattering ($\nu_\mu$--$q$),
neutrino-electron scattering ($\nu_\mu$--$e$), atomic parity violation
(APV), and polarized electron-deuteron scattering ($e$--D).
Theoretical predictions are given for all model-independent parameters
\cite{nuq_dat,old_ncfit,beyer} characterizing the electroweak low
energy neutral current experiments.
They are the effective $\nu_\mu$--$q$ coupling factors\cite{nuq_dat}
\bea
 g_L^2\,,\pthree g_R^2\,,\pthree \delta_L^2\,,\pthree \delta_R^2,
\label{nuq_para}
\eea
for the $\nu_\mu$--$q$ scattering experiments,
the effective neutral current parameters\cite{beyer}
\bea
 \rho_{\nu e}\,,\pthree s^2_{\nu e}
\label{nue_para}
\eea
for the $\nu_\mu$--$e$ scattering experiments,
the weak charge of nuclei\cite{apv_dat}
\bea
 Q_W^{}(A,Z)\,,
\label{apv_para}
\eea
for parity violation in atoms, and the effective neutral current
couplings\cite{old_ncfit}
\bea
 2C_{1u}-C_{1d}\,,\pthree 2C_{2u}-C_{2d}\,,
\label{eq_para}
\eea
for the $e$--D polarization asymmetry. Definitions of these
model-independent parameters are given below and re-expressed in
terms of the helicity amplitudes of eq.~(\ref{m_nc}).
 
In this subsection terms of order $\alpha\cdot(q^2/\mmw)$ are neglected,
while keeping terms of order $q^2/\mmw$ and $\alpha\cdot(m_f^2/q^2)$.
The generic amplitude for the process $ij \to ij$ follows then
from eq.~(\ref{m_nc})~:
\bea
     M_{ij}^{NC} &=&
     \frac{1}{q^2}
       \biggl\{  (Q_i\,Q_j)\,[\ebar^2(q^2)
               +\ehat^2\,\Gamma_1^{i}(q^2) +\ehat^2\,\Gamma_1^{j}(q^2)]
     \nonumber \\
     &&  \hspace{21mm}
               + (Q_i\,{I_3}_j)\,\ehat^2\, \ol{\Gamma}_2^j(q^2)
               + (Q_j\,{I_3}_i)\,\ehat^2\, \ol{\Gamma}_2^i(q^2) \biggr\}
     \nonumber \\
     && +\frac{1}{q^2-\mmz}
       \biggl\{ ({I_3}_i -Q_i\shat^2)\,({I_3}_j -Q_j\shat^2)\gzbar^2(0)
     \nonumber \\
     &&  \hspace{21mm}  -({I_3}_i -Q_i\shat^2)\, Q_j\,
                        \gzhat^2\, [\sbar^2(q^2)-\shat^2]
     \nonumber \\
     &&  \hspace{21mm}  -({I_3}_j -Q_j\shat^2)\, Q_i\,
                        \gzhat^2\, [\sbar^2(q^2)-\shat^2]\,
       \biggr\}
     \nonumber \\
     &&  +B_{ij}^{NC}(0,0) +O\Bigl(\ehat^2\frac{q^2}{\mmw}\Bigr)\,.
\label{m_nc_le}
\eea
All electroweak observables of the low energy neutral current sector
are calculated by using the above approximation.
Contributions from the neglected terms are completely negligible.
The numerical predictions for all observables
(\ref{nuq_para})--(\ref{eq_para}) depend on just
the two universal charge form factors $\sbar^2(0)$ and $\gzbar^2(0)$,
since the running of the charge form factors
$1/\bar{\alpha}(q^2)-1/\alpha$ and
$\sbar^2(q^2)/\bar{\alpha}(q^2)-\sbar^2(0)/\alpha$
at low energies $|q^2|\ll\mmz$ are governed completely by known physics
only and are hence accurately calculable (see appendix B).
Although the expression (\ref{m_nc_le}) with the $\msbar$ coupling
normalization (\ref{hatcouplings}) is used in all numerical calculations
presented below, we often quote below a slightly more compact expression
that is obtained from eq.~(\ref{m_nc_le}) by dropping the terms
proportional to $[\sbar^2(q^2)-\shat^2]$ and replacing $\shat^2$ by
$\sbar^2(q^2)$ in the term multiplying the $Z$ propagator factor. This
is a valid approximation to eq.~(\ref{m_nc_le}) differing only by terms
of order $\gzhat^2[\sbar^2(q^2)-\shat^2]^2$.
 
\subsubsection*{\normalsize \bf 3.2.1
Neutral currents in $\nu_\mu-q$ scattering}
\pr
The neutral current data from the $\nu$--$q$ scattering experiments 
can be conveniently parametrized by the four model-independent
parameters\cite{nuq_dat}
\bsub
\label{g2_d2}
\bea
     g_\alpha^2 &\equiv& u_\alpha^2 + d_\alpha^2\,,
     \\
     \delta_\alpha^2 &\equiv& u_\alpha^2 - d_\alpha^2\,,
\eea
\esub
for $\alpha=L$ or $R$. The effective chiral couplings $q_\alpha\,
(=\uL,\dL,\uR,\dR)$ can be directly expressed in terms of the helicity
amplitudes of eq.~(\ref{m_nc_le}) by
\bea
     q_\alpha 
              = -\frac{\MnqLa}{2 \sqrt{2} G_{\rm c.c.}}
     \qquad (\,q=u,d\:;\:\alpha=L,R)\,,
\label{q_alpha}
\eea
with the notation
$M_{\alpha \beta}^{ij} \equiv M_{i_\alpha j_\beta}^{NC}$. The
amplitudes (\ref{m_nc_le}) can then be written in compact form~:
\bea
     \MnqLa &=&
       \frac{1}{2}\,Q_q\,\ehat^2\,\frac{\ol{\Gamma}_2^\nu(t)}{t}\,
      +\frac{1}{2}\,\Bigl[{I_3}_{q_\alpha} -Q_q\sbar^2(t) \Bigr]
         \frac{\gzbar^2(0)}{t-\mmz}
      +\BnqLa(0,0)\,,
\label{m_nuq}
\eea
and the charged current factor is approximated by
\bea
     G_{\rm c.c.} = G_F \,\frac{\;[1+\delta_{\rm c.c.}]^{1/2}\;}
     {1+ \displaystyle{\frac{\langle -t \rangle _{\rm c.c.}}{\mmw}}}
     \,.
\label{gcc_nuq}
\eea
The QED correction factor
$\delta_{\rm c.c.}$ is accounted for (following Sirlin and
Marciano\cite{nuq_cc}) by,
\bsub
\label{del_cc}
\bea
     \delta_{\rm c.c.} &=& \frac{\alpha}{\pi}\,
          \biggl[\,\ln \frac{\mmz}{2\,\langle -t\rangle_{\rm c.c.}}
                   +2 \,\biggr]
      \\
      &\approx& 0.017 \qquad
      \mbox{ for }\langle -t \rangle _{\rm c.c.}=20\GeV^2\,.
\label{del_cc_num}
\eea
\esub
Note that the leading logarithm approach of ref.\cite{4f_rc} gives
\bsub
\label{del_cc_ll}
\bea
     \delta_{\rm c.c.} &=& \frac{\alpha}{\pi} \,
         \ln \frac{\mmw}{\langle -t \rangle _{\rm c.c.}}
      \\
      &\approx& 0.013 \qquad
      \mbox{ for }\langle -t \rangle _{\rm c.c.}=20\gev^2\,,
\eea
\esub
for the above correction factor. In our numerical calculation we adopt
the factor (\ref{del_cc_num}). The $\nu_\mu$ charge radius factor
$\ol{\Gamma}_2^\nu(t)/t$ and the box form factors $\BnqLa(0,0)$ in the
amplitude (\ref{m_nuq}) can easily be read off from the generic
expressions in appendix~A~:
\bea
     \frac{\ol{\Gamma}_2^\nu(t)}{t}
        = \frac{1}{\mmw} \frac{\ghat^2}{16 \pi^2} \ol{J}_W(t;m_\mu) \,,
\label{nu_cr}
\eea
with
\bsub
\label{jw_mu}
\bea
     \ol{J}_W(t;m_\mu) &=& 4F_3(t;m_\mu,m_\mu) -\frac{2}{3}\ln \mmw -1
\\
     &=& \frac{2}{3} \ln \frac{-t}{\mmw} -\frac{19}{9}
           + O \biggl(\frac{t}{\mmw},\, \frac{m_\mu^2}{t} \biggr)\,,
\label{jw_mu_approx}
\eea
\esub
from eq.~(\ref{eq:jw}) and
\bsub
\label{box_nuq}
\bea
     \BnuLL (0,0) &=& -\frac{\ghat^4}{64\, \pi^2 \mmw}
                +\frac{3\, \gzhat^4}{64\, \pi^2 \mmz}
           \Bigl(\,\frac{1}{2} -\frac{2}{3} \shat^2 \Bigr)^2\,,
     \\
     \BnuLR (0,0) &=& \hphantom{-\frac{\ghat^4}{64\, \pi^2 \mmw}}
                -\frac{3\, \gzhat^4}{64\, \pi^2 \mmz}
           \Bigl( -\frac{2}{3} \shat^2 \Bigr)^2\,,
     \\
     \BndLL (0,0) &=& \hphantom{+}\frac{\ghat^4}{16\, \pi^2 \mmw}
                +\frac{3\, \gzhat^4}{64\, \pi^2 \mmz}
           \Bigl(-\frac{1}{2} +\frac{1}{3} \shat^2 \Bigr)^2\,,
     \\
     \BndLR (0,0) &=& \hphantom{+\frac{\ghat^4}{16\, \pi^2 \mmw}}
                -\frac{3\, \gzhat^4}{64\, \pi^2 \mmz}
           \Bigl(\,\frac{1}{3} \shat^2 \Bigr)^2\,,
\eea
\esub
from eq.~(\ref{eq:box_le}). These expressions are sufficient to evaluate
the helicity amplitudes (\ref{m_nuq}) as functions of $\sbar^2(0)$ and
$\gzbar^2(0)$, for the $\msbar$ coupling normalization of
eq.~(\ref{hatcouplings}). We set $\mw=80.24$~GeV and $\mz=91.187$~GeV in
all numerical calculations.
 
At $\langle -t \rangle _{\rm n.c.}=20\GeV^2$,
\bea
     \frac{\ehat^2}{16\,\pi^2}\,\ol{J}_W(t=-20\gev^2;m_\mu)
     &\approx&  -0.0037,
\label{jw_nuq_num}
\eea
and the $q_\alpha$'s are approximated as
  \begin{eqnarray}
     q_\alpha &\approx& 0.9923\,\bar{\rho}\,
     \Bigl[\,I_{3q_\alpha} -Q_q\,\sbar^2(t)\, \Bigr]+
     \left\{
     \begin{array}{ll}
       +0.0031 & (q_\alpha=u_L^{})\\
       +0.0026 & (q_\alpha=u_R^{})\\
       -0.0074 & (q_\alpha=d_L^{})\\
       -0.0012 & (q_\alpha=d_R^{})
     \end{array}
     \right.\,,
\label{q_alpha_approx}
\end{eqnarray}
for the $\msbar$ coupling normalization of eq.~(\ref{hatcouplings}).
Here the universal $\bar{\rho}$ parameter is defined by
\bea
 \ol{\rho} &\equiv& \frac{\gzbar^2(0)}{4\sqrt{2}\,G_F\, \mmz}
 = \frac{1}{1+\delg-\alpha T}
 \approx \frac{\gzbar^2(0)}{0.54864} \,.
\label{rhobar}
\eea
The relation between the form factor $\gzbar^2(0)$ and the $T$
parameter is seen in eq.~(\ref{gzbarfromt}). The running of 
$\sbar^2(t)$ is estimated as
\bea
     \sbar^2(t=-20\gev^2)
     &\approx& \frac{\bar{\alpha}(t=-20\gev^2)}{\alpha}
 [\sbar^2(0)  -0.0097]
\nonumber\\
 &\approx& 1.0295\, \sbar^2(0) -0.0100 \,.
\label{s2b_nuq_num}
\eea
The approximations (\ref{q_alpha_approx})--(\ref{s2b_nuq_num}) are found
to give excellent numerical predictions for all $q_\alpha$ as functions
of the two charge form factors, $\sbar^2(0)$ and $\gzbar^2(0)$.
 
The major effects of radiative corrections can be made transparent by
parametrizing the model-independent coupling factors of
eq.~(\ref{q_alpha}) in terms of the effective couplings $\rho_{\nu q}$
and $s^2_{\nu q}$ of ref.\cite{nuq_ew}:
\bsub
\label{q_alpha_ms}
\bea
     \uL &=& \rho_{\nu q}
         \Bigl(\hphantom{-} \frac{1}{2} -\frac{2}{3} s^2_{\nu q} \Bigr)
        +\Delta_{\uL}\,,
     \\
     \dL &=& \rho_{\nu q}
         \Bigl(-\frac{1}{2} +\frac{1}{3} s^2_{\nu q} \Bigr)
        +\Delta_{\dL}\,,
     \\
     \uR &=& \rho_{\nu q}
         \Bigl(-\frac{2}{3} s^2_{\nu q} \Bigr) \hspace{8mm}
        +\Delta_{\uR}\,,
     \\
     \dR &=& \rho_{\nu q}
         \Bigl(\hphantom{-}\frac{1}{3} s^2_{\nu q} \Bigr) \hspace{8mm}
        +\Delta_{\dR}\,.
\eea
\esub
The extra terms $\Delta_{q_\alpha}$ are fixed such that they do not
interfere with the leading terms in the most accurately measured
quantities, that is, $g_L^2$ and $g_R^2$.  One finds
\bsub
\label{del_q_alpha}
\bea
     \Delta_{\uL}
       &=& \frac{\gzhat^2}{8\, \pi^2}
           \Bigl(\,\chat^2 +\frac{\shat^2}{3}\,\Bigr)\,a_{\beta L}^{}\,,
     \\[1mm]
     \Delta_{\dL}
       &=& \frac{\gzhat^2}{8\, \pi^2}
           \Bigl(\,\chat^2 -\frac{\shat^2}{3}\,\Bigr)\,a_{\beta L}^{}\,,
     \\[1mm]
     \Delta_{\dR} &=& 2\,\Delta_{\uR}
                = \frac{\gzhat^2}{8\, \pi^2}\, \frac{\shat^4}{5}\,,
\eea
\esub
with
\bea
     a_{\beta L}^{} &=& -\frac{1}{2\chat^2}
       \biggl(
        \frac{9}{8} -\frac{3}{2}\,\shat^2 +\frac{8}{15}\, \shat^4
       \biggr)\,.
\eea
The radiatively corrected amplitudes can then be expressed approximately
in terms of the effective strengths `$\rho_{\nu q}$' of the neutral
current and the effective weak mixing factor `$s^2_{\nu q}$' in the
$\nu_\mu$--$q$ scattering process. In terms of the two universal charge
form factors $\gzbar^2(0)$ and $\sbar^2(t)$ they are given by
\bsub
\label{rho_s2_nuq}
\bea
     \rho_{\nu q} &=&
             \frac{\ol{\rho}}{\;[1+\delta_{\rm c.c.}]^{1/2}}\,
             \frac{ 1 +\frac{\langle -t \rangle_{\rm c.c.}}{\mmw} }
                  { 1 +\frac{\langle -t \rangle_{\rm n.c.}}{\mmz} }
        + \frac{\gzhat^2}{8\pi^2} \, a_Z^{} \,,
\label{rho_nuq}
\\
     s^2_{\nu q}(t) &=& \sbar^2(t) + \frac{\ehat^2}{16 \pi^2} \ol{J}_W(t;m_\mu)
        - \frac{\ehat^2}{8\pi^2\chat^2} \, a_\gamma \,.
\label{s2_nuq}
\eea
\esub
The box factors in eq.~(\ref{rho_s2_nuq}) are obtained from
eq.~(\ref{box_nuq}):
\bsub
\label{av_box}
\bea
     \frac{\gzhat^2}{8\pi^2} \, a_Z^{} &=&
        \frac{\gzhat^2}{16 \pi^2 \chat^2}
          \biggl(\,\frac{5}{2} -\frac{15}{4}\shat^2 -\frac{1}{5}\shat^4
                 +\frac{14}{9} \shat^6 \biggr)\,
        \approx 0.0074\,,
\label{az_box}
\\
     \frac{\ehat^2}{8\pi^2\chat^2} \, a_\gamma &=&
        \frac{\ehat^2}{16 \pi^2 \chat^4}
          \biggl(\,\frac{5}{2} -\frac{61}{20}\shat^2-\frac{9}{10}\shat^4
                 +\frac{14}{9} \shat^6 \biggr)\,
        \approx  0.0018\,,
\label{aa_box}
\eea
\esub
where $\mmw/\mmz$ is replaced by $\chat^2$ in order to reproduce the
expressions in ref.\cite{nuq_ew}. With the estimates
(\ref{jw_nuq_num}) and (\ref{s2b_nuq_num}),
we find
\bsub
\label{rho_s2_nuq_num}
\bea
 \rho_{\nu q} &\approx& 0.9923\,\bar{\rho} +0.0074\,,
\label{rho_nuq_num}
\\
 s^2_{\nu q} &\approx& 1.0295\,\sbar^2(0) -0.0155 \,.
\label{s2_nuq_num}
\eea
\esub
These equations are useful in understanding qualitatively the effect
of the $\nu_\mu$--$q$ scattering experiments off isoscalar targets,
but we find that they give slightly inaccurate approximations to
the quantities $q_\alpha$ (\ref{q_alpha}).
 
In the following table, we compare the numerical predictions for
the basic quantities $q_\alpha$ and the model-independent parameters
of eq.~(\ref{g2_d2}) by using the exact matrix elements (\ref{q_alpha})
and by using the approximation (\ref{q_alpha_ms}),
for $\gzbar^2(0)=0.5492$ and $\sbar^2(-20\gev^2)=0.2359$
(the SM predictions for $m_t=175$~GeV and $\mh=100$~GeV)~:
  \begin{eqnarray}
  \begin{array}{|c|r|r|}
     \hline
                & {\rm eq.~(\protect\ref{q_alpha})}
                & {\rm eq.~(\protect\ref{q_alpha_ms})} \\
                & {\rm exact} & {\rm approx.} \\
     \hline
     u_L^{}     &  0.3435 &  0.3343 \\
     u_R^{}     & -0.1537 & -0.1537 \\
     d_L^{}     & -0.4260 & -0.4336 \\
     d_R^{}     &  0.0769 &  0.0769 \\
     \hline
     g_L^2      &  0.2995 &  0.2998 \\
     g_R^2      &  0.0295 &  0.0295 \\
     \delta_L^2 & -0.0634 & -0.0763 \\
     \delta_R^2 &  0.0177 &  0.0177 \\
     \hline
  \end{array}
  \end{eqnarray}
It is clearly seen that the formulae (\ref{q_alpha_ms}), although
reproducing $u_L^{}$ and $d_L^{}$ rather poorly, give, as expected,
an excellent approximation for the most precisely measured parameter
$g_L^2$. They give, however, a rather poor approximation for the parameter
$\delta_L^2$ being off by 20\%, which is unsatisfactory in view of the
experimental uncertainty (see section~4.2.1).
 
Fig.~\ref{figureofnuqparameters} illustrates the relation between the
model-independent parameters ($g_L^2$, $g_R^2$) and the two universal
form factors ($\sbar^2(0)$, $\gzbar^2(0)$). The present data
\cite{nuq_dat} (see section~4) constrain the 2-dimensional parameter
space to the ellipse drawn in the same figure. The dashed line is the
$\bar{\rho}=1$ ($\delg-\alpha T=0$) curve~:
$\gzbar^2(0) =4\sqrt{2}G_F\mmz = 0.5486$. The thinness of the ellipse
in the ($g_L^2,\,g_R^2$) plane implies a strong correlation between
$\sbar^2(0)$ and $\gzbar^2(0)$. It is worth noting that the effective
charge $\sbar^2(0)$ derived from $\nu_\mu$--$q$ scattering experiments
at $q^2\approx -20$~GeV$^2$ is larger than the process-dependent
effective mixing factor $s^2_{\nu q}$ by as much as 0.01: 
see eq.~(\ref{s2_nuq_num}).
 
\subsubsection*{\normalsize \bf 3.2.2
Neutral currents in $\nu_\mu-e$ scattering}
 
\pr
The total cross section for the processes $\nu_\mu e \to \nu_\mu e $
and $\bar{\nu}_\mu e \to \bar{\nu}_\mu e $ in terms of the helicity
amplitudes $\MneLL$ and $\MneLR$ are given by
\bsub
\label{sig_nue}
\bea
     \sigma^{\nu e} &=& \frac{m_e E_\nu}{4 \pi} \int^1_0\,dz\,
             \biggl\{\,\squared{\MneLL}
                     +(1-z)^2\,\squared{\MneLR} \,\biggr\}\,,
\\
     \sigma^{\bar{\nu} e} &=& \frac{m_e E_\nu}{4 \pi} \int^1_0\,dz\,
             \biggl\{\, (1-z)^2\,\squared{\MneLL}
                     +\squared{\MneLR} \,\biggr\}\,,
\eea
\esub
where the variable $z$ is related to the momentum transfer $t$ by
\bea
     z = -t/t_{\rm max}\,,\quad
     t_{\rm max}=\frac{(s-m_e^2)^2}{s} \approx 2 m_e E_\nu\,,
\label{z_nue}
\eea
with the approximation $s \equiv (p_\nu +p_e)^2 \approx 2 m_e E_\nu$.
The amplitudes in eq.\,(\ref{sig_nue}) are obtained from
eq.~(\ref{m_nc_le})
\bsub
\label{m_nue}
\bea
     \MneLL &=& -\frac{1}{2}\ehat^2\frac{\ol{\Gamma}_2^\nu(t)}{t}
                +\frac{1}{2}\biggl[-\frac{1}{2} +\sbar^2(t) \biggr]
                \frac{\gzbar^2(0)}{t-\mmz}
                +\BneLL(0,0) \,,
\\
     \MneLR &=& -\frac{1}{2}\ehat^2\frac{\ol{\Gamma}_2^\nu(t)}{t}
                \hspace{15mm}
                +\frac{1}{2} \sbar^2(t)
                 \frac{\gzbar^2(0)}{t-\mmz}
                +\BneLR(0,0) \,,
\eea
\esub
where the $\nu_\mu$ charge radius factor $\ol{\Gamma}_2^\nu(t)/t$
is given by eq.~(\ref{jw_mu}) and the box form factors $\BneLL$,
$\BneLR$ by
\bsub
\label{box_nue}
\bea
     \BneLL (0,0) &=& \frac{\ghat^4}{16 \pi^2 \mmw}
               +\frac{3\gzhat^4}{16 \pi^2 \mmz}
                \Bigl(\frac{1}{2}\Bigr)^2
                \Bigl(-\frac{1}{2}+\shat^2\Bigr)^2 \,,
\\
     \BneLR (0,0) &=& \hphantom{\frac{\ghat^4}{16 \pi^2 \mmw}}
                -\frac{3\gzhat^4}{16 \pi^2 \mmz}
                \Bigl(\frac{1}{2}\Bigr)^2
                \Bigl(\shat^2\Bigr)^2 \,,
\eea
\esub
see eq.~(\ref{eq:box_le}). It is then straightforward to express the
cross sections (\ref{sig_nue}) in terms of the universal charge form
factors $\sbar^2(0)$ and $\gzbar^2(0)$. Our results (\ref{m_nue}) and
(\ref{box_nue}) agree with ref.~\cite{nue_ew}.
 
As in the case of the $\nu_\mu$--$q$ scattering analysis it is useful
to introduce the process-dependent effective couplings $\rho_{\nu e}$
and $s^2_{\nu e}$\cite{old_ncfit}:
\bsub
\label{m_nue_eff}
\bea
     \MneLL &=& 2 \sqrt{2} \,G_F\, \rho_{\nu e}
                       \biggl[\,\frac{1}{2} -s_{\nu e}^2(t) \biggr] \,,
\\
     \MneLR &=& 2 \sqrt{2} \,G_F\, \rho_{\nu e}
                       \biggl[ -s^2_{\nu e}(t) \biggr] \,.
\eea
\esub
{}From eqs.~(\ref{m_nue}) and (\ref{m_nue_eff}) one finds
\bsub
\label{rho_s2_nue}
\bea
     \rho_{\nu e}
     &=& \bar{\rho} +\frac{\gzhat^2}{16 \pi^2}
                     \biggl[ \frac{19}{4} -7 \shat^2 +6 \shat^4 \biggr] \,,
     \\
     s^2_{\nu e}(t)
     &=& \sbar^2(t) +\frac{\ehat^2}{16 \pi^2} \ol{J}_W(t;m_\mu)
                    -\frac{\gzhat^2 \shat^2}{16 \pi^2}
                     \biggl[ \frac{19}{4} -\frac{17}{2} \shat^2
                            +6 \shat^4 \biggr] \,,
 \label{s2_nue}
\eea
\esub
by neglecting higher order terms and by setting $\mmw/\mmz=\chat^2$.
Here $\bar{\rho}$ and $\ol{J}_W(t;m_\mu)$ are given by (\ref{rhobar})
and (\ref{jw_mu}), respectively. The cross sections can then be
expressed in terms of the model-independent parameters $\rho_{\nu e}$
and $s^2_{\nu e}$ by
\bsub
\label{sig_nue_eff_int}
\bea
     \frac{\sigma^{\nu e}}{E_\nu}
        &=& \frac{2\,m_e\,G_F^2}{\pi} \,\rho_{\nu e}^2 \,
            \int_0^1 dz \,
           \biggl\{\, \Bigl[\, \frac{1}{2}-s^2_{\nu e}(t) \Bigr]^2
                  +(1-z)^2 \Bigl[s^2_{\nu e}(t)\Bigr]^2 \,
           \biggr\}\,,
     \\
     \frac{\sigma^{\bar{\nu} e}}{E_\nu}
        &=& \frac{2\,m_e\,G_F^2}{\pi} \,\rho_{\nu e}^2 \,
            \int_0^1 dz \,
           \biggl\{\,(1-z)^2\,
                   \Bigl[\, \frac{1}{2}-s^2_{\nu e}(t) \Bigr]^2
                  +\Bigl[s^2_{\nu e}(t) \Bigr]^2 \,
           \biggr\}\,,
\eea
\esub
where $t=-2\,m_e\,E_\nu z$ (\ref{z_nue}).
For $E_\nu = 25.7 \gev$ (CHARM-II\cite{nue_dat}), we find
\bea
     t_{\rm max} = 2\, m_e \,E_\nu
              \sim 2\, m_\mu^2 \,.
\label{tmax_nue}
\eea
In this momentum region the running of $\sbar^2(t)$ is negligible:
\bea
     \sbar^2(t) &=& \left\{
       \begin{array}{ll}
            1.0072\, \sbar^2(0) -0.0018 & (t = -m_\mu^2) \\
            1.0080\, \sbar^2(0) -0.0020 & (t = -2\,m_\mu^2)
       \end{array} \right. \,.
\label{s2b_nue_num}
\eea
Also the $\nu$-charge radius factor $\ol{J}_W(t;m_\mu)$ has little
$t$-dependence:
\bea
     \frac{\ehat^2}{16\,\pi}\, \ol{J}_W(t;m_\mu)  &=& \left\{
           \begin{array}{ll}
              -0.0061 & (t=0) \\
              -0.0060 & (t=-m_\mu^2) \\
              -0.0059 & (t=-2\,m_\mu^2)
           \end{array}  \right. \,.
\label{jw_nue}
\eea
Thus, the $t$-dependence of the effective mixing factor $s^2_{\nu e}(t)$
(\ref{s2_nue}) 
is negligibly small. From eqs.~(\ref{rho_s2_nue}), (\ref{s2b_nue_num})
and (\ref{jw_nue}) follows
\bsub
\label{rho_s2_nue_num}
\bea
 \rho_{\nu e} &\approx& \bar{\rho} +0.0121\,,
\\
 s^2_{\nu e}(0) \approx s^2_{\nu e}(-m_\mu^2)
 &\approx& \,  1.0072 \sbar^2(0)  -0.0103\,.
\eea
\esub
In the limit of negligible $t$-dependence of $s^2_{\nu e}$, 
eq.~(\ref{sig_nue_eff_int}) becomes~:
\bsub
\label{sig_nue_eff}
\bea
     \frac{\sigma^{\nu e}}{E_\nu}
        &=& \frac{2\,m_e\,G_F^2}{\pi} \,\rho_{\nu e}^2 \,
           \biggl[\, \Bigl(\, \frac{1}{2}-s^2_{\nu e} \Bigr)^2
                  +\frac{1}{3} \Bigl(s^2_{\nu e}\Bigr)^2 \,
           \biggr]\,,
\\
     \frac{\sigma^{\bar{\nu} e}}{E_\nu}
        &=& \frac{2\,m_e\,G_F^2}{\pi} \,\rho_{\nu e}^2 \,
           \biggl[\,\frac{1}{3}\,
                   \Bigl(\, \frac{1}{2}-s^2_{\nu e} \Bigr)^2
                  +\Bigl(s^2_{\nu e}\Bigr)^2 \,
           \biggr]\,,
\eea
\esub
with $s^2_{\nu e} = s^2_{\nu e}(0)$. This is the form entering the
analysis of ref.\cite{beyer}: they combined the three experiments
\cite{nue_dat} and expressed the result in terms of the
model-independent parameters $\rho_{\nu e}$ and $s^2_{\nu e}$
(\ref{nue_para}).
In our analysis the above parametrization (\ref{sig_nue_eff})
is used to
reproduce the combined measured cross sections from
the fit\cite{beyer} in terms of $\rho_{\nu e}$ and $s^2_{\nu e}$.
These cross sections are then analysed in our framework
by using the defining equation (\ref{sig_nue}).
 
Fig.~\ref{figureofnueparameters} illustrates the constraint by the data
similarly to Fig.~\ref{figureofnuqparameters}. The approximation
(\ref{rho_s2_nue_num}) is found to reproduce our results accurately.
The dashed line denotes the curve $\bar{\rho}=1$ ($\delg-\alpha T=0$).
The ratio of the $\nu_\mu e$ and $\bar{\nu}_\mu e$ cross sections is
measured accurately, and hence the form factor $\sbar^2(0)$ is
constrained fairly independently of $\gzbar^2(0)$ from the $\nu_\mu$--$e$
scattering experiments.
 
\subsubsection*{\normalsize \bf 3.2.3
Neutral currents in $e$--$q$ interactions }
 
\pr
The effective Lagrangian of the parity-violating $e$--$q$ interaction
\cite{old_ncfit}
\bea
 {\cal L}_{PV} &=& -\frac{G_F}{\sqrt{2}}\sum_q\,
         \left[C_{1q}\, \ol{\psi}_e\gamma^\mu\gamma_5 \psi_e
               \cdot \ol{\psi}_q\gamma_\mu \psi_q
              +C_{2q}\, \ol{\psi}_e\gamma^\mu \psi_e
               \cdot \ol{\psi}_q\gamma_\mu\gamma_5 \psi_q \right]\,,
\label{lpv_eq}
\eea
can be rewritten in terms of left- and right-handed currents as follows~:
\bea
 {\cal L}_{PV} &=& -\frac{G_F}{\sqrt{2}} \sum_q
         \Bigl[ C_{1q}\left(J_R^{e} -J_L^{e} \right) \cdot
                      \left(J_R^{q} +J_L^{q} \right)
               +C_{2q}\left(J_R^{e} +J_L^{e} \right) \cdot
                      \left(J_R^{q} -J_L^{q} \right)   \Bigr]\,.
\eea
The effective couplings $C_{1q}$, $C_{2q}$ expressed in terms of the
helicity amplitudes read~:
\bsub
\label{c1c2_vs_m}
  \bea
     C_{1q} &=&  C_{1q}^M +C_{1q}^{(\gamma)}
           = \frac{1}{2\sqrt{2}\,G_F}
                \Bigl[\MeqLL -\MeqRL +\MeqLR -\MeqRR\Bigr]
               + C_{1q}^{(\gamma)}\,,
     \\
     C_{2q} &=&  C_{2q}^M +C_{2q}^{(\gamma)}
           = \frac{1}{2\sqrt{2}\,G_F}
                \Bigl[\MeqLL -\MeqLR +\MeqRL -\MeqRR\Bigr]
               + C_{2q}^{(\gamma)}\,.
\eea
\esub
Here $C_{1q}^{(\gamma)}$ and $C_{2q}^{(\gamma)}$ denote the sum of the
contributions from the photonic correction to the axial vector $Zee$
vertex and the $Z\gamma$ box correction\cite{apv_ew}, which are not
included in our helicity amplitudes (\ref{m_nc_le}). They are found in
refs.\cite{ed_ew,apv_ew}~:
\bsub
\label{c1c2_gamma}
\bea
     \Bigl(4 \sqrt{2} G_F\,\mmz\Bigr)\,C_{1q}^{(\gamma)}
     &=& \frac{\gzhat^2\, \ehat^2}{16 \pi^2}\,
          \biggl\{ -2\Bigl( I_{3q}-2\,Q_q\,\shat^2 \Bigr)
                   +6 I_{3q} Q_q \,( 1-4\,\shat^2)
                    \Bigl(\ln\frac{\mmz}{M^2}+\frac{3}{2}\Bigr)
          \biggr\}\,,\hspace{12mm}
     \\
     \Bigl(4 \sqrt{2} G_F\,\mmz\Bigr)\,C_{2q}^{(\gamma)}
     &=& \frac{\gzhat^2\, \ehat^2}{16 \pi^2}\,
          \biggl\{ -2 I_{3q} Q_q^2\,( 1-4\,\shat^2)
                   +6\,Q_q \Bigl(I_{3q}-2\,Q_q\,\shat^2\Bigr)
                    \Bigl(\ln\frac{\mmz}{M^2}+\frac{3}{2}\Bigr)
          \biggr\}\,. \quad
     \nonumber \\[-1mm]
\eea
\esub
$M^2\!\approx \langle -t \rangle \approx\! 1.5 \gev^2$ is used in the
analysis of the SLAC $eD$ scattering experiments\cite{ed_dat}. By
inserting eq.~(\ref{m_nc_le}) into $C_{1q}^M$ and $C_{2q}^M$ defined
above, one finds
\bsub
\label{c1c2_m}
\bea
     2\sqrt{2} G_F\cdot C_{1q}^M &=&
        \frac{\ehat^2}{t}(-Q_q)
          \left[\, 2\left(\Gamma_1^{e_L}-\Gamma_1^{e_R}\right)
                 +\ol{\Gamma}_2^{e_L} \right](t)
      +\frac{\gzbar^2(0)}{t-\mmz}\Bigl(-\frac{1}{2}\,\Bigr)
         \left(I_{3q}-2Q_q\sbar^2(t)\right)
     \nonumber \\
     & &    +\BeqLL -\BeqRL +\BeqLR -\BeqRR\,,
     \\
     2\sqrt{2}\,G_F\cdot C_{2q}^M &=&
        \frac{\ehat^2}{t}
           \left[\,(-Q_q)\, 2\left(\Gamma_1^{q_L}-\Gamma_1^{q_R}\right)
                 -2I_{3q}\ol{\Gamma}_2^{q_L} \right](t)
      +\frac{\gzbar^2(0)}{t-\mmz}\Bigl(-\frac{1}{2}+2\sbar^2(t)\Bigr)
         \left(I_{3q}\right)
     \nonumber \\
     & &    +\BeqLL -\BeqLR +\BeqRL -\BeqRR \,.
\eea
\esub
By adopting the SM predictions for the vertex and box form factors of
appendix A, the model-independent parameters $C_{iq}$ of the low energy
effective Lagrangian (\ref{lpv_eq}) are readily evaluated as functions
of $\sbar^2(t)$ and $\gzbar^2(0)$. More explicitly, one finds
\bsub
\label{c1c2_m_2}
\bea
     \lefteqn{4 \sqrt{2} G_F\,\mmz\,C_{1q}^M}
     \nonumber \\
     &=&  \frac{\gzbar^2(0)}{1 - t/\mmz}
             \Bigl[\,I_{3q} -2Q_q \sbar^2(t)\, \Bigr]
         +\frac{\gzhat^2\,\ehat^2}{16 \pi^2}\, Q_q\,
             \Bigl[\,(1-4\shat^2) J_Z + 2\,(J_W-\ol{J}_W)\, \Bigr]
\nonumber \\
     & & +\frac{\gzhat^4}{16 \pi^2}
            \biggl\{\,\frac{3}{4}\,I_{3q}\Bigl(I_{3q}-2\,Q_q\shat^2\Bigr)
                    \Bigl[1+(1-4\shat^2)^2\, \Bigr]
                  +2\, \chat^2(q=u) -\frac{\chat^2}{2}(q=d)
            \biggr\}\,,
\\
     \lefteqn{4 \sqrt{2} G_F\,\mmz\,C_{2q}^M}
     \nonumber \\
     &=&  \frac{\gzbar^2(0)}{1 - t/\mmz} \,I_{3q}\,
             \Bigl[\,1 -4 \sbar^2(t)\, \Bigr]
         +\frac{\gzhat^2\,\ehat^2}{16 \pi^2}\,
             \Bigl[\,2\,I_{3q}\,Q_q(2\,I_{3q}-4\,Q_q\,\shat^2) J_Z
                   + 2\,Q_q\,J_W -4\,I_{3q}\,\ol{J}_W)\, \Bigr]
\nonumber \\
     & & +\frac{\gzhat^4}{16 \pi^2}
            \biggl\{\,\frac{3}{2} \,(1-4\,\shat^2)
                   \Bigl[ (I_{3q}-Q_q\,\shat^2)^2
                          +(Q_q\,\shat^2)^2\Bigr]
                   +2\, \chat^2(q=u) -\frac{\chat^2}{2}(q=d)
            \biggr\}\,,
\eea
\esub
where the factors $J_Z \equiv J_Z(t;m_e)$,
$J_W \equiv J_W(t;m_{\nu_e})$ and
$\ol{J}_W\equiv \ol{J}_W(t;m_{\nu_e})$ are given in appendix A.
The sum of (\ref{c1c2_m_2}) and (\ref{c1c2_gamma}) agrees with
ref.\cite{apv_ew}.
 
At $t=-1.5\gev^2$, $\sbar^2(t)$ is calculated from
$\sbar^2(0)$ as
  \begin{eqnarray}
     \sbar^2(t) &\approx& 1.0183\, \sbar^2(0)-0.0058\,,
     \label{sbar_running_ed}
  \end{eqnarray}
and the numerical values for $J_Z$, $J_W$ and $\ol{J}_W$ are
$ J_Z      \approx -6.97$,
$ J_W      \approx -6.80$,
$ \ol{J}_W \approx -7.69$.
The non-universal (vertex and box) corrections
for $C_{iq}$ are estimated numerically as
  \begin{subequations}
     \label{vertex_and_box_eq}
  \begin{eqnarray}
     C_{1u} &\approx& \bigl[C_{1u}^M\bigr]_{\rm IBA} +0.0061 +0.0007,\\
     C_{1d} &\approx& \bigl[C_{1d}^M\bigr]_{\rm IBA} -0.0011 +0.0009,\\
     C_{2u} &\approx& \bigl[C_{2u}^M\bigr]_{\rm IBA} +0.0082 +0.0048,\\
     C_{2d} &\approx& \bigl[C_{2d}^M\bigr]_{\rm IBA} -0.0070 +0.0043,
  \end{eqnarray}
  \end{subequations}
where the second terms in the r.h.s. 
denote the electroweak vertex/box corrections
for $C_{iq}^M$, and the last terms denote the external photonic
corrections, $C_{iq}^{(\gamma)}$. The improved Born expressions
$\bigl[C_{iq}^M\bigr]_{\rm IBA}$'s can be expressed by
  \begin{subequations}
  \begin{eqnarray}
     \bigl[C_{1q}^M\bigr]_{\rm IBA} &=& \frac{\bar{\rho}}{1-t/\mmz}\,
                            \Bigl[I_{3q} -2\,Q_q\,\sbar^2(t) \Bigr]\,,
     \\
     \bigl[C_{2q}^M\bigr]_{\rm IBA} &=& \frac{\bar{\rho}}{1-t/\mmz}\,
                            I_{3q}\,\Bigl[1 -4\,\sbar^2(t) \Bigr]\,,
  \end{eqnarray}
  \end{subequations}
with $\bar{\rho} \equiv{\gzbar^2(0)}/{(4\sqrt{2}\,G_F\,\mmz)}$
as in eq.~(\ref{rhobar}).
 
In the polarized $eD$ experiment only the combinations
$2C_{1u}-C_{1d}$ and $2C_{2u}-C_{2d}$\cite{old_ncfit} are well measured.
A model-independent determination of these two combinations is performed
in section~4.  Fig.~\ref{figureofedparameters} shows the relation
between the model-independent parameters ($2C_{1u}-C_{1d}$,
$2C_{2u}-C_{2d}$) and the two universal parameters
($\sbar^2(0)$, $\gzbar^2(0)$), together with the 1-$\sigma$ contour of
the result of the analysis obtained in section~4 from the experimental
data\cite{ed_dat}. Note that the vertex and box corrections (especially
the $WW$ box contributions) in eqs.~(\ref{c1c2_m}) are important in
these combinations yielding~:
\bsub
\label{c_ed_num}
\bea
   2C_{1u}-C_{1d} &\approx& [2C_{1u}^M-C_{1d}^M]_{\rm IBA}
   + 0.0134 +0.0005\,,
\\
   2C_{2u}-C_{2d} &\approx& [2C_{2u}^M-C_{2d}^M]_{\rm IBA}
   + 0.0234 +0.0052\,.
\eea
\esub
As before, the second terms denote the vertex/box corrections in
$C_{iq}^M$, while the last terms denote contributions from
$C_{iq}^{\gamma}$. The majority of the non-universal contributions
above come from the $WW$ box diagram. Since the typical contribution
of the improved Born approximation to these factors are
$[2C_{1u}^M-C_{1d}^M]_{\rm IBA} \approx 0.7089$ and
$[2C_{2u}^M-C_{2d}^M]_{\rm IBA} \approx 0.0751$ for
$\gzbar^2(0)=0.5492$ and $\sbar^2(-1.5$GeV$^2)=0.2375$ (the SM
predictions for $m_t=175$~GeV and $\mh=100$~GeV), the non-propagator
correction terms are appreciable in these observables.
 
In the case of atomic parity violation the momentum transfer is so
small that the matrix elements for nucleons should be calculated.
Marciano and Sirlin\cite{apv_ew} introduced effective couplings
$C_{1p}$ and $C_{1n}$ for nucleons, which may be separated as in
eq.~(\ref{c1c2_vs_m})
\bsub
\label{c1_pn}
\bea
     C_{1p} &=&  C_{1p}^M +C_{1p}^{(\gamma)}\,,
\\
     C_{1n} &=&  C_{1n}^M +C_{1n}^{(\gamma)}\,.
\eea
\esub
Here $C_{1p}^M$ and $C_{1n}^M$ are the contributions from the neutral
current amplitudes (\ref{m_nc_le}), which can be expressed in terms of
$C_{1u}^M$ and $C_{1d}^M$ by
\bsub
\label{c1_pn_m}
\bea
     C_{1p}^M = 2\,C_{1u}^M+C_{1d}^M\,,
\\
     C_{1n}^M = C_{1u}^M+2\,C_{1d}^M\,,
\eea
\esub
or more explicitly,
\bsub
\label{c1_pn_m_2}
\bea
     \Bigl(4 \sqrt{2} G_F\,\mmz\Bigr) \,C_{1p}^M
     &=& \gzbar^2(0) \biggl[\,\frac{1}{2} -2 \sbar^2(0)\, \biggr]
         +\frac{\gzhat^2\, \ehat^2}{16 \pi^2}\,
            \biggl\{\, (1-4\shat^2) J_Z + 2\,(J_W-\ol{J}_W)\, \biggr\}
\nonumber \\
     & & +\frac{\gzhat^4}{16 \pi^2}
            \biggl\{ \frac{9}{16}\, \Bigl(1-\frac{20}{9}\shat^2\Bigr)
                    \left[1+(1-4\shat^2)^2\, \right]
                    +\frac{7}{2}\, \chat^2
            \biggr\}\,,
\label{c1_p_m}
\\
     \Bigl( 4 \sqrt{2} G_F\,\mmz\Bigr) \,C_{1n}^M
     &=& \gzbar^2(0) \biggl[-\frac{1}{2}\biggr]
        +\frac{\gzhat^4}{16 \pi^2}
            \biggl\{ \frac{9}{16}\, \Bigl(1-\frac{16}{9}\shat^2\Bigr)
                    \left[1+(1-4\shat^2)^2\, \right] +\,\chat^2
            \biggr\}\,,\qqquad
\label{c1_n_m}
\eea
\esub
with
\bsub
\label{jv_0}
\bea
    & & J_Z = \frac{2}{3}\,\ln \frac{m_e^2}{\mmz} -\frac{1}{9}\,,
\\
    & & J_W - \ol{J}_W = \frac{8}{9}\,,
\eea
\esub
which can be obtained from eqs.~(\ref{eq:j}) by taking the 
$q^2\to 0$ limit.
$C_{1p}^{(\gamma)}$ and $C_{1n}^{(\gamma)}$ are the contributions
from the photonic correction to the axial vector $Zee$ vertex and the
$Z\gamma$ box correction\cite{apv_ew}~:
\bsub
\label{c1_pn_gamma}
\bea
     4 \sqrt{2} G_F\,\mmz\,C_{1p}^{(\gamma)}
     &=& \frac{\gzhat^2\, \ehat^2}{16 \pi^2}\,
         \biggl\{-(1-4\,\shat^2)^2
             +5\,(1-4\,\shat^2)\Bigl[K +\frac{4}{5}(\xi_1)_B^p\Bigr]
         \biggr\}\,,
\\
     4 \sqrt{2} G_F\,\mmz\,C_{1n}^{(\gamma)}
     &=& \frac{\gzhat^2\, \ehat^2}{16 \pi^2}\,
         \biggl\{-(1-4\,\shat^2)^2
             +4\,(1-4\,\shat^2)\Bigl[K +(\xi_1)_B^n\Bigr]
         \biggr\}\,.
\eea
\esub
The last terms on the right-hand sides of eqs.~(\ref{c1_pn_gamma})
denote the $\gamma Z$-box corrections which are sensitive to the nucleon
structure. The constants $K$, $(\xi_1)_B^p$ and $(\xi_1)_B^n$ have been
estimated in ref.\cite{apv_ew} to be
\bea
      K = 9.6\pm 1\,,\;  (\xi)_B^p = 2.55\,,\;  (\xi)_B^n = 1.74\,.
\label{apv_pn}
\eea
 
By estimating numerically the vertex/box corrections in
eqs.~(\ref{c1_pn_m_2}) and (\ref{c1_pn_gamma}), we find
  \begin{subequations}
  \begin{eqnarray}
    C_{1p} &\approx& \Bigl[C_{1p}^M\Bigr]_{\rm IBA} +0.0107 +0.0027,
    \\
    C_{1n} &\approx& \Bigl[C_{1n}^M\Bigr]_{\rm IBA} +0.0038 +0.0023,
  \end{eqnarray}
  \end{subequations}
where the second terms denote the weak vertex/box corrections for
$C_{1p}^M$ and $C_{1n}^M$, and the last terms denote the photonic
corrections, $C_{1p}^{(\gamma)}$ and $C_{1n}^{(\gamma)}$.
The improved Born approximations
$\bigl[C_{1p}^M\bigr]_{\rm IBA}$ and $\bigl[C_{1n}^M\bigr]_{\rm IBA}$
are given simply by
  \begin{subequations}
\label{c1_pn_iba}
  \begin{eqnarray}
    \bigl[C_{1p}^M\bigr]_{\rm IBA} &=&
    \bar{\rho}\,\biggl[\,\frac{1}{2} -2\sbar^2(0)\biggr] \,,
    \\
    \bigl[C_{1n}^M\bigr]_{\rm IBA} &=&
    \bar{\rho}\,\biggl[-\frac{1}{2} \,\biggr] \,.
  \end{eqnarray}
  \end{subequations}
Their typical numerical values are found to be
$\bigl[C_{1p}^M\bigr]_{\rm IBA} \approx 0.0223$ and
$\bigl[C_{1n}^M\bigr]_{\rm IBA} \approx -0.5005$,
for $\gzbar^2(0)=0.5492$ and $\sbar^2(0)=0.2389$
(the SM predictions for $m_t=175$~GeV and $\mh=100$~GeV).
Note that the non-universal corrections are important especially for
$C_{1p}$, where the effect comes mainly from the $WW$ box contribution,
or the term with the factor $\frac{7}{2}\chat^2$ in eq.~(\ref{c1_p_m}).
 
The weak charge $Q_W(A,Z)$ of an atom is given in terms of
$C_{1p}$ an $C_{1n}$ by
\bea
     Q_W(A,Z) & = & 2(A-Z)C_{1n}+2ZC_{1p}\,,
\label{qw_az}
\eea
which in the case of cesium is
\bea
     Q_W(^{133}_{\hphantom{1}55}{\rm Cs}) = 156\,C_{1n} +110\,C_{1p}\,.
\label{qw_cs}
\eea
Numerically they are estimated as
  \begin{eqnarray}
    Q_W({\rm Cs}) \approx \gzbar^2(0)\,[\,-41.92 -400.99\sbar^2(0)\,]
     +1.77+0.65\,.
\label{qw_cs_num}
  \end{eqnarray}
where the first term comes from the IBA approximation
to $C_{1p}^M$ and $C_{1n}^M$ (\ref{c1_pn_iba}),
the second term comes from the electroweak vertex/box contributions to
them, and the last term from the external photonic corrections of
eq.~(\ref{c1_pn_gamma}). It is clear from the above result that the
vertex and box corrections should be carefully taken account of
in extracting the electroweak parameters from the $Q_W$ measurements.
 
In Fig.~\ref{figureofapvparameters},  the parameter $Q_W({\rm Cs})$ of
eq.~(\ref{qw_cs}) is shown as a function of the two universal parameters
($\sbar^2(0)$, $\gzbar^2(0)$) in the range $0.20<\sbar^2(0)<0.26$
and $0.52 <\gzbar^2(0)<0.57$ together with the 1-$\sigma$ contour of
the data\cite{apv_dat} (dashed lines). The horizontal straight dashed
line denotes the line $\bar{\rho}=1$ ($\delg-\alpha T=0$). It is worth
noting that the correlation in the $\sbar^2(0)$ and $\gzbar^2(0)$ form
factors obtained from the cesium weak charge $Q_W(133,55)$ in the 
figure is opposite to that obtained from the $\nu_\mu$--$q$ scattering 
experiments (see Fig.~\ref{figureofnuqparameters}). The cesium $Q_W$
measurement implies an anti-correlation between $\sbar^2(0)$ and
$\gzbar^2(0)$, or $\bar{\rho}$. This is opposite to the trend observed
for the constraints from the $\nu_\mu$-$q$ experiments. For further
discussion, see section~4.
 
\vspace*{5mm}
\subsection*{\normalsize \bf 3.3  Charged current experiments }
 
\pr
In the charged current sector we consider two precision experiments:
the muon lifetime\cite{pdg92} and the $W$ boson mass measurements
\cite{pdg92,mw93}.
 
{}From the matrix element (\ref{m_cc}) one finds for the muon decay
constant
\bea
     G_F = \frac{\gwbar^2(0) +\ghat^2 \delg}{4\sqrt{2}\,\mmw}\,,
\label{gf_vs_gw0}
\eea
where the factor $\delg$ denotes the sum of the vertex and the box
contributions. It has been calculated within the SM in
ref.\cite{del_gf}~:
\bsub
\label{delg_cc}
\bea
     \delg &=& \frac{\ghat^2}{8\,\pi^2}
             \biggl[\,1 +\biggl(\frac{1}{4\,\shat^2}-1\biggr)
                     \ln\frac{1}{\chat^2} \biggr]\,,
\\[1mm]
          &\approx& 0.0055\,,
\eea
\esub
where the pinch term\cite{pt2} has been subtracted as explained
in section~2: see eqs.~(\ref{delg}) and (\ref{delg_old}).
 
The expression (\ref{gf_vs_gw0}) enables one to predict the physical
$W$ mass in terms of the charge form factor $\gwbar^2(0)$.
Numerically, one finds~:
\bsub
\label{mw_from_gf}
\bea
 \mmw &=& \frac{\gwbar^2(0) +\ghat^2\delg}{4\sqrt{2}\,G_F} \\
 &\approx& \Bigl[15155.9\,\gwbar^2(0)
 +46.7\,\frac{\delg -0.0055}{\alpha}
 +35.2\Bigr]~{\rm GeV}^2\,.
\eea
\esub
Once the numerical value of $\delg$ factor is known, the measurement
of the $\mw$ mass determines directly the charge form factor
$\gwbar^2(0)$.
 
The form factor $\gwbar^2(0)$ can be calculated in terms of the
$S$, $T$ and $U$ parameters in the ${\rm SU(2)_L \times U(1)_Y}$
models. Insertion of the expansion (\ref{gwbar_approx}) leads to
\bea
\label{mw_stu}
 \mw({\rm GeV}) =
 79.840 -0.291\,S +0.417\,T +0.332\,U -0.136\,\delta_\alpha \,,
\eea
in excellent agreement with eq.~(\ref{mw_from_gf}) for $\delg =0.0055$.
The prediction for a different $\delg$ value follows from the above
expression by simply making the substitution (\ref{t_subst}).
 
Fig.~\ref{figureofmw} shows the SM predictions for $\mw$ in
the plane of $m_t$ and $\mh$, for $\delta_\alpha=0$ and $\delg=0.0055$.
In the O($\alpha \alpha_s$) corrections to the SM contributions to
the $S$, $T$, $U$ parameters $\alpha_s(\mz)$ is set to 0.12. Changing
$\alpha_s(\mz)$ by $\pm 0.01$ affects the prediction of $\mw$ by about
$\mp 0.004$~GeV. The mean and standard deviation of the present $\mw$
measurement (see section~4.3) are indicated by dashed lines.
 
Note that among the electroweak observables examined in this paper, only
$\mw$ is sensitive to the $U$ parameter. Hence, when performing a
general fit to the $S$, $T$, $U$ parameters, the mean 
($\langle U\rangle$) and
standard deviation ($\Delta U$) of the $U$ parameter are determined 
solely by the mean ($\langle\mw\rangle$) 
and standard deviation ($\Delta\mw$)
of $\mw$~:
\bsub
\label{u_mw}
\bea
 \langle U \rangle &=& [\,\langle m_W^{}({\rm GeV}) \rangle -79.840
 +0.291\,\langle S \rangle -0.417\,\langle T \rangle 
 +0.136\,\delta_\alpha\,]/0.332 \,,
\\
 \Delta U &\approx& \Delta \mw({\rm GeV})/0.332 \,.
\label{delu_delmw}
\eea
\esub
Here $\langle S \rangle$ and $\langle T\rangle$ 
denote the best-fit values from other
experiments. The present experimental error of $\Delta \mw$ = 0.16~GeV
induces $\Delta U=0.48$, while $\Delta \mw$ = 0.05~GeV, the precision
anticipated in future LEP200 experiments, would give $\Delta U=0.15$.
The full error $\Delta U$ should be slightly larger than the above
estimate, since $S$ and $T$ were fixed and set at their best values in
deriving (\ref{delu_delmw}).


\newpage
\section*{\large \bf 4  Experimental data and the electroweak parameters}
\renewcommand{\theequation}{4.\arabic{equation}}
\cleqn
 
\pr
Based on the formalism introduced in the previous sections the values
for the form factors are inferred from fits to the data of electroweak
precision experiments~: $\gzbar^2(\mmz)$, $\sbar^2(\mmz)$,
$\delb(\mmz)$ from the LEP/SLC experiments on the $Z$-pole,
$\gzbar^2(0)$, $\sbar^2(0)$ from the low energy neutral current
experiments at $q^2\approx 0$, and $\gwbar^2(0)$ from the $W$ mass
measurements at $p\bar{p}$ colliders.
 
\subsection*{\normalsize \bf 4.1  $Z$ boson parameters}
 
\pr
The analysis is based on the data from the LEP and SLC experiments
published up to the year 1993\cite{zdata93,lep93,alr92}. 
Discussions
of the recent update from LEP\cite{lep94} and 
the precision measurement of the left-right asymmetry at
SLC\cite{alr94} are postponed to section~6.
 
The $Z$ line-shape parameters resulting from a combined fit performed
by the LEP electroweak group\cite{lep93} are~:
%
\dataofzlineshape
The other electroweak data used in our fit are\cite{zdata93,lep93}:
%
\dataofzothers
%
Definitions of all the above observables and their theoretical
expressions have been given in section~3.1.
 
The $Z$ mass, $\mz\!=\!91.187\gev$, is treated as an input parameter 
neglecting its error.
This is justified because of the smallness of the
experimental uncertainty and correlations.
For the fits to be described below a few general conditions are
anticipated~:
(a) only three neutrinos ($N_\nu = 3$) contribute to the invisible
width of $Z$,
(b) the perturbative QCD corrections with the finite quark mass effects
are taken as given explicitly in section~3.1,
(c) the vertex and box corrections are calculated in the SM and given
in Table~\ref{tableofvertexformfactors} and \ref{tableofboxformfactors},
(d) the $\zbb$ vertex is taken into account by the quantity
$\delb(\mmz)$, which is treated in the fit as a free parameter just as
the universal parameters $\gzbar^2(\mmz)$ and $\sbar^2(\mmz)$.
 
Various methods to determine the QCD coupling constant have led to
consistent results with a typical uncertainty of $\Delta \alpha_s(m_Z)$
$\approx$ 0.01. However, this is far from making it 
precise enough to be used as a fixed
input parameter, since
the fitted electroweak parameters are found to be rather sensitive to
the assumed value of $\alpha_s(\mz)$: see, for instance,
eq.~(\ref{fitofzparameter}) below. For this reason, and also for the
convenience of GUT studies, $\alpha_s \equiv \alpha_s(\mz)_{\msbar}$ is
treated throughout our fits as an external input parameter and,
consequently, the best-fit values of the fit parameters and the minimum
$\chi^2$ are always presented as functions of $\alpha_s$.
Once a precise determination of $\alpha_s$ from independent data is
available, it is straightforward to get the correspondingly adjusted
best-fit values without
repeating the fit. It is also easy to infer from our results the
quantitative consequences of a particular GUT model predicting the
relationship between $\alpha_s$ and $\sin^2\theta_W^{}(\mz)_{\msbar}$.
 
The overall fit to all $Z$ parameters listed above gives the following
result~:
%
\fitofzparameter
%
where the errors and correlations are nearly independent of $\alpha_s$.
The above parametrization for the $\alpha_s$ dependences of the mean
values and $\chi^2_{\rm min}$ are accurate interpolations of our fit
results (at the level of 1\%) in the range $0.09 <\alpha_s <0.15$. The
bottom and charm quark masses were set to $m_b=4.7\gev$ and $m_c=1.4\gev$.
A shift of the bottom mass by $\pm 0.2 \gev$ implies only the fitted
$\delb(\mmz)$ value to be displaced by $\pm 0.0002$, which is
negligibly small compared to its error ($\pm 0.0034$). Similarly,
shifting the charm quark mass by $\pm 0.2 \gev$  does not affect the
above results, as expected. In particular, in the favored range
$0.11 \simlt \alpha_s(\mz) \simlt 0.13$ the quality of the fit is good,
e.g. $\chi^2_{\rm min}=4.6$ at $\alpha_s=0.12$ for $9-3=6$ degrees of
freedom.
 
Results of the fit (\ref{fitofzparameter}) are displayed in
Fig.~\ref{figureofzparameterfit} by the 1-$\sigma$ allowed contours in
the three projections, ($\sbar^2(\mmz),\gzbar^2(\mmz)$),
($\sbar^2(\mmz),\delb(\mmz)$), and ($\delb(\mmz),\gzbar^2(\mmz)$).
The contours are shown for three representative $\alpha_s$-values~:
$\alpha_s$=0.11\,(dashed), 0.12\,(solid) and 0.13\,(dash-dotted). Also
shown by the lattices are the SM predictions for $m_t$=(100--200)$\gev$
and $\mh$=(50--1000)$\gev$. In these SM predictions, all known two-loop
corrections of the $O(m_t^4)$ and $O(\alpha\alpha_s)$
level\cite{rho_ew1,rho_ew2,zbb_qcd,rho_zbb_ew,rho_ew3,
rho_pqcd,pivv_pqcd,rho_pqcd_mh2}
are
included, as explained in detail in appendix C. Hence, the predictions
depend weakly on $\alpha_s$
due to the $O(\alpha\alpha_s)$ corrections, as well as on $\delg$ and
$\delta_\alpha$ which are needed to predict the charge form factors
from the known ($\alpha$,$G_F$,$\mz$) values. The SM predictions in
Fig.~\ref{figureofzparameterfit} are calculated for $\alpha_s$=0.12,
$\delg$=0.0055 and $\delta_\alpha$=0. Changing $\alpha_s$ by $\pm$0.01
has little effect, but changing $\delta_\alpha$ by $\pm$0.10 leads to a
shift in the SM predictions for $\sbar^2(\mmz)$ by $\mp 0.00026$,
which is as large as
40\% of its uncertainty: see eq.~(\ref{fitofzparameter}).
 
It is clearly seen from the figure and eq.~(\ref{fitofzparameter})
that the weak mixing form factor 
$\sbar^2(\mmz)$ is determined almost independently of $\alpha_s$,
while the $Z$ coupling strength $\gzbar^2(\mmz)$ is anti-correlated
with the assumed $\alpha_s$ value as a reflection of its sensitivity to
the total $Z$ width. This anti-correlation leads in the SM to a
preference of larger $m_t$  for smaller $\alpha_s$, since
$\gzbar^2(\mmz)$ grows with $m_t$ 
(see Fig.~\ref{figureofchargeformfactors}).
Furthermore, the fitted   $\delb(\mmz)$ value depends strongly on the
$\alpha_s$ value assumed. The minimum of $\chi^2$ is reached at
$\alpha_s=0.1029$ in eq.~(\ref{chisqofzparameter}), a value slightly
outside the range $0.11 \simlt \alpha_s(\mz) \simlt 0.13$ expected 
from various QCD analyses\cite{alpha_s}. 
 
It is instructive to elucidate the properties of the fit to the $Z$
parameters in three steps. First, the relatively small
sensitivity of the parameter $\sbar^2(\mmz)$ to $\alpha_s$ can be
understood easily, since it is derived essentially from the asymmetry
parameters being either completely or nearly insensitive to QCD
corrections. Indeed, the fitted values of $\sbar^2(\mmz)$ as determined
from each asymmetry measurement (see also Fig.~\ref{figureofasymmetry})
turn out to be~:
%
\fitofzasymmetries
%
almost independent of $\gzbar^2(\mmz)$ and $\delb(\mmz)$.
Note that although the quark ($q=b,c$)
forward-backward asymmetries have mild $\alpha_s$-dependences due to
the perturbative QCD corrections\cite{qcd_afb}, they still can be
neglected compared to the experimental uncertainties. From the above
asymmetry data alone one finds
%
\fitofzasymmetriesall
%
The precision of the above determination of $\sbar^2(\mmz)$ from the
asymmetry data alone is almost as good as that of the global
fit to all the $Z$ parameters. 
These asymmetry measurements are particularly important for GUT studies,
since the parameter $\sbar^2(\mmz)$ is directly related to the
unifying coupling $\shat^2(\mu)\equiv \sin^2\theta_W^{}(\mu)_{\msbar}$
via eq.~(\ref{rg}).

Next, the best-fit value $\sbar^2(\mmz)\approx 0.2313$ is taken to
probe the sensitivity of the remaining four observables to the
parameters $\gzbar^2(\mmz)$ and $\delb(\mmz)$. As explained in
section~3.1, three of the remaining four observables, $\Gamma_Z$,
$\sigma_h^0$ and $R_\ell$, are sensitive to the $\alpha_s$ value
assumed, but only through the combination $\delb(\mmz)+0.6\alpha_s$
(\ref{delb_alps}), or equivalently $\alpha_s+1.6\delb(\mmz)$. $\Gamma_Z$
is also sensitive to $\gzbar^2(\mmz)$. Hence, a 2-parameter fit to the
above three observables for $\sbar^2(\mmz)=0.2313$ leads to~:
\bea
\label{fit_gzsighrl}
    \left.
    \begin{array}{lll}
 \gzbar^2(\mmz)           &\mm=& 0.5547 \pm 0.0017 \\
 \alpha_s+1.6\delb(\mmz)  &\mm=& 0.106 \pm 0.007
    \end{array}
    \right\}\;
 \rho_{\rm corr} = -0.46 \,.
\eea
The above result is found to be insensitive to the $\alpha_s$ value in
the range $0.10<\alpha_s<0.14$.
The above result for $\gzbar^2(\mmz)$ is consistent with the global
fit (\ref{fitofzparameter}), as may be verified by evaluating
$\gzbar^2(\mmz)$ at the minimum of $\chi^2$ ($\alpha_s=0.1029$). The
anti-correlation above reflects the fact that $\Gamma_Z$ remains
unaltered, while increasing $\gzbar^2(\mmz)$ and decreasing
$\alpha_s+1.6\delb(\mmz)$ simultaneously.
 
Only one $Z$ observable is now left, namely $R_b$. In section
3.1 $R_b$ was found to be sensitive to the parameter $\delb(\mmz)$
alone. A 1-parameter fit to $R_b$ yields~:
\bea
 \delb(\mmz)=0.0012 \pm 0.0068 \,,
\label{fit_rb}
\eea
keeping the other parameters fixed at $\sbar^2(\mmz)=0.2313$,
$\gzbar^2(\mmz)=0.5542$
                        and $\alpha_s=0.12$.
However, this fit is insensitive to variations around the values of
the fixed parameters. Note, the SM predicts a negative value of
$\delb(\mmz)$ for large $m_t$ (see Fig.~\ref{figureofstusm}). Thus,
there is poor agreement with the expected large $m_t$ behavior of the
$\zbb$ vertex correction from the present $R_b$ measurement alone.
Since the parameter $\alpha_s$ enters the fit only in the combination
$\alpha_s+1.6\delb(\mmz)$, the fitted $\delb(\mmz)$ can be interpreted
as a constraint of $\alpha_s$. From eqs.~(\ref{fit_gzsighrl}) and
(\ref{fit_rb}) follows
\bea
 \alpha_s = 0.104 \pm 0.013 \,.
\label{fit_alps_approx}
\eea
The fit to the $Z$ shape parameters with both $\delb$ and $\alpha_s$
left free yields~:
%
\fitofdelbalps
%
The large errors and the strong anti-correlation among them show
that it makes little sense to extract $\alpha_s$ model-independently
from the electroweak experiments on the $Z$-pole,
as also noted in ref.\cite{aziz}. The low best-fit value of $\alpha_s$
reflects essentially the actual value of $R_b$, which is larger
than the SM prediction in the range $150~\gev <m_t < 200~\gev$ 
(see Fig.~\ref{figureofzparametersvsmt}).
It is therefore necessary to assume the SM contributions to
$\delb(\mmz)$, and to a lesser extent those to $\gzbar^2(\mmz)$, in
order to extract $\alpha_s$ from the electroweak $Z$ parameters. The
result of such an analysis is given in section~5.4, where consequences
of the minimal SM are studied.
 
Finally, we present the result of 1-parameter fits to four observables,
$\Gamma_Z$, $\sigma_h^0$, $R_\ell$ and $R_b$, respectively, in terms of
the parameter $\delb(\mmz)$, for 
various values of $\gzbar^2(\mmz)$, $\sbar^2(\mmz)$
and $\alpha_s$.
Here, we neglect correlations in the errors and find:
%
\fitofdelb
%
The above fits clearly confirm quantitatively our observations that
$\Gamma_Z$, $\sigma_h^0$ and $R_\ell$ measure the combination 
$\delb +0.6\,\alpha_s$ 
(\ref{delb_alps}), that $\Gamma_Z$ is also sensitive to $\gzbar^2(\mmz)$, 
and that $R_b$ is sensitive only to $\delb(\mmz)$. 
At present the data $\Gamma_Z$, $\sigma_h^0$ and 
$R_\ell$ favor a negative $\delb(\mmz)$ value consistent 
with the SM prediction for $150\gev <m_t<200\gev$, 
while $R_b$ data gives a $\delb(\mmz)$ value consistent with zero, 
at $\alpha_s\approx 0.12$. 
The combination of all the above measurements
together with all the asymmetry data, and properly accounting for the
correlations in the errors, yields
%
\fitofdelball
%
in accordance with the result (\ref{fitofzparameter}).
Note that the coefficient in front of $\alpha_s$ in eq.~(\ref{delb_all})
is smaller than 0.6 in the combination
(\ref{delb_alps}) as a consequence of including
the additional information due to $R_b$.
 
\subsection*{\normalsize \bf 4.2  Low energy neutral current experiments}
 
\pr
The two universal parameters $\sbar^2(0)$ and $\gzbar^2(0)$ can be
extracted from four types of low energy neutral current experiments~:
the neutrino-nucleon scattering ($\nu_\mu$--$q$), the neutrino-electron
scattering ($\nu_\mu$--$e$), atomic parity violation (APV) and the
polarized electron-deuteron scattering experiments ($e$--D). Effects due
to small, but finite, momentum transfer in these processes are accounted
for by assuming the running of these form factors to be governed by
the SM particles only (see Fig.~\ref{figureofchargeformfactors}), which,
at low energies, is an excellent approximation. Vertex and box
corrections are calculated by assuming that they are dominated by the
SM contributions. For details of the theoretical predictions, see
section~3.2. For each sector, first a model-independent parametrization
of the data is given, and then the fit result in the 
($\sbar^2(0),\,\gzbar^2(0)$) plane.
 
\subsubsection*{\normalsize \bf 4.2.1 Neutral currents in
$\nu_\mu$--$q$ scattering }
 
\pr
For the $\nu_\mu$--$q$ data, the results of the analysis of
ref.\cite{nuq_dat} are adopted. In terms of the model-independent
parameters ($g_L^2$,$g_R^2$,$\delta_L^2$,$\delta_R^2$), the following
fit has been obtained:
%
\dataofnuqorig
%
where the former and the latter errors denote the experimental and the
parametrization errors. 
The correlation matrices for the two types of
uncertainties also quoted in ref.\cite{nuq_dat} are respectively
%
\dataofnuqorigcorr
%
The fitted parameters depend
on the assumed value of the charm quark mass 
($m_c$ in GeV units)\cite{fogli} entering the
slow-rescaling formula\cite{slow_rescaling}
for the charged current cross sections.
The data\cite{nuq_dat} constrain the charm quark mass to
  \begin{eqnarray}
     m_c =1.54 \pm 0.33\gev.
     \label{mc_dat}
  \end{eqnarray}
After summing the experimental and the parametrization errors in
quadrature, and integrating out the $m_c$ dependence of the above
parametrization under the constraint (\ref{mc_dat}), the new
model-independent parametrization of the $\nu_\mu$--$q$ data gives~:
%
\dataofnuq
%
which properly accounts for the uncertainty in $m_c$. The
parametrization (\ref{dataofnuq}) serves as input to our analysis.

By using the theoretical formulae (\ref{g2_d2}) and (\ref{q_alpha})
of section~3.2.1 the data (\ref{dataofnuq}) can be confronted with
the predictions in terms of $\sbar^2(0)$ and $\gzbar^2(0)$. Corrections
due to small, but finite, momentum transfer are evaluated at
  \begin{eqnarray}
    \langle -t \rangle_{n.c.} =\langle -t \rangle_{c.c.} =20\gev^2,
  \end{eqnarray}
in eqs.~(\ref{gcc_nuq}) and (\ref{jw_mu}) and in the running of
$\sbar^2(t)$: see eq.~(\ref{s2b_nuq_num}). The fit result is~:
%
\fitofnuqnotgauss
%
Asymmetric errors are quoted. The non-gaussian behaviour of the $\chi^2$
function reflects the non-linear transformation between the charge form
factors ($\gzbar^2(0),\sbar^2(0)$) and the model-independent parameters
($g_L^2,g_R^2,\delta_L^2,\delta_R^2$), as seen in
Fig.~\ref{figureofnuqparameters}. The strong positive correlation
between the fitted values of $\sbar^2(0)$ and $\gzbar^2(0)$ is a
consequence of eq.~(\ref{dataofnuq}):
the precisely measured combination
$g_L^2 +g_R^2$ in (\ref{dataofnuq}) dominates the total neutral current
cross section off isoscalar targets. The 1-$\sigma$ contour of the
above fit is shown in Fig.~\ref{figureoflencfit}. It can be reproduced
rather accurately by the following parametrization:
%
\fitofnuq
%
which serves merely for estimating the constraints from the
$\nu_\mu$--$q$ experiments. We stress that all the quantitative
analyses in the following sections are performed by fitting directly
to the original parametrization of the data (\ref{dataofnuq}).
 
\subsubsection*{\normalsize \bf 4.2.2 Neutral currents in
$\nu_\mu$--$e$ scattering }
 
\pr
The $\nu_\mu$--$e$ data from the three experiments: CHARM, BNL~E374 and
CHARM-II\cite{nue_dat}, have been summarized in ref.\cite{beyer} in
terms of the model-independent parameters
$s^2_{\nu e}$ and $\rho_{\nu e}$:
%
\dataofnue
%
As explained in detail in section~3.2.2, first the total cross section
$\sigma^{\nu e}$ and $\sigma^{\bar{\nu} e}$ is reconstructed by using
the formula (\ref{sig_nue_eff}), and then the fit is performed by using
the theoretical expressions (\ref{sig_nue}). The
reconstructed cross sections are found to be
  \begin{eqnarray}
    \label{sig_nue_dat}
    \left.
    \begin{array}{lll}
       \sigma^{\nu e}/E_\nu(10^{-42}{\rm cm}^2/\gev)
        &\mm=& 1.56 \pm 0.10
       \\
       \sigma^{\bar{\nu} e}/E_{\bar{\nu}}(10^{-42}{\rm cm}^2/\gev)
        &\mm=& 1.36 \pm 0.09
    \end{array}
    \right\}\quad
    \rho_{\rm corr} = 0.51 \,,
  \end{eqnarray}
and the 2-parameter fit to the above data gives
%
\fitofnue
%
The same result follows if we use the approximation 
(\ref{rho_s2_nue_num}) 
directly to fit the parametrization (\ref{dataofnue}). 
Here $\chi^2_{\rm min} =0$, 
since the number of degrees is $2-2=0$. 
The
result is shown in Fig.~\ref{figureoflencfit} by the 1-$\sigma$ contour.
The weak mixing form factor $\sbar^2(0)$ is measured more
accurately in the $\nu_\mu$--$e$ experiments than in the $\nu_\mu$--$q$
experiments, whereas for $\gzbar^2(0)$ it is the other way around.
 
\subsubsection*{\normalsize \bf 4.2.3  Atomic parity violation }
 
\pr
As for the APV experiments the result of the analysis\cite{apv_dat} on
the parity violating transitions in the cesium atom $(A,Z)=(133,55)$
are used~:
%
\dataofapv
%
The quoted uncertainty is the quadratic sum of experimental and
theoretical errors. After correcting for the vertex and box
corrections\cite{apv_ew} as explained in detail in section~3.2.3, one
finds
%
\fitofapv
%
Here the value $\gzbar^2(0)$=0.5486 stands for the prediction at
$\bar{\rho}$=1 or $T=\delg/\alpha$. The result is shown in
Fig.~\ref{figureoflencfit} by 1-$\sigma$ contours. As anticipated in
the previous section, the correlation between the fitted $\sbar^2(0)$
and $\gzbar^2(0)$ values is opposite to that from $\nu$--$q$ fit.
As a consequence, the constraints on both $\sbar^2(0)$ and $\gzbar^2(0)$
are improved significantly by combining the two types of experiments.
 
\subsubsection*{\normalsize \bf 4.2.4 Polarization asymmetry in 
$e-$D scattering }
 
\pr
Finally, for the SLAC $e$D polarization asymmetry experiment
\cite{ed_dat} a model-independent fit is performed to the original data
by using the two combinations, $2C_{1u}-C_{1d}$ and $2C_{2u}-C_{2d}$ of
the coefficients of the effective parity violating $e$--$q$ neutral
current operators\cite{old_ncfit}: see eq.~(\ref{lpv_eq}). In the quark
parton model with the valence quark approximation the observed
polarization asymmetry is expressed in terms of the above parameters by
  \begin{eqnarray}
    -\frac{A}{Q^2} = \frac{6\,G_F}{5\,\sqrt{2} \ebar^2(-Q^2)}\,
    \biggl\{(2\,C_{1u}-C_{1d}) + (2\,C_{2u}-C_{2d})\,
      \frac{1-(1-y)^2}{1+(1-y)^2} \biggr\},
     \label{a_eq_qpm}
  \end{eqnarray}
which depends on the scaling variable $y$, but not on $x$. The mild
$Q^2$ dependence due to the running of the effective QED charge
$\ebar^2(-Q^2)$ is accounted for.
There have been extensive studies\cite{ed_qcd,ed_ht1}, which show that
the above approximation is in fact valid on more general grounds, but
that it may suffer from higher-twist contributions. We therefore
perform a new model-independent fit to the original data\cite{ed_dat},
and obtain quantitatively the theoretical
uncertainty in the fitted parameters.
 
By taking account of the sea-quark contributions and finite
$R=\sigma_L/\sigma_T$\cite{ed_qcd}, as well as possible higher twist
contributions\cite{ed_ht1,ed_ht2}, the above simple expression for the
asymmetry (\ref{a_eq_qpm}) is modified as follows~:
  \begin{eqnarray}
     -\frac{A}{Q^2} &=&
      \frac{6 G_F}{5\,\sqrt{2} \ebar^2(-Q^2)}
      \biggl\{ (2C_{1u}-C_{1d}) \Bigl(1-\frac{3}{4}c\Bigr)
              + (2C_{2u}-C_{2d}) \Bigl(b +\frac{5}{12}c\Bigr)\biggr\},
     \label{a_eq_corr}
  \end{eqnarray}
with
  \begin{eqnarray}
     b &=& \frac{1 -(1-y)^2}{1 +(1-y)^2 -y^2 \frac{R}{1 +R}}
           \Bigl(1 +8.35 \delta -\epsilon_u -\frac{\epsilon_s}{5}\Bigr),
     \label{b_eq}
    \\
     c &=& 1.34 \delta -\epsilon_s/5\,.
     \label{c_eq}
  \end{eqnarray}
Here the $x$-dependent parameters $\epsilon_u$ and $\epsilon_s$
denote the relative contribution of the sea $u$-quark and that of
$s$ and $\bar{s}$ quarks, respectively, which are parametrized by
  \begin{eqnarray}
     \epsilon_u &=& \epsilon \frac{(1 -x)^4}{\sqrt{x}}\,,
     \\
     \epsilon_s &=& \frac{\epsilon}{3}\frac{(1 -x)^4}{\sqrt{x}}\,.
  \end{eqnarray}
The uncertainty in the factor $\epsilon$ above is estimated to be
  \begin{eqnarray}
     \epsilon &=& 0.1 \pm 0.03\,.
  \end{eqnarray}
The effects of introducing sea-quark contributions in the fit is shown
in Fig.~\ref{figureofed}(a). As found in ref.~\cite{ed_qcd}, the effect
is very small along the tree level SM prediction as shown by the
straight line in the figure. Some representative values of
$\sin^2\theta_W^{}$ in the SM are denoted by blobs.
The longitudinal to transverse cross section ratio $R=\sigma_L/\sigma_T$
is allowed to vary within the rather conservative limits
  \begin{eqnarray}
     R &=& 0.2 \pm 0.2 \,.
  \end{eqnarray}
The effect of introducing the $R$ parameter alone is shown in
Fig.~\ref{figureofed}(b) and the result turns out to be insensitive
to its uncertainty, especially along the tree-level SM trajectory,
confirming the earlier observation of ref.\cite{ed_qcd}.
Finally, the parameter $\delta$ in the factors $b$ and $c$ parametrizes
the higher twist effects as expected in the MIT bag model\cite{ed_ht2}.
Taking as the magnitude of the uncertainty the largest value of the MIT bag
model estimate of ref.\cite{ed_ht2} yields
  \begin{eqnarray}
     \delta &=& (1.58 \pm 1.58) \times 10^{-3}.
  \end{eqnarray}
The effects of introducing the $\delta$ parameter alone are shown
in Fig.~\ref{figureofed}(c). As in the case of the sea-quark
contributions (Fig.~\ref{figureofed}(a)), the effect is negligibly
small along the line representing the tree-level SM prediction. Note
that the higher-twist effects are found to be rather model
dependent\cite{nuq_ht2}. The MIT-Bag model estimates\cite{ed_ht2}
adopted here lead to quite small corrections, as in the neutrino
scattering off isoscalar targets\cite{nuq_ht1}. Further study on the
higher twist effects may be needed to achieve precision measurements of
the electroweak parameters in these reactions.
 
After allowing for all of the above uncertainties, one finds
%
\dataofed
%
with $\chi^2_{\rm min}=9.95$ for 11 data points, that is, a good fit.
The above result is shown in Fig.~\ref{figureofed}(d). Because of the
strong correlation, only a linear combination of the two coupling
factors is measured well.
 
By using the theoretical formulae (\ref{c1c2_vs_m}), a fit is made to
the data (\ref{dataofed}) in terms of the  two parameters $\sbar^2(0)$
and $\gzbar^2(0)$. In order to fix the $q^2$-dependent factors
($Q^2\equiv -q^2$) $\Gamma_1(-Q^2)$, $\gambar_2(-Q^2)$ and $\sbar^2(-Q^2)$
in the amplitudes we choose
$\langle Q^2 \rangle =1.5\gev^2$.
Note, however, that $Q^2$-dependence of each data point\cite{ed_dat}
and that of the QED running coupling $\ebar^2(-Q^2)$ in
eq.~(\ref{a_eq_corr}) have been respected in the model-independent fit
(\ref{dataofed}).
The result is~:
%
\fitofed
%
and shown in Fig.~\ref{figureoflencfit}. Note that the parametrization
(\ref{chisqofed}) is valid only in the vicinity of the SM predictions
$\gzbar^2(0)\sim 0.55$ (but is valid in the whole region of
Fig.~\ref{figureoflencfit}), and that the global $\chi^2_{\rm min}$ is
zero, since the two parameter parametrization (\ref{dataofed}) is
adopted as the original data of our fit.
 
\subsubsection*{\normalsize \bf 4.2.5
Summary of low energy neutral current experiments }
 
\pr
In this section the fits to the electroweak observables in the four
low energy neutral current experiments are summarized. The fit results
are illustrated in Fig.~\ref{figureoflencfit} by 1-$\sigma$ allowed
regions in the ($\sbar^2(0),\,\gzbar^2(0)$) plane. Since all four pieces
of information are consistent with each other, a combined fit can been
performed~:
%
\fitoflenc
%
The fit with $7=9-2$ degrees of freedom is good and its result is shown
in Fig.~\ref{figureoflencfit} by the ellipse with the thick
1-$\sigma$ contour.
 
It is sometimes useful to analyse the neutral current sector with and
without inclusion of the neutrino data, since in some models they
receive different new physics contributions. To this end the fit
is done separately for $\nu_\mu$--$q$ and $\nu_\mu$--$e$ experiments~:
%
\fitofnuonly
%
The fit for the APV and $e$D experiments gives~:
%
\fitofeqonly
%
These two fits are again consistent and their combination reproduces,
of course, the above global fit (\ref{fitoflenc}).
 
\subsection*{\normalsize \bf 4.3  Charged current experiments }
\pr
The $W$ mass measurements have been updated recently by the CDF and D0
collaborations. By combining the most recent measurements\cite{mw93} and
the previous result of PDG\cite{pdg92} one obtains
%
\dataofmw
%
Note that in this analysis the $W$ mass definition follows the LEP
convention\cite{lep1}, as opposed to the pole mass definition: see
eq.~(\ref{mv_lep}). The pole mass should be smaller by about
$0.03\gev$. 
The difference is still negligibly small as compared to the error of
$0.16\gev$. It is worth noting that the $W$ propagator with running
width factor gives a more accurate description of the scattering
amplitudes when no imaginary parts are introduced outside the propagator
factor.
 
The electroweak parameter $\gwbar^2(0)$ is now obtained by combining
the $\mw$ measurement with the $\mu$ life-time parameter $G_F$
(\ref{mw_from_gf})~:
we find
%
\fitofmw
%
where $\delg =0.0055$ is the SM estimate for the process specific correction
to the $\mu$ life-time: see eq.~(\ref{delg_cc}).  No other experiment in
the charged current sector is accurate enough to provide adequate
information for our electroweak analysis.
Precise measurements of the $W$ shape parameters\cite{wwidth_ew} 
would improve 
our knowledge in this sector considerably.
 

\section*{\large \bf 5  Systematic analysis}
\renewcommand{\theequation}{5.\arabic{equation}}
\cleqn
 
\pr
In this section, first the $q^2$-dependence of the two charge form
factors $\gzbar^2(q^2)$ and $\sbar^2(q^2)$ is examined between
$q^2=0$ and $q^2=\mmz$. Next a combined fit in terms of the $S$, $T$
and $U$ parameters is made assuming the $q^2$-dependence of these
charge form factors to be governed by the SM. Finally, only the SM
particles are assumed to contribute to the radiative effects and the
preferred range of the two mass parameters $m_t$ and $\mh$ is searched
for. Also the $\alpha_s$ and $\delta_\alpha$ dependences of the fits are
discussed in detail.
 
\subsection*{\normalsize \bf
5.1  Summary of all experimental constraints on the electroweak
     parameters}
 
\pr
The information on all electroweak precision data has been represented
in the previous sections in terms of the charge form
factor values 
(see eqs.~(\ref{fitofzparameter}),
(\ref{fitoflenc}) and (\ref{fitofmw}))
and is, for convenience, collected in
Table~\ref{tableoffitsummary}.
In addition, the fine structure constant $\alpha$ determining the
charge form factor $\ebar^2(0)=4\pi\alpha$ (see
Tables~\ref{tableofchargeformfactors} and \ref{tableofstepsinanalysis})
has been used as an input parameter.
In calculating $\chi^2$ the model-independent parametrizations of the
original data are used as inputs for the fit~:
eqs.~(\ref{dataofzlineshape})--(\ref{dataofzothers}) for the $Z$ parameters
(section~4.1),
eq.~(\ref{dataofnuq}) for the $\nu_\mu$--$q$ scattering experiments
(section~4.2.1),
eq.~(\ref{dataofnue}) for the $\nu_\mu$--$e$ scattering experiments
(section~4.2.2),
eq.~(\ref{dataofapv}) for the atomic parity violation experiments
(section~4.2.3),
eq.~(\ref{dataofed}) for the $e$--D polarization asymmetry measurements
(section~4.2.3),
and eq.~(\ref{dataofmw}) for the $W$ mass measurements
(section~4.3).
The $\chi^2$ fits in each of the various sectors look all fine and
it is concluded that the whole body of data is consistent with the
assumption of the ${\rm SU(2)_L\times U(1)_Y}$ universality and the
SM dominance of the vertex and box corrections.
 
\subsection*{\normalsize \bf
5.2  Testing the running of the charge form factors}
 
\pr
If there are new particles coupled to the weak gauge bosons with
masses near or below $\mw$ and $\mz$, their signal can be identified as an
anomalous running of the charge form factors\cite{susy_thr,hisz}. 
In principle, the running of all four charge form factors provides 
us with information on new physics contributions via 
eq.~(\ref{runningalpha}) for $1/\bar{\alpha}(q^2)$, 
eq.~(\ref{runningsb2}) for $\sbar^2(q^2)/\bar{\alpha}(q^2)$, 
eq.~(\ref{runninggzb}) for $1/\gzbar^2(q^2)$ 
and via eq.~(\ref{runninggwb}) for $1/\gwbar^2(q^2)$. 
At present, only two of the four form factors, $\sbar^2(q^2)$ and
$\gzbar^2(q^2)$, have been determined with sufficient accuracy at two
different energy scales, $q^2 = 0$ and $\mmz$.
 
The results collected in Table~\ref{tableoffitsummary} yield~:
%
\fitofrunning
%
In the absence of a precise value for $\bar{\alpha}(\mmz)$ the SM 
prediction $\bar{\alpha}(\mmz)_{\rm SM}=1/128.72$ (or, more generally,
$\delta_\alpha \equiv 1/\bar{\alpha}(\mmz)-128.72 = 0$) is used above.
 
Fig.~\ref{figureofrunningvsmh} illustrates SM running of the charge
form factor $\gzbar^2(q^2)$,
  \begin{eqnarray}
    \frac{4\pi}{\gzbar^2(\mmz)} -\frac{4\pi}{\gzbar^2(0)}
     &=&  \frac{1}{4} \Bigl[S_Z(\mmz) -S_Z(0)\Bigr]
     \label{gzbar_running}
  \end{eqnarray}
as a function of $\mh$, together with the experimental constraint
(\ref{fitofrunning}). The $q^2$-dependent $S_Z$ function is defined
in terms of the gauge boson two-point functions in eq.~(\ref{eq:sz_def})
of appendix B. The difference (\ref{gzbar_running}) takes the form
(\ref{runninggzb}) of section~2. The $m_t$ dependence of the SM
prediction is very small compared to the experimental error for
$m_t>100 \gev$.  The SM is consistent with the data as long as the Higgs
boson mass is not too small. Note that the 1-$\sigma$ constraint on
$\mh$, $\mh > 2.9\gev$  $(67\% {\rm CL})$, is obtained merely by
comparing the $Z$ boson coupling strengths at $q^2$=0 and
$q^2\!=\!\mmz$. These values are, however, 
obtained by neglecting the $Z \to Hf\bar{f}$ contribution to 
$\Gamma_Z$, and are 
anyway excluded by direct
searches at LEP ($\mh>63\gev$)\cite{mh_limit}.
 
The Higgs boson does not contribute to the running of the other neutral
current form factors, $1/\bar{\alpha}(q^2)$ and
$\sbar^2(q^2)/\bar{\alpha}(q^2)$. They are affected by loops of charged
particles only, and, for instance, the top quark contributions to the
running of these form factors are parametrized in appendix~B, in
eqs.~(\ref{eq:alphamz_sm}) and (\ref{eq:sbar_running_sm}):
\bsub
\label{running_vs_mt}
\bea
     \frac{1}{\bar{\alpha}(\mmz)}_{\rm SM} \qquad
     &=&\; 128.71 +\delta_{\rm had}
      +0.024 \Bigl(1+5\frac{\alpha_s}{\pi}\Bigr)
      \biggl(\frac{100 \gev}{m_t}\biggr)^2,
\label{alphamz_mt}
\\
     \bigg[ \frac{\sbar^2(\mmz)}{\bar{\alpha}(\mmz)}
           -\frac{\sbar^2(0)}{\alpha}
     \biggr]_{\rm SM}
     &=& -3.09 + \frac{\delta_{\rm had}}{2}
      +0.009\Bigl(1+5\frac{\alpha_s}{\pi}\Bigr)
            \biggl(\frac{100\gev}{m_t}\biggr)^2\,,
\label{sbar_running_mt}
\eea
\esub
The $m_t$ dependences of these runnings are very small for
$m_t \simgt 100\gev$.
 
The running may be appreciable, if there is a charged fermion with mass
near to half the $Z$ mass\cite{susy_thr}. The case of a light wino,
the fermionic partner of the $W$ in the supersymmetric SM,
is shown in Fig.~\ref{figureofrunningfit}:
(a)~$4\pi/\gzbar^2(\mmz) -4\pi/\gzbar^2(0)\,$,
(b)~$\sbar^2(\mmz)/\bar{\alpha}(\mmz) -\sbar^2(0)/\alpha\,$,
and (c)~$\delta_\alpha \equiv 1/\bar{\alpha}(\mmz)-128.72\,$.
The singularity at $m_{\rm wino}^{}=\mz/2$ of the charge form factor
$4\pi/\gzbar^2(\mmz)$ in (a) reflects\cite{mz_thr} the deviation
of the $Z$ line-shape from the Breit-Wigner form assumed both
in the experimental fit and the corresponding theoretical formulae,
and is unphysical.
The 1-$\sigma$ bound on the wino mass, $m_{\rm wino}^{}>46.1\gev$,
as read off from Fig.~\ref{figureofrunningfit} is unrealistic, since the
threshold $2m_{\rm wino}^{}=92.2\gev$ is less than a half width away
from the $Z$-pole. In order to derive constraints on particles very near
to the threshold, one should look for deviations of the $Z$ line shape
from the simple Breit-Wigner form\cite{mz_pole,mz_thr}. When
calculating the predictions for (b) and (c) the hadronic vacuum
polarization contribution to the running of these form factors is set
to $\delta_{\rm had}=0$, 
while the present estimate\cite{piqq_h_latest} is
$\delta_{\rm had}=0\pm0.1$ (\ref{eq:delta_had}). Wino of masses around
$50\gev$ may shift $\delta_\alpha =1/\bar{\alpha}(\mmz) -128.72$ from
its canonical value $\delta_\alpha =0$ by about 0.1, which is of the
same order as the present uncertainty in the SM prediction.
 
It is clearly seen from Fig.~\ref{figureofrunningvsmh} and from
eq.~(\ref{running_vs_mt}) that the results (\ref{fitofrunning})
are consistent with the SM predictions in the range
$\mh>60\gev$ and $m_t>100\gev$. The study of
the two examples, a very light Higgs boson and a supersymmetric wino,
demonstrates that more accurate values of $\sbar^2(0)$ and
$\gzbar^2(0)$ are needed to detect effects of new physics through the
running of the charge form factors. Accurate measurements of the charge
form factor $1/\bar{\alpha}(q^2)$ at $|q^2| \sim \mmz$ should also
provide independent information.
 
Fig.~\ref{figureofncfit} shows the above results in the
($\sbar^2(\mmz),\,\gzbar^2(\mmz)$) plane, where the $Z$ parameter fit
(`LEP$+$SLC') is taken from Fig.~\ref{figureofzparameterfit} for
$\alpha_s=0.12$, and the combined low energy fit of
Fig.~\ref{figureoflencfit} has been rescaled to the $\mz$ scale by
assuming SM running of the two charge form factors, $\sbar^2(q^2)$
and $\gzbar^2(q^2)$. The combined low energy neutral current data
(see Table~\ref{tableoffitsummary} and Fig.~\ref{figureoflencfit}) are
displayed for various choices of $m_t$ and $\mh$ in order to put in
evidence their small, but finite, effects on the running of these form
factors. The four contours are obtained for $m_t$=100, 200$\gev$ and
$\mh$=60, 1000$\gev$. 
At $m_t=175\gev$ and $\mh=100\gev$, the fit 
(\ref{fitoflenc_a}) for the low energy neutral current data can 
be re-parametrized as
%
\fitoflencatmz
%
It is seen from the figure that the low energy
neutral current fit and the $Z$ parameter fit in terms of the charge
form factors $\sbar^2(q^2)$ and $\gzbar^2(q^2)$ are 
in accordance with 
the running of these form factors as predicted by the SM.
 
The thick solid contour marks the result of the fit to all neutral
current experiments as summarized in Table~\ref{tableoffitsummary}
assuming the SM for the running of $\gzbar^2(q^2)$ and $\sbar^2(q^2)$~:
%
\fitofallnc
%
In the global fit the uncertainty due to $m_t$ and $\mh$ in the running
of the form factors is negligible in the range $m_t$=100--200$\gev$
and $\mh$=60--1000$\gev$. The $\chi^2_{\rm min}$ value of 6.6 for
$\alpha_s=0.12$ is acceptable for 15 (=$18-3$) degrees of freedom. In
conclusion, there is no indication of new particles with mass $\simlt
\mz$ in the running of the charge form factors.
 
Note that the errors in (\ref{fitofrunning}) are dominated by those of
the low energy experiments. Further improvements in the low energy
precision experiments are required to detect a signal of relatively
light new particles, should they exist, through anomalous running of
the charge form factors. In comparing the global fit of 
Fig.~\ref{figureofncfit} with the
individual fit to the low energy $NC$ data in 
Fig.~\ref{figureoflencfit}, the fit from the
$\nu_\mu$--$q$ and $\nu_\mu$--$e$ experiments (\ref{fitofnuonly}) are
remarkably consistent with the $Z$ parameter fit of
Table~\ref{tableoffitsummary}, whereas the fit of the $e$--$q$ sector
(\ref{fitofeqonly}) based on the APV and the $e$--D asymmetry measurements
are about 1.5 standard deviations away.
For $m_t=175\gev$ and $\mh=100\gev$, the fit (\ref{fitofnuonly_a}) for 
$\nu_\mu$--$q$ and $\nu_\mu$--$e$ scattering is re-parametrized as 
%
\fitofnuonlyatmz
%
while the fit (\ref{fitofeqonly_a}) for the APV 
and polarized $e$--D experiments gives
%
\fitofeqonlyatmz
%
Further studies of polarization asymmetries in the $e$--$q$ sector as
well as studies of the neutral current processes at
TRISTAN energies might be potentially rewarding.
 
\subsection*{\normalsize \bf
5.3  Testing the 3 parameter universality}
 
\pr
Once the $q^2$-dependence of the charge form factors is assumed to be
governed by SM physics alone, all radiative effects to the gauge
bosons depend on three universal parameters~: $S$, $T$, $U$. They
include the SM radiative effects as well as new physics contributions,
as opposed to the original definitions of ref.\cite{stu}. While the
charge form factors $\gzbar^2(\mmz)$, $\sbar^2(\mmz)$, $\gwbar^2(0)$
can be directly confronted with experiments, the $S,T,U$ parameter fit
suffers from uncertainty in the QED effective coupling
$\bar{\alpha}(\mmz)$, the reason being the fact that the charge form
factors are determined
by the $S,T,U$ parameters under the
($\alpha,G_F,\mz$) constraints (see discussion in section~2.3). The
magnitude of $\bar{\alpha}(\mmz)$ is
controlled by the external parameter
$\delta_\alpha \equiv 1/\bar{\alpha}(\mmz)-128.72$.
 
A 4-parameter fit yields~:
%
\fitofstu
%
from the result of the
global fit summarized in Table~\ref{tableoffitsummary}.
The best-fit values
of $S,T,U$ and $\delb$ are weakly dependent upon $\alpha_s$ and 
$\delta_\alpha$ as quoted explicitly in eq.~(\ref{fitofstu_a}). 
The minimum of $\chi^2$ turns out to be practically independent 
of $\delta_\alpha$. 
We therefore add to the fit the independent knowledge 
$\delta_\alpha=0.0\pm 0.1$\cite{piqq_h_latest}
leading then to~: 
%
\fitofstuchisqmin
%
The correlation between $S$ and $T$ is strong, since they are
constrained by the precisely measured weak mixing form
factor $\sbar^2(\mmz)$ via eq.~(\ref{sbar_approx}).
 
The above results are shown in Fig.~\ref{figureofstufit} by 1-$\sigma$
contours as projections onto the ($S,\,T$), ($S,\,U$), and ($U,\,T$)
planes. The contours are drawn for three $\alpha_s$ values,
$\alpha_s$=0.11 (dashed lines), 0.12 (solid lines) and 0.13 (dash-dotted
lines), and for $m_t=150, 200\gev$ and $\mh=100, 1000\gev$ in the
running of the charge form factors $\sbar^2(q^2)$ and $\gzbar^2(q^2)$
between $q^2=0$ and $q^2=\mmz$. The fit results depend slightly on
$m_t$ and $\mh$ in the above range. The numerical values of
eq.~(\ref{fitofstu}) are obtained for $m_t=175\gev$ and $\mh=100\gev$.
The SM predictions of appendix~C are drawn in Fig.~\ref{figureofstufit}
by lattices in the region $m_t$=100--200$\gev$ and $\mh$=50--1000$\gev$.
 
The fitted $T$ parameter depends only slightly on $\alpha_s$, 
when the parameter $\delb$ is allowed to 
vary freely within the experimental constraints. 
If we fix $\delb$ by a theoretical model, then the $T$ parameter
should have stronger $\alpha_s$ dependence due to the correlation
$-0.31$ between the errors of $T$ and $\delb$ (see section~6.3
for more discussions).
The $S$ parameter depends on $\delta_\alpha$. 
The fitted $S$ value is shifted
by about 0.07 (that is, 20\% of its present uncertainty of 0.33)
for $[\delta_\alpha]_{\rm SM} \approx \delta_{\rm had} =0\pm0.10$.

The parameters $S,T$ and $U$ measure electroweak radiative effects 
in the gauge boson propagators. 
The fit (\ref{fitofstu}) shows that the data favor negative $S$ 
and positive $T$ at $\alpha_s=0.12$ and $\delta_\alpha=0$. 
The point $S=T=U=\delb=0$, which 
represent no electroweak 
radiative effects in the gauge boson propagators nor in the 
$\zbb$ vertex, 
is about 4.5 standard deviations away from the  
minimal for $\alpha_s=0.12$ and $\delta_\alpha=0$. 
However, if in addition the electroweak radiative effects 
are dropped 
in the muon decay by setting $\delg=0$ in eq.~(\ref{gf}),  
then according to the substitution rule (\ref{t_subst}) the 
`no-radiative effects' point becomes $T=0.0055/\alpha=0.75$, 
$S=U=\delb=0$ in the fit (\ref{fitofstu}), 
which is only 2.6 standard deviations away from the minimal. 
Although this result still assumes the SM radiative corrections 
for the remaining vertex/box corrections, it is essentially 
the mechanism that led the authors of ref.~\cite{okun} to 
state that there had not yet been an evidence for genuine 
electroweak radiative effects. 
Our analysis makes it clear that it is more natural 
to interpret significant
radiative effects in the $T$ parameter 
which are approximately cancelled by
the effect of the  radiative effect $\delg$ in the prediction of the
electroweak observables.

The resulting $\chi^2_{\rm min}$ of eq.~(\ref{chisqofstu}) agrees
nearly with that of eq.~(\ref{chisqofallnc}). The effective number of
degrees of freedom is in both cases 15, namely 
$19-4$ respectively $18-3$.
The fit to the $NC$ data contains actually only three parameters, $S$,
$T$ and $\delb(\mmz)$, corresponding to the charge form factors
$\sbar^2(\mmz)$, $\gzbar^2(\mmz)$ and $\delb(\mmz)$ in the global fit. 
The present fit depends in addition upon $U$, 
when the charged current 
data (and thus the forth form factor, $\gwbar^2(0)$) are included.
 
\subsection*{\normalsize \bf
5.4  Testing the Minimal Standard Model}
 
\pr
In the minimal SM, all the parameters $\gzbar^2(\mmz)$,
$\sbar^2(\mmz)$, $\gzbar^2(0)$, $\sbar^2(0)$, $\gwbar^2(0)$ and
$\delb(\mmz)$ depend uniquely upon the two mass parameters $m_t$
and $\mh$. Consequently, the results of the fits summarized in
Table~\ref{tableoffitsummary} are constraining $m_t$ and $\mh$.
We should repeat here that the SM contributions from the 
top-bottom doublet to the form factors are calculated by 
using the simple $O(\alpha\alpha_s)$ two-loop 
formula\cite{rho_pqcd,pivv_pqcd,rho_pqcd_mh2}. 
Non-perturbative $t\bar{t}$ threshold 
effects\cite{rho_thr,rho_thr2,rho_thr3} will affect these corrections 
and the predicted $m_t$ value will shift upwards by as much as 
a few GeV\cite{rho_thr3} from the effect in the $T$ parameter. 
Our approach separates clearly the data analysis in terms of 
the generic form factors and the analysis of the SM 
contributions to these form factors. 
Uncertainties in the latter process can hence be 
studied separately. 
In fact if the SM $m_t$-dependence of the fit is dictated by 
the $m_t$-dependence of the $T$ parameter alone, 
then the sole effect of the non-perturbative threshold 
corrections can be expressed as a rescaling of the $m_t$ parameter 
in the following analysis. 
 
Fig.~\ref{figureofmtmhconstraints} shows the result of the global SM
fit to all electroweak data in the ($\mh,\,m_t$) 
plane\cite{ellisfogli,okun_sm} 
for three representative $\alpha_s$ values. 
The ``$\times$'' indicate  the
minimum of $\chi^2\,$; 7.4, 6.6, 10.3 for $\alpha_s=0.11$, 0.12, 0.13, 
respectively, 
the inner contours
correspond to 1-$\sigma$, the outer to $\chi^2=\chi^2_{\rm min}+4.61$
(that is, 90\,\%~CL). Dashed lines show the best $m_t$ values
for a given $\mh$. Note the positive correlation between the
preferred values of $m_t$ and $\mh$, which is found to be independent
of the assumed $\alpha_s$ value. On the other hand, the preferred range
of $\mh$ depends rather sensitively on $\alpha_s$. For the cases
$\alpha_s(\mz) =0.11$ and $0.12$ smaller $\mh$ values are preferred,
whereas for $\alpha_s(\mz) = 0.13$ larger $\mh$ is slightly favored.
If the lower bound for $\mh$, $\mh>63\gev$ at 95\%~CL measured by the
LEP experiments\cite{mh_limit}, 
is imposed, $m_t$ below 100~$\gev$ is clearly
disfavored for all $\alpha_s$, in agreement with the directly
established lower top mass limit\cite{top_cdf_limit,top_d0}.
 
The $\chi^2$ function in the  global fit to all electroweak data can be
represented in terms of the four parameters $m_t$, $\mh$,
$\alpha_s(\mz)$ and $\delta_\alpha$ 
together with the constraint
$\delta_\alpha=0.0 \pm 0.1$\cite{piqq_h_latest} by~:
  \chisqsm
where
  \fitofmt
and
  \chisqminsm
Here $m_t$ and $\mh$ are measured in GeV. This parametrization
reproduces the exact $\chi^2$ within a few \% accuracy in the range
$100\gev<m_t<250\gev$, $60\gev<\mh<1000\gev$ and 
$0.10<\alpha_s(\mz)<0.13$. The best-fit value of $m_t$ for a given set
of $\mh$, $\alpha_s$ and $\delta_\alpha$ is readily obtained from
eq.~(\ref{fitofmtbest}) with its approximate error of
(\ref{fitofmterror}), mutatis mutandis for $\mh$. 
Due to the quadratic
form it is easy to get the $\alpha_s$ or $\delta_\alpha$ independent
results. 
Also additional constraints on the external parameters 
$\alpha_s$ and $\delta_\alpha$, such as those from their improved 
measurements, can be 
discussed without difficulty.
As explained in section~4.1, the SM does not fit well the ratio $R_b$.
If we remove from our global fit the data on $R_b$, we find that
the best-fit $m_t$ value above becomes larger by 3.9~GeV, almost
independent of $\mh$ and $\alpha_s$, and the $\chi^2_{\rm min}$
decreases by 2.4.
 
Fig.~\ref{figureofchisqvsmt} displays the overall $\chi^2$ of the SM
fit, $\chi^2_{\rm SM}$, as function of $m_t$ for $\mh=60$, 300,
1000~GeV and $\alpha_s(\mz)=0.11$, 0.12, 0.13. Also the uncertainty
due to $\delta_\alpha$ is shown for three cases, $\delta_\alpha=-0.1$
(a), $0$ (b), $+0.1$ (c). The results of the parametrization
eq.~(\ref{total_chisqsm}) is shown by the dotted line.
It is remarkable to see that the present knowledge of $\delta_\alpha$ to
$\pm 0.10$ affects the best-fit value of $m_t$ by about 5 GeV,
while the uncertainty in $\alpha_s$ of $\pm 0.01$ affects
it by about 2 GeV. This observation emphasizes the importance of the
asymmetry measurements for the prediction of $m_t$ through
$\sbar^2(\mmz)$, where the dependence on $\delta_\alpha$ in the SM is
not negligible: see eq.~(\ref{sbar_approx}).
On the other hand, the $\alpha_s$-dependence of the fitted $m_t$ comes
from the constraint due to the $Z$ total width, $\Gamma_Z$, which in
turn is sensitive to $m_t$ mainly through $\gzbar^2(\mmz)$.
We come back to this point in the next section when discussing the
new left-right asymmetry measurement\cite{alr94}.
 
In Fig.~\ref{figureofchisqvsmh} the overall  $\chi^2$ is plotted
as functions of $\mh$ for $m_t=$120, 140, 160, 180, 200~GeV and
$\alpha_s(\mz)=0.11$, 0.12, 0.13 setting $\delta_\alpha=0$.
The dotted lines indicate our approximation $\chi^2_{\rm SM}$ of
(\ref{total_chisqsm}). Obviously, the best-fit value of
$\mh$ depends very sensitively on the $m_t$ and $\alpha_s$ values.
A small value of the Higgs mass is favored for $m_t<140\gev$, values
of a few hundred GeV for $m_t$ around $160\gev$ and large values for
$m_t>180\gev$.
The preference of lighter $\mh$ is
more pronounced for small $\alpha_s$, while heavier $\mh$ for
larger $\alpha_s$. However, the $\mh$ dependence of $\chi^2$ is very
mild and  meaningful upper bound on $\mh$ can only be obtained for
small $\alpha_s$ and small $m_t$. The upper and lower bounds on $\mh$
will be discussed more quantitatively in section~6 after inclusion of
the new left-right asymmetry data\cite{alr94}.
 
For given $m_t$ and $\mh$
the QCD coupling $\alpha_s(\mz)$ may be
extracted within the SM from the electroweak data alone with the result~:
\bsub
\label{alphas_sm}
\bea
    \alpha_s &=& \langle\alpha_s\rangle \pm 0.0060\,,
    \\
    \langle\alpha_s\rangle &=&
      0.1165 -0.00085\,\Bigl(\frac{m_t}{100}\Bigr)^2
             +0.00031\,\Bigl(\ln \frac{\mh}{100}+2.6\Bigr)^2
             -0.0006\,\frac{\delta_\alpha}{0.10}\,,
\label{alphas_sm_bar}
\eea
\esub
where $m_t$ and $\mh$ are measured in GeV. The above
parametrization
reproduces well the $\alpha_s$ dependence of the $\chi^2$ 
function (\ref{total_chisqsm}) in the range $100\gev <m_t<200\gev$ and
$60\gev <\mh<1000\gev$. The error on $\alpha_s$ determined
from the electroweak data is found to be approximately $0.0060$,
almost independently of the assumed $m_t$, $\mh$ and
$\delta_\alpha$,
while the mean value $\langle\alpha_s\rangle$ 
is slightly sensitive to them;
\bea
 \langle\alpha_s\rangle =
 \left\{\begin{array}{lll}
             0.1159  & \mbox{for} & (m_t,\,\mh) = (150,\,60)\gev \\
             0.1153  & \mbox{for} & (m_t,\,\mh) = (175,\,60)\gev \\
             0.1145  & \mbox{for} & (m_t,\,\mh) = (200,\,60)\gev \\
             0.1220  & \mbox{for} & (m_t,\,\mh) = (150,\,1000)\gev \\
             0.1214  & \mbox{for} & (m_t,\,\mh) = (175,\,1000)\gev \\
             0.1206  & \mbox{for} & (m_t,\,\mh) = (200,\,1000)\gev
            \end{array} \right.\,,
\eea
for $\delta_\alpha=0$.  There is a tendency in the SM fit to prefer
larger $\alpha_s$ for larger $\mh$.
 
Furthermore, if all radiative effects are assumed to be dominated
by the SM contributions, the present electroweak data have some
sensitivity to the parameter
$\delta_\alpha \equiv 1/\bar{\alpha}(\mmz)-128.72$.
By excluding the last term in eq.~(\ref{chisqminsm}), 
$(\delta_\alpha/0.1)^2$\cite{piqq_h_latest}, 
the electroweak data alone provide the constraint~:
\bsub
\label{dela_fit}
\bea
  \delta_\alpha &=& \langle\delta_\alpha\rangle \pm 0.24\,,
  \\
  \langle\delta_\alpha\rangle
    &=& 0.010 - 0.139\, \frac{m_t -150}{10}
              + 0.246\, \ln\frac{\mh}{100}
              - 0.112\, \frac{\alpha_s -0.12}{0.01}\,,
\label{dela_fit_bar}
\eea
\esub
where $m_t$ and $\mh$ are measured in GeV. The above
parametrization is valid in the range $120\gev<m_t<200\gev$,
$60\gev<\mh<1000\gev$ and $0.11<\alpha_s<0.13$. For some representative
$m_t$ and $\mh$ values the exact evaluation of the $\chi^2$
function leads to~:
\bea
  \langle\delta_\alpha\rangle =
    \left\{\begin{array}{rll}
           -0.09  & \mbox{for} & (m_t,\,\mh) = (150,\,60)~\gev \\
           -0.45  & \mbox{for} & (m_t,\,\mh) = (175,\,60)~\gev \\
           -0.87  & \mbox{for} & (m_t,\,\mh) = (200,\,60)~\gev \\
            0.59  & \mbox{for} & (m_t,\,\mh) = (150,\,1000)~\gev \\
            0.25  & \mbox{for} & (m_t,\,\mh) = (175,\,1000)~\gev \\
           -0.12  & \mbox{for} & (m_t,\,\mh) = (200,\,1000)~\gev
    \end{array} \right.\,,
\label{delabar_fit}
\eea
for $\alpha_s$=0.12. The above fit is consistent with the direct
measurement $[\delta_\alpha]_{\rm SM} \approx \delta_{\rm had} =
0\pm0.10$ when $m_t$ and $\mh$ are in the preferred range in
Fig.~\ref{figureofmtmhconstraints}.
This confirms the importance of the direct $\delta_{\rm had}$
measurement in constraining the model parameters from the electroweak
precision measurements.
 
Considering the $\chi^2_{\rm min}$ per degree of freedom (see
parametrization (\ref{total_chisqsm}) and
Figs.~\ref{figureofmtmhconstraints}--\ref{figureofchisqvsmh})
the SM predictions provide a good description of the data over a
still wide range of $m_t$ and $\mh$ for the values of
$\alpha_s$ and $\delta_\alpha$ in the ranges: $0.11 \simlt \alpha_s
\simlt 0.13$ and $-0.1 \simlt \delta_\alpha \simlt 0.1$. In conclusion,
the analysis of the present precision experiments does not show a
signal of new physics beyond the SM.
 

\section*{\large \bf 6  Discussion }
\renewcommand{\theequation}{6.\arabic{equation}}
\cleqn
 
\pr
In this section, the consequences of the update of LEP data, the new
precision measurement of the left-right asymmetry at SLC\cite{alr94}
and the impact of a direct top mass measurement are considered.
Finally, the predictions of all electroweak observables within the
SM are discussed.
 
\subsection*{\normalsize \bf 6.1 Update of LEP data}
 
\pr
Recently the LEP Electroweak Working Group has published a
report\cite{lep94} summarizing the combination of preliminary LEP data
for the 1994 La Thuile and Moriond conferences. During 1993 the four
LEP experiments have performed a high precision scan roughly 1.8 GeV
above and below the $Z$ resonance and within 200~MeV of $\mz$.
The new $Z$ shape parameters agree with the ones quoted in section~4.1
within one standard deviation. The $Z$ mass moved to 91.1895 $\pm$
0.0044 GeV with improved uncertainty. 
Changing of the `constant' $\mz$ from $91.187\gev$ to $91.1895\gev$ 
does not lead to noticeable effects in the analysis.
The total $Z$ width increased to
2.4969 $\pm$ 0.0038 GeV with considerably reduced uncertainty, also the
forward-backward lepton asymmetry increased to 0.0170 $\pm$ 0.0016.
Other parameters, $\sigma_h^0$, $R_\ell$, $R_b$, 
have changed very little.
The correlations in the $Z$ line-shape 
parameter fit have become slightly smaller.
 
For the time being no attempt has been made to incorporate the
updated values, since the analyses of the 1993 data are still
preliminary.
 
\subsection*{\normalsize \bf 6.2 %
The new left-right asymmetry data at SLC}
 
\pr
As emphasized in sections~3 and 4, the left-right asymmetry as well as
the other asymmetry measurements at LEP have the advantage of
determining the universal parameter $\sbar^2(\mmz)$ almost independently
of the other form factors, $\gzbar^2(\mmz)$ and $\delb(\mmz)$, and
almost unaffected by uncertainty in $\alpha_s$. Since the parameter
$\sbar^2(\mmz)$ is directly related to the $\msbar$ coupling
$\shat^2(\mu)$, these asymmetry measurements are particularly important
for the GUT studies.
 
The new measurement of the left-right asymmetry\cite{alr94},
\dataofnewalr
implies
\fitofsbarfromnewalr
This value is 2.5 standard deviations smaller than 
(\ref{fitofsbarfromasymmetries}). Excluding the possibility of a shift
caused by a systematic effect this measurement may be considered as a
statistical fluctuation and then be combined with the other asymmetry
data on the $Z$-pole, that is, the lepton ($e$, $\mu$, $\tau$)
forward-backward asymmetry\cite{lep93}, the $\tau$ polarization
asymmetry\cite{lep93}, the left-right asymmetry\cite{alr94} and the
quark ($b$,$c$) forward-backward asymmetries\cite{lep93}, 
as well as with the old left-right asymmetry data 
from SLD\cite{alr92}. 
The result is
  \begin{eqnarray}
     \sbar^2(\mmz) = 0.2302 \pm 0.0005\,.
     \label{fitofsbarfromasymmetrieswithnewalr}
  \end{eqnarray}
The new average (denoted by ``ALL'')
is shown in Fig.~\ref{figureofsbarfromasymmetries} together
with the individual contributions.%
\footnote{
\label{fn:new_and_old_alr}
\normalsize \baselineskip 18pt
In Fig.~\ref{figureofsbarfromasymmetries} and 
in the following analysis, we use the combined
result of \protect\cite{alr92} and \protect\cite{alr94} 
as the data for $A_{\rm LR}^0$: 
$A_{\rm LR}^0 = 0.1637 \pm 0.0075$ 
gives $\sbar^2(\mmz) = 0.2284 \pm 0.0010$.
} 
Note that $\sbar^2(\mmz)$ derived
from the $\tau$ forward-backward asymmetry is as small as
(\ref{fitofsbarfromnewalr}) from the new left-right asymmetry.
Although the inclusion of the new left-right asymmetry lowers the
$\sbar^2(\mmz)$ fit value by about 1.5 standard deviations, the
quality of the fit ($\chi^2$ = 6.6 for 5 degrees of freedom) does not
indicate an inconsistency with the other data, as may be seen also from
the histogram of the distribution in the figure.
 
With the proviso of excluding a shift due to systematic error sources
we include the data (\ref{dataofnewalr}) into our global analysis,
and discuss its effect by comparing the results with those obtained in
section~4.
The 3 parameter fit to the $Z$ parameters only gives
%
\fitofzparameterwithnewalr
%
The result is shown in the $(\sbar^2(\mmz),\,\gzbar^2(\mmz))$ plane by
the thick lines of Fig.~\ref{figureofzparameterfitwithnewalr}(a) for
three values $\alpha_s=0.11, 0.12, 0.13$ along with the old
fits (thin lines) copied from Fig.~\ref{figureofzparameterfit}.
The SM prediction for $\delta_\alpha = 0$ is also shown in the range
$100\gev$\,$<m_t<$\,$240\gev$ and $1\gev$\,$<\mh<$\,$1000\gev$.
It can be seen that the new $A_{\rm LR}$ measurement by itself
implies large $m_t$
($m_t\simgt 200\gev$) for $\mh>50\gev$. The combined fit, however,
favors $m_t\sim 180\gev$ for $\mh\sim 100 \gev$.
The remaining two parameters $\gzbar^2(\mmz)$ and $\delb(\mmz)$
are less affected.
The $\chi^2_{\rm min}$ per degree of freedom is
$8.4/6$ for $\alpha_s=0.12$, which is fine.
 
Next, the 4-parameter fit in terms of $S,T,U$ and $\delb$ is 
performed analogously to the one in section~5.3.
Combining the above result with
eq.~(\ref{fitoflenc}) from the low energy neutral current experiments
and eq.~(\ref{fitofmw}) from the $W$ mass measurements leads to
%
\fitofstuwithnewalr
%
where $m_t = 175 \GeV$ and $\mh=100 \GeV$ are used to calculate 
SM running of the form factors between $q^2=0$ and $q^2=\mmz$.
Fig.~\ref{figureofzparameterfitwithnewalr}(b) shows the 1-$\sigma$
contours in the ($S,\,T$) plane for the three values $\alpha_s =
0.11, 0.12, 0.13$ for $\delta_\alpha=0$. The old fits
(\ref{fitofstu}) are also shown by thin lines.
The results are insensitive to the above ($m_t$, $\mh$) values assumed
in the running of the charge form factors in the region
$m_t>100\GeV$ and $\mh>50\GeV$, although they are considerably
modified for $\mh \simlt 50 \gev$ (see Fig.~\ref{figureofrunningvsmh}).
It is worth noting this qualitative
difference between the fit to $\sbar^2(\mmz)$ and $\gzbar^2(\mmz)$ and
that to $S$ and $T$. As a matter of fact, the experiments on the $Z$
resonance are far more precise than those from the low energy neutral
current experiments implying that the global fit to all the electroweak
measurements in the neutral current sector measures essentially
$\sbar^2(\mmz)$ and $\gzbar^2(\mmz)$. In the SM the two charge form
factors can be calculated for arbitrary $m_t$ and $\mh$, as shown in
the figure for $m_t$=100--240$\gev$ and $\mh$=1--1000$\gev$. On the
other hand, in our definition, the $T$ parameter determines
$\gzbar^2(0)$ rather than $\gzbar^2(\mmz)$. Hence, only if the running
of the $\gzbar^2(q^2)$ between $q^2=0$ and $q^2=\mmz$ is small, can we
make the global fit to the $S,T$ parameters. For this reason we
restrict the SM predictions to the region $\mh=$50--1000$\gev$ in the
($S,T$) figure. It is remarkable that the electroweak data including
the new left-right asymmetry measurement clearly favor negative $S$,
thus putting severe constraints on technicolor models\cite{stu}. Note
that in the ($S,\,T$) plane only the $S$ parameter is strongly affected
by the new $A_{\rm LR}$ data, while the $T$ parameter is constrained,
independent of the $S$ parameter, by $\gzbar^2(\mmz)$ from $\Gamma_Z$.
 
Next, the impact of the left-right asymmetry measurement on the SM
fit is discussed using all electroweak data.
Fig.~\ref{figureofmtmhconstraintswithnewalr} shows the results of the
SM fit in the ($m_t,\,\mh$) 
plane for $\alpha_s(\mz)\,$=$\,0.11,0.12,0.13$,
and for $\delta_\alpha=-0.1$~(a), 0~(b), and  $+0.1$~(c).
The contours of $\chi^2=\chi^2_{\rm min}+1$
and $\chi^2=\chi^2_{\rm min}+4.61$ are shown by thick lines.
The minima of $\chi^2$ in the figure are marked by crosses~:
12.1, 11.4, 15.7 for $\alpha_s=0.11$, 0.12, 0.13, respectively, 
for $\delta_\alpha=0$. 
The 1-$\sigma$ contour for each $\alpha_s$ value is now clearly outside
the physical region allowed by the direct Higgs searches at LEP
($\mh>63\gev$, denoted by ``LEP limit'' in the figures), although the
$\mh$ dependence of the $\chi^2$ is very mild for $\alpha_s\simgt 0.12$.
The result favoring a light Higgs boson reflects the fact that the new
left-right asymmetry measurement shifts the $S$ parameter to negative
values.
 
Finally, the status of the SM fit is studied in detail as in section~5.
To this end the representation of the $\chi^2$ of the SM fit
including the new left-right asymmetry data is obtained 
(analogous to section~5.4)~:
%
\chisqsmwithnewalr
%
where
%
\fitofmtwithnewalr
%
and
%
\chisqminsmwithnewalr
 
Fig.~\ref{figureofchisqvsmtwithnewalr} (in analogy to the previous
results of Fig.~\ref{figureofchisqvsmt}) shows the total $\chi^2$ of the
SM fit as functions of $m_t$ for $\mh=$60, 300, 1000~GeV and
$\alpha_s(\mz)=0.11$, 0.12, 0.13. The uncertainty $\delta_\alpha$ is
shown for three cases~: $\delta_\alpha=-0.1$~(a), 0~(b), $+0.1$~(c).
The dotted lines are obtained by the approximate formulae
(\ref{total_chisqsmwithnewalr}). It is obvious
from Fig.~\ref{figureofchisqvsmtwithnewalr} and
Fig.~\ref{figureofchisqvsmt}, or from 
eq.~(\ref{fitofmtbestwithnewalr}) and
eq.~(\ref{fitofmtbest}), that the best-fit value of $m_t$ is shifted by
about $+17\gev$ for given $\mh$, $\alpha_s$ and $\delta_\alpha$ values.
Here again the uncertainty of $\delta_\alpha$ is important for the top
mass prediction, as observed from (\ref{fitofmtbestwithnewalr}) and
Fig.~\ref{figureofchisqvsmtwithnewalr}~:
$\delta_\alpha=\pm 0.1$ causes a shift $\mp 5\gev$ in the best-fit
value $\langle m_t\rangle$. 
The $\alpha_s$-dependence of the $\langle m_t\rangle$ values is
considerably weakened.
 
Fig.~\ref{figureofchisqvsmhwithnewalr} (in correspondence to
Fig.~\ref{figureofchisqvsmh} in the previous fit) shows the total
$\chi^2$ of the SM fit including the new left-right asymmetry
data\cite{alr94}, as functions of $\mh$ for $m_t=$100--200~GeV.
Three $\alpha_s$ cases are displayed;
$\alpha_s(\mz)=0.11$~(a), 0.12~(b), and 0.13~(c), all for
$\delta_\alpha=0$. The dotted lines show our approximation
(\ref{total_chisqsmwithnewalr}), valid only in the
`physical' region $63\gev < \mh < 1000\gev$. As seen, the best-fit
value of $\mh$ is as low as $10\gev$ for $m_t \simlt 150\gev$, while it
increases with $m_t$ for $m_t \simgt 150\gev$. This trend can also be
appreciated from the global fit of
Fig.~\ref{figureofzparameterfitwithnewalr}(a) in the
($\sbar^2(\mmz),\,\gzbar^2(\mmz)$) plane.
 
\subsection*{\normalsize \bf 6.3
The impact of the top mass measurement }
 
\pr
The top quark searches of the two collaborations CDF and D0 at the
Tevatron entered in their decisive phase\cite{top_cdf,top_d0}. 
The range of values for the
top quark mass coming out of the fits to the electroweak precision data
is within reach for direct observation in the detectors at the Tevatron.
In view of the recent publication by the CDF collaboration\cite{top_cdf}
it is instructive to examine the impact of the constraint
  \begin{eqnarray}
     m_t=174\pm 16\gev.
     \label{mt_dat}
  \end{eqnarray}

First, the $m_t$-dependence of the global fit to the electroweak
data in terms of the charge form factors $\sbar^2(\mmz)$ and
$\gzbar^2(\mmz)$ is considered, now assuming SM dominance to the
$\delb(\mmz)$ form factor. Using the $Z$ parameters 
including the new $A_{\rm LR}$ measurement\cite{alr94} 
one obtains
%
\fitofzparameterwithmtfixed
%
which is a good approximation in the region $150\gev <m_t<200\gev$.
Here the errors and the correlations are almost independent of the
$m_t$ value. The fit to all electroweak data gives
%
\fitofstuwithmtfixed
%
The appearance of essentially the same combination 
 \begin{eqnarray}
    \alpha_s -0.12-0.0022\,\frac{m_t-174}{16}
 \end{eqnarray}
in eqs.~(\ref{fitofzparameterwithmtfixed}) 
and (\ref{fitofstuwithmtfixed})
is the expected consequence of the strong correlation between
$\delb(\mmz)$ and $\alpha_s$ as discussed in detail in section~4.
 
Next, the above constraint on the top quark mass (\ref{mt_dat}) is
imposed on the $\chi^2$ function of the SM fit in the previous
subsection. The result displayed in
Fig.~\ref{figureofmtmhconstraintswithnewalrandmt}
shows the improvement over
Fig.~\ref{figureofmtmhconstraintswithnewalr}.
Now, light Higgs boson masses are moderately favored, as a consequence
of the constraint (\ref{mt_dat}) being somewhat larger than the
best-fit value of $m_t$ obtained by freely fitting the two parameters,
$m_t$ and $\mh$ without the $\mh$ constraint from LEP.
 
It is instructive to anticipate the impact a precise measurement of
the top mass would have in the context of the present electroweak data. 
The top quark mass is expected to be measured eventually 
with an uncertainty of 
about 5~GeV at Tevatron by the end of this decade\cite{mt_tevatron},
which may be improved to about 3~GeV at an upgraded Tevatron\cite{mt_b0}.
The uncertainty is expected to be reduced 
by an order of magnitude to a few hundred MeV 
at next linear $e^+e^-$ colliders\cite{mt_lc}.
The top mass acts then like an external parameter and the only remaining 
free parameter is the Higgs mass. 
Fig.~\ref{figureofmhconfidencelevelwithnewalr} 
shows the 95\%~CL constraints for three values
of $\alpha_s(\mz)=0.11, 0.12, 0.13$, and for $\delta_\alpha=0$.
For small $m_t$ values, rather strict upper bounds on $\mh$ are
found. 
On the other hand no strict upper bound is obtained for
$m_t\simgt 180\gev$.
In the region $160\gev <m_t< 190\gev$, the upper bound on $\mh$
at the 95\%~CL is approximately expressed as
\bea
    \ln\frac{\mh}{100} < \left\{
      \begin{array}{lll}
        1.20 +1.12\frac{m_t-174}{16} &\mbox{for}& \alpha_s = 0.11 \\[1mm]
        1.55 +1.25\frac{m_t-174}{16} &\mbox{for}& \alpha_s = 0.12 \\[1mm]
        1.95 +1.45\frac{m_t-174}{16} &\mbox{for}& \alpha_s = 0.13
      \end{array} \right. \,,
\eea
where $m_t$ and $\mh$ are measured in GeV.
The upper bound is lower for smaller $m_t$. Since these bounds are very
sensitive to the $m_t$ value as well as the assumed $\alpha_s$ value,
further accurate measurements of $m_t$ are needed to obtain more
stringent limits on $\mh$. Nevertheless, it is remarkable that the
constraint on the top quark mass (\ref{mt_dat}) would favor a
relatively light Higgs boson, $\mh=O(100\gev)$, 
which may exist in the minimal SUSY-SM.
Also, the direct $\mh$ bound from LEP\cite{mh_limit} 
$\mh>63\gev\,$(95\%CL) implies that the top quark should be 
heavier than about $145\gev$, since otherwise the Higgs boson 
should have been discovered at the 95\%CL. 
This lower $m_t$ bound changes by about $\mp 5\gev$ for 
$\delta_\alpha=\pm 0.1$. 

One comment is in order.
Though our approximate formulae of the $\chi^2$ for the SM fit,
(\ref{total_chisqsmwithnewalr}), reproduce the
exact result within about 1\% accuracy in the Higgs mass range 
$63\gev<\mh <1000\gev$ 
as seen Figs.~\ref{figureofchisqvsmtwithnewalr}--%
\ref{figureofchisqvsmhwithnewalr}, one should not use them in
finding the confidence levels of $\mh$ for small $m_t$, since the
neighborhood of the minimum of the $\chi^2$ is outside the above range,
where the exact $\chi^2$ and the approximate formulae are fairly
different as seen from Fig.~\ref{figureofchisqvsmhwithnewalr}.
 
\subsection*{\normalsize \bf 6.4  Summary of the data and the SM fit}
 
\pr
Table~\ref{tableofsummary} collects the complete list of all input
data (except for $\alpha$, $G_F$ and $\mz$) and the corresponding
minimal SM predictions for several sets of $(m_t, \mh, \alpha_s)$
values. The total $\chi^2$ of each sector is also given in the table.
The correlations between the errors (given in the text) are properly
taken into  account. The numbers demonstrate that the present
electroweak experiments are well described by the SM, perhaps except
for a combination of a light top and a heavy Higgs, see the case
$(m_t,\mh)=(150,1000)$GeV in the last column of the table.
Its total $\chi^2$ at $\alpha_s=0.12$ is 
30.22 for 19 data points, whose 
$\chi^2$-probability corresponds to 95\%.
In Table~\ref{tableofsummary} also the results of two approximations
are listed. The `no-EW' column is obtained by dropping all electroweak
corrections to the two-point functions ($S=T=U=0$) as well as vertex/box
corrections ($\delg=\delb=\Gamma_i=B_{ij}=0$), while retaining the QED
running of the charge form factors $\bar{\alpha}(q^2)$ and
$\sbar^2(q^2)/\bar{\alpha}(q^2)$ due to light particles
(excluding the $W$ and $t$ contributions). The `IBA' column shows the
result of the improved Born approximation, where all the gauge boson
propagator corrections are retained and hence all the four charge form
factors are kept exact, but all vertex/box corrections
($\delb=\Gamma_i=B_{ij}=0$) dropped, except for $\delg$ in
the $\mu$ decay.
 
It is amazing to note that the `no-EW' hypothesis is, from a
statistical point of view, not completely unacceptable. The comparison
between the `no-EW' and the `IBA' hypothesis is surprising, since in
the `IBA' prediction all the most important electroweak corrections
are supposed to be contained, including the dominant $m_t^2$
corrections in the $T$ parameter. It is even more striking, if
$\delg$ in IBA is set to 0 (this may be called a genuine IBA), to
obtain $\sbar^2(\mmz)=0.2286$ for $m_t=175\gev$ and the total $\chi^2$
jumps nearly to 100.
The measurement of the $Z$ parameters are equally well described
by the `no-EW' and the full calculation for $m_t=175\gev$. 
This confirms the observation of ref.~\cite{okun,kang} that there is no
evidence of the genuine electroweak correction in the present
electroweak precision experiments.
As explained in sections~2.3 and 5.3, 
this is because of the accidental 
cancellation between the propagator corrections and the remaining 
vertex/box corrections.
The no-EW calculation for 
all the asymmetries on the $Z$-pole 
give almost the same values
with the predictions of the exact calculation for $m_t=175\gev$ and
$\mh=100\gev$. As discussed in section~4.1, $R_b$ also gives a large
contribution to $\chi^2$ in the full calculation. For a large top
quark mass, the $\zbb$ vertex from factor $\delb(\mmz)$ decreases 
(see
Fig.~\ref{figureofstusm}), and hence it gives smaller $R_b$. For this
reason the present data of $R_b$ agree better with the no-EW and the
IBA calculations, where $\delb(\mmz)$ is set to 0.
 
The most significant differences between the no-EW prediction
and the full SM predictions in ($m_t=175\gev\,,\mh=100\gev$) column
appear actually in the predictions for the low energy $\nu_\mu$--$q$
scattering and the atomic parity violation experiments. When evaluating
the no-EW and IBA predictions, all the external photonic corrections
and the tree-level propagator effects are retained, as explained in
section~3.2. The difference between the full SM predictions and the
no-EW or IBA predictions is mainly caused by the absence of the $WW$
box contribution in the latter.
 
Another significant difference appears in the predictions for $\mw$,
where the no-EW prediction (79.95$\gev$) is much smaller than the
observed value, $80.24\pm0.16\gev$. This observation has also 
been made in
refs.~\cite{hioki,okun2,kang}. In contrast to the low energy neutral
current experiments above, the difference here is due to $S$ and $U$
contributing to $\mw$ proportional to $-0.294S+0.332U$ (c.f.
eq.~(\ref{mw_stu})). For instance, the full SM for $m_t=175~\gev$ and
$\mh=100~\gev$ predicts $S=-0.2323$ and $U=0.3577$, which implies for
$\mw$ a shift by $0.19~\gev$ corresponding to more than one standard
deviation.
 
Finally, Fig.~\ref{figureofchisqvsmhwithnewalrwideeachsector} shows
separately for each sector the $\chi^2$ of the SM fit as functions of
$\mh$ for $m_t=100-200\gev$. In all sectors, the preferred Higgs mass
range is strongly correlated with the assumed top mass. For
$m_t$=170--180$\gev$, a light Higgs boson is favored by the
$Z$ parameter measurements and by the low energy neutral current
experiments, while the data of $\mw$ alone prefer a rather heavy Higgs
boson. Although the overall trend of the total $\chi^2$ shown
in Fig.~\ref{figureofchisqvsmhwithnewalr}
is dominated by the contribution from the Z parameter measurements,
also the $W$ mass measurement plays an important role for some
$m_t$, $\mh$ ranges. For instance, a relatively light Higgs boson
($\mh\simlt 100\gev$) appears incompatible with a heavy top quark
($m_t\sim 200\gev$) by the $\mw$ measurement alone.
 

\section*{\large \bf 7  Conclusions}
\renewcommand{\theequation}{7.\arabic{equation}}
\cleqn
 
\pr
A novel method to confront electroweak data with theory at the quantum
level has been proposed and a comprehensive analysis has been carried
out. The electroweak observables were first expressed in terms of
model-independent parameters, which in turn were expressed in terms of
$S$-matrix elements of processes with four light fermions and
factorized into the short-distance part and the part related with the
external QED/QCD corrections for neutral current processes.
Only two quantities, the Fermi coupling constant $G_F$ and the $W$ mass
are considered for charged current processes.
Since all electroweak observables were expressed in terms of helicity
amplitudes, they can be evaluated in an arbitrary model on and off
the $Z$ resonance. Our formalism is hence useful to study effects of
tree-level deviations from the SM, arising, for instance, from an
additional $Z$ boson.
After careful evaluation of the external QED/QCD corrections, the
theoretical predictions were confronted with experiment in three
steps of increasing theoretical stringency.
First, in the class of theories respecting the electroweak gauge group
${\rm SU(2)_L \times U(1)_Y}$ broken spontaneously to ${\rm U(1)_{EM}}$
the radiative effects were classified into process-independent and
process-dependent ones. Apart from the $\zbb$ vertex, all vertex and box
corrections were assumed to be given by the SM, while new physics
contributions were studied in the most general way by four universal
charge form factors. Next, by assuming the running of the charge form
factors to be governed by SM physics alone, the electroweak parameters
$S$, $T$, $U$ were determined. Finally, the SM itself was confronted
with experiments.
 
It was our aim to render this analysis as transparent as possible
by developing the theoretical formalism in full detail and by
presenting the results in figures and parametrizations in a form
useful for appreciating consequences of future improvements in the
experimental data.
 
The analysis proceeded in two steps. First, the information in the
whole body of electroweak precision data has been condensed in the 9
electroweak parameters~:
$\mw$ and $\mz$, $\ebar^2(0)$, $\sbar^2(0)$, $\gzbar^2(0)$ and
$\gwbar^2(0)$, $\sbar^2(\mmz)$, $\gzbar^2(\mmz)$ and $\delb^2(\mmz)$.
At the present time no direct information exists for $\ebar^2(\mmz)$.
In order to keep the analysis flexible $\ebar^2(\mmz)$ and also
the QCD coupling constant $\alpha_s$ have been treated as external
parameters in the fit procedure. Second, this universal set of
quantities with the complete covariance matrix has been interpreted
within the electroweak theory at three qualitatively distinct levels.
 
The main result is that the data can be consistently interpreted
at all levels, in particular there is nowhere evidence against the SM.
This conclusion is not affected, when the new precision measurements of
the left-right asymmetry from SLD\cite{alr94} is included.
The fits to the universal charge form factors or that to the universal
$S$, $T$, $U$ parameters work well and do not hint at a violation
of the ${\rm SU(2)_L \times U(1)_Y}$ universality, nor at an anomalously
large non-standard vertex/box corrections. Generally speaking, the
inclusion of the SM vertex/box corrections improves the fit to the
data, while the improved Born approximation gives a poor fit to
experiments.
The ratio $R_b \equiv \sigma_b^0/\sigma_h^0$ measured by the LEP
experiments turned out to be in poor agreement with the large $\zbb$
vertex correction predicted by the SM.
The fit to the $S$, $T$, $U$ parameters gives us information on
spontaneous symmetry breaking. The $T$ parameter is essentially
determined by the charge form factor $\gzbar^2(\mmz)$, and positive
value is favored. The $S$ parameter is then fixed mainly via
$\sbar^2(\mmz)$, and hence its best-fit value is affected by the
asymmetry data. A negative $S$ value is favored by the new left-right
asymmetry from SLD, and the naive technicolor models are disfavored
\cite{stu}. 
Due to strong correlation between the fitted $S$ and $T$ values, 
the region of the ($S,\,T$) plane with relatively large $S$ and $T$ 
($-0.3 \simlt S \simlt -0.1$ and $0.5 \simlt T \simlt 1$) 
is consistent with the SM prediction for 
$150\gev \simlt m_t \simlt 200 \gev$ and
$ 50\gev \simlt \mh \simlt 200 \gev$ 
(see Figs.~\ref{figureofzparameterfitwithnewalr}(b) 
and \ref{figureofzparameterfitwithmt}(b)).
The $U$ parameter is measured only via $\gwbar^2(0)$, and
it is consistent with zero.
 
The analysis showed that the experimental precision required to
detect a deviation from the SM is still insufficient. For instance, the
running of the charge form factors can be tested presently only for
$\sbar^2(q^2)$ and $\gzbar^2(q^2)$ and is limited by the precision of
the low energy neutral current experiments.
Nevertheless, the data are precise enough to show that their consistent
description within the SM is only guarantied, if the top quark mass
exceeds about 145~GeV. 
This low mass bound of $m_t$ is
nearly independent of $\alpha_s$,
but changes by about $\mp 5\gev$ due
to the uncertainty $\pm 0.1$ in the hadronic vacuum polarization
contribution to $\delta_\alpha \equiv 1/\bar{\alpha}(\mmz)-128.72$. 
Note that the SM top-bottom contribution to the 
form factors have been calculated by using the $O(\alpha\alpha_s)$ 
two-loop formula\cite{rho_pqcd,pivv_pqcd,rho_pqcd_mh2}. 
Perturbative $t\bar{t}$ 
threshold effects\cite{rho_thr,rho_thr2,rho_thr3} will affect 
these corrections, 
and the predicted $m_t$ value may shift upwards by as much as a 
few GeV\cite{rho_thr3}. 

The near future promises a clarification of the value of the
left-right asymmetry published by the SLD group and the ratio $R_b$
from LEP experiments. The precision scan around the $Z$ resonance
performed 1993 by the four LEP experiments will further improve
substantially the $Z$ resonance parameters. It would be advantageous
to publish the data without the subtraction of the $Z$--$\gamma$
interference contribution. Eagerly awaited is the definitive
observation of the top quark. If its mass turns out to be compatible
with the electroweak analysis of the 1-loop effects there is hope to
constrain for the first time the elusive Higgs sector.
 
By introducing the QCD coupling strength $\alpha_s(\mz)_{\msbar}$ and
the shift $\delta_\alpha \equiv 1/\bar{\alpha}-128.72$ as external
parameters in the fit, we have made clear the significance of their
precise measurements. Unless these parameters are accurately measured,
the search for effects beyond the SM through the electroweak radiative
effects gets increasingly limited.

 
\section*{Acknowledgements}
\pr 
We thank K.~Hara,  H.~Masuda and T.~Mori 
for their help 
in understanding the experimental data.
We also thank B.K.~Bullock, S.~Ishihara, B.~Kniehl, K.~Kondo, 
P.~Langacker, J.~Schneps, R.~Szalapski, Y.~Yamada 
and D.~Zeppenfeld for clarifying discussions.
The work of CSK was supported 
in part by the Korean Science and Engineering Foundation, 
in part by Non-Direct-Research-Fund, Korea Research Foundation 1993, and
in part by the Basic Science Research Institute Program, 
Ministry of Education, 1994, Project No.~BSRI-94-2425.
 
\vfill
 

\newpage
\section*{\large \bf App.~A  $\:\:$
SM radiative corrections at one-loop order }
\renewcommand{\theequation}{A.\arabic{equation}}
\cleqn
\pr
In this appendix the propagator, vertex and box corrections of the
standard model (SM) are presented, all at one-loop level and partly
at two-loop level for the $O(\alpha \alpha_s)$ terms of the gauge
boson propagators. All the Green's functions are calculated in the
't~Hooft Feynman gauge in the dimensional regularization and
renormalized in the $\msbar$ scheme. Definitions of the scalar one-loop
integrals, $A$, $B$, $C$, $D$ functions, are given in Appendix D.
Vector boson propagators are given in A.1, the vector boson fermion
vertex functions and the fermion wave function corrections follow in
A.2, while the box corrections are listed in A.3.
All the one-loop calculations are done independently 
and we reproduce the known results of 
ref.~\cite{aoki82,pv,kl} for two-point functions and those of 
ref.~\cite{hollik90,fujimoto90,kniehlstuart,hollik93} for 
the three- and four-point functions. 

\subsection*{\normalsize \bf A.1  Propagator corrections}
\pr
There are four vector boson two-point functions contributing to
processes with external light quarks and leptons at one-loop order.
They can be parametrized by\cite{kl}
  \begin{subequations}
     \label{eq:pi_t}
  \begin{eqnarray}
     \pibar^{\gamma\gamma}_{T}(q^2)
       & = &  \ehat^2\,\pibar^{QQ}_{T}(q^2),
     \label{eq:pigg_t}
     \\
     \pibar^{Z\gamma}_{T}(q^2)
       & = &  \ehat\gzhat^{}\,
        \left\{\pibar^{3Q}_{T}(q^2)-\shat^2\pibar^{QQ}_{T}(q^2)\right\},
     \label{eq:pigz_t}
     \\
     \pibar^{ZZ}_{T}(q^2)
       & = & \gzhat^2\,
          \left\{\pibar^{33}_{T}(q^2)-2\shat^2\pibar^{3Q}_{T}(q^2)
          +\shat^4\pibar^{QQ}_{T}(q^2) \right\},
     \label{eq:pizz_t}
     \\
     \pibar^{WW}_{T}(q^2) & = & \hat{g}^2\,\pibar^{11}_{T}(q^2),
    \label{eq:piww_t}
  \end{eqnarray}
  \end{subequations}
with the coupling factors
  \begin{equation}
     \gzhat^{} =\frac{\hat{g}}{\chat}=\frac{\ehat}{\shat\chat}\,,
      \label{eq:gzhat}
  \end{equation}
and the use of the compact notation
  \begin{equation}
       \shat^2=1-\chat^2=\sin^2\hat{\theta}_{_W}\,,\label{eq:shat}
  \end{equation}
throughout the appendix. These two-point functions and the coupling
factors are renormalized in the $\overline{\rm MS}$ (the modified
minimal subtraction) scheme, and hence they  depend on the 't~Hooft
unit of mass $\mu$ which appears explicitly in the $B$ functions as
defined in appendix~D. The coupling factors of (\ref{eq:gzhat}) and
(\ref{eq:shat}) also depend implicitly on the unit of mass $\mu$.
The subscripts $T$ in eqs.~(\ref{eq:pi_t}) denote the transverse part
of the polarization tensor
  \begin{equation}
     \Pi_{\mu\nu}(q^2)
     = \biggl(-g_{\mu\nu} +\frac{q_\mu q_\nu}{q^2} \biggr)
      \Pi_T(q^2)
     +\frac{q_\mu q_\nu}{q^2} \Pi_L(q^2).
     \label{eq:pi}
  \end{equation}
The longitudinal parts $\Pi_L(q^2)$ do not contribute to processes
with light external fermions.
 
With the help of the four $B$ functions, $B_0, B_3, B_4$ and $B_5$
(see appendix~D), all SM contributions to the above two-point functions
are expressed compactly. $\ol{\Pi}_T^{AB}(q^2)$'s is decomposed into the
bosonic and the fermionic contributions,
  \begin{equation}
     \ol{\Pi}_T^{AB}(q^2)
     = \ol{\Pi}_T^{AB}(q^2)_B +\Pi_T^{AB}(q^2)_F\,,
     \label{eq:piab_b_and_f}
  \end{equation}
and the expressions are given separately.
 
\subsubsection*{\normalsize \bf A.1.1  Bosonic contributions}
\pr
The bosonic contributions with pinch terms are given by\cite{pt2}
  \begin{subequations}
     \label{eq:bpi_t_pt}
  \begin{eqnarray}
     \ol{\Pi}^{QQ}_{T} (q^2)_B
     &=& \Pi^{QQ}_{T} (q^2)_B - \frac{1}{4\,\pi^2}\,q^2\,B_0(q^2;W,W)\,,
     \label{eq:bpiqq_t_pt}
     \\
     \ol{\Pi}^{3Q}_{T} (q^2)_B
     &=& \Pi^{3Q}_{T} (q^2)_B
         - \frac{1}{4\,\pi^2}\,\biggl(q^2 -\frac{1}{2} \mmw\biggr)
             B_0(q^2;W,W)\,,
     \label{eq:bpi3q_t_pt}
     \\
     \ol{\Pi}^{33}_{T} (q^2)_B
     &=& \Pi^{33}_{T} (q^2)_B
         - \frac{1}{4\,\pi^2}\,(q^2 -\mmw) B_0(q^2;W,W)\,,
     \label{eq:bpi33_t_pt}
     \\
     \ol{\Pi}^{11}_{T} (q^2)_B
     &=& \Pi^{11}_{T} (q^2)_B
        - \frac{1}{4\,\pi^2}\,(q^2 -\mmw) \left[ \chat^2 B_0(q^2;W,Z)
                              +\shat^2 B_0(q^2;W,\gamma) \right]\,,
     \label{eq:bpi11_t_pt}
  \end{eqnarray}
  \end{subequations}
where the short-hand notation
  \begin{eqnarray}
     B_{n}(q^2;A,B) & = & B_{n}(q^2;m_A^{},m_B^{}),
    \label{eq:b_notation}
  \end{eqnarray}
is introduced for the $B$ functions. Each $\Pi_T(q^2)$ function without
overline is calculated in the 't~Hooft-Feynman gauge, whereas the
$\ol{\Pi}_T^{}(q^2)$ functions with pinch terms are gauge invariant
\cite{pt2}. The explicit expressions are
  \begin{subequations}
     \label{eq:bpi_t}
  \begin{eqnarray}
     \ol{\Pi}^{QQ}_{T} (q^2)_B
       &=& -\frac{q^2}{16\,\pi^2}\,
            \biggl\{ \biggl[5 B_0 +12 B_3\biggr](q^2;W,W)
                                       +\frac{2}{3} \biggr\}\,,
     \label{eq:bpiqq_t}
     \\[1mm]
    \ol{\Pi}^{3Q}_{T} (q^2)_B
      &=& -\frac{q^2}{16\,\pi^2}\,
           \biggl\{ \biggl[\frac{11}{2} B_0 +10 B_3 \biggr](q^2;W,W)
                                 +\frac{2}{3} \biggr\}\,,
     \label{eq:bpi3q_t}
    \\[1mm]
    \ol{\Pi}^{33}_{T} (q^2)_B
     &=& \quad  \frac{1}{16\,\pi^2}\,
            \biggl[\mmz B_0 +\frac{1}{4} B_5 \biggr](q^2;Z,H)
         \nonumber \\
     & &       - \frac{1}{16\,\pi^2}
            \biggl[ \biggl(\frac{23}{4} q^2 -2 \mmw \biggr) B_0
                              + 9 q^2 B_3 \biggr] (q^2;W,W)
                     - \frac{q^2}{24\,\pi^2}\,,
     \label{eq:bpi33_t}
    \\[2mm]
    \ol{\Pi}^{11}_{T} (q^2)_B
     &=&  \quad   \frac{1}{16\,\pi^2}\,
                \biggl[ \mmw B_{0} +\frac{1}{4}B_{5} \biggr](q;W,H)
         \nonumber \\  \quad
     & &       - \frac{1}{16\,\pi^2}\,
                  \biggl[ \biggl(8 \chat^2 q^2 -(1- 4 \shat^2) \mmw
                      -\mmz \biggr) B_0
              -\biggl(\frac{9}{4} -2\shat^2 \biggr) B_5\biggr](q^2;W,Z)
        \nonumber \\
     & &        -\frac{\shat^2}{8 \,\pi^2}
                 \biggl[ (4 q^2 -2 \mmw) B_0 -B_5 \biggr]
                        (q^2;W,\gamma)
                -\frac{q^2}{24\,\pi^2}\,.
     \label{eq:bpi11_t}
  \end{eqnarray}
  \end{subequations}
At one-loop order of the minimal SM, the first terms in
eqs.~(\ref{eq:bpi33_t}) and (\ref{eq:bpi11_t}) are the only ones in the
transverse component of the vector boson propagators being dependent on
the Higgs boson mass $(\mh)$.
 
\subsubsection*{\normalsize \bf A.1.2  Fermionic contributions}
\pr 
The fermionic contributions to the gauge boson propagators
are known to $O(\alpha\alpha_s)$ two-loop level~:
  \begin{subequations}
\label{eq:fpi_t}
  \begin{eqnarray}
     \Pi^{QQ}_{T}(q^2)_F & = &
         \frac{q^2}{16\,\pi^2}
         \sum_{i}\, Q^{2}_{\ell}\, 8\,B_{3}(q^2;\ell_i,\ell_i)
         \nonumber \\
       &&\!+  \frac{q^2}{16\,\pi^2} \,C_q \!\!\sum_{f=u_i,d_i}  Q^{2}_{f}
         \biggl\{
         8\,B_{3}(q^2;f,f)
         + \frac{4}{3} \,\frac{\alpha_s}{\pi}\,B'_{V}(q^2;f,f)
         \biggr\} ,
      \label{eq:fpiqq_t}\\
     \Pi^{3Q}_{T}(q^2)_F & = &
         \frac{q^2}{16\,\pi^2}
         \sum_{i} Q_{\ell}\,I_{3\ell}\,4\,B_{3}(q^2;\ell_i,\ell_i),
         \nonumber \\
       &&\!+ \frac{q^2}{16\,\pi^2} \,C_q\!\! \sum_{f=u_i,d_i} Q_f\,I_{3f}
          \biggl\{
          4\,B_{3}(q^2;f,f)
          + \frac{2}{3} \,\frac{\alpha_s}{\pi}\,B'_{V}(q^2;f,f)
          \biggr\} ,
      \label{eq:fpi3q_t}\\
     \Pi^{33}_{T}(q^2)_F & = &
         \frac{1}{16\,\pi^2}
         \sum_{f=\ell_i,\nu_i} (I_{3f})^{2}
                 \Bigl[4q^{2}B_{3} -2m_f^{2}B_{0}\Bigr](q^2;f,f)
        \nonumber \\
      &&\!\!\!\!+ \frac{1}{16\,\pi^2}  C_q \!\!
         \sum_{f=u_i,d_i} \!(I_{3f})^{2}
          \biggl\{
                 \Bigl[ 4q^{2}B_{3}\! -2m_f^{2}B_{0}\Bigr](q^2;f,f)
          + \frac{1}{3} \,\frac{\alpha_s}{\pi}
                 \Bigl[ B_V +B_A \Bigr] (q^2;f,f)
          \biggr\} ,
       \nonumber \\[-2mm]
      \label{eq:fpi33_t}\\
     \Pi^{11}_{T}(q^2)_F & = &
         \frac{1}{16\,\pi^2}
         \sum^{}_{i} \Bigl[2q^{2}B_{3} -B_{4}\bigr](q^2;\nu_i,\ell_i)
         \nonumber \\
       &&\!\!\!\!+ \frac{1}{16\,\pi^2} C_q
         \sum^{}_{i,j}\Bigl|V_{u_{i}d_{j}}\Bigr|^{2}
          \biggl\{
               \Bigl[2q^{2}B_{3} -B_{4}\Bigr](q^2;u_{i},d_{j})
          + \frac{1}{6} \,\frac{\alpha_s}{\pi}
                 \Bigl[ B_V +B_A \Bigr] (q^2;u_i,d_j)
          \biggr\} ,
       \nonumber \\[-2mm]
      \label{eq:fpi11_t}
  \end{eqnarray}
  \end{subequations}
%
The summation over $i$, $j$ extends over the three generations of
lepton and quark flavors,
$(\nu_1, \nu_2, \nu_3) =(\nu_e, \nu_{\mu}, \nu_\tau)$,
$(\ell_1,\ell_2,\ell_3)=(e,\mu,\tau)$, $(u_1,u_2,u_3)=(u,c,t)$
and $(d_1,d_2,d_3)=(d,s,b)$.
$C_q=3$ is the color factor,
$Q_f$ the electric charge of the fermion $f$ in units of the proton
charge, $I_{3f}$ the weak isospin
  \begin{equation}
     I_{3f} =
     \left\{
        \begin{array} {ll}
          +\frac{1}{2} & \mbox{for $f$ $=$ $\nu_i$ or $u_{iL}$},\\
          -\frac{1}{2} & \mbox{for $f$ $=$ $\ell_{iL}$ or $d_{iL}$},\\
          \quad  0     & \mbox{for $f$ $=$ $\ell_{iR}$,
                               $u_{iR}$ or $d_{iR}$},
         \end{array}
     \right.     \label{eq:i3f}
  \end{equation}
while $V_{u_{i}d_{j}}$ are the Kobayashi-Maskawa quark mixing matrix
elements.
The $O(\alpha\alpha_s)$ corrections in perturbative 
QCD\cite{pivv_pqcd,rho_pqcd_mh2,kniehl} are given by 
the functions $B_V$ and $B_A$~:
\begin{subequations}
\label{eq:b_2loop}
\begin{eqnarray}
\label{eq:bv_mm}
   B_V(q^2;m,m) &=& q^2 B_V'(q^2;m,m)\,,
\\
\label{eq:ba_mm}
   B_A(q^2;m,m) &=& q^2 B_A'(q^2;m,m) +B_A(0;m,m)\,,
\\
\label{eq:bv_m0}
   B_V(q^2;m,0) &=& q^2 B_V'(q^2;m,0) +B_V(0;m,0)\,,
\\
\label{eq:ba_m0}
   B_A(q^2;m,0) &=& q^2 B_A'(q^2;m,0) +B_A(0;m,0)\,,
\end{eqnarray}
\end{subequations}
where
\begin{subequations}
\label{eq:bp_2loop}
\begin{eqnarray}
 B_V'(q^2;m,m) &=& \ln \frac{\mu^2}{m^2}
        +\frac{55}{12} -4\,\zeta_3
        +\frac{4 m^2}{q^2}\, V_1 (\frac{q^2}{4m^2})\,,
\label{eq:bvp_mm}
\\
 B_A'(q^2;m,m) &=& \ln \frac{\mu^2}{m^2}
        +\frac{55}{12} -4\,\zeta_3
        +\frac{4 m^2}{q^2}\, [A_1(\frac{q^2}{4m^2})-A_1(0)]\,,
\label{eq:bap_mm}
\\
 B_V'(q^2;m,0) &=& B_A'(q^2;m,0)
\nonumber\\
 &=& \ln \frac{\mu^2}{m^2}
        +\frac{55}{12} -4\,\zeta_3
        +\frac{4 m^2}{q^2}\, \biggl[ F_1(\frac{q^2}{m^2})-F_1(0) \biggr]\,,
\label{eq:bvp_m0}
\end{eqnarray}
\end{subequations}
and
\begin{subequations}
\label{eq:b_2loop_le}
\begin{eqnarray}
 B_A(0;m,m) &=& m^2 \left[
 12 \,\ln^2 \frac{\mu^2}{m^2}
        +22 \,\ln \frac{\mu^2}{m^2}
        +\frac{31}{2} \right] \,,
\label{eq:ba_le}
\\
 B_V(0;m,0) &=& B_A(0;m,0)
\nonumber\\
 &=& m^2 \left[
 3 \,\ln^2 \frac{\mu^2}{m^2}
        +\frac{11}{2}  \,\ln \frac{\mu^2}{m^2}
        +\zeta_2 +\frac{35}{8} \right]\,.
\label{eq:bv_le}
\end{eqnarray}
\end{subequations}
Here $\zeta_2 = \pi^2/6$, $\zeta_3 =1.2020569$, and the complex functions
$V_1$, $A_1$ and $F_1$ are given in refs.\cite{pivv_pqcd,rho_pqcd_mh2}. 
The following limits are useful~:
\begin{subequations}
\label{eq:bp_2loop_limits}
\begin{eqnarray}
 B_V'(M^2;0,0) &=& B_A'(M^2;0,0)
\nonumber\\
 &=& \ln\frac{\mu^2}{M^2} +\frac{55}{12} -4\zeta_3 +i\pi \,,
\\
 B_V'(M^2;m,m) &=&
 \ln\frac{\mu^2}{m^2} +\frac{15}{4} +O(\frac{M^2}{m^2}) \,,
\\
 B_A'(M^2;m,m) &=&
 \ln\frac{\mu^2}{m^2} +\frac{67}{36} +O(\frac{M^2}{m^2}) \,,
\\
 B_V'(M^2;m,0) &=& B_A'(M^2;m,0)
\nonumber\\
 &=& \ln\frac{\mu^2}{m^2} +\frac{115}{36}
 -\frac{4}{9}\zeta_2 +O(\frac{M^2}{m^2}) \,.
\end{eqnarray}
\end{subequations}
 
\subsection*{\normalsize \bf A.2  Vertex correction }
 
\pr
The vertex form factors $\Gamma_1^{f}(q^2)$, $\ol{\Gamma}_2^{f}(q^2)$, 
$\Gamma_3^{f}(q^2)$ and $\Gamma_4^{f}(q^2)$ appearing in the helicity 
amplitudes (\ref{m_nc}) contribute to the $\gamma ff$ and $Zff$ vertices
as follows~:
\bsub
\label{eq:gamma_vff}
\bea
\Gamma^{\gamma ff}(q^2) &=& -\ehat \Bigl\{ Q_f [\,1+\Gamma_1^f(q^2)\,]
                                     +I_{3f}\gambar_2^f(q^2) \Bigr\}\,,
\label{eq:gamma_aff}
\\
\Gamma^{Zff}(q^2) &=& -\gzhat^{} \Bigl\{ 
          (I_{3f}-Q_f\shat^2) \,[\,1+\Gamma_1^f(q^2)\,]
                  +I_{3f} [\,\chat^2\gambar_2^f(q^2) +\Gamma_3^f(q^2)\,] 
                      +\Gamma_4^f(q^2) \Bigr\}\,.\hspace{1.2cm}
\label{eq:gamma_zff}
\eea
\esub
It should be noted that the functions $\Gamma_1^f(q^2)$ and $\gambar_2^f(q^2)$
are common to the $\gamma ff$ and $Zff$ vertices, and that 
$\Gamma_3^{f}(q^2)$ and $\Gamma_4^{f}(q^2)$ are additional contributions
to the $Zff$ vertex.  
These vertex functions depend on the chirality of $f$ and their explicit
forms at one-loop level of the SM are
  \begin{subequations}
     \label{eq:gamma_f}
  \begin{eqnarray}
     \Gamma_1^{f_R}(q^2) &=&
        \biggl( \frac{g_R^{Zff}}{4 \pi} \biggr)^2
         \Gamma_{1Z}^f(q^2)\,,
     \label{eq:gamma1_fr}
     \\
     \ol{\Gamma}_2^{f_R}(q^2)
	 &=& \Gamma_3^{f_R}(q^2) = \Gamma_4^{f_R}(q^2) =  0 \,,
     \label{eq:gamma234_fr}
     \\[1mm]
     \Gamma_1^{f_L}(q^2) &=&
        \biggl( \frac{g_L^{Zff}}{4 \pi} \biggr)^2
         \Gamma_{1Z}^f(q^2)
      + \sum_{f'} \biggl| \frac{g_L^{Wff'}}{4 \pi} \biggr|^2
         \Gamma_{1W}^{f'}(q^2)\,,
     \label{eq:gamma1_fl}
     \\[1mm]
     \ol{\Gamma}_2^{f_L}(q^2) &=&
          -2\,\sum_{f'} \biggl| \frac{g_L^{Wff'}}{4 \pi} \biggr|^2
          \gambar_{2W}^{f'}(q^2)\,,
     \label{eq:gamma2_fl}
     \\[1mm]
     \Gamma_3^{f_L}(q^2) &=&
          \sum_{f'} \biggl| \frac{g_L^{Wff'}}{4 \pi} \biggr|^2
          \Gamma_{mW}^{f'}(q^2)\,,
     \label{eq:gamma3_fl}
     \\[1mm]
     \Gamma_4^{f_L}(q^2) &=& 0 \,,
  \end{eqnarray}
  \end{subequations}
with the gauge boson coupling convention
  \begin{eqnarray}
    \label{eq:vff_couplings}
    \begin{array}{cc}
     g_L^{\gamma ff} = g_R^{\gamma ff} = \ehat\,Q_f\,, \quad &
     g_R^{Zff}  = -\gzhat^{}\,Q_f\,\shat^2\,,
     \\[3mm]
     g_L^{Zff}  = \gzhat^{}\,(I_{3f} -Q_f\,\shat^2)\,, \quad &
     g_L^{Wff'} = \displaystyle{\frac{\ghat}{\sqrt{2}}}\,V_{ff'}\,,
    \end{array}
  \end{eqnarray}
where
  \begin{subequations}
      \label{eq:gamma_fv}
  \begin{eqnarray}
      \Gamma_{1Z}^f(q^2)
       &=& \Gamma_1(q^2;f,Z,f) -\Sigma' (m_f^2;f,Z) \,,
      \label{eq:gamma1z}
      \\
      \Gamma_{1W}^{f'}(q^2)
       &=& \left(\Gamma_1+\Gamma_{1m}\right) (q^2;f',W,f')
                -\Sigma'(m_f^2;f',W)\,,
      \label{eq:gamma1w}
       \\
      \gambar_{2W}^{f'}(q^2)
       &=&  \left(\Gamma_1+\Gamma_{1m}\right)(q^2;f',W,f')
                  -\Gamma_2 (q^2;W,f',W)
            +2 {\rm Re}\, B_0(q^2;W,W)\,, \quad
      \label{eq:gamma2w}
       \\
          \Gamma_{mW}^{f'}(q^2)
       &=&  \Gamma_{1m} (q^2;f',W,f') + \Gamma_{2m} (q^2;W,f',W) \,.
      \label{eq:gammamw}
  \end{eqnarray}
  \end{subequations}
Here
  \begin{eqnarray}
     \Sigma'(q^2;m,M) &=&
       -\biggl( 2 +\frac{m^2}{M^2} \biggr) B_1(q^2;m,M) -1\,,
     \label{eq:sigmap}
  \end{eqnarray}
is the external light fermion self energy correction, and the last term
in eqs.~(\ref{eq:gamma2w}) of $\ol{\Gamma}_{2W}^{f'}$ is the pinch
term\cite{kl,pt2} which is subtracted from the vertex functions as
calculated in the 't~Hooft-Feynman gauge. The remaining vertex
functions in eqs.~(\ref{eq:gamma_fv}) are
  \begin{subequations}
\label{eq:gamma's}
  \begin{eqnarray}
     &\Gamma_1(q^2;m,M,m) &=
        \biggl[\, 2 q^2 (C_{11} +C_{23}) +4C_{24}
              -\frac{m^4}{M^2}C_{0} \biggr] (q^2;m,M,m) -2\,,
     \label{eq:gamma_1}
     \\
     &\Gamma_{1m}(q^2;m,M,m) &= \frac{m^2}{M^2}
        \left\{ \left[ q^2 (C_{12} +C_{23}) +2C_{24}
              -2 M^2 C_{0} \right] (q^2;m,M,m) -\frac{1}{2}
        \right\},
     \label{eq:gamma_1m}
     \\
     &\Gamma_{2}(q^2;M,m,M) &=
        2\bigg[ q^2 (C_{11} +C_{23})
               +\biggl(6 +\frac{m^2}{M^2} \biggr) C_{24}
               + (q^2 -m^2) C_{0} \biggr] (q^2;M,m,M) -2,
     \nonumber \\[-2mm]
     \label{eq:gamma_2}
     &&\\
     &\Gamma_{2m}(q^2;M,m,M) &= 2 \frac{m^2}{M^2}
        \left[\, 2 M^2 C_{0} -C_{24} \right] \!(q^2;M,m,M)\,,
     \label{eq:gamma_2m}
  \end{eqnarray}
  \end{subequations}
with the shorthand notation for the $C$ functions of appendix D:
  \begin{eqnarray}
     C_i(q^2;m_1,m_2,m_3) &\equiv& C_i(0,0,q^2;m_1,m_2,m_3)\,.
     \label{eq:ci_def}
  \end{eqnarray}

In the limit of the diagonal KM matrix elements $V_{{u_i}{d_j}} =
\delta_{ij}$, which is assumed in all our numerical results, the
internal fermion mass $m=m_{f'}$ is non-negligible only for
$f=b_L^{}$($f'=t$). Otherwise we can set $m=0$ at high energies
($m^2/q^2\sim 0$) and find
\bsub
\label{eq:gamma's_limits}
  \begin{eqnarray}
     \Gamma_1(q^2;0,M,0)
      &=& \ln \frac{\mu^2}{M^2} -4 -2\frac{M^2}{q^2}
          +\Bigl(3+2\frac{M^2}{q^2}\Bigr)\ln\frac{-q^2-i\epsilon}{M^2}
         \nonumber \\
      & &   +2\Bigl(1+\frac{M^2}{q^2}\Bigr)^2
          \Bigl[{\rm Sp}\Bigl(1+\frac{q^2}{M^2}+i\epsilon\Bigr)
                -{\rm Sp}(1)\Bigr]\,,
     \\
     \Gamma_2(q^2;M,0,M)
      &=& 3\ln \frac{\mu^2}{M^2} +2 -2\frac{M^2}{q^2}
          +\Bigl(1+2\frac{M^2}{q^2}\Bigr)\beta L
          +2\frac{M^2}{q^2}\Bigl(2+\frac{M^2}{q^2}\Bigr) L^2\,,\quad
     \\
     \Gamma_{1m}(q^2;0,M,0) &=& \Gamma_{2m}(q^2;M,0,M) = 0\,,
  \end{eqnarray}
\esub
and
  \begin{eqnarray}
     \Sigma'(0;0,M) &=& \ln \frac{\mu^2}{M^2}-\frac{1}{2}\,,
\label{eq:sigmap_00m}
     \\
     {\rm Re} B_0(q^2;M,M) &=& \ln\frac{\mu^2}{M^2} +2 -\beta\,L\,.
\label{eq:pt_mm}
  \end{eqnarray}
Here ${\rm Sp}(z) = -\int_0^z \frac{1-t}{t}\,dt$ is the complex Spence
(dilogarithm) function, and
  \begin{subequations}
  \begin{eqnarray}
     \beta &=& \sqrt{1-4(M^2-i\epsilon)/q^2}\,,
     \\
      L    &=& \ln\frac{\beta+1}{\beta-1}\,.
  \end{eqnarray}
  \end{subequations}

At low energies, light fermion masses may not be neglected as compared
to the momentum transfer $q^2$. In the limit of 
$|q^2|/\mmz\!\ll\! 1$ and
$m_f^2/\mmz\! \ll\! 1$, but at fixed $m_f^2/q^2$, the vertex functions
reduce to
\bsub
\label{eq:gammafv_le}
  \begin{eqnarray}
     \Gamma_{1Z}^f(q^2)&=&\frac{-q^2}{\mmz}
           \Bigl[J_Z(q^2;f) +O\Bigl(\frac{q^2}{\mmz}\Bigr)\Bigr]\,,
     \\
     \Gamma_{1W}^{f'}(q^2)&=&\frac{-q^2}{\mmw}
           \Bigl[J_W(q^2;f') +O\Bigl(\frac{q^2}{\mmz}\Bigr)\Bigr]\,,
     \\
     \gambar_{2W}^{f'}(q^2)&=&\frac{-q^2}{\mmw}
          \Bigl[\ol{J}_W(q^2;f') +O\Bigl(\frac{q^2}{\mmz}\Bigr)\Bigr]\,,
  \end{eqnarray}
\esub
where
\bsub
\label{eq:j}
  \begin{eqnarray}
     J_V(q^2;f)
     &=& 4 F_3(q^2;f,f)-\frac{2}{3} \ln m_V^2 -\frac{1}{9}\,,
\label{eq:jv}     \\
     \ol{J}_W(q^2;f)
     &=& 4 F_3(q^2;f,f)-\frac{2}{3} \ln \mmw
         -\frac{2}{3}-\frac{1}{3}\,.
\label{eq:jw}
  \end{eqnarray}
\esub
The function $F_3$ is defined in appendix D. The last $1/3$ factor is
the pinch term.
 
\subsection*{\normalsize \bf A.3  Box correction }
\pr 
Box corrections for the process 
$e_\lambda \ol{e_\lambda} \to f_\sigma
\ol{f_\sigma}$ are expressed by $\boxfactor{e}{f}{\lambda}{\sigma}$,
where $\lambda,\sigma=-1$ is used for left-handed fermions and
$\lambda,\sigma=+1$ for right-handed fermions.
  \begin{subequations}
     \label{eq:b_box}
  \begin{eqnarray}
     &B_{\lambda,\lambda}^{\,e\,f}(s,t) &=
         \frac{1}{16 \pi^2}
         \left| g_\lambda^{Zee} g_\lambda^{Zff} \right|^2
         \left[ I_1(u,s;\mz,m_f) - I_2(t,s;\mz,m_f) \right]
     \\
     &&+\frac{\delta_{\lambda L}}{16 \pi^2}
        \left| g_L^{We\nu} g_L^{Wff'}  \right|^2
        \times
         \left\{
         \begin{array}{ll}
            \! +I_1(u,s;\mw,m_{f'})
            & \mbox{ for }\; {I_3}_f= +\frac{1}{2} \;\; (f=\nu_\ell,u_i)
            \\[1mm]
            \! -I_2(t,s;\mw,m_{f'})
            & \mbox{ for }\; {I_3}_f= -\frac{1}{2} \;\; (f=\ell,d_i)
  \end{array}
         \right. ,
     \nonumber \\[-1mm]
     \nonumber\\
     &B_{\lambda,-\lambda}^{\,e\,f}(s,t) &=
        \frac{1}{16 \pi^2}
        \left| g_\lambda^{Zee} g_{-\lambda}^{Zff} \right|^2
        \left[ I_2(u,s;\mz,m_f) - I_1(t,s;\mz,m_f) \right]\,,
  \end{eqnarray}
  \end{subequations}
with
  \begin{subequations}
\label{eq:stu_kin}
  \begin{eqnarray}
    s &=& (p_e +p_{\bar{e}})^2 = (p_f +p_{\bar{f}})^2,
    \\
    t &=& (p_e -p_f)^2 = (p_{\bar{e}} -p_{\bar{f}})^2,
    \\
    u &=& (p_e -p_{\bar{f}})^2 = (p_{\bar{e}} -p_f)^2,
  \end{eqnarray}
  \end{subequations}
and $\:p_i\:$ being the 4-momentum of particle $\:i\:$.
The internal fermion mass $m_{f'}$ is non-negligible only for $f=b_L^{}$,
for which the top quark contributes in the limit of the diagonal KM
matrix elements.
The functions $\:I_1(u,s\,;m_V^{},m_f)\:$ and $I_2(u,s\,;m_V^{},m_f)$
are expressed in terms of the $D$ functions of appendix D:
  \begin{subequations}
     \label{eq:i_box}
  \begin{eqnarray}
     I_1(u,s;m_V^{},m_f) &=&
      -2u \left( D_{11} +D_{12} -D_{13} +2 D_{24} \right)
       -4t D_{25} -4s D_{26} -16 D_{27}\,, \qquad
     \label{eq:i1_box}
     \\
     I_2(u,s;m_V^{},m_f) &=&
      -2u \left( D_{11} +D_{24} -D_{25} \right) -4 D_{27}\,,
     \label{eq:i2_box}
  \end{eqnarray}
  \end{subequations}
where
  \begin{eqnarray}
     D_i &\equiv& D_i(0,0,0,0,u,s;0,m_V^{},m_f,m_V^{})\,,
       \qquad i=0, 11\!-\!13, 21\!-\!27\,.
\label{eq:di_def}
  \end{eqnarray}
After reduction of the higher $D$ functions
  \begin{subequations}
     \label{eq:i_box_red}
  \begin{eqnarray}
     I_1(u,s;M,m) &=&
      -2\,C_0^{(124)} - 2\,C_0^{(234)}
      + 2\left(\,u-\,m^2\right)D_0^{(1234)} \,,
     \label{eq:i1_box_red}
     \\
     I_2(u,s;M,m) &=&
      \frac{2}{s+u}\left[\,B_0^{(13)} - B_0^{(24)} \right]
     \nonumber \\
     &&
      + \frac{u}{(s+u)^2}\left(s +2\,u+2\,M^2 -m^2\right)
       \left[ C_0^{(123)}+C_0^{(134)} \right]
     \nonumber \\
     &&
      - \frac{1}{(s+u)^2}
        \biggl( s\left(s +2\,u -2\,M^2+m^2\right)+2u^2 \biggr)
       \left[ C_0^{(124)}+C_0^{(234)} \right]
     \nonumber \\
     && + \frac{1}{(s+u)^2}
         \biggl\{\left(u-m^2\right)s^2
        - \left(2\,M^4-2\,M^2\,m^2+m^4+m^2\,u-2\,u^2\right)s
     \nonumber \\
     &&\hspace{2cm}
        + 2\,\left(M^2+u\right)\left(M^2-m^2+u\right)u
        \biggr\} D_0^{(1234)}\,,
     \label{eq:i2_box_red}
  \end{eqnarray}
  \end{subequations}
is obtained. For the case $f \ne b_L^{}$ the limit $m \to 0$ can be
carried out~:
  \begin{subequations}
     \label{eq:i_nonb}
  \begin{eqnarray}
     I_1(u,s;M,0) &=& -4\,C_0^{(124)} + 2\,u\,D_0^{(1234)}\,,
     \label{eq:i1_nonb}
     \\
     I_2(u,s;M,0) &=&
      \frac{2}{s+u} \left[B_0^{(13)} -B_0^{(24)}\right]
     \nonumber\\
     && +\frac{2}{(s+u)^2}
        u\left(s + 2\,u +2\,M^2 \right)  C_0^{(123)} \!
     \nonumber\\
     && -\frac{2}{(s+u)^2}
           \left( s(s +2\,u-2\,M^2) +2\,u^2\right) C_0^{(124)}
     \nonumber\\
     && +\frac{1}{(s+u)^2}
          \biggl\{ s^2\,u -2\,s \left(M^4-u^2\right)
                  + 2\,u \left(M^2+u\right)^2 \biggr\} D_0^{(1234)}\,.
      \quad
     \label{eq:i2_nonb}
  \end{eqnarray}
  \end{subequations}
Eqs.~(\ref{eq:i_nonb}) agree exactly with ref.\cite{hollik90}.
 
In the low energy limit, only $D_{27}$ survives:
  \begin{subequations}
\label{eq:d27_le}
  \begin{eqnarray}
     D_{27}(0,0,0,0,0,0;0,m_V^{},0,m_V^{}) &=& \frac{1}{4\,m_V^2}\,,
     \\
     D_{27}(0,0,0,0,0,0;0,\mw,m_t,\mw)
     &=& \frac{1}{4(m_t^2 -\mmw)}
        \left[ \frac{m_t^2}{m_t^2-\mmw} \ln \frac{m_t^2}{\mmw} -1
        \right]\,,\qquad
  \end{eqnarray}
  \end{subequations}
and hence
  \begin{subequations}
     \label{eq:box_le}
  \begin{eqnarray}
     &B_{\lambda,\lambda}^{\,e\,f}(0,0) &=
        \frac{1}{16 \pi^2}
        \left| g_\lambda^{Zee} g_\lambda^{Zff} \right|^2
        \frac{3}{\mmz}
     \nonumber \\
     &&\;+\;\frac{\delta_{\lambda L}}{16 \pi^2}
            \left| g_L^{We\nu} g_L^{Wff'}  \right|^2
        \times
         \left\{
         \begin{array}{ll}
            \!\displaystyle{+\frac{4}{\mmw}}
            & \mbox{ for }\; {I_3}_f= +\frac{1}{2} \;\; (f=\nu_\ell,u_i)
            \\
            \!\displaystyle{-\frac{1}{\mmw}}
            & \mbox{ for }\; {I_3}_f= -\frac{1}{2} \;\; (f=\ell,d_i)
  \end{array}
         \right., \qquad
     \\[1mm]
     &B_{\lambda,-\lambda}^{\,e\,f}(0,0) &=
        -\frac{1}{16 \pi^2}
         \left| g_\lambda^{Zee} g_{-\lambda}^{Zff} \right|^2
         \!\frac{3}{\mmz}\,,
  \end{eqnarray}
  \end{subequations}
for $f \ne b_L$, and
  \begin{eqnarray}
     &B_{L,L}^{\,e\,b}(0,0) &=
        \frac{\left| g_L^{Zee} g_L^{Zbb} \right|^2}{16 \pi^2}
          \frac{3}{\mmz}
       -\frac{\left| g_L^{We\nu} g_L^{Wbt} \right|^2}{16 \pi^2}
        \frac{1}{m_t^2 -\mmw}
        \left[ \frac{m_t^2}{m_t^2-\mmw} \ln \frac{m_t^2}{\mmw} -1
        \right] \,,\qquad
\label{eq:box_le_b}
  \end{eqnarray}
for $f=b_L$.
 
\newpage
\section*{\large \bf App.~B $\:\:$
Renormalization group and hadronic contributions }
\renewcommand{\theequation}{B.\arabic{equation}}
\cleqn
\pr 
The effective charges of the $SU(2) \times U(1)$ theory are expressed
in terms of the $\overline{\rm{MS}}$ couplings by
  \begin{eqnarray}
     \frac{1}{\ebar^2(q^2)} & = & \frac{1}{\ehat^2(\mu)}
     + {\rm Re}\ol{\Pi}^{QQ}_{T,\gamma}(q^2)\,,
     \label{eq:ebar_def}
     \\
     \frac{1}{\bar{g}^2(q^2)} & = & \frac{1}{\ghat^2(\mu)}
      +{\rm Re} \ol{\Pi}^{3Q}_{T,\gamma}(q^2)\,,
     \label{eq:gbar_def}
  \end{eqnarray}
where the $SU(2)$ effective charge
  \begin{equation}
    \bar{g}^2(q^2) \equiv \frac{\ebar^2(q^2)}{\sbar^2(q^2)}
    \label{eq:sbar_def}
 \end{equation}
is introduced for convenience. The expressions (\ref{eq:ebar_def}) and
(\ref{eq:gbar_def}) are explicit solutions of the renormalization
group (RG) equation in the $\overline{\rm{MS}}$ scheme~:
  \begin{equation}
     D[\mbox{effective charges}] = 0\,, \label{eq:rge_msbar}
  \end{equation}
with the RG operator
  \begin{eqnarray}
     D & \equiv & \Bigl[\mu^2\frac{d^2}{d\mu^2}\Bigr]_{_{\bf{Bare}}}
     \nonumber \\
     & = & \mu^2\frac{\partial}{\partial\mu^2} +
    \hat{\beta}_e \Bigl(\frac{\ehat^2}{16\pi^2}\Bigr)
    \frac{\partial}{\partial(\ehat^2/16\pi^2)}+
    \hat{\beta}_g \Bigl(\frac{\ghat^2}{16\pi^2}\Bigr)
    \frac{\partial}{\partial(\ghat^2/16\pi^2)}.
    \label{eq:rge_op}
  \end{eqnarray}
The $\overline{\rm{MS}}$ $\hat{\beta}$-functions
in the minimal SM read at one-loop order
  \begin{eqnarray}
     \hat{\beta}_e\Bigl(\frac{\ehat^2}{16\pi^2}\Bigr)
      &=& \Bigl[-7+\frac{4}{3}\Sigma_{f} C_fQ^2_f\Bigr]
          \Bigl(\frac{\hat{e}^2}{16\pi^2}\Bigr)^2,
     \label{eq:beta_e}
     \\
    \hat{\beta}_g\Bigl(\frac{\hat{g}^2}{16\pi^2}\Bigr)
      &=& \Bigl[-\frac{43}{6}+\frac{2}{3}\Sigma_{f}C_f I_{3f}Q_f\Bigr]
          \Bigl(\frac{\hat{g}^2}{16\pi^2}\Bigr)^2,\quad
     \label{eq:beta_g}
  \end{eqnarray}
where $C_f=1(3)$ for $f=\ell(q)$. The two-loop $O(\alpha\alpha_s$)
contributions are accounted for by replacing
$C_q \to C_q(1+\frac{\hat{\alpha}_s}{\pi})$
in eqs.~(\ref{eq:beta_e}) and (\ref{eq:beta_g}). Note that the effective
charges $\bar{e}^2(q^2)$ and $\bar{g}^2(q^2)$ behaves similarly to the
$\msbar$ couplings at asymptotically high energies, $|q^2|\gg \mmw$,
since the functions $\ol{\Pi}_{T,\gamma}^{QQ}(q^2)$ and
$\ol{\Pi}_{T,\gamma}^{3Q}(q^2)$ do not have large logarithms at
$\mu^2 \sim |q^2|\gg \mmw$\cite{kl}. This is enabled by adding the
pinch terms\cite{kl,pt2} in the self energy $\ol{\Pi}(q^2)$, and our
$\ebar^2(q^2)$ and $\sbar^2(q^2)$ are equivalent to the corresponding
$*$-charges\cite{kl} up to the imaginary parts and the two-loop
corrections.
 
Although the $\overline{\rm{MS}}$ couplings $\hat{e}$ and $\hat{g}$
could be adopted directly in our analysis, we prefer the effective
charges of (\ref{eq:ebar_def}) and (\ref{eq:gbar_def}) as 
quantities to be used when confronting theory with experiment. 
We give two reasons,
one being associated with the non-decoupling of heavy particles
in the $\overline{\rm{MS}}$ scheme, the other being related
with the treatment of non-perturbative hadronic contributions to the
electroweak parameters.
 
Traditionally, the appearance of large logarithms\, of heavy particle
masses (non-decoupling) in the $\overline{\rm{MS}}$ scheme is avoided
by adopting the effective field theory \cite{elft1,elft2}, where the
heavy particle fields are integrated out in the action. The couplings
of the effective theories are then related to each others by matching
conditions ensuring that all effective theories give identical results
at zero momentum transfer, since the effects of heavy particles in the
effective light field theory
must be proportional to $q^2/m^2_{\rm heavy}$.
 
In general, the two $\overline{\rm{MS}}$ couplings $\hat{e}^2(\mu)_{eff}
$ and $\hat{g}^2(\mu)_{eff}$ of the effective light particle theory can
be obtained by the matching conditions
  \begin{eqnarray}
     \frac{1}{\bar{e}^2(0)}
       &=& \frac{1}{\hat{e}^2(\mu)_{eff}}
          +\Bigl[{\rm Re}\ol{\Pi}^{QQ}_{T,\gamma}(0)\Bigr]_{eff},
     \label{b.8}\\
     \frac{1}{\bar{g}^2(0)}
       &=& \frac{1}{\hat{g}^2(\mu)_{eff}}
          +\Bigl[{\rm Re}\ol{\Pi}^{3Q}_{T,\gamma}(0)\Bigr]_{eff},
     \label{b.9}
  \end{eqnarray}
where only the light particles at the scale $\mu$ contribute to the
two-point functions at the right-hand side. In the minimal SM, one may,
for instance, employ an effective theory of particles of mass up to the
scale $\mu$:
  \begin{eqnarray}
     \frac{16\pi^2}{\bar{e}^2(0)}
      &=& \frac{16\pi^2}{\hat{e}^2(\mu)_{eff}}
         +\frac{4}{3}\Sigma_f Q^2_f \ln \frac{\mu^2}{m^2_f}
           \theta(\mu-m_f)
         -\Bigl(7\ln \frac{\mu^2}{\mmw}+\frac{2}{3}\Bigr)
           \theta(\mu-\mw),
    \label{eq:e2_eff}
    \\
    \frac{16\pi^2}{\bar{g}^2(0)}
     &=& \frac{16\pi^2}{\hat{g}^2(\mu)_{eff}}
         +\frac{2}{3}\Sigma_f I_{3f}Q_f \ln \frac{\mu^2}{m^2_f}
            \theta(\mu-m_f)
         -\Bigl(\frac{43}{6}\ln \frac{\mu^2}{\mmw}
                 +\frac{2}{3}\Bigr)\theta(\mu-\mw),\quad
    \label{eq:g2_eff}
  \end{eqnarray}
Such a scheme is often adopted in quantum chromodynamics (QCD), but
leads to a discontinuity at $\mu = \mw$ of the effective $\msbar$
coupling constants. The appearance of the discontinuity in the
unphysical $\overline{\rm{MS}}$ couplings is not really a problem%
\footnote{\normalsize \baselineskip 18pt
In fact the discontinuity can be evaded
by using yet another unphysical effective coupling, the
so called dimensional reduction $\ol{\rm DR}$ scheme\cite{dr}.
},
but the appearance of many quark and lepton mass scales renders
the use of these effective couplings impractical at the scale
$\mu<\mz$.
Furthermore, direct use of the effective $\msbar$ couplings at lower
energies leads to expressions with light-quark masses suffering from
large non-perturbative QCD corrections.
 
These two problems of the $\overline{\rm{MS}}$ scheme can be overcome
simultaneously by adopting the effective charges (\ref{eq:ebar_def})
and (\ref{eq:gbar_def}) as expansion parameters at
  \begin{equation}
     0 \leq |q^2| < O(\mmz) \label{eq:below_z}
  \end{equation}
when confronting with experiments. The connection with a high energy
theory, e.g. at $q^2 = \mmz$, can then be made free from light quark
mass ambiguities by the use of the manifestly RG invariant expressions
(\ref{eq:ebar_def}) and (\ref{eq:gbar_def}). In the region
(\ref{eq:below_z}) the effective charges at two different $q^2$ are
related by dispersion relations.
 
The light hadron (first 5-quark, or "$5q$") contributions to the
differences
  \begin{eqnarray}
     \frac{1}{\bar{\alpha}(q^2)} \:-\:\:\frac{1}{\alpha} \quad
      &=&\frac{4\pi}{\ebar^2(q^2)} -\frac{4\pi}{\ebar^2(0)}
       = 4\,\pi \Bigl[ {\rm Re}\ol{\Pi}^{QQ}_{T,\gamma}(q^2)
                      -\ol{\Pi}^{QQ}_{T,\gamma}(0) \Bigr],
     \label{eq:alpha_running}\\
     \frac{\sbar^2(q^2)}{\bar{\alpha}(q^2)}
        -\frac{\sbar^2(0)}{\alpha}
      &=& \frac{4\pi}{\bar{g}^2(q^2)}
        - \frac{4\pi}{\bar{g}^2(0)}
       =  4\,\pi\,\Bigl[ {\rm Re}\ol{\Pi}^{3Q}_{T,\gamma}(q^2)
          -\ol{\Pi}^{3Q}_{T,\gamma}(0) \Bigr],
     \label{eq:sbar_running}
  \end{eqnarray}
have been parametrized in the region $0<|q^2|<\mmz$ as follows. 
For the photon vacuum polarization function, we use 
  \begin{subequations}
     \label{eq:piqq_5q}
  \begin{eqnarray}
     && \hspace{-10mm}
        4\pi\Bigl[\,{\rm Re}\ol{\Pi}_{T,\gamma}^{QQ}(q^2)
                           -\ol{\Pi}_{T,\gamma}^{QQ}(0) \Bigr]_{5q}
     \nonumber  \\
     && \quad -f(q^2) \hspace{65.2mm}
                  \quad \mbox{for}\quad -\mmz<q^2<0,
     \label{eq:piqq_5q_spacelike}
        \\[-6mm]
     &=& \left\{\vphantom{\begin{array}{l}a\\[4mm]a\end{array}}\right.
        \nonumber
        \\[-9mm]
     && \quad -f(q^2)
                 +4\pi\sum_{q=c,b}
            \Bigl[ {\rm Re} \ol{\Pi}_{T,\gamma}^{QQ}(q^2)
                          - \ol{\Pi}_{T,\gamma}^{QQ}(-q^2) \Bigr]_{q}
                  \quad \mbox{for}\quad \,0<q^2<\mmz.
        \label{eq:piqq_5q_timelike}
  \end{eqnarray}
  \end{subequations}
Here the results of the dispersion integral 
analyses\cite{piqq_h,piqq_h_latest} are
parametrized by
  \begin{eqnarray}
     f(q^2) = \left\{
     \begin{array}{lll}
     1.096 \ln(1+|q^2|)
                &  \!\mbox{for}\!& 0.0\le \sqrt{|q^2|}(\gev) \le 0.3
     \\
     0.3261\ln(1+3.927|q^2|)
                &  \!\mbox{for}\!& 0.3\le \sqrt{|q^2|}(\gev) \le 3
     \\
     0.2486+0.4009\ln(1+|q^2|)
                &  \!\mbox{for}\!& 3.0\le \sqrt{|q^2|}(\gev) \le 50
     \\
     \displaystyle{
     3.878 +0.4084\Bigl\{\ln \frac{|q^2|}{s_0}
                            +0.005696(\frac{|q^2|}{s_0}-1) \Bigr\}
     }
                & \!\mbox{for}\!&\,50\,\le \sqrt{|q^2|}(\gev)
                  \le\mbox{$\mz$}
    \end{array}
    \right.
    \nonumber \\
    \label{eq:f_5q}
  \end{eqnarray}
with $s_0=(91.176\gev)^2$. 
The parametrization (\ref{eq:f_5q}) is copied from 
ref.~\cite{piqq_h} for $0\gev<\sqrt{|q^2|}<50\gev$
and smoothly connected to the most recent estimates
of ref.~\cite{piqq_h_latest} at $q^2=\mmz$; 
$f(\mmz)=(0.0283 \pm 0.0007)/\alpha$.
In the
time-like region ($0<q^2<\mmz$), the second term in 
(\ref{eq:piqq_5q_timelike}) is added in order to account approximately 
for the threshold contributions of the charm and bottom quarks.
 
Hadronic contribution to the photon-$Z$ mixing two-point function can
then be estimated as\cite{pi3q_h}
  \begin{equation}
     4\pi \Bigl[{\rm Re}\ol{\Pi}^{3Q}_{T,\gamma}(q^2)
                       -\ol{\Pi}^{3Q}_{T,\gamma}(0)\Bigr]_{5q}
     =
     2\pi \Bigl[{\rm Re}\ol{\Pi}^{QQ}_{T,\gamma}(q^2)
                       -\ol{\Pi}^{QQ}_{T,\gamma}(0)\Bigr]_{5q}
     +\Delta_{\omega\phi}(q^2) +\Delta_c(q^2) +\Delta_b(q^2),
     \label{eq:pi3q_5q}
  \end{equation}
where
  \begin{eqnarray}
     \Delta_{\omega\phi}(q^2) & = & \frac{3 q^2}{4\alpha^2}\,
     \Biggl\{\,
          \frac{2\Gamma(\omega\to e^+e^-)}{m_\omega(m^2_\omega+|q^2|)}
         -\frac{\Gamma(\phi\to e^+e^-)}{m_\phi(m^2_\phi+|q^2|)}\,
     \Biggr\} \,,
     \label{eq:delwp}
  \end{eqnarray}
is an estimate\cite{pi3q_h} for the extra contribution from the $u$,
$d$, $s$ quarks, and
  \begin{eqnarray}
     \Delta_q(q^2) &=& \frac{1}{6\pi} C_q\,Q_q
        \biggl\{ B_3(  0;m_q,m_q) -B_3(q^2;m_q,m_q)
      \nonumber \\
        & & \hspace{17mm} +\frac{\alpha_s}{6\pi}
                  \Big[B_V'(0;m_q,m_q)-B_V'(q^2;m_q,m_q)\Bigr]
        \biggr\}\,,
     \label{eq:delq}
  \end{eqnarray}
for $q=c$ and $b$, are calculated perturbatively. Note that in the
$m_u = m_d = m_s$ limit, the identity
$\Delta_u + \Delta_d + \Delta_s = 0$ holds. Thus, the term
$\Delta_{\omega\phi}$ gives an estimate \cite{pi3q_h} of the flavor
SU(3) violation effect. Contributions of leptons, the top quark and any
other new particles, as well as the light 5-quark contributions
at $\sqrt{|q^2|}> \mz$ are treated perturbatively.
 
The light quark masses to be used
in the region $|q^2|>\mmz$ are determined by requiring continuity of
the two effective charges at $q^2 = m^2_Z$.  The left-hand sides of
eqs.~(\ref{eq:piqq_5q}) and (\ref{eq:pi3q_5q}) are evaluated
perturbatively, and equated with the estimate at $q^2 =m^2_Z$~:
  \begin{eqnarray}
     4\pi\, \Bigl[Re \ol{\Pi}^{QQ}_{T,\gamma}(m^2_Z)
                   - \ol{\Pi}^{QQ}_{T,\gamma}(0)\Bigr]_{5q}
     &=& -f(m^2_Z)\,,
     \label{eq:piqq_5q_matching}
     \\
     4\pi\Bigl[{\rm Re}\ol{\Pi}^{3Q}_{T,\gamma}(\mmz)
                  -\ol{\Pi}^{3Q}_{T,\gamma}(0)\Bigr]_{5q}
     &=& -\frac{1}{2}f(\mmz)+\Delta_{\omega\phi}(\mmz)
         +\Delta_c(\mmz)+\Delta_b(\mmz)\,,\qquad
    \label{eq:pi3q_5q_matching}
  \end{eqnarray}
where the mean value of the estimate\cite{piqq_h_latest}
  \begin{eqnarray}
     -f(m^2_Z) 
      = \frac{-0.0283 \pm 0.0007}{\alpha}
      &\equiv& -3.88 +\delta_{\rm had}\; ;\quad
     \delta_{\rm had} = 0\pm 0.1
    \label{eq:delta_had}
  \end{eqnarray}
is taken at $\mz = 91.187$ GeV. Note that the additional term at the
right-hand side of (\ref{eq:piqq_5q_timelike}) is less than $0.001$ and
the discontinuity at $q^2\!=\!\mmz$ is negligibly small. With the use
of the expressions (\ref{eq:fpiqq_t}) and (\ref{eq:fpi3q_t}),
the two matching conditions can be approximated by:
  \begin{equation}
      \sum_{5q} Q_q^2
      \biggl[ \frac{1}{6} \ln\frac{\mmz}{m_q^2} -\frac{5}{18}\biggr]
      +O\Bigl(\frac{m_q^2}{\mmz},\frac{\alpha_s}{\pi}\Bigr)
     = \frac{\pi}{6}f(\mmz),
     \label{eq:qmass_matching1}
  \end{equation}
  \begin{equation}
     \ln\frac{m_d m_s}{m^2_u}
      +O(\frac{\alpha_s}{\pi},\frac{m^2_s}{\mmz})
      =18\pi\Delta_{\omega\phi}(\mmz)
      =0.152\,.
     \label{eq:qmass_matching2}
  \end{equation}
Taking the charm and bottom quark masses
  \begin{subequations}
     \label{eq:mcmb}
  \begin{eqnarray}
     m_c & = & 1.4 \gev,
     \label{eq:mc} \\
     m_b & = & 4.7 \gev,
     \label{eq:mb}
  \end{eqnarray}
  \end{subequations}
and including $O(\alpha_s)$ corrections one finds for $\delta_{\rm
had}=0$~:
 \begin{equation}
   \label{eq:tableofqmass}
 \begin{array}{|c|c|c|c|c|}
    \hline
    \alpha_s(\mz)  \:&\: 0     \:&\: 0.11  \:&\: 0.12  \:&\: 0.13  \\
    \hline
    m_u=m_d(\gev)  \:&\: 0.055 \:&\: 0.089 \:&\: 0.093 \:&\: 0.097 \\
    \hline
    m_s(\gev)      \:&\: 0.064 \:&\: 0.104 \:&\: 0.108 \:&\: 0.113 \\
    \hline
 \end{array}
 \end{equation}
Our program finds appropriate light quark masses for arbitrary
$\alpha_s(\mz)$, $\delta_{\rm had}$, $m_c$ and $m_b$ input values by
solving the continuity conditions (\ref{eq:piqq_5q_matching}) and
(\ref{eq:pi3q_5q_matching}). It should be pointed out here that these
light quark masses are fixed merely to ensure the continuity of the
effective charges at $q^2\!=\!\mmz$ and that they do not have a direct
physical significance. At $|q^2| > m^2_Z$, where those quark masses are
used, the mass effects are suppressed by $m^2_q/m^2_Z$ and never become
significant. Whenever the light quark mass values play a physically
significant role, their values must be chosen independent of those of
eq.~(\ref{eq:tableofqmass}) by appropriate physics
arguments.
 
In Fig.~\ref{figureofchargeformfactors} the SM predictions for the
effective charge $4\pi/\bar{e}^2(q^2)$ and the effective weak mixing
angle $\bar{s}^2(q^2)$ are shown in the region 1~MeV
$<\sqrt{|q^2|}<1$~TeV for $m_t =100$, $150$, $200 \gev$ and
$\mh = 100, 1000\gev$ with $\delta_{\rm had} = 0$. The solid lines show
the space-like $(q^2 < 0)$ effective charge, whereas the dashed lines
the time-like $(q^2 > 0)$ effective charge. The top-quark effect at
$q^2=\mmz$ can be parametrized by
  \begin{eqnarray}
     \frac{1}{\bar{\alpha}(\mmz)}_{\rm SM} \qquad
     &=&\; 128.71 +\delta_{\rm had}
      +0.024 \Bigl(1+5\frac{\alpha_s}{\pi}\Bigr)
      \biggl(\frac{100 \gev}{m_t}\biggr)^2,
     \label{eq:alphamz_sm}
     \\
     \bigg[ \frac{\sbar^2(\mmz)}{\bar{\alpha}(\mmz)}
           -\frac{\sbar^2(0)}{\alpha}
     \biggr]_{\rm SM}
     &=& -3.09 + \frac{\delta_{\rm had}}{2}
      +0.009\Bigl(1+5\frac{\alpha_s}{\pi}\Bigr)
            \biggl(\frac{100\gev}{m_t}\biggr)^2\,,
     \label{eq:sbar_running_sm}
  \end{eqnarray}
for $m_t >100\gev$ 
representing typical contributions of a heavy particle
to the running of the effective charge form factor $\bar{\alpha}(q^2)$
and $\bar{s}^2(q^2)$ between $q^2 = 0$ and $\mmz$.
 
When constraining new physics contributions the value of
$\bar{\alpha}(\mmz)$
is required, but only $\alpha=\bar{\alpha}(0)$, that is, the fine 
structure
constant, is precisely measured. When new physics is contributing
significantly to the running of the effective charge form factors
between $q^2=0$ and $q^2=\mmz$, its value can deviate from the SM
prediction (\ref{eq:alphamz_sm}). In order to account for
both such new physics contributions and future improvements in the
measurement of $\delta_{\rm had}$, the parameter
\begin{equation}
\label{eq:delta_alpha_def}
 \delta_\alpha \equiv \frac{1}{\bar{\alpha}(\mmz)} -128.72\,.
\end{equation}
is introduced. For instance, in the SM one finds
from (\ref{eq:alphamz_sm})
\begin{equation}
\label{eq:delta_alpha_sm}
 [\delta_\alpha]_{\rm SM} =
  \delta_{\rm had}
  +0.024 \Bigl(1+5\frac{\alpha_s}{\pi}\Bigr)
        \biggl(\frac{100 \gev}{m_t}\biggr)^2
  -0.01\,.
\end{equation}
The last two terms are close to zero for $m_t =150$--$200$GeV, such that
within the SM~:
\begin{equation}
\label{eq:delta_alpha_sm2}
 [\delta_\alpha]_{\rm SM} \approx \delta_{\rm had} \,.
\end{equation}
In general, new physics contributions can be accounted for by
\begin{equation}
\label{eq:delta_alpha}
 \delta_\alpha = [\delta_\alpha]_{\rm SM}
 +4\pi\Bigl[ {\rm Re}\Pi^{QQ}_{T,\gamma}(\mmz)
      -\Pi^{QQ}_{T,\gamma}(    0)
      \Bigr]_{\rm New Physics} \,.
\end{equation}
An example of the extra term is found in ref.\cite{hisz}, where
consequences of the gauge-invariant dimension six operators\cite{gw}
have been studied in detail.
 
The $\msbar$ couping constants $\hat{e}^2(\mu)$ and $\ghat^2(\mu)$ are
determined from the identities eqs.~(\ref{eq:ebar_def}) and
(\ref{eq:gbar_def}) evaluated at large $|q^2|$, say at $q^2 = m^2_Z$.
The magnitude of $\ehat^2(\mz)$ depends on $m_t$ and the assumed
$\alpha_s(\mz)$ value, and that of $\shat^2(\mz)$ depends also on the
$\sbar^2(\mmz)$ value as observed at LEP/SLC. For $\alpha_s = 0.12$
one obtains
  \begin{eqnarray}
     \frac{1}{\hat{\alpha}(\mz)_{\rm SM}}
 \equiv \frac{4\pi}{\hat{e}^2(\mz)_{\rm SM}} =
        128.00 + \delta_{\rm had} +
     \left\{\begin{array}{lll}
             -0.12  & \mbox{for} & m_t = 100 \gev \\
             +0.00  & \mbox{for} & m_t = 150 \gev \\
             +0.08  & \mbox{for} & m_t = 200 \gev \\
             +0.15  & \mbox{for} & m_t = 250 \gev
            \end{array} \right.\,,
    \label{eq:alphahat_vs_mt}
  \end{eqnarray}
and
  \begin{equation}
     \hat{s}^2(\mz)_{\rm SM}
 = \sbar^2(\mmz) +\frac{\alpha}{2}\delta_{\rm had}
 + \left\{
       \begin{array}{ccc}
         0.0007 & \mbox{for} & m_t= 100 \gev \\
         0.0009 & \mbox{for} & m_t= 150 \gev \\
         0.0010 & \mbox{for} & m_t= 200 \gev \\
         0.0011 & \mbox{for} & m_t= 250 \gev
       \end{array}
      \right. \,,
    \label{eq:shat_vs_mt}
  \end{equation}
The relatively large $m_t$ dependences above, as opposed to those of
eqs.~(\ref{eq:alphamz_sm}) and (\ref{eq:sbar_running_sm}), result from
the non-decoupling of the heavy top quark due to the logarithmic $m_t$
dependence of the $\msbar$ renormalized two-point functions in
eqs.~(\ref{eq:ebar_def}) and (\ref{eq:gbar_def}), as explained earlier.
In the presence of many new particles at the TeV scale, such as in the
supersymmetric standard model, all new particle contributions are
suppressed by their inverse mass-squared as demonstrated for a heavy
top quark in (\ref{eq:alphamz_sm}) and (\ref{eq:sbar_running_sm})
for the effective charges, while
the magnitude of $\ehat^2(\mz)$ and
$\shat^2(\mz)$ are affected strongly.
One should then either adopt the effective light particle theory 
for the $\msbar$ couplings\cite{stu_mod,marciano,llm,sgut93} 
or use the above effective
charges below TeV scale.
 
Finally, it is worth mentioning that the expressions for the running
of the remaining 
two charge form factors:
  \begin{eqnarray}
     \frac{1}{\gzbar^2(q^2)} &=& \frac{1}{\gzhat^2(\mu)}
     +{\rm Re} \ol{\Pi}_{T,Z}^{33}(q^2)
     -2\shat^2 {\rm Re} \ol{\Pi}_{T,Z}^{3Q}(q^2)
     +\shat^4  {\rm Re} \ol{\Pi}_{T,Z}^{QQ}(q^2)\,,
     \label{eq:gzbar_def}
     \\
     \frac{1}{\gwbar^2(q^2)} &=& \frac{1}{\ghat^2(\mu)}
     +{\rm Re} \ol{\Pi}_{T,W}^{11}(q^2)\,,
     \label{eq:gwbar_def}
  \end{eqnarray}
are not the exact solution of the one-loop RG equations of the $\msbar$
couplings, but that the $O(1)$ terms at the right-hand sides remain
small at all $q^2$, provided the renormalization condition
  \begin{subequations}
     \label{eq:r_cond}
  \begin{eqnarray}
     \ehat^2(\mu) &=& \ebar^2(Q^2)\,,
     \\
     \shat^2(\mu) &=& \sbar^2(Q^2)\,,
  \end{eqnarray}
  \end{subequations}
is chosen with
  \begin{eqnarray}
     Q^2 = \left\{
       \begin{array}{lll}
          \mmz &\mbox{if} & |q^2|\:\simlt\: O(\mmz)\,,\\[2mm]
          q^2   &\mbox{if} & |q^2|     >       \mmz \,. \\
       \end{array}
     \right.
     \label{eq:Q2}
  \end{eqnarray}
Therefore, $\gzbar^2(q^2)$ and $\gwbar^2(q^2)$ can be safely calculated
from
  \begin{eqnarray}
     \frac{4\,\pi}{\gzbar^2(q^2)}
      &=& \frac{\sbar^2(Q^2)\,\cbar^2(Q^2)}{\bar{\alpha}(Q^2)}
          +\frac{1}{4}\,S_Z(q^2)\,,
     \label{eq:gzbar_vs_sz}
     \\
     \frac{4\,\pi}{\gwbar^2(q^2)}
      &=& \frac{\sbar^2(Q^2)}{\bar{\alpha}(Q^2)}
          +\frac{1}{4}\,S_W(q^2)\,,
     \label{eq:gwbar_vs_sw}
  \end{eqnarray}
where the two quantities
  \begin{subequations}
     \label{eq:szsw_def}
  \begin{eqnarray}
     S_Z(q^2)
     &\equiv&16\,\pi\,{\rm Re}\biggl\{
         \ol{\Pi}_{T,\gamma}^{3Q}(Q^2) -\ol{\Pi}_{T,Z}^{33}(q^2)
           -2\sbar^2(Q^2)\Bigl[
         \ol{\Pi}_{T,\gamma}^{3Q}(Q^2) -\ol{\Pi}_{T,Z}^{3Q}(q^2)
                    \Bigr]
        \nonumber \\
     & & \hspace{2cm} + \sbar^4(Q^2)\Bigl[
         \ol{\Pi}_{T,\gamma}^{QQ}(Q^2) -\ol{\Pi}_{T,Z}^{QQ}(q^2)\Bigr]
      \biggr\}\,,
     \label{eq:sz_def}
     \\[2mm]
     S_W(q^2)
       &\equiv& 16\,\pi\,{\rm Re}\Bigl[
           \ol{\Pi}_{T,\gamma}^{3Q}(Q^2)
          -\ol{\Pi}_{T,W}^{11}(q^2)_{\shat^2=\sbar^2(Q^2)} \Bigr]\,,
     \label{eq:sw_def}
  \end{eqnarray}
  \end{subequations}
remain small (free of large logarithm) at all $q^2$, $0<|q^2|<\infty$.

In principle, the parametrization (\ref{eq:f_5q}) can be used
to account for the hadronic contributions to the
$\ol{\Pi}_{T,Z}^{33}(q^2)$ and $\ol{\Pi}_{T,W}^{11}(q^2)$ terms 
at $|q^2|<\mmz$ 
with the help of the CVC and PCAC hypotheses.
However, we find that the contribution of light hadrons are negligible 
at low momentum transfers $|q^2| \ll \mmz$,
and hence the perturbative expressions (\ref{eq:fpi_t}) with the light 
quark masses as obtained by the matching conditions 
(\ref{eq:piqq_5q_matching}) and (\ref{eq:pi3q_5q_matching})
are used when evaluating these functions.
 
It is important to note that the expressions (\ref{eq:gzbar_vs_sz}) and
(\ref{eq:gwbar_vs_sw}) are valid in the sense of a  perturbative
expressions, and therefore the scale $Q^2$ has been chosen such that the
$S_Z(q^2)$ and $S_W(q^2)$ terms remain small. The typical scale of the
charge form factors $\gzbar^2(q^2)$ and $\gwbar^2(q^2)$ are $Q^2=\mmz$
rather than $Q^2=q^2$ for $|q^2|\ll \mmz$. Our definitions of the $S$ and
$U$ parameters then follow
  \begin{eqnarray}
     S   &=& S_Z(0) = 16\pi\,{\rm Re}
             \Bigl[ \ol{\Pi}_{T,\gamma}^{3Q}(\mmz)
                   -\ol{\Pi}_{T,Z}^{33}(0) \Bigr]\,,
\label{eq:s_def}\\
     S+U &=& S_W(0) = 16\pi\,{\rm Re}
          \Bigl[ \ol{\Pi}_{T,\gamma}^{3Q}(\mmz)
                -\ol{\Pi}_{T,W}^{11}(0)_{\shat^2=\sbar^2(\mmz)} \Bigr]\,.
    \label{eq:u_def}
  \end{eqnarray}

\newpage
\section*{\large \bf App.~C  $\:\:$
SM contributions to $S$, $T$, $U$ and $\protect\delb(\mmz)$ }
\renewcommand{\theequation}{C.\arabic{equation}}
\cleqn
 
\pr
This Appendix deals with the SM contributions to the universal
electroweak parameters $S$, $T$, $U$ and the $Zb_Lb_L$ vertex
form factor $\delb(\mmz)$ used as free fit parameters. The complete
analytic formulae are given at one-loop level and the two-loop
corrections are also included as far as they are known.
We adopt the perturbative order 
$\alpha\alpha_s$\cite{kniehl,rho_pqcd,pivv_pqcd,rho_pqcd_mh2,zbb_qcd} 
corrections at $\alpha_s =\alpha_s(\mz)_{\msbar}$ in evaluating 
the $S$, $T$, $U$ and $\delb(\mmz)$ parameters, 
since it allows the readers 
to reproduce our results unambiguously and straightforwardly.
The effects due to non-perturbative threshold 
corrections\cite{rho_thr,rho_thr2,rho_thr3} should be evaluated carefully,
and one can obtain more precise predictions of the SM from our formulae
by adjusting the effective top-quark mass to produce the same $S$, $T$,
$U$ and $\delb(\mmz)$ values.
 
\subsection*{\normalsize \bf C.1  $S_{\rm SM}$ }
 
\pr
The $S$ parameter in the SM can be expressed as a sum of three pieces:
\bea
      S_{\rm SM} &=& S_\ell +S_q +S_B \,,
\label{eq:s_sm}
\eea
where the indices denote contributions from 
the leptonic, hadronic and the bosonic
(that is, $W,Z,H$) sectors of the SM, respectively. Each term is separately
finite. $S_\ell$ and $S_B$ are given at one-loop order, whereas the
hadronic contribution $S_q$ with the two-loop $O(\alpha\alpha_s)$
correction\cite{kniehl,pivv_pqcd,rho_pqcd_mh2}.
 
The leptonic contribution is a sum of three terms
\bea
 S_\ell = \frac{1}{\pi}\,\sum_{i=1}^3 \,
 G_S^\ell \biggl( \frac{m_{\ell_i}^2}{\mmz} \biggr)\,,
\label{eq:s_l}
\eea
where each generation ($\nu_i, \ell_i$) contributes
\bea
      G_S^\ell(x) = -\frac{1}{6}
           \Bigl\{\,\ln x +(1+5\,x)\,A(x)  -10\,x\, \Bigr\} \,.
\label{eq:gs_l}
\eea
The real function $A$ is
\bea
       A(x) &=&
 \left\{ \begin{array}{lll}
                  2\sqrt{1-4x}\; \ln \,
                    \displaystyle{\frac{1+\sqrt{1-4x}}{2\sqrt{x}}} &
                    \mbox{ for} & 0<x<\displaystyle{\frac{1}{4}}\,,
                  \\[3mm]
                  2\sqrt{4x-1}\; \tan^{-1}\,
                    \displaystyle{\frac{1}{\sqrt{4x-1}}} &
                    \mbox{ for} & x>\displaystyle\frac{1}{4}\,.
                  \end{array}
        \right.
\label{eq:ax}
\eea
For the case of charged lepton masses much smaller than the $Z$ mass
one finds
\bea
      G_S^\ell(x) &=& (\,2 +\frac{1}{2}\ln x \Bigr)\, x +O(x^2) \,.
\label{eq:gs_l_light}
\eea
The ($\nu_\tau\,,\,\tau$) doublet contribution is hence
$S_\ell\approx -0.0002$.
 
The hadronic contribution calculated up to $O(\alpha\alpha_s)$ 
two-loop level is\cite{kniehl,pivv_pqcd,rho_pqcd_mh2}:
\bea
      S_q = S_q^{(0)} +S_q^{(1)} \,.
\label{eq:s_q}
\eea
The one-loop contribution is again a sum of the three terms
\bea
 S_q^{(0)} &=& \frac{C_q}{\pi}\,\sum_{i=1}^3 \,
 G_S^q \biggl( \frac{m_{u_i}^2}{\mmz} \,,\,
        \frac{m_{d_i}^2}{\mmz} \biggr)\,,
\label{eq:s_q0}
\eea
with $C_q = 3$, where each quark generation contributes
\bea
 G_S^q(x,y) = \frac{1}{18} \Bigl\{\,
 \ln \frac{y}{x} -(1+11x)A(x) +(1-7y)A(y) +22x +14y \Bigr\}\,.
\label{eq:gs_q}
\eea
When both quarks are light as compared to $Z$, one finds
\bea
      G_S^q(x,y) =
  \Bigl(\,\frac{4}{3}+{\frac{1}{2}\,\ln x}\Bigr) x
      +\Bigl(\,\frac{2}{3}+{\frac{1}{2}\,\ln y}\Bigr) y
      +O(x^2,y^2) \,.
\label{eq:gs_q_light}
\eea
When only the down-type quark is light ($y\ll 1$), one finds
\bea
 G_S^q(x,y) &=& \frac{1}{18} \Bigl\{\,
 22x -\ln x -(1+11x)A(x) +3y(4+3\ln y) +O(y^2) \Bigr\}\,,
\label{eq:gs_q_tb1}\\
 &=& \frac{1}{18} \Bigl\{\,
 -\ln x -\frac{1}{6} +\frac{7}{20\,x} +3y(4+3\ln y)
 +O(\frac{1}{x^2},y^2) \Bigr\}\,.
\label{eq:gs_q_tb2}
\eea
For large $m_t$, the quark contribution $S_q$ becomes negative and
its magnitude grows logarithmically.
 
The two-loop $O(\alpha\alpha_s)$ 
correction\cite{kniehl,pivv_pqcd,rho_pqcd_mh2} can be expressed
in terms of the $B_V'$ and $B_A'$ functions (\ref{eq:bp_2loop})
of appendix A~:
\bea
 S_q^{(1)} &=& C_q\,\frac{\alpha_s}{\pi^2}\,\sum_{i=1}^3 \,\Bigl\{\,
    \frac{5}{36}\,B_V'(\mmz;u_i,u_i)
   +\frac{1}{36}\,B_V'(\mmz;d_i,d_i)
\nonumber\\
 && \hspace{20mm}
     -\frac{1}{12}\,B_A'(\mmz;u_i,u_i)
     -\frac{1}{12}\,B_A'(\mmz;d_i,d_i)
 \Bigr\}\,,
\label{eq:s_q1}
\eea
where the quark label stands for its mass as in appendix A.
It is easily seen that the right-hand side of the above equation is
independent of the unit-of-mass $\mu$ for each generation, and that
they are in fact a function of the ratios $m_{u_i}^2/\mmz$ and
$m_{d_i}^2/\mmz$. The contributions from the first two quark
generations are again negligible. The two-loop term $S_q^{(1)}$ is
hence dominated by the ($t$, $b$) doublet contribution, which can be
approximated by
\bea
 S_q^{(1)} &\approx&
C_q\,\frac{\alpha_s}{\pi^2}\,
  \biggl\{
 -\frac{1}{18}\ln\,x_t
 +\frac{5x_t}{9}V_1(\frac{1}{4x_t})
 -\frac{x_t}{3}[A_1(\frac{1}{4x_t}) -A_1(0)] \biggr\}
\nonumber\\
 &=&
C_q\,\frac{\alpha_s}{\pi^2}\, \biggl\{
 -\frac{1}{18}\ln\,x_t
 +\frac{2}{9}\zeta_3
 +\frac{1}{9} +O(\frac{1}{x_t}) \biggr\}\,,
\label{eq:s_q1_tb}
\eea
with $x_t\equiv m_t^2/\mmz$. 
The expression (\ref{eq:s_q1_tb}) agrees with ref.\cite{kniehl}. 
The following table shows the full
hadronic contribution $S_q$ for several values of $m_t$ in lowest
order ($\alpha_s=0$) and with the $O(\alpha\alpha_s)$ corrections for
$\alpha_s=0.12$:
\bea
     \begin{array}{|c|r|r|}
     \hline
      m_t\,(\GeV)  &
     \multicolumn{2}{|c|}{S_q}\\ \cline{2-3}
     &   \alpha_s=0 & \alpha_s=0.12
     \\
     \hline
      100  &  -0.008 &   0.010 \\
      120  &  -0.033 &  -0.017 \\
      140  &  -0.052 &  -0.038 \\
      160  &  -0.069 &  -0.055 \\
      180  &  -0.083 &  -0.070 \\
      200  &  -0.095 &  -0.082 \\
     \hline
     \end{array}
\label{table_sq}
\eea
using for the bottom mass $m_b=4.7 \GeV$.  The two-loop correction
is important for relatively small $m_t$ values.
 
The bosonic contribution is expressed as
\bea
     S_B = \frac{1}{\pi}\,
           \biggl\{ F_S\Bigl(\frac{\mmw}{\mmz}\Bigr)
                  + H_S\Bigl(\frac{\mmh}{\mmz}\Bigr) \biggr\}\,,
\label{eq:s_b}
\eea
where
\bsub
\label{eq:s_b2}
\bea
   \ F_S(c^2)
     &=& -\frac{1}{12}\ln c^2 - \frac{7}{8} - \frac{14}{3}\,c^2
         + \biggl(\frac{7}{3}\,c^2-\frac{1}{12}\biggr)\,A(c^2)
\label{eq:fs}
\\[2mm]
     &=& -1.451 \quad  
          \mbox{for} \quad c^2 = \biggl(\frac{80.24}{91.187}\biggr)^2
          \,,
\nonumber
\\[2mm]
     H_S(x)
     &=&
          \,\frac{3}{8}\,x -\frac{1}{12}\,x^2
         +\biggl[\frac{3-x}{4} +\frac{x^2}{24} +\frac{3}{4(1-x)} \biggr]
           x \ln x  +\biggl[1-\frac{x}{3} +\frac{x^2}{12}\biggr]B(x).
     \hspace{12mm}
\label{eq:hs}
\eea
\esub
Here $A(x)$ is given by (\ref{eq:ax}), and
\bea
     && B(x)
       =  \left\{ \begin{array}{lll}
                  \sqrt{x(4-x)}\; \tan^{-1}\,
                    \displaystyle{\sqrt{\frac{4}{x}-1}} &
                    \mbox{ for } & 0<x<4 \,,
                  \\[3mm]
                  \sqrt{x(x-4)}\; \ln \,
                    \displaystyle{\frac{2}{\sqrt{x}+\sqrt{x-4}}} &
                    \mbox{ for } & x>4 \,.
                  \end{array}
          \right.
\label{eq:bx}
\eea
For large $\mh$ one has
\bea
     H_S\Bigl(\frac{\mmh}{\mmz}\Bigr) &=&\;
        \frac{1}{12}\ln \frac{\mmh}{\mmz}
               +\frac{37}{36} - \frac{17}{48}\,\frac{\mmz}{\mmh}
       + O\!\left(\frac{m_Z^4}{m_H^4}\right) \,.
\label{eq:hs_heavy}
\eea
The total bosonic contribution $S_B$ is tabulated below 
for several $\mh$ values~:
\begin{eqnarray}
\begin{array}{|c|c|}
\hline
\mh\,(\GeV) & S_B \\
\hline
  50  & -0.234 \\
 100  & -0.166 \\
 200  & -0.107 \\
 400  & -0.061 \\
1000  & -0.008 \\
\hline
\end{array}
\label{table_sb}
\end{eqnarray}
 
\subsection*{\normalsize \bf C.2  $T_{\rm SM}$ }
 
\pr
The $T$ parameter in the SM can be expressed as a sum of
three individually finite pieces:
\bea
      T_{\rm SM} &=& T_\ell +T_q +T_B \,,
\label{eq:t_sm}
\eea
where the indices denote the leptonic, the hadronic and the bosonic
(that is, $\gamma,W,Z,H$) contributions, 
respectively. $T_\ell$ and $T_B$ are
evaluated at one-loop order, whereas $T_q$ contains irreducible
two-loop contributions in $\alpha\alpha_s$ 
order\cite{kniehl,rho_pqcd,pivv_pqcd,rho_pqcd_mh2} and in
the $m_t^4$ order\cite{rho_zbb_ew,rho_ew2,rho_ew3}. Reducible
higher-oder contributions\cite{rho_ew1} are taken account of by the
identity (\ref{gzbarfromt}).
 
The leptonic contribution is a sum of three terms
\bea
 T_\ell = \frac{G_F\, \mmz}{2\sqrt{2}\,\pi^2\, \alpha}\,\sum_{i=1}^3\,
 G_T\biggl(0\,,\, \frac{m_{\ell_i}^2}{\mmz} \biggr)\,,
\label{eq:t_l}
\eea
where
\bea
      G_T(x,y) = \frac{x+y}{4} +\frac{x y}{2\,(x-y)}\ln \frac{y}{x} \,.
\label{eq:gt}
\eea
The leptonic contribution of the first three generations is hence
negligible; even the ($\nu_\tau\,,\,\tau$) doublet contributes to
$T_\ell$ only about 0.00005.
 
The hadronic contribution is calculated including the
$O(\alpha\alpha_s)$\cite{kniehl,rho_pqcd,pivv_pqcd,rho_pqcd_mh2} 
and the irreducible
$O(m_t^4)$\cite{rho_zbb_ew,rho_ew2,rho_ew3} two-loop corrections:
\bea
      T_q = T_q^{(0)} + T_q^{(1)} +T_q^{(2)}\,.
\label{eq:t_q}
\eea
The one-loop contribution is a sum of the nine terms
\bea
 T_q^{(0)} &=& \frac{G_F\, \mmz}{2\sqrt{2}\,\pi^2\, \alpha}\,C_q\,
 \sum_{i,j=1}^3 \, |V_{u_i,d_j}|^2\,
 G_T\biggl( \frac{m_{u_i}^2}{\mmz} \,,\,
        \frac{m_{d_j}^2}{\mmz} \biggr)\,,
\label{eq:t_q0}
\eea
with $C_q = 3$. The function $G_T(x,y)$ is found in (\ref{eq:gt}). In
the limit of the diagonal KM matrix elements $V_{ij}=\delta_{ij}$,
the contributions from the light quarks of the first and second
generations can be neglected.
 
The two-loop contributions are only important for the $t$-$b$ doublet~:
\bea
      T_q^{(1)} &=& -C_q\,\frac{\alpha_s}{\pi}\cdot\frac{3+\pi^2}{18}\,
 \biggl(\frac{m_t^2}{\mmz}\biggr)
 \biggl(\frac{G_F\,\mmz}{2\,\sqrt{2}\,\pi^2\,\alpha}\biggr)\,,
\label{eq:t_q1}
\\
      T_q^{(2)} &=& C_q\,\alpha\,
 \biggl(\frac{m_t^2}{4\,\mmz}\biggr)^2\,
 \biggl(\frac{G_F\,\mmz}{2\,\sqrt{2}\,\pi^2\,\alpha}\biggr)^2\,
 \rho^{(2)}(\mh/m_t)\,,
\label{eq:t_q2}
\eea
where terms of order $m_b^2/m_t^2$ are neglected. 
The function $\,\rho^{(2)}\,(\mh/m_t)\,$ gives\cite{rho_zbb_ew}~: 
$\rho^{(2)}(\mh/m_t)= -0.74, -4.72, -6.95, -11.70, -10.74$,
for $\mh/m_t = 0,\frac{1}{2},1,5,10$. The numerical value of the
'expansion parameter' in the above expressions is
$G_F\mmz/2\sqrt{2}\pi^2\alpha =0.4761$.
 
The following table shows the contributions from each term in
eq.~(\ref{eq:t_q}) for several values of $m_t$, the lowest order
contribution $T_q^{(0)}$ and the $O(\alpha\alpha_s)$ contribution
$T_q^{(1)}$ with $\alpha_s = 0.12$,
and $O(m_t^4)$ contribution $T_q^{(2)}$ with $\mh=100,1000 \gev$:
\bea
     \begin{array}{|c|c|c|c|c|}
     \hline
      m_t\,(\GeV)  & \vphantom{\Bigl]}
      T_q^{(0)} & T_q^{(1)} &
     \multicolumn{2}{|c|}{T_q^{(2)}} \\
      \cline{3-5}
     &   & \alpha_s=0.12 & \mh=100 & \mh=1000
     \\
     \hline
      100  & \quad 0.419 \quad& -0.047  &  -0.003 &  -0.005 \\
      120  & \quad 0.607 \quad& -0.068  &  -0.006 &  -0.011 \\
      140  & \quad 0.830 \quad& -0.092  &  -0.010 &  -0.020 \\
      160  & \quad 1.087 \quad& -0.120  &  -0.016 &  -0.035 \\
      180  & \quad 1.379 \quad& -0.152  &  -0.024 &  -0.055 \\
      200  & \quad 1.705 \quad& -0.188  &  -0.034 &  -0.084 \\
     \hline
     \end{array}
\label{table_tq}
\eea
using for the bottom mass $m_b=4.7 \gev$ in $T_q^{(0)}$.
 
The bosonic contribution is
\bea
      T_B = \frac{G_F\,\mmz}{2\,\sqrt{2}\,\pi^2\,\alpha}\,
 \biggl[\,F_T\Bigl(\frac{\mmw}{\mmz}\Bigr)
               + H_T\Bigl(\frac{\mmh}{\mmz}\Bigr) \biggr]\,,
\label{eq:t_b}
\eea
where
\bea
     F_T(c^2) &=& \biggl( \frac{1}{4}+2 \shat^2 \biggr)\,
                  \frac{c^2\, \ln c^2}{1-c^2}
     -\frac{3}{4}\,c^2 +1-\shat^2
\label{eq:ft}
\\
     &=& -0.4371 \quad
          \mbox{for} \quad c^2 = \biggl(\frac{80.24}{91.187}\biggr)^2
                               \mbox{ and }\; \shat^2 =0.2312\,,
\nonumber
\\
     H_T(x) &=& \frac{3}{4}\,x
       \biggl[ \frac{\ln x}{1-x}
              -\frac{\ln(x/c^2)}{1-x/c^2} \biggl]\;;
     \quad c^2 \equiv\frac{\mmw}{\mmz}\,.
\label{eq:ht}
\eea
For a heavy Higgs boson ($\mmh \gg \mmz$) one finds
\bea
     H_T\biggl(\frac{\mmh}{\mmz}\biggr) &=&
     -\frac{3}{4}\,\biggl[(1-c^2)\ln \frac{\mmh}{\mmz}
      +c^2 \ln c^2
      +\biggl\{(1-c^4)\ln \frac{\mmh}{\mmz} +c^4\ln c^2\biggr\}
       \frac{\mmz}{\mmh}
      + O\biggl(\frac{m_Z^4}{m_H^4}\biggr)\biggr] .
  \nonumber \\
\label{eq:ht_heavy}
\eea
The total bosonic contribution $T_B$ is tabulated below for several $\mh$
values~:
\begin{eqnarray}
\begin{array}{|c|c|}
\hline
\mh\,(\GeV) & T_B \\
\hline
  50  & -0.227 \\
 100  & -0.257 \\
 200  & -0.314 \\
 400  & -0.396 \\
1000  & -0.529 \\
\hline
\end{array}
\label{table_tb}
\end{eqnarray}
 
\subsection*{\normalsize \bf C.3  $U_{\rm SM}$ }
 
\pr
The $U$ parameter in the SM can be expressed as a sum of three pieces:
\bea
      U_{\rm SM} &=& U_\ell +U_q +U_B \,,
\label{eq:u_sm}
\eea
where the indices denote the leptonic, the hadronic and the bosonic
(that is, $\gamma,W,Z,H$) contributions. Each term is separately finite.
$U_\ell$ and $U_B$ are given at one-loop order, whereas the hadronic
contribution $U_q$ is given with the 
two-loop $O(\alpha\alpha_s)$ correction
\cite{kniehl,rho_pqcd,pivv_pqcd,rho_pqcd_mh2}.
 
The leptonic contribution is a sum of three terms
\bea
 U_\ell = \frac{1}{\pi}\,\sum_{i=1}^3 \,
 G_U \biggl(0\,,\,\frac{m_{\ell_i}^2}{\mmz} \biggr)\,,
\label{eq:u_l}
\eea
where the contribution of each generation ($\nu_i, \ell_i$) is 
\bea
     G_U(x,y) &=&
         -\frac{x +y}{3} -\frac{1-x}{6}A(x) -\frac{1-y}{6}A(y)
                  +f_U^{}(x/c^2,y/c^2) \,.
\label{eq:gu}
\eea
Here $A(x)$ is given in (\ref{eq:ax}), and
\bea
     f_U^{}(x,y) &=& \frac{x+y}{12} +\frac{(x-y)^2}{6}
               + \frac{(x-y)^4-3\,(x^2+y^2)}{12\,(x-y)\,}
                \ln\frac{y}{x}
\nonumber \\
                 && + \frac{(x-y)^2 +x+y -2}{6}\,
                \beta(x,y)L(x,y)\,,
\label{eq:f_u}
\eea
with
\bea
      \beta(x,y) = \sqrt{ \bigl| 1-2(x+y)^2 +(x-y)^2 \bigr|}\,,
\label{eq:beta_xy}
\eea
and
\bea
     L(x,y)
       &=&\left\{
        \begin{array}{lll}
           \displaystyle{\frac{1}{2}} \ln
           \displaystyle{\frac{1-x-y+\beta(x,y)}{1-x-y-\beta(x,y)}} &
           \mbox{for} & |\sqrt{x}-\sqrt{y}|>1 \mbox{ or }
                           \sqrt{x}+\sqrt{y} <1 \,,\\
           \displaystyle{\tan^{-1}\frac{1-x+y}{\beta(x,y)}}
            +\displaystyle{\tan^{-1}\frac{1+x-y}{\beta(x,y)}} &
           \mbox{for} & |\sqrt{x}-\sqrt{y}|<1<\sqrt{x}+\sqrt{y}\,.
       \end{array}
       \right.
\nonumber \\[-5mm] &&
\label{eq:l_xy}
\eea
In the limit of vanishing lepton mass one has
\bea
 G_U(0,0) = \frac{1}{3}\,\ln c^2\,,
\label{eq:gu_light}
\eea
for $c^2\equiv \mmw/\mmz$, and hence the contribution of the first
three lepton generations is
\bea
      U_\ell &\approx& \frac{1}{\pi}\,\ln c^2
      =-0.0814 \quad
      \mbox{for} \quad c^2 = \biggl(\frac{80.24}{91.187}\biggr)^2\,.
\label{eq:u_l_light}
\eea
 
The hadronic contribution is calculated up to two-loop
$O(\alpha\alpha_s)$ level\cite{kniehl,pivv_pqcd,rho_pqcd_mh2}:
\bea
      U_q = U_q^{(0)} +U_q^{(1)} \,.
\label{eq:u_q}
\eea
The one-loop contribution is again a sum of the three terms in the
limit of diagonal KM matrix elements $V_{ij}=\delta_{ij}$:
\bea
 U_q^{(0)} &=& \frac{C_q}{\pi}\,\sum_{i=1}^3 \,
 G_U \biggl( \frac{m_{u_i}^2}{\mmz} \,,\,
        \frac{m_{d_i}^2}{\mmz} \biggr)\,,
\label{eq:u_q0}
\eea
with $C_q = 3$, where $G_U(x,y)$ has been given above
in (\ref{eq:gu}). For the first two quark generations the approximation
(\ref{eq:gu_light}) holds. For the contribution of the ($t\,,\,b$)
doublet the following approximation is useful:
\bea
     G_U(x,y) &=&
        \frac{1}{6} \ln x +\frac{1}{12} -\frac{1}{2}\,y\,\ln y
      + \biggl( \frac{1}{40}-\frac{1}{8}\,c^2 +\frac{2}{3}\,y \biggr)
        \frac{1}{x} +O\biggl(\frac{1}{x^2},y^2 \biggr)\,.
     \quad
\label{eq:gu_tb}
\eea
 
The two-loop $O(\alpha\alpha_s)$ 
correction\cite{kniehl,pivv_pqcd,rho_pqcd_mh2} can be expressed by
\bea
 U_q^{(1)} = C_q\,\frac{\alpha_s}{\pi^2}\sum_{i=1}^3\,
 G_U^{(1)}(m_{u_i},m_{d_i})\,,
\label{eq:u_q1}
\eea
where the two-loop function $G_U^{(1)}$ is given in terms of the $B_V$
and $B_A$ functions (\ref{eq:b_2loop}) of appendix A:
\bea
 G_U^{(1)}(m_u,m_d) &=&
 {\rm Re}\biggl\{
  \frac{1}{12}[B_V'+B_A'](\mmz;m_u,m_u)
 +\frac{1}{12}[B_V'+B_A'](\mmz;m_d,m_d)
 \nonumber \\
 && \hspace{1cm}
 -\frac{1}{6} [B_V'+B_A'](\mmw;m_u,m_d) \biggr\}\,.
\label{eq:gu_1}
\eea
It is readily seen that the function $G_U^{(1)}$ is independent of the
unit-of-mass $\mu$. The contribution of the first two quark generations
can be approximated by
\bea
 G_U^{(1)}(0,0) = \frac{1}{3} \ln c^2 \,,
\label{eq:gu_1_light}
\eea
just like in eq.~(\ref{eq:gu_light}) for the one-loop contribution. The
top-bottom contribution can be approximated by
\bsub
\label{eq:gu_1_tb}
\bea
 G_U^{(1)}(m_t,0) &=&
 \frac{1}{6}\ln\frac{m_t^2}{\mmz}
 +\frac{m_t^2}{3\mmz}\biggl[ V_1(\frac{\mmz}{4m_t^2})
               +A_1(\frac{\mmz}{4m_t^2}) -A_1(0) \biggr]
\nonumber\\&&
 -\frac{4m_t^2}{3\mmw}\biggl[F_1(\frac{\mmw}{m_t^2}) -F_1(0) \biggr]
\label{eq:gu_1_tb1}
\\
 &=& \frac{1}{6}\ln\frac{m_t^2}{\mmz}
 -\frac{2}{3}\,\zeta_3
 +\frac{4}{27}\,\zeta_2
 +\frac{1}{6}
 +O\bigg(\frac{\mmz}{m_t^2}\biggr)\,.
\label{eq:gu_1_tb2}
\eea
\esub
The expressions (\ref{eq:gu_1_light}) and (\ref{eq:gu_1_tb}) 
agree with ref.\cite{kniehl}. 
The following table shows the total hadronic contribution $U_q$ for
several values of $m_t$ in lowest order ($\alpha_s=0$) and with
$O(\alpha\alpha_s)$ corrections for $\alpha_s=0.12$:
\bea
     \begin{array}{|c|r|r|}
     \hline
      m_t\,(\GeV)  &
     \multicolumn{2}{|c|}{U_q}\\ \cline{2-3}
     &   \alpha_s=0 & \alpha_s=0.12
     \\
     \hline
      100  &  -0.118  & -0.148  \\
      120  &  -0.034  & -0.057  \\
      140  &   0.029  &  0.009  \\
      160  &   0.079  &  0.063  \\
      180  &   0.122  &  0.108  \\
      200  &   0.159  &  0.147  \\
     \hline
     \end{array}
\label{table_uq}
\eea
using for the bottom mass $m_b=4.7\gev$ in evaluating the lowest order
$G_U^{(0)}$, while the $O(\alpha\alpha_s)$ correction $G_U^{(1)}$
is calculated in the limit of vanishing bottom quark mass.
 
The bosonic contributions are given as
\bea
     U_B &=& \frac{1}{\pi}
          \biggl\{ F_U\Bigl(\frac{\mmw}{\mmz}\Bigr)
                 + H_U\Bigl(\frac{\mmh}{\mmz}\Bigr)\biggr\}
\label{eq:u_b}
\eea
where the constant term $F_U(\mmw/\mmz)$ is found to be
\bea
     F_U(c^2) &=&
       \left\{
              -2\,\biggl( \frac{1}{1-c^2} +\frac{1}{c^4}
                        -\frac{1}{6\,c^6}  \biggr)\, \shat^2
              +\frac{1}{4}\,
                 \biggl( \frac{1}{3}  -\frac1{1-c^2} +\frac{9}{c^2}
                         +\frac{7}{c^4}  - \frac{3}{2\,c^6} \biggr)
       \right\}
       \ln c^2
\nonumber \\ &&
      + \biggl(c^2+\frac{29}{4}\biggr) A(c^2)
      + \biggl\{\left(6+\frac{4}{c^2}-\frac{1}{c^4}\right)
                \frac{2\,\shat^2}{3}
           -\biggl(7+\frac{2}{c^2}-\frac{3}{4\,c^4}\biggr)\biggr\}
         B\biggl(\frac{1}{c^2}\biggr)
\nonumber \\
      && - \biggl(3-\frac{2}{3\,c^2}\biggr) \frac{\shat^2}{c^2}
     - \biggl(2\,c^2-\frac{1}{8}-\frac{19}{8\,c^2}+\frac{3}{4\,c^4}\biggr)
\label{eq:fu}
\\[2mm]
      &=& 1.043  \quad
          \mbox{for} \quad c^2 = \biggl(\frac{80.24}{91.187}\biggr)^2
                               \mbox{ and }\; \shat^2 =0.2312\,,
\nonumber
\eea
and the $\mh$ dependence is given by
\bea
 H_U(x_H^{}) &=& -H_S(x_H^{})
   +H_S\bigg(\frac{x_H^{}}{c^2}\biggr)\;;
 x_H^{} \equiv \frac{\mmh}{\mmz}\,.
\label{eq:hu}
\eea
The function $H_S(x_H^{})$ is defined in eq.~(\ref{eq:hs}).
In the large mass limit ($\mh \gg \mz$) the leading logarithm
($\ln\mh$) of the function $H_S$ (see eq.~(\ref{eq:hs_heavy}))
cancels in $H_U$ of eq.~(\ref{eq:hu}), and hence one has
\bea
     H_U\biggl(\frac{\mmh}{\mmz}\biggr)
     &=& \frac{1}{12}\ln c^2 +\frac{17}{48}(1-c^2) \frac{\mmz}{\mmh}
     +O\biggl(\frac{m_Z^4}{m_H^4}\biggr)\,.
\label{eq:hu_heavy}
\eea
Note, however, that the $\mh$ dependence is very small as seen from the
table below showing the total bosonic contribution $U_q$ for several
$\mh$ values:
\begin{eqnarray}
\begin{array}{|c|c|}
\hline
\mh\,(\GeV) & U_B \\
\hline
 50   &  0.345  \\
 100  &  0.344  \\
 200  &  0.341  \\
 400  &  0.340  \\
 1000 &  0.339  \\
\hline
\end{array}
\end{eqnarray}
 
\subsection*{\normalsize \bf C.4  $\delb(\mmz)_{\rm SM}$ }
 
\pr
The $\zbb$ vertex form factor $\delb(\mmz)_{\rm SM}$ in the
SM is expressed by:
  \begin{eqnarray}
     \delb(\mmz)
       &=& \delb^{(0)}(\mmz) +\delb^{(1)}(\mmz) +\delb^{(2)}(\mmz) \,.
      \qquad
      \label{eq:delta_b}
  \end{eqnarray}
The one-loop contribution
  \begin{eqnarray}
     \delb^{(0)}(\mmz) &=&
             \Gamma_1^{b_L}(\mmz)
            +\chat^2 \ol{\Gamma}_2^{b_L}(\mmz)
            +\Gamma_3^{b_L}(\mmz)
     \label{eq:delb_0}
  \end{eqnarray}
is calculated using the vertex functions of appendix~A, which can be
approximated by
  \begin{eqnarray}
      \delb^{(0)} &\approx&
         -0.00076 -0.00217 \biggl(\frac{m_t+36}{100}\biggr)^2\,,
     \label{eq:delb_approx}
  \end{eqnarray}
for $100\gev <m_t <250\gev$. The second term at the right-hand side of
eq.~(\ref{eq:delta_b}) is the $O(\alpha_s m_t^2)$ two-loop contribution
\cite{zbb_qcd}:
  \begin{eqnarray}
     \delb^{(1)}(\mmz) &=& \frac{\alpha_s}{\pi}\cdot
            2\, \Bigl(\frac{\pi^2}{3}-1\Bigr)
             \frac{G_F\, m_t^2}{8\sqrt{2}\, \pi^2 } \,.
     \label{eq:delb_1}
  \end{eqnarray}
The last term is the $O(m_t^4)$ two-loop contribution
\cite{rho_zbb_ew,zbb_ew}:
  \begin{eqnarray}
     \delb^{(2)}(\mmz) &=&
      - 2\,\biggl(\frac{G_F\, m_t^2}{8\sqrt{2}\, \pi^2 } \biggr)^2
           \,\tau^{(2)}(\mh/m_t) \,,
     \label{eq:delb_2}
  \end{eqnarray}
where the function $\tau^{(2)}(\mh/m_t)$ is given in
ref.~\cite{rho_zbb_ew}. For $\mh/m_t = 0,\frac{1}{2},1,5,10$,
it gives $\tau^{(2)}(\mh/m_t)=5.71,2.46,1.47,3.69,7.92$.
 
The following table shows the contributions from each term in
eq.~(\ref{eq:delta_b}), $\delb^{(0)}(\mmz)$,
$\frac{\alpha_s}{\pi}\delb^{(1)}(\mmz)$ with $\alpha_s = 0.12$,
and $\delb^{(2)}(\mmz)$ with $\mh=100,1000 \gev$, for several $m_t$
values:
\bea
     \begin{array}{|c|c|c|c|c|}
     \hline
      m_t\,(\GeV)  & \vphantom{\biggl]}
      \delb^{(0)}(\mmz) & \delb^{(1)}(\mmz) &
     \multicolumn{2}{|c|}{\delb^{(2)}(\mmz)} \\
      \cline{3-5}
     &   & \alpha_s=0.12 & \mh=100 & \mh=1000
     \\
     \hline
      100  &  -0.00481 & 0.00018 & -0.00000 & -0.00002  \\
      120  &  -0.00603 & 0.00026 & -0.00001 & -0.00003  \\
      140  &  -0.00746 & 0.00036 & -0.00002 & -0.00005  \\
      160  &  -0.00908 & 0.00047 & -0.00003 & -0.00007  \\
      180  &  -0.01089 & 0.00059 & -0.00005 & -0.00010  \\
      200  &  -0.01285 & 0.00073 & -0.00009 & -0.00013  \\
     \hline
     \end{array}
\label{table_delb}
\eea
 
\newpage
\section*{\large \bf App.~D  $\:\:$ One loop scalar functions }
\renewcommand{\theequation}{D.\arabic{equation}}
\cleqn
 
\pr
In this appendix explicit analytic expressions for the $B$ functions
are given, as well as the reduction of higher $C$ and $D$ functions
to $C_0$ and $D_0$ functions.
 
\subsection*{\normalsize \bf D.1  $A$, $B$, $C$ and $D$ functions}
\pr 
Following Passarino and Veltman \cite{pv} the $A$, $B$, $C$ and $D$
functions are defined by~:
  \begin{eqnarray}
     A(m_i)
       &=& \int {\ol{d^Dk} \over i \pi^2}
                {1 \over N_i}\,,
     \\
     \left[B_0,\,B^\mu,\,B^{\mu\nu}\right]\,(ij)
       &=& \int {\ol{d^Dk} \over i \pi^2}
                {\left[1,\,k^\mu,\,k^{\mu}k^{\nu}\right]
                 \over N_i N_j}\,,
     \\
     \left[C_0,\,C^\mu,\,C^{\mu\nu}\right]\,(ijk)
       &=& \int {\ol{d^Dk} \over i \pi^2}
                {\left[1,\,k^\mu,\,k^{\mu}k^{\nu}\right]
                 \over N_i N_j N_k}\,,
     \\
     \left[D_0,\,D^\mu,\,D^{\mu\nu}\right]\,(ijk\ell)
       &=& \int {\ol{d^Dk} \over i \pi^2}
                {\left[1,\,k^\mu,\,k^{\mu}k^{\nu}\right]
                 \over N_i N_j N_k N_\ell}\,,
  \end{eqnarray}
where $D=4-2\,\epsilon$,
    \begin{equation}
      \ol{d^Dk} = \Gamma(1-\epsilon)\,(\pi\mu^2)^\epsilon\,d^Dk\,
    \end{equation}
is the $\msbar$ regularization\cite{beta_qcd,msbreg},
and the propagator factors are
  \begin{subequations}
  \begin{eqnarray}
     N_1 &=& k^2 -m_1^2 +i \varepsilon\,,
     \\
     N_2 &=& (k+p_1)^2 -m_2^2 +i \varepsilon\,,
     \\
     N_3 &=& (k+p_1+p_2)^2 -m_3^2 +i \varepsilon\,,
     \\
     N_4 &=& (k+p_1+p_2+p_3)^2 -m_4^2 +i \varepsilon\,.
  \end{eqnarray}
  \end{subequations}
The vector/tensor functions are reduced to scalar functions
as
\begin{subequations}
\begin{eqnarray}
  B^\mu(12) &=& p_1^\mu B_1(12) \,,
  \\
  B^{\mu\nu}(12) &=& p_1^\mu p_1^\nu B_{21}(12)
                    +g^{\mu\nu} B_{22}(12)\,,
\end{eqnarray}
\end{subequations}
for the two-point functions,
\begin{subequations}
\begin{eqnarray}
  C^\mu(123) &=& p_1^\mu C_{11}(123) +p_2^\mu C_{12}(123) \,,
  \\
  C^{\mu\nu}(123) &=& p_1^\mu p_1^\nu C_{21}(123)
                     +p_2^\mu p_2^\nu C_{22}(123)
                     +p_1^{\{\mu} p_2^{\nu\}} C_{23}(123)
                     +g^{\mu\nu} C_{24}(123)\,,
\end{eqnarray}
\end{subequations}
for the three-point functions, and
\begin{subequations}
\begin{eqnarray}
  D^\mu(1234) &=& p_1^\mu D_{11}(1234) +p_2^\mu D_{12}(1234)
                 +p_3^\mu D_{13}(1234)\,,
  \\
  D^{\mu\nu}(1234) &=&  p_1^\mu p_1^\nu D_{21}(1234)
                       +p_2^\mu p_2^\nu D_{22}(1234)
                       +p_3^\mu p_3^\nu D_{23}(1234)
  \nonumber \\
                   & & +p_1^{\{\mu} p_2^{\nu\}} D_{24}(1234)
                       +p_1^{\{\mu} p_3^{\nu\}} D_{25}(1234)
                       +p_2^{\{\mu} p_3^{\nu\}} D_{26}(1234)
  \nonumber \\
                   & & +g^{\mu\nu} D_{27}(1234)\,,
\end{eqnarray}
\end{subequations}
for the four-point functions. Higher rank tensor functions do not appear
in our applications in the 't~Hooft-Feynman gauge.
 
The scalar functions $B_i$, $C_i$, $D_i$, are defined by
  \begin{eqnarray}
     B_i(12) &=& B_i(p_1^2;m_1,m_2)
  \end{eqnarray}
for $i=0,1,21$ and $22$,
  \begin{eqnarray}
     C_i(123) &=& C_i(p_1^2,p_2^2,(p_1+p_2)^2;m_1,m_2,m_3)
  \end{eqnarray}
for $i=0,11,12$ and $21$--$24$, and
  \begin{eqnarray}
     D_i(1234)
     &=& D_i(p_1^2,p_2^2,p_3^2,(p_1+p_2+p_3)^2,(p_1+p_2)^2,(p_2+p_3)^2
                  ;m_1,m_2,m_3,m_4) \hspace{7mm}
  \end{eqnarray}
for $i=0,11$--$13$ and $21$--$27$.
The basic scalar functions $B_0$, $C_0$ and $D_0$ were obtained by
't~Hooft and Veltman\cite{tv}. The Fortran code FF\cite{ff} is used
for the general form of $C_0$ and $D_0$ functions. Reductions of higher
$B$, $C$, $D$ functions are given in the following subsections.

\subsection*{\normalsize \bf D.2  $B$ functions }
\pr 
It is convenient to introduce the following four scalar $B$ functions
in addition to $B_0$ and $B_1$ above~:
  \begin{subequations}
  \begin{eqnarray}
     B_2(q^2;m_1,m_2) &=&  B_{21}(q^2;m_1,m_2) ,
     \label{d.5}\\
     B_3(q^2;m_1,m_2) &=& -B_1(q^2;m_1,m_2) -B_2(q^2;m_1,m_2) ,
     \label{d.6}\\
     B_4(q^2;m_1,m_2) &=& -m_1^2 B_1(q^2;m_2,m_1)
                          -m_2^2 B_1(q^2;m_1,m_2) ,
     \label{d.7}\\
     B_5(q^2;m_1,m_2) &=& A(m_1) +A(m_2) -4\,B_{22}(q^2;m_1,m_2) .
     \label{d.8}
  \end{eqnarray}
  \end{subequations}
All two-point functions of the standard model and its supersymmetric
extension~\cite{dhy} are expressed compactly in terms of the above six
$B_{\rm n}$ functions (${\rm n}=0,1,..,5$) being only logarithmically
divergent. The ultra-violet singular factor is parametrized by~:
  \begin{eqnarray}
     \Delta = \frac{1}{\epsilon} +\ln\mu^2.
     \label{d.9}
  \end{eqnarray}
In the $\msbar$ (modified minimal subtraction) renormalization scheme
the singular piece $\Delta$ in these functions is simply replaced
by a logarithm of the unit of mass $\mu$:
  \begin{eqnarray}
     \Delta \stackrel{\overline{\rm MS}}{\longrightarrow}\ln \mu^2.
     \label{d.10}
  \end{eqnarray}
The six $B_{\rm n}$ functions are then expressed by
  \begin{subequations}
  \begin{eqnarray}
     B_0(q^2;m_1,m_2) &=&             \Delta -F_0(q^2;m_1,m_2),
     \label{d.11}\\
     B_1(q^2;m_1,m_2) &=& -\frac{1}{2}\Delta +F_1(q^2;m_1,m_2),
     \label{d.12}\\
     B_2(q^2;m_1,m_2) &=&  \frac{1}{3}\Delta -F_2(q^2;m_1,m_2),
     \label{d.13}\\
     B_3(q^2;m_1,m_2) &=&  \frac{1}{6}\Delta -F_3(q^2;m_1,m_2),
     \label{d.14}\\
     B_4(q^2;m_1,m_2) &=& \frac{m^2_1+m^2_2}{2}\Delta -F_4(q^2;m_1,m_2),
     \label{d.15}\\
     B_5(q^2;m_1,m_2) &=& \frac{q^2}{3}\Delta -F_5(q^2;m_1,m_2),
     \label{d.16}
  \end{eqnarray}
  \end{subequations}
where the finite parts $F_{\rm n}$ have the following
Feynman parametrizations~:
  \begin{subequations}
  \begin{eqnarray}
      F_0(q^2;m_1,m_2) &=&  \int_0^1 dx\,\ln H,
      \\
      F_1(q^2;m_1,m_2) &=&  \int_0^1 dx\,\,x \,\ln H,
      \\
      F_2(q^2;m_1,m_2) &=&  \int_0^1 dx\,\,x^2\,\ln H,
      \\
      F_3(q^2;m_1,m_2) &=& \int_0^1 dx\,\, x(1-x) \,\ln H,
      \\
      F_4(q^2;m_1,m_2) &=& \int_0^1 dx\,[(1-x)m^2_1+xm^2_2]\,\ln H,
      \\
      F_5(q^2;m_1,m_2) &=&  \int^1_0 dx\,
                          [(1-2x)(m^2_1-m^2_2)+(1-2x)^2q^2]\,\ln H,
  \end{eqnarray}
  \end{subequations}
with
  \begin{equation}
    H \equiv \left[(1-x)m^2_1 +xm^2_2 -x(1-x)q^2 -i\epsilon \right]\,.
  \end{equation}
Among the six $F_{\rm n}$ functions four (n = 0, 3, 4, 5) are
symmetric under the exchange of the two masses. It is useful to
introduce the antisymmetric $F$ function
  \begin{eqnarray}
     F_A(q^2;m_1,m_2) \equiv F_1(q^2;m_2,m_1) - F_1(q^2;m_1,m_2).
     \label{d.21}
  \end{eqnarray}
In terms of the two symmetric functions $F_0$ and $F_3$ and the
antisymmetric function $F_A$ all the remaining $F_{\rm n}$ functions
can be expressed compactly~:
  \begin{subequations}
  \begin{eqnarray}
     F_1(q^2;m_1,m_2)
     &=& \frac{1}{2} \left[ F_0-F_A \right](q^2;m_1,m_2),
 \label{d.22}\\
     F_2(q^2;m_1,m_2)
     &=& \left[ \frac{1}{2}(F_0-F_A)-F_3 \right](q^2;m_1,m_2)
 ,\label{d.23}\\
     F_4(q^2;m_1,m_2)
     &=& \left[ \frac{m^2_1+m^2_2}{2}F_0
                     +\frac{m^2_1-m^2_2}{2}F_A \right](q^2;m_1,m_2),
     \label{d.24}\\
     F_5(q^2;m_1,m_2) &=& \left[ q^2(F_0-4F_3) +(m^2_1-m^2_2)F_A \right]
                          (q^2;m_1,m_2).
     \label{d.25}
  \end{eqnarray}
  \end{subequations}
Therefore it is convenient to give closed analytic expressions for
the three functions, $F_0, F_3$ and $F_A$:
  \begin{subequations}
  \begin{eqnarray}
     F_0(q^2;m_1,m_2) &=&
      \ln(m_1m_2)-\delta \ln\frac{m_2}{m_1} -2 +\beta L,        \label{d.26}
     \\
     F_3(q^2;m_1,m_2) &=&
      \frac{1}{6}\ln(m_1m_2) -\frac{3\sigma-2\delta^2}{6}
      \delta \ln\frac{m_2}{m_1} -\frac{5}{18} -\frac{\sigma-\delta^2}{3}
      +\frac{1+\sigma-2\delta^2}{6} \beta L, \nonumber\\  \label{d.27}
     \\
     F_A(q^2;m_1,m_2) &=&
      -(\sigma-\delta^2)\ln\frac{m_2}{m_1} +\delta(1-\beta L), \label{d.28}
  \end{eqnarray}
  \end{subequations}
where
  \begin{eqnarray}
    \sigma &=& \frac{m^2_1+m^2_2}{q^2},   \label{d.29}\\
    \delta &=& \frac{m^2_1-m^2_2}{q^2},   \label{d.30}\\
    \beta  &=&
      \left\{ \begin{array}{ll}
          (1-2\sigma+\delta^2)^{\frac{1}{2}} &\:\:\:
         \mbox{for $q^2<(m_1-m_2)^2$ or $q^2>(m_1+m_2)^2$ },\\
          i(2\sigma-\delta^2-1)^{\frac{1}{2}} &\:\:\:
         \mbox{for $(m_1-m_2)^2 < q^2<(m_1+m_2)^2$ },
              \end{array}\right. \label{d.31}
  \end{eqnarray}
and the function $L$ is defined as
  \begin{eqnarray}
     L(q^2;m_1,m_2) = \left\{
       \begin{array}{ll}
         \frac{1}{2} \ln\frac{1+\beta-\sigma}{1-\beta-\sigma} -i\pi &
         \mbox{for $q^2>(m_1+m_2)^2$}, \\
         \frac{1}{2} \ln\frac{1+\beta-\sigma}{1-\beta-\sigma} &
         \mbox{for $q^2<(m_1-m_2)^2$}, \\
         \frac{1}{i}(\tan^{-1}\frac{1-\delta}{|\beta|}
          +\tan^{-1}\frac{1+\delta}{|\beta|}) \:\:\: &
         \mbox{for $(m_1-m_2)^2<q^2<(m_1+m_2)^2$}.
       \end{array} \right.      \label{d.32}
  \end{eqnarray}
 
Also the derivatives $F'_0$, $F'_3$ and $F'_A$ are needed for certain
applications. One finds
  \begin{subequations}
  \begin{eqnarray}
     F'_0(q^2;m_1,m_2) &=&
     \frac{1}{q^2}
     \biggl\{\,1+\delta\ln\frac{m_2}{m_1}-(\delta^2-\sigma)\frac{L}{\beta}
     \biggr\},
     \label{d.49}
     \\
     F'_3(q^2;m_1,m_2) &=&
      \frac{1}{q^2}
     \biggl\{\frac{1}{6}-\delta^2+\frac{\sigma}{2}+\delta(\sigma-\delta^2)
      \ln\frac{m_2}{m_1}+
        \Bigl(\frac{\sigma^2+\delta^2}{2}+\delta^2(\delta^2-2\sigma)\Bigr)
      \frac{L}{\beta}\biggr\}, \hspace{0.5cm}
     \nonumber \\ &&
     \label{d.50}
     \\
     F'_A(q^2;m_1,m_2) &=&
       \frac{1}{q^2}\biggl\{\,
        (\sigma-2\delta^2)\ln\frac{m_2}{m_1} -2\delta
       +\delta(1-3\sigma+2\delta^2)\frac{L}{\beta}\biggr\}.
     \label{d.51}
  \end{eqnarray}
  \end{subequations}
The phase factors in $\beta$ and $L$ in eqs.~(\ref{d.31},\ref{d.32})
are required to obtain correctly the ratio $L/\beta$. In terms of the
above three functions all the other $F'_{\rm n}$ functions are
expressed compactly:
  \begin{subequations}
  \begin{eqnarray}
     F'_1(q^2;m_1,m_2) &=& \frac{1}{2}[F'_0-F'_A](q^2;m_1,m_2),
     \label{d.52}
     \\
     F'_2(q^2;m_1,m_2) &=&[\frac{F'_0-F'_A}{2}-F'_3](q^2;m_1,m_2),
     \label{d.53}
     \\
     F'_4(q^2;m_1,m_2) &=&[\frac{m^2_1+m^2_2}{2}F'_0
      +\frac{m^2_1-m^2_2}{2}F'_A](q^2;m_1,m_2),
     \label{d.54}
     \\
     F'_5(q^2;m_1,m_2) &=&
       [F_0-4F_3+q^2(F'_0-4F'_3)+(m^2_1-m^2_2)F'_A](q^2;m_1,m_2),
    \label{d.55}
  \end{eqnarray}
  \end{subequations}
The derivative of the $B_{\rm n}$ functions is found to be~:
  \begin{subequations}
  \begin{eqnarray}
     B'_{\rm n}(q^2;m_1,m_2) &=& -F'_{\rm n}(q^2;m_1,m_2)
        \,\,\,\,\, \mbox{for}\,\,\,{\rm n}=0,2,3,4
     \label{d.56}
     \\
     B'_1(q^2;m_1,m_2) &=& F'_1(q^2;m_1,m_2),
     \label{d.57}
     \\
     B'_5(q^2;m_1,m_2) &=& \frac{1}{3}\Delta-F'_5(q^2;m^2_1,m^2_2).
     \label{d.58}
  \end{eqnarray}
  \end{subequations}
 
\subsection*{\normalsize \bf D.3  $C$ functions }
\pr
The higher $C$ functions
  \begin{equation}
     C_i\equiv C_i(p_1^2, p_2^2, (p_1+p_2)^2; m_1, m_2, m_3)
  \end{equation}
for $i=11,12$, and 21--24
are given in terms of the $C_0$, $B_0$ and $B_1$ functions as
  \begin{subequations}
     \label{eq:c_fn}
  \begin{eqnarray}
        \left(\begin{array}{c}
           \! C_{11}\!\! \\ \! C_{12}\!\!
        \end{array} \right)
      &=& X^{-1}
        \left(
          \begin{array}{c}
             B_0^{(13)} -B_0^{(23)} +f_1 C_0 \\
             B_0^{(12)} -B_0^{(13)} +f_2 C_0
          \end{array}
        \right)
    \\[2mm]
     C_{24} \:\: &=& \frac{1}{4} + \frac{1}{4}B_0^{(23)}
     + \frac{m_1^2}{2}\,C_0
     - \frac{f_1}{4}\,C_{11}  - \frac{f_2}{4}\,C_{12}
    \\[2mm]
        \left(\begin{array}{c}
            \! C_{21}\!\! \\ \! C_{23}\!\!
        \end{array}\right)
      &=& X^{-1}
          \left(
           \begin{array}{l}
              \! B_1^{(13)} +B_0^{(23)} +f_1 C_{11} -2\,C_{24}\!\\
              \! B_1^{(12)} -B_1^{(13)} +f_2 C_{11} \!
           \end{array}
         \right)
    \\[2mm]
         \left(\begin{array}{c}
             \! C_{23}\!\! \\ \! C_{22}\!\!
         \end{array}\right)
      &=& X^{-1}
          \left(
           \begin{array}{l}
              \! B_1^{(13)} -B_1^{(23)} +f_1 C_{12} \!\\
              \hphantom{B_1^{(12)}}
                -B_1^{(13)} +f_2 C_{12} -2\,C_{24}\!
           \end{array}
         \right)
  \end{eqnarray}
  \end{subequations}
where
  \begin{equation}
     X \equiv
       \left(
         \begin{array}{cc}
            \! 2 p_1^2  & 2 p_1p_2 \! \\
            \! 2 p_1p_2 & 2 p_2^2  \!
         \end{array}
       \right) \,,
  \end{equation}
and
  \begin{subequations}
  \begin{eqnarray}
     f_1 &=& m_2^2 -m_1^2 -p_1^2 \,,
     \\
     f_2 &=& m_3^2 -m_2^2 -(p_1+p_2)^2 +p_1^2 \,.
  \end{eqnarray}
  \end{subequations}
Here the shorthand notations
  \begin{subequations}
  \begin{eqnarray}
     B_i^{(12)} &\equiv& B_i(p_1^2;m_1,m_2)\,, \\
     B_i^{(13)} &\equiv& B_i((p_1+p_2)^2;m_1,m_3)\,, \\
     B_i^{(23)} &\equiv& B_i(p_2^2;m_2,m_3)\,,
  \end{eqnarray}
  \end{subequations}
are used for $B_0$ and $B_1$ functions.
 
\subsection*{\normalsize \bf D.4  $D$ functions }
\pr 
The higher $D$ functions
  \begin{equation}
    D_i \equiv
        D_i(p_1^2,p_2^2,p_3^2,(p_1+p_2+p_3)^2,(p_1+p_2)^2,(p_1+p_3)^2;
            m_1, m_2, m_3, m_4) \hspace{5mm}
  \end{equation}
for $i=0,11$--$13$ and $21$--$27$
are expressed in terms of the $D_0$, $C_0$ and $B_0$ functions
as follows:
  \begin{subequations}
     \label{eq:d_fn}
  \begin{eqnarray}
         \left(\!\!
               \begin{array}{l} D_{11} \\ D_{12} \\ D_{13} \end{array}
          \!\! \right)
     &=& X^{-1}
        \left(\!
        \begin{array}{l}
           C_0^{(134)} - C_0^{(234)} + f_1\,D_0 \\
           C_0^{(124)} - C_0^{(134)} + f_2\,D_0 \\
           C_0^{(123)} - C_0^{(124)} + f_3\,D_0
        \!\end{array}
       \right)
        \,,
    \\
      D_{27} \;\;\; &=& \;
         m_1^2 D_0 + \frac{1}{2}C_0^{(234)}
       - \frac{1}{2} \Bigl[f_1\,D_{11} + f_2\,D_{12} + f_3\,D_{13} \Bigr]
        \,,
    \\
        \left(\!\!
              \begin{array}{l} D_{21} \\ D_{24} \\ D_{25} \end{array}
        \!\! \right)
     &=& X^{-1}
     \left(\! \begin{array}{l}
      C_{11}^{(134)}+C_{ 0}^{(234)}+f_1\,D_{11} -2\,D_{27} \\
      C_{11}^{(124)}-C_{11}^{(134)}+f_2\,D_{11} \\
      C_{11}^{(123)}-C_{11}^{(124)}+f_3\,D_{11}
     \end{array}\! \right)
     \,,
    \\
        \left(\!\!
              \begin{array}{l} D_{24} \\ D_{22} \\ D_{25} \end{array}
        \!\! \right)
     &=& X^{-1}
     \left(\! \begin{array}{l}
      C_{11}^{(134)}-C_{11}^{(234)}+f_1\,D_{12} \\
      C_{12}^{(124)}-C_{11}^{(134)}+f_2\,D_{12} -2\,D_{27} \\
      C_{12}^{(123)}-C_{12}^{(124)}+f_3\,D_{12}
     \end{array}\! \right)
     \,,
    \\
        \left(\!\!
              \begin{array}{l} D_{25} \\ D_{26} \\ D_{23} \end{array}
        \!\! \right)
     &=& X^{-1}
     \left(\! \begin{array}{l}
      C_{12}^{(134)}-C_{12}^{(234)}+f_1\,D_{13} \\
      C_{12}^{(124)}-C_{12}^{(134)}+f_2\,D_{13} \\
      \hphantom{C_{12}^{(124)}}   -C_{12}^{(124)}+f_3\,D_{13} -2\,D_{27}
      \end{array}\! \right)
      \,,
  \end{eqnarray}
  \end{subequations}
where
  \begin{equation}
     X \equiv
         \left(
           \begin{array}{ccc}
              2 p_1^2   & 2 p_1 p_2 & 2 p_1 p_3 \\
              2 p_1 p_2 & 2 p_2^2   & 2 p_2 p_3 \\
              2 p_1 p_3 & 2 p_2 p_3 & 2 p_3^2
           \end{array}
         \right) \,,
  \end{equation}
and
  \begin{subequations}
  \begin{eqnarray}
       f_1  &=& m_2^2 -m_1^2 -p_1^2 \,, \\
       f_2  &=& m_3^2 -m_2^2 -(p_1+p_2)^2 +p_1^2 \,, \\
       f_3  &=& m_4^2 -m_3^2 -(p_1+p_2+p_3)^2 +(p_1+p_2)^2 \,.
  \end{eqnarray}
  \end{subequations}
The higher $C$ functions in eqs.~(\ref{eq:d_fn}) are written in terms
of the $C_0$ and $B_0$ functions in analogy to the previous
subsection~:
  \begin{subequations}
  \begin{eqnarray}
        \left(\begin{array}{l}
           \! C_{11}^{(123)}\!\! \\ \! C_{12}^{(123)}\!\!
        \end{array} \right)
      &=&\left(\begin{array}{cc}
            \!\! 2 p_1^2  &\! 2 p_1p_2 \!\! \\
            \!\! 2 p_1p_2 &\! 2 p_2^2  \!\!
         \end{array}\right)^{-1}
         \left(\begin{array}{l}\!\!
                  B_0^{(13)} -B_0^{(23)} +f_1 C_0^{(123)} \!\!\\
             \!\! B_0^{(12)} -B_0^{(13)} +f_2 C_0^{(123)} \!\!
         \end{array}\right) ,
    \\[2mm]
        \left(\begin{array}{l}
           \! C_{11}^{(124)}\!\! \\ \! C_{12}^{(124)}\!\!
        \end{array} \right)
      &=&\left(\begin{array}{cc}
            \!\! 2 p_1^2          &\! 2 p_1\,(p_2+p_3) \!\! \\
            \!\! 2 p_1\,(p_2+p_3) &\! 2 (p_2+p_3)^2  \!\!
         \end{array}\right)^{-1}
         \left(\begin{array}{l}\!\!
                 B_0^{(14)} -B_0^{(24)} +f_1 C_0^{(124)} \!\!\\
            \!\! B_0^{(12)} -B_0^{(14)} +(f_2+f_3)\, C_0^{(124)} \!\!
         \end{array}\right) , \qqquad
    \\[2mm]
        \left(\begin{array}{l}
           \! C_{11}^{(134)}\!\! \\ \! C_{12}^{(134)}\!\!
        \end{array} \right)
      &=&\left(\begin{array}{cc}
            \!\! 2 (p_1+p_2)^2    &\! 2 (p_1+p_2)\,p_3 \!\! \\
            \!\! 2 (p_1+p_2)\,p_3 &\! 2 p_3^2  \!\!
         \end{array}\right)^{-1}
         \left(\begin{array}{l}\!\!
                 B_0^{(14)} -B_0^{(34)} +(f_1+f_2)\, C_0^{(134)} \!\!\\
            \!\! B_0^{(13)} -B_0^{(14)} +f_3\, C_0^{(134)} \!\!
         \end{array}\right) ,\qqquad
    \\[2mm]
        \left(\begin{array}{l}
           \! C_{11}^{(234)}\!\! \\ \! C_{12}^{(234)}\!\!
        \end{array} \right)
      &=&\left(\begin{array}{cc}
            \!\! 2 p_2^2  &\! 2 p_2p_3 \!\! \\
            \!\! 2 p_2p_3 &\! 2 p_3^2  \!\!
         \end{array}\right)^{-1}
         \left(\begin{array}{l}\!\!
                 B_0^{(24)} -B_0^{(34)} +(f_2+2p_1p_2) C_0^{(234)} \!\!\\
            \!\! B_0^{(23)} -B_0^{(24)} +(f_3+2p_1p_3) C_0^{(234)} \!\!
         \end{array}\right) ,
  \end{eqnarray}
  \end{subequations}
with
  \begin{subequations}
  \begin{eqnarray}
     B_0^{(12)} &\equiv& B_0(p_1^2;m_1,m_2) \,,
     \\
     B_0^{(13)} &\equiv& B_0((p_1+p_2)^2;m_1,m_3) \,,
     \\
     B_0^{(14)} &\equiv& B_0((p_1+p_2+p_3)^2;m_1,m_4) \,,
     \\
     B_0^{(23)} &\equiv& B_0(p_2^2;m_2,m_3) \,,
     \\
     B_0^{(24)} &\equiv& B_0((p_2+p_3)^2;m_2,m_4) \,,
     \\
     B_0^{(34)} &\equiv& B_0(p_3^2;m_3,m_4) \,.
  \end{eqnarray}
  \end{subequations}

\def\tcaptionofchargeformfactors{%
Universal electroweak parameters of the spontaneously broken generic
${\rm SU(2)_L \times U(1)_Y}$ theory. 
Column 2 list the 10 universal parameters, the masses of the weak bosons 
and the 4 charge form factors at two $q^2$ scales, 0 and $\mmz$. 
Column 3 contains three precisely measured quantities (the fine structure 
constant $\alpha$, 
the Fermi coupling constant $G_F$ and the $Z$ boson mass 
$m_Z$) together with their relation to the universal charge form factors. 
The factor $\delg$ is explained in the text: see eq.~(\ref{delg}). 
The last column lists those parameters which are used in fits. 
The `star' marks parameters for which no direct experimental information 
is available.
}%
 
\def\tcaptionofstepsinanalysis{%
Three types of fits are considered. 
For each sector the free parameters are listed. 
External parameters in the fits in addition to the precisely 
known fine structure constant $\alpha$, the Fermi coupling constant 
$G_F$ and the $Z$ boson mass $m_Z^{}$ are listed separately. 
The quantity $\delta_{\alpha}$ is defined as 
$\delta_\alpha \equiv 1/\bar{\alpha}(m_Z^2) -128.72$ 
(\ref{delta_alpha}). 
The parameters $S$, $T$, $U$ are defined 
in eqs.~(\ref{stu_def}) and $\delb$ in eq.~(\ref{delb}).
}%
 
\def\tcaptionofvertexformfactors{%
Vertex form factors $\Gamma_1^f(q^2)$, $\gambar_2^f(q^2)$ and 
$\Gamma_3^f(q^2)$ in the SM at $q^2=m_Z^2$. 
The definitions of the form factors are given in eq.~(\ref{gamma_f}) 
and their explicit forms in appendix~A.2.
}
 
\def\tcaptionofboxformfactors{%
Box form factors $B(e_\alpha,f_\sigma) \equiv B^{ef}_{\alpha\sigma}(s,t)$
for the process $e_\alpha \ol{e_\alpha} \to f_\sigma \ol{f_\sigma}$
in the SM at $s=-2t=m_Z^2$.
The definitions of the form factors are given in eq.~(\ref{box})
and their explicit forms in appendix~A.3.
}
 
\def\tcaptionofwidths{%
Partial and total $Z$ widths in $\mev$ units in the minimal SM 
for $\mz=91.197\gev$, $\alpha_s(\mz)=0.12$, $\delta_\alpha=0$, 
$\delg =0.0055$, $m_b=4.7\gev$ and $m_c=1.4\gev$.
See eqs.~(\ref{gamma_z})--(\ref{m_zff}) in section~3.1 for details.
}
 
\def\tcaptionoffitsummary{%
Summary of all the electroweak data used in the fit, and the fit results.
The $Z$ boson parameters are studied in section~4.1, the low energy
neutral current experiments are studied in section~4.2, and the charged
current experiments are studied in section~4.3.  
In addition, we use the fine structure constant $\alpha$ datum which 
fixes the charge form factor $\ebar^2(0)$.
$\chi^2$ has been calculated by taking the model-independent parametrizations
of the original data as the inputs of our analysis:
eqs.~(\ref{dataofzlineshape})--(\ref{dataofzothers}) for the $Z$ parameters
(section~4.1), 
eq.~(\ref{dataofnuq}) for the $\nu_\mu$--$q$ scattering experiments 
(section~4.2.1), 
eq.~(\ref{dataofnue}) for the $\nu_\mu$--$e$ scattering experiments 
(section~4.2.2), 
eq.~(\ref{dataofapv}) for the atomic parity violation experiments
(section~4.2.3), 
eq.~(\ref{dataofed}) for the $e$--D polarization asymmetry measurements
(section~4.2.3), 
and eq.~(\ref{dataofmw}) for the $W$ mass measurements
(section~4.3).
}
 
\def\tcaptionofsummary{%
The SM predictions for the electroweak parameters.
The column `no-EW' is obtained by dropping all radiative 
corrections except in
the running of $\bar{\alpha}(q^2)$ and $\sbar^2(q^2)$ due 
to light quarks and leptons.  
The column `IBA' is obtained by dropping all vertex and box corrections
except $\delg$.  
In both `no-EW' and `IBA' cases, corrections due to the tree-level
propagator effects and the external QED/QCD corrections are kept.
When the predictions depend on $\alpha_s(m_Z)$, we show
three representative cases for $\alpha_s(m_Z)=$0.11, 0.12 and 0.13
from top to bottom.  The $\chi^2$ values are obtained by taking
account of the correlations among the errors that are presented in
the text (see section~4). 
The total number of the data is 22 by counting also ($\alpha$, $G_F$, $\mz$),
while the above three parameters are used as inputs of the SM analysis.
The degree of freedom of the fit is hence $22-3=19$.
}
 
\def\propagator#1#2{%
    \begin{picture}(60,16)(-5,-4)
       \Text( 6,0)[r]{#1}
       \Text(54,0)[l]{#2}
       \Photon(10,0)(24,0){3}{3}
       \Photon(36,0)(50,0){3}{3}
       \GCirc(30,0){6}{0.5}
    \end{picture}
}
\def\tableofchargeformfactors{%
\begin{center}
{\Large Table 1}
\end{center}
\begin{tabular}{|c|c|c|c|}
\hline
     \begin{tabular}{c} Electroweak     \\ propagators \end{tabular}
   & \begin{tabular}{c} Universal       \\ parameters  \end{tabular}
   & \begin{tabular}{c} Precisely known \\ parameters  \end{tabular}
   & \begin{tabular}{c} Fit             \\ parameters  \end{tabular}
  \\
  \hline
   \propagator{$\gamma$}{$\gamma$}  \vphantom{$\Big/$}
   & \hphantom{$m_W$} $\;\bar{e}^2(0)$ $\:\:\:\bar{e}^2(m_Z^2)$
   & $\alpha = \bar{e}^2(0)/4\pi$
   & $\hphantom{\bar{g}_W^2(m_Z^2)}$  $\;*$
  \\
  \hline
   \propagator{$\gamma$}{$Z$} \vphantom{$\Big/$}
   & \hphantom{$m_W$} $\;\bar{s}^2(0)$ $\;\;\bar{s}^2(m_Z^2)$
   &
   & $\bar{s}^2(0)$  $\;\;\bar{s}^2(m_Z^2)$
  \\
  \hline
   \propagator{$Z$}{$Z$} \vphantom{$\Big/$}
   & $m_Z\,$ $\;\bar{g}_Z^2(0)$  $\;\bar{g}_Z^2(m_Z^2)$
   & $m_Z$
   & $\bar{g}_Z^2(0)$  $\;\;\bar{g}_Z^2(m_Z^2)$
  \\
  \hline
   \propagator{$W$}{$W$} \vphantom{$\Big/$}
   & $m_W$  $\:\bar{g}_W^2(0)$  $\bar{g}_W^2(m_Z^2)$
   & $4\sqrt{2}G_F ={\bar{g}_W^2(0) \over m_W^2}\,(1+\bar{\delta}_G)$
   & $\bar{g}_W^2(0)$  $\hspace{9mm}*\quad$
  \\
  \hline
\end{tabular}
}

\def\dashed{$\hspace{-14pt}\mbox{
\begin{tabular}{p{0.5pt}}
\rule[6pt]{0.4pt}{2pt}\\[-12pt]
\rule[3pt]{0.4pt}{2pt}\\[-12pt]
\rule[0pt]{0.4pt}{2pt}\\[-12pt]
\rule[-3pt]{0.4pt}{2pt}\\[-12pt]
\rule[-6pt]{0.4pt}{2pt}
\end{tabular}
}\hspace{-12pt}$
} 
\def\tableofstepsinanalysis{%
\begin{center}
{\Large Table 2}
\end{center}
\begin{center}
\begin{tabular}{|c|ccclc|cccclcc|cclcc|}
\hline \vphantom{$\Big/$}
\begin{tabular}{c}
Experimental \\ inputs
\end{tabular}
&
\multicolumn{5}{|c|}{6-parameter fit} & 
\multicolumn{7}{|c|}{4-parameter fit} & 
\multicolumn{5}{|c|}{2-parameter fit}
\\
\hline \vphantom{$\Big/$}
$\alpha$, $G_F$, $m_Z^{}$ &
\multicolumn{5}{|c|}{(input)} &
\multicolumn{7}{|c|}{(input)} &
\multicolumn{5}{|c|}{(input)}
\\
\hline
\begin{tabular}{c}
low energy \\ neutral currents
\end{tabular}
&
$\!\sbar^2(0)\!$  & $\!\gzbar^2(0)\!$ & &\dashed & &
$\!S\!$   & $\!T\!$  &  &  &\dashed & $\!\delta_\alpha\!$ &  &
$\!m_t\!$ & $\!m_H^{}\!$ &\dashed & $\!\delta_\alpha\!$ & 
\\
\hline \vphantom{$\Big/$}
$Z$ parameters &
$\!\sbar^2(m_Z^2)\!$ & $\!\gzbar^2(m_Z^2)\!$ & 
$\!\!\delb(m_Z^2)\!$ &\dashed & $\!\alpha_s\!$      &
$\!S\!$ & $\!T\!$ &  & $\!\!\!\!\!\delb(m_Z^2)\!$ &\dashed
&$\!\delta_\alpha\!$  & $\!\!\!\alpha_s\!$        &
$\!m_t\!$ & $\!m_H^{}\!$ &\dashed 
& $\!\delta_\alpha\!$ & $\!\!\!\!\alpha_s\!$ 
\\
\hline \vphantom{$\Big/$}
$m_W^{}$ & 
$\!\gwbar^2(0)\!$ & & & \dashed & &
$\!S\!$ &  $\!T\!$ & $\!U\!$ &  &\dashed & $\!\delta_\alpha\!$ &  &
$\!m_t\!$ & $\!m_H^{}\!$ &\dashed & $\!\delta_\alpha\!$ &
\\
\hline
\begin{tabular}{c}
number of \\
fit parameters
\end{tabular}
&
\multicolumn{3}{c}{6} &\dashed & & 
\multicolumn{4}{c}{4} &\dashed & & &
\multicolumn{2}{c}{2} &\dashed & &
\\
\dline{1-18}
\begin{tabular}{c}
number of \\$\!\!\!\!$external parameters$\!\!\!\!$
\end{tabular}
& & &   &\dashed& \multicolumn{1}{c|}{1}
& & & & &\dashed& \multicolumn{2}{c|}{2}
& &     &\dashed& \multicolumn{2}{c|}{2} 
\\
\hline
\end{tabular}
\end{center}
}
 
\def\non{-----------$\;\;\;\;\;\;\;\;\;\;$}              
\def\tableofvertexformfactors{%
\begin{center}
{\Large Table 3}
\end{center}
\begin{center}
\begin{tabular}{|c|r|r|r|}
\hline
$ f $ &
$ \Gamma_1^f       $ \hspace{16mm} &
$ \gambar_2^f $ \hspace{16mm} &
$ \Gamma_3^f       $ \hspace{16mm} \\
\hline 
             $\nu_L$
 & $ 0.00252 \, +0.00431 \,i $ & $-0.00680 \, -0.00565 \,i $ & \non \\
            $\ell_L$
 & $ 0.00185 \, +0.00325 \,i $ & $-0.00680 \, -0.00565 \,i $ & \non \\
            $\ell_R$
 & $ 0.00020 \, +0.00032 \,i $ & \non 			     & \non \\
               $u_L$
 & $ 0.00203 \, +0.00354 \,i $ & $-0.00680 \, -0.00565 \,i $ & \non \\
               $u_R$
 & $ 0.00009 \, +0.00014 \,i $ & \non 			     & \non \\
               $d_L$
 & $ 0.00225 \, +0.00389 \,i $ & $-0.00680 \, -0.00565 \,i $ & \non \\
               $d_R$
 & $ 0.00002 \, +0.00004 \,i $ & \non 			     & \non \\
  $b_L\,(m_t = 100)$
 & $ 0.00176 \, +0.00107 \,i $ & $-0.00402 \, +0.00000 \,i $ 
 & $-0.00347 \, +0.00000 \,i $\\
  $b_L\,(m_t = 150)$
 & $ 0.00141 \, +0.00107 \,i $ & $-0.00261 \, +0.00000 \,i $ 
 & $-0.00763 \, +0.00000 \,i $\\
  $b_L\,(m_t = 200)$
 & $ 0.00126 \, +0.00107 \,i $ & $-0.00179 \, +0.00000 \,i $
 & $-0.01270 \, +0.00000 \,i $\\
\hline
\end{tabular}
\end{center}
}

\def\tableofboxformfactors{%
\begin{center}
{\Large Table 4}
\end{center}
\begin{center}
\begin{tabular}{|c|r|r|}
\hline
$ f $ &
$ sB(e_L,f_\sigma) $ \hspace{10mm} &
$ sB(e_R,f_\sigma) $ \hspace{10mm} \\
\hline 
             $\nu_L$
 & $ 0.00109 \, +0.00000 \,i $ & $-0.00006 \, +0.00000 \,i $\\
            $\ell_L$
 & $-0.00005 \, +0.00000 \,i $ & $-0.00002 \, +0.00000 \,i $\\
            $\ell_R$
 & $-0.00002 \, +0.00000 \,i $ & $ 0.00001 \, +0.00000 \,i $\\
               $u_L$
 & $ 0.00104 \, +0.00000 \,i $ & $-0.00003 \, +0.00000 \,i $\\
               $u_R$
 & $-0.00001 \, +0.00000 \,i $ & $ 0.00001 \, +0.00000 \,i $\\
               $d_L$
 & $-0.00001 \, +0.00000 \,i $ & $-0.00005 \, +0.00000 \,i $\\
               $d_R$
 & $ 0.00000 \, +0.00000 \,i $ & $ 0.00000 \, +0.00000 \,i $\\
  $b_L\,(m_t = 100)$
 & $-0.00002 \, +0.00000 \,i $ & $-0.00005 \, +0.00000 \,i $\\
  $b_L\,(m_t = 150)$
 & $-0.00001 \, +0.00000 \,i $ & $-0.00005 \, +0.00000 \,i $\\
  $b_L\,(m_t = 200)$
 & $ 0.00001 \, +0.00000 \,i $ & $-0.00005 \, +0.00000 \,i $\\
\hline
\end{tabular}
\end{center}
}

\def\tableofzwidths{%
\begin{center}
{\Large Table 5}
\end{center}
\begin{center}
\begin{tabular}{|c|rrrrrr|}
 \hline
$m_t\,(\gev) $
 &  150 $\;\;$ &  150 $\;\;$ &  175 $\;\;$
 &  175 $\;\;$ &  200 $\;\;$ &  200 $\;\;$\\
$m_H^{}\,(\gev) $
 &  100 $\;\;$ & 1000 $\;\;$ &  100 $\;\;$
 & 1000 $\;\;$ &  100 $\;\;$ & 1000 $\;\;$\\
 \hline
$\bar{g}_Z^2(m_Z^2)    $
 &  0.55516 &  0.55405 &  0.55641 &  0.55523 &  0.55784 &  0.55656\\
$\bar{s}^2(m_Z^2)      $
 &  0.23119 &  0.23245 &  0.23040 &  0.23170 &  0.22952 &  0.23086\\
$\delb(m_Z^2)          $
 &$-$0.00789&$-$0.00792&$-$0.00994&$-$0.00999&$-$0.01226&$-$0.01230\\
 \hline
$\Gamma_\nu            $
 &   166.95 &   166.61 &   167.32 &   166.97 &   167.75 &   167.37\\
$\Gamma_e=\Gamma_\mu$
 &    83.81 &    83.59 &    84.04 &    83.80 &    84.30 &    84.04\\
$\Gamma_\tau           $
 &    83.62 &    83.40 &    83.85 &    83.61 &    84.11 &    83.85\\
$\Gamma_u              $
 &   299.20 &   297.94 &   300.41 &   299.09 &   301.76 &   300.35\\
$\Gamma_c              $
 &   299.14 &   297.88 &   300.35 &   299.03 &   301.70 &   300.30\\
$\Gamma_d =\Gamma_s    $
 &   382.65 &   381.28 &   383.77 &   382.34 &   385.09 &   383.56\\
$\Gamma_b              $
 &   376.90 &   375.51 &   376.25 &   374.79 &   375.55 &   374.01\\
$\Gamma_h              $
 &  1740.54 &  1733.87 &  1744.55 &  1737.59 &  1749.18 &  1741.78\\
 \hline
$\Gamma_Z              $
 &  2492.63 &  2484.27 &  2498.44 &  2489.70 &  2505.15 &  2495.82\\
 \hline
\end{tabular}
\end{center}
}
 
\def\tableoffitsummary{%
\begin{center}
{\Large Table 6}
\end{center}
\begin{center}
\begin{tabular}{|c|c|c|}
\hline 
\multicolumn{2}{|c}{$Z$ parameters measurements} &
(section 4.1)
\vphantom{$\Big/$}
\\ \hline \vphantom{$\Big/$}
data &
\multicolumn{2}{c|}{$
   \mz,\; \Gamma_Z,\; \sigma_h^0,\; R_\ell,\;  A_{\rm FB}^{0,\ell},\;
     P_\tau,\; A_{\rm LR},\;A_{\rm FB}^{0,b},\;
        A_{\rm FB}^{0,c},\;R_b \,
$}
\\ \hline
fit parameters & 
$\mz\,$(input), $\sbar^2(\mmz)$, $\gzbar^2(\mmz)$, $\delb(\mmz)$ &
d.o.f. = 10$-$4
\vphantom{$\Big/$}
\\ \hline
external parameter & \multicolumn{1}{c}{$\alpha_s$} & \vphantom{$\Big/$}
\\ \hline
\multicolumn{3}{|c|}{}\\[-9pt]
\multicolumn{3}{|c|}{$\displaystyle{
     \left.
     \begin{array}{lll}
        \gzbar^2(m_Z^2)&\mm =&
           0.5542 -0.00030\,\frac{\alpha_s-0.12}{0.01} \pm 0.0017 \\[1mm]
        \sbar^2(m_Z^2) &\mm =&
           0.2313 +0.00008\,\frac{\alpha_s-0.12}{0.01} \pm 0.0007 \\[1mm]
        \delb(m_Z^2)   &\mm =&\mfive
          -0.0061 -0.00430\,\frac{\alpha_s-0.12}{0.01} \pm 0.0034
     \end{array}
     \right. 
        \rho_{\rm corr} =
        \left(
        \begin{array}{rrr}
          1 &  0.14 & -0.36 \\[1mm]
            &  1    &  0.20 \\[1mm]
            &       &  1
       	\end{array}
        \right)
}$}
\\
\multicolumn{3}{|c|}{}\\[-9pt]
\multicolumn{3}{|c|}{$\displaystyle{
	\chi^2_{\rm min}/{\rm (d.o.f.)} =
          \biggl[ 
              2.48 +\biggl(\frac{\alpha_s-0.1017}{0.0127}\biggr)^2
          \biggr] \bigg/6
         \hspace{35mm}
}$}
\\[-9pt]
\multicolumn{3}{|c|}{}\\
\hline
\hline
\multicolumn{2}{|c}{Low energy neutral current experiments} &
(section 4.2)
\vphantom{$\Big/$}
\\ \hline \vphantom{$\Big/$}
data &
\multicolumn{2}{c|}{$
 (g_L^2\,,\,g_R^2\,,\,\delta_L^2\,,\,\delta_R^2),\,
 (\rho_{\nu e}\,,\,s^2_{\nu e}),\,
 Q_W^{},\,
 (2C_{1u}\!-C_{1d}\,,\, 2C_{2u}\!-C_{2d})
$}
\\ \hline \vphantom{$\Big/$}
fit parameters & 
$\sbar^2(0)$, $\gzbar^2(0)$ &
d.o.f. = 9$-$2
\\ \hline
\multicolumn{3}{|c|}{}\\[-9pt]
\multicolumn{3}{|c|}{
  $\displaystyle{
       \left.
	\begin{array}{lll}
	\gzbar^2(0)  \!\! &=&  \!\!  0.5462 \pm 0.0036 \\[1mm]
	\sbar^2(0)   \!\! &=&  \!\!  0.2353 \pm 0.0044
	\end{array}
	\right\} \qquad
	\,\rho_{\rm corr}=0.53
  }$}
\\
\multicolumn{3}{|c|}{}\\[-9pt]
\multicolumn{3}{|c|}{$\displaystyle{
     \chi^2_{\rm min}/{\rm (d.o.f.)}= 2.22/7
         \hspace{35mm}
}$}
\\[-9pt]
\multicolumn{3}{|c|}{}\\
\hline
\hline
\multicolumn{2}{|c}{Charged current experiments} &  (section 4.3)
\vphantom{$\Big/$}
\\ \hline \vphantom{$\Big/$}
data &
\multicolumn{2}{c|}{$
G_F\,,\, \mw
$}
\\ \hline \vphantom{$\Big/$}
fit parameters & 
$G_F\,$(input), $\gwbar^2(0)$ &
d.o.f. = 2$-$2
\\ \hline
\multicolumn{3}{|c|}{}\\[-9pt]
\multicolumn{3}{|c|}{$\displaystyle{
\gwbar^2(0) = 0.4225 -0.0031\,\mbox{$\frac{\delg-0.0055}{\alpha}$}
\pm 0.0017 \hspace{35mm}
}$}
\\
\multicolumn{3}{|c|}{}\\[-9pt]
\multicolumn{3}{|c|}{$\displaystyle{
     \chi^2_{\rm min}/{\rm (d.o.f.)}= 0/0
         \hspace{35mm}
}$}
\\[-9pt]
\multicolumn{3}{|c|}{}\\
\hline
\end{tabular}
\end{center} 
} 
 
\def\tableofsummary{%
 \def\equnit{$\times 10^{-5}$}       
 \def\afb{A_{\rm FB}}                
 \def\non{$\;\;$------}              
 \footnotesize
\begin{center}
\vspace{-1cm}
 \begin{tabular}{|l|r|l|l|l|llll|} 
\multicolumn{9}{c}{\Large Table 7}\\
\multicolumn{9}{c}{}\\
 \hline                              
  & data \hspace{7mm} & no-EW &      
  \hspace{2mm}IBA  & Exact&  &       
  \hspace{4mm}  SM & &  \\           
 \hline
         $m_t$ (GeV)   &               
& \non  &    175 &    175 &    175 &    175 &    150 &    150\\
         $m_H$ (GeV)   &               
& \non  &    100 &    100 &     60 &   1000 &     60 &   1000\\
 \hline
$S$                    &               
& \non  &-0.2323 &-0.2323 &-0.2832 &-0.0749 &-0.2638 &-0.0555\\
$T$                    &               
& \non  & 0.8869 & 0.8869 & 0.9174 & 0.5875 & 0.6136 & 0.3002\\
$U$                    &               
& \non  & 0.3577 & 0.3577 & 0.3591 & 0.3529 & 0.2993 & 0.2931\\
$\bar{\delta}_G$       &               
& \non  & 0.0055 & 0.0055 & 0.0055 & 0.0055 & 0.0055 & 0.0055\\
$1/\bar{\alpha}(m_Z^2)$&               
 & 128.85 & 128.71 & 128.71 & 128.71 & 128.71 & 128.72 & 128.72\\
$\bar{s}^2(m_Z^2)$     &               
 & 0.2312 & 0.2304 & 0.2304 & 0.2301 & 0.2317 & 0.2309 & 0.2325\\
$\bar{g}_Z^2(m_Z^2)$   &               
 & 0.5486 & 0.5564 & 0.5564 & 0.5564 & 0.5552 & 0.5552 & 0.5540\\
$\bar{\delta}_b(m_Z^2)$&               
& \non & \non  &-0.0099 &-0.0100 &-0.0100 &-0.0079 &-0.0079\\
$\bar{s}^2(0)$         &               
 & 0.2388 & 0.2389 & 0.2389 & 0.2386 & 0.2401 & 0.2394 & 0.2408\\
$\bar{g}_Z^2(0)$       &               
 & 0.5486 & 0.5492 & 0.5492 & 0.5493 & 0.5480 & 0.5481 & 0.5468\\
$\bar{g}_W^2(0)$       &               
 & 0.4218 & 0.4242 & 0.4242 & 0.4245 & 0.4224 & 0.4229 & 0.4208\\
 \hline
$\Gamma_Z$(GeV) &   2.489 $\pm$  0.007 
 &  2.481 &  2.519 &  2.493 &  2.494 &  2.484 &  2.488 &  2.479\\
                &                      
 &  2.487 &  2.524 &  2.498 &  2.499 &  2.490 &  2.493 &  2.484\\
                &                      
 &  2.493 &  2.530 &  2.504 &  2.505 &  2.495 &  2.499 &  2.490\\
$\sigma_h^0$(nb)&   41.56 $\pm$   0.14 
 &  41.53 &  41.53 &  41.52 &  41.52 &  41.52 &  41.50 &  41.51\\
                &                      
 &  41.47 &  41.47 &  41.46 &  41.46 &  41.47 &  41.45 &  41.46\\
                &                      
 &  41.42 &  41.42 &  41.41 &  41.41 &  41.42 &  41.39 &  41.40\\
$R_\ell$        &  20.763 $\pm$  0.049 
 & 20.734 & 20.747 & 20.689 & 20.693 & 20.665 & 20.701 & 20.673\\
                &                      
 & 20.801 & 20.814 & 20.756 & 20.760 & 20.732 & 20.769 & 20.741\\
                &                      
 & 20.869 & 20.880 & 20.823 & 20.827 & 20.799 & 20.836 & 20.808\\
$\afb^{0,\ell}$ &  0.0158 $\pm$ 0.0018 
 & 0.0167 & 0.0182 & 0.0167 & 0.0171 & 0.0144 & 0.0157 & 0.0132\\
$P_\tau$        &  -0.139 $\pm$  0.014 
 & -0.149 & -0.156 & -0.148 & -0.150 & -0.138 & -0.144 & -0.132\\
$A_{\rm LR}$    &  0.1637 $\pm$ 0.0075 
 & 0.1494 & 0.1557 & 0.1480 & 0.1500 & 0.1378 & 0.1438 & 0.1318\\
$R_b$           &  0.2203 $\pm$ 0.0027 
 & 0.2183 & 0.2182 & 0.2157 & 0.2156 & 0.2157 & 0.2165 & 0.2165\\
                &                      
 & 0.2183 & 0.2182 & 0.2157 & 0.2157 & 0.2157 & 0.2165 & 0.2166\\
                &                      
 & 0.2183 & 0.2182 & 0.2157 & 0.2157 & 0.2157 & 0.2165 & 0.2166\\
$\afb^{0,b}$    &   0.099 $\pm$  0.006 
 &  0.105 &  0.109 &  0.104 &  0.105 &  0.096 &  0.101 &  0.092\\
                &                      
 &  0.105 &  0.109 &  0.104 &  0.105 &  0.096 &  0.101 &  0.092\\
                &                      
 &  0.105 &  0.109 &  0.104 &  0.105 &  0.097 &  0.101 &  0.092\\
$\afb^{0,c}$    &   0.075 $\pm$  0.015 
 &  0.075 &  0.078 &  0.074 &  0.075 &  0.069 &  0.072 &  0.065\\
                &                      
 &  0.075 &  0.078 &  0.074 &  0.075 &  0.069 &  0.072 &  0.065\\
                &                      
 &  0.075 &  0.078 &  0.074 &  0.075 &  0.069 &  0.072 &  0.066\\
    $\chi^2$        &$(\alpha_s=0.11)\quad $
 &   7.65 &  26.38 &  11.16 &  11.00 &  19.88 &  10.78 &  29.21\\
                    &$(\alpha_s=0.12)\quad $
 &   7.40 &  35.10 &  10.71 &  10.94 &  16.35 &  10.15 &  25.10\\
                    &$(\alpha_s=0.13)\quad $
 &  12.87 &  49.38 &  15.76 &  16.39 &  18.31 &  15.09 &  26.55\\
 \hline
$g_L^2$         &  0.2980 $\pm$ 0.0044 
 & 0.2887 & 0.2893 & 0.2995 & 0.2998 & 0.2973 & 0.2979 & 0.2955\\
$g_R^2$         &  0.0307 $\pm$ 0.0047 
 & 0.0302 & 0.0303 & 0.0295 & 0.0295 & 0.0297 & 0.0295 & 0.0298\\
$\delta_L^2$    & -0.0589 $\pm$ 0.0237 
 &-0.0588 &-0.0589 &-0.0634 &-0.0634 &-0.0634 &-0.0633 &-0.0632\\
$\delta_R^2$    &  0.0206 $\pm$ 0.0160 
 & 0.0181 & 0.0182 & 0.0177 & 0.0177 & 0.0178 & 0.0177 & 0.0178\\
$\chi^2 $       &                      
 &   6.91 &   6.09 &   0.24 &   0.29 &   0.25 &   0.19 &   0.78\\
 \hline
$s^2_{eff}$     &   0.233 $\pm$  0.008 
 &  0.239 &  0.239 &  0.230 &  0.230 &  0.231 &  0.231 &  0.232\\
$\rho_{eff}$    &   1.007 $\pm$  0.028 
 &  1.000 &  1.001 &  1.013 &  1.013 &  1.011 &  1.011 &  1.009\\
$\chi^2$        &                      
 &   0.61 &   0.60 &   0.18 &   0.21 &   0.06 &   0.11 &   0.02\\
 \hline
$Q_W$           &  -71.04 $\pm$   1.81 
 & -74.89 & -74.98 & -73.21 & -73.17 & -73.31 & -73.17 & -73.30\\
$\chi^2$        &                      
 &   4.52 &   4.74 &   1.43 &   1.39 &   1.57 &   1.38 &   1.57\\
 \hline
$2C_{1u}-C_{1d}$&   0.938 $\pm$  0.264 
 &  0.709 &  0.709 &  0.723 &  0.724 &  0.717 &  0.720 &  0.713\\
$2C_{2u}-C_{2d}$&  -0.659 $\pm$  1.228 
 &  0.081 &  0.080 &  0.104 &  0.105 &  0.096 &  0.101 &  0.092\\
$\chi^2$        &                      
 &   1.96 &   1.94 &   1.27 &   1.23 &   1.51 &   1.40 &   1.69\\
 \hline
$m_W$           &   80.24 $\pm$   0.16 
 &  79.95 &  80.39 &  80.39 &  80.42 &  80.22 &  80.27 &  80.08\\
$\chi^2$        &                      
 &   3.23 &   0.91 &   0.91 &   1.28 &   0.02 &   0.03 &   1.06\\
 \hline
  $\chi^2_{\rm tot}$&$(\alpha_s=0.11)\quad $
 &  24.87 &  40.66 &  15.20 &  15.40 &  23.29 &  13.88 &  34.33\\
                    &$(\alpha_s=0.12)\quad $
 &  24.62 &  49.38 &  14.74 &  15.34 &  19.76 &  13.26 &  30.22\\
                    &$(\alpha_s=0.13)\quad $
 &  30.10 &  63.65 &  19.79 &  20.78 &  21.72 &  18.20 &  31.66\\
 \hline
 \end{tabular}
\end{center}
\normalsize
}
 
\newpage
\noindent
\section*{\large Table captions }
\renewcommand{\labelenumi}{Table \arabic{enumi}}
\begin{enumerate}
 
\item \tcaptionofchargeformfactors \label{tableofchargeformfactors}
\item \tcaptionofstepsinanalysis   \label{tableofstepsinanalysis}
\item \tcaptionofvertexformfactors \label{tableofvertexformfactors}
\item \tcaptionofboxformfactors    \label{tableofboxformfactors}
\item \tcaptionofwidths            \label{tableofzwidths}
\item \tcaptionoffitsummary        \label{tableoffitsummary}
\item \tcaptionofsummary           \label{tableofsummary}
 
\end{enumerate}
 
\newpage
\tableofchargeformfactors
\vspace{3cm}
\tableofstepsinanalysis
\newpage
\tableofvertexformfactors
\vspace{3cm}
\tableofboxformfactors
\newpage
\tableofzwidths
\newpage
\vspace{2cm}
\tableoffitsummary
\newpage
\tableofsummary

\textwidth  15.5cm
\textheight 23.0cm
\topmargin -0.65cm  
\oddsidemargin  0.25cm 
\evensidemargin 0.25cm 
\baselineskip 18pt
\raggedbottom

 
\def\fcaptionofstusm{%
The SM predictions for the ($S$, $T$, $U$, $\protect\delb$) parameters
defined in eqs.~(\protect\ref{stu}) and (\protect\ref{delb}) are shown as
functions of $m_t$ for selected $\mh$ values. 
Their closed analytic expressions are given in appendix~C. 
$\alpha_s(\mz)$ is set to 0.12 in the two-loop 
$O(\alpha\alpha_s)$ corrections 
for $S$, $T$, $U$\protect\cite{rho_pqcd} and 
$\protect\delb(\mmz)$\protect\cite{zbb_qcd}.
}
 
\def\fcaptionofchargeformfactors{%
The four charge form factors in the minimal SM as functions of the
momentum transfer scale. 
The SM predictions are given for $m_t=100,\,150,\,200\gev$ and 
$\mh=100,\,1000\gev$. 
The parametrization\protect\cite{piqq_h} of the hadronic vacuum polarization 
contribution is used in the space-like region ($-\mmz<q^2<0$). 
In the time-like region ($0<q^2<\mmz$) 
only the heavy quark ($c$, $b$) threshold corrections
are taken into account. 
The light quark contributions at $|q^2|>\mmz$ are calculated in 
perturbative QCD by requiring continuity at $q^2=\mmz$.
See appendix B for details. 
}
 
\def\fcaptionofgammaz{%
The $Z$ total width $\Gamma_Z$ as function of the universal charge form
factors $\protect\gzbar^2(\mmz)$ and $\protect\sbar^2(\mmz)$
for $\protect\delb(\mmz)=0$\,(a), $\protect\delb(\mmz)=-0.01$\,(b)
and $\protect\delb(\mmz)=-0.02$\,(c). Three cases of $\alpha_s(\mz)$
($0.11, 0.12, 0.13$) are shown for each $\delb(\mmz)$.
}
 
\def\fcaptionofsigmah{%
The hadronic cross section on the $Z$-pole $\sigma_h^0$(LEP)
as function of
$\protect\sbar^2(\mmz)$ and $\protect\delb(\mmz)$
for $\alpha_s(\mz) = 0.11, 0.12, 0.13$.
The solid (dashed) lines are obtained for
$\protect\gzbar^2(\mmz) = 0.55$ ($0.57$).
Here $\sigma^0_h$(LEP) is defined by 
$12\pi\Gamma_e\Gamma_h/\mmz\Gamma_Z^2$ in ref.~\cite{lep93};
see discussions in the text. 
}
 
\def\fcaptionofrl{%
The ratio $R_\ell\equiv\sigma^0_h{\rm (LEP)}/\sigma^0_\ell{\rm (LEP)}$ 
of the on-$Z$-pole cross sections 
as function of
$\protect\sbar^2(\mmz)$ and $\protect\delb(\mmz)$
for $\alpha_s(\mz) = 0.11, 0.12, 0.13$.
The solid (dashed) lines are obtained for
$\protect\gzbar^2(\mmz) = 0.55$ ($0.57$).
}
 
\def\fcaptionofrb{%
The ratio $R_b \equiv \sigma_b^0/\sigma_h^0$
of the $b$-quark production cross section to the hadronic 
cross section on the $Z$-pole 
as function of
$\protect\sbar^2(\mmz)$ and $\protect\delb(\mmz)$
for $\alpha_s(\mz) = 0.11, 0.12, 0.13$.
The solid (dashed) lines are obtained for
$\protect\gzbar^2(\mmz) = 0.55$ ($0.57$).
}
 
\def\fcaptionofasymmetry{%
The asymmetries on the $Z$-pole as functions of the effective charge
$\protect\sbar^2(\mmz)$:
the leptonic forward-backward asymmetry $A_{\rm FB}^{0,\ell}$\,(a),
the left-right beam polarization asymmetry $A_{\rm LR}^0$\,(b),
the $b$-quark forward-backward asymmetry $A_{\rm FB}^{0,b}$\,(c),
the forward-backward asymmetry of the $c$-quark $A_{\rm FB}^{0,c}$\,(d).
The solid lines (`Full') are obtained from the full helicity 
amplitudes (\ref{m_nc}) 
including the $\gamma$ and $Z$ exchange as well as the box
contributions (which are negligibly small). The long dashed lines
(`Full $-\gamma$') are obtained by subtracting from the full 
amplitudes (\ref{m_nc})
the real and imaginary parts of the $\gamma$-exchange contribution 
$Q_iQ_j\,[\,\ebar^2(\mmz) -i\,\ehat^2\Delta_{\gamma\gamma}(\mmz)\,]/s$. 
The thick dashed lines
(`$Z$-only') are obtained by retaining only the $Z$-pole term, 
the term multiplying the $Z$-propagator factor in eq.~(\ref{m_nc}). 
The dotted lines 
(`$Z$-only(IBA)') are obtained by using the improved Born
approximation to the $Z$-exchange amplitudes. The thick dashed lines
(`$Z$-only') are used in the present analysis. QCD corrections to
$A_{\rm FB}^{0,b}$ and $A_{\rm FB}^{0,c}$ are calculated for
$\alpha_s = 0.12$.
}
 
\def\fcaptionofzparametersvsmt{%
$m_t$ dependence of the SM predictions for the electroweak $Z$ boson
parameters.
Predictions for three values of $\mh$ are shown by dashed lines
(60~GeV), solid lines (300~GeV) and by dash-dotted lines (1000~GeV),
all calculated for $\alpha_s(\mz)=0.12$.
Also shown by straight lines are the mean (dotted lines) and
the 1-$\sigma$ allowed ranges of the experimental 
data\cite{lep93,alr94} (see sections~4.1 and 6.2).
}
 
\def\fcaptionofzparametersvsalphas{%
$\alpha_s(\mz)$ dependence of the SM predictions for the electroweak
$Z$ boson parameters. Predictions for three values of $m_t$ are shown
by dashed lines (100~GeV), solid lines (150~GeV) and by dash-dotted
lines (200~GeV), all for $\mh=100$~GeV.
Also shown by straight lines are the mean (dotted lines) and
the 1-$\sigma$ allowed ranges of the experimental 
data\cite{lep93} (see section~4.1).
}
 
\def\fcaptionofnuqparameters{%
Relation between the model-independent parameters of
the $\nu_\mu$--$q$ scattering experiments ($g_L^2$, $g_R^2$) and
the two universal form factors
($\protect\sbar^2(0)$, $\protect\gzbar^2(0)$).
The 1-$\sigma$ contour of the present data\cite{nuq_dat} is also
shown: 
see eq.~(\ref{dataofnuq}) in section~4.2.1.
The $\bar{\rho}=1$ line corresponds to 
$\gzbar^2(0) = 4\,\sqrt{2}\,G_F\,\mmz = 0.5486$. 
}
 
\def\fcaptionofnueparameters{%
Relation between the $\nu(\bar{\nu})$--$e$ scattering cross sections
per neutrino energy ($\sigma(\nu e)/E_\nu$, $\sigma(\bar{\nu} e)/E_\nu$)
and the two universal parameters
($\protect\sbar^2(0)$, $\protect\gzbar^2(0)$).
The 1-$\sigma$ contour shows the experimental constraint: see
eq.~(\ref{dataofnue})\cite{nue_dat} and 
eq.~(\ref{sig_nue_dat}) in section~4.2.2.
The $\bar{\rho}=1$ line corresponds to 
$\gzbar^2(0) = 4\,\sqrt{2}\,G_F\,\mmz = 0.5486$. 
}
 
\def\fcaptionofedparameters{%
Relation between the model-independent parameters 
($2C_{1u}-C_{1d}$, $2C_{2u}-C_{2d}$)\cite{old_ncfit} of
the $e$--${\rm D}$ polarization asymmetry experiments 
and the two universal form factors
($\protect\sbar^2(0)$, $\protect\gzbar^2(0)$).
The 1-$\sigma$ contour of the present data\cite{ed_dat} is also
shown:
see eq.~(\ref{dataofed}) in section~4.2.4.
The $\bar{\rho}=1$ line corresponds to 
$\gzbar^2(0) = 4\,\sqrt{2}\,G_F\,\mmz = 0.5486$. 
}

\def\fcaptionofapvparameters{%
The weak charge ($Q_W$) of 
the cesium atom $^{133}_{55}{\rm C_s}$ in the 
atomic parity violation experiments 
as function of the two universal parameters
($\protect\sbar^2(0)$, $\protect\gzbar^2(0)$).
The 1-$\sigma$ contour of the present data\cite{apv_dat} is
shown by dashed lines: 
see eq.~(\ref{dataofapv}) in section~4.2.3. 
The $\bar{\rho}=1$ line corresponds to 
$\gzbar^2(0) = 4\,\sqrt{2}\,G_F\,\mmz = 0.5486$. 
}
 
\def\fcaptionofmw{
The SM predictions for $\mw$ as functions of $m_t$ and $\mh$
for $\delta_\alpha=0$, $\delg=0.0055$ and $\alpha_s=0.12$.
The 1-$\sigma$ allowed range of the present data\cite{mw93} 
is shown by thick dashed lines:
see eq.~(\ref{dataofmw}) in section~4.3.
}
 
\def\fcaptionofzparameterfit{%
3-parameter fit to the $Z$ boson parameters:
the $Zb_Lb_L$ vertex form factor $\protect\delb(\mmz)$ is introduced
as the third parameter of the fit in addition to the two universal
charge form factors $\protect\sbar^2(\mmz)$ and
$\protect\gzbar^2(\mmz)$: see eq.~(\ref{fitofzparameter}).
The 1-$\sigma$ contours are shown for three
representative $\alpha_s(\mz)$ values, 0.11 (dashed lines), 0.12 (solid
lines), 0.13 (dot-dashed lines). Also shown are the SM predictions in
the range 100~GeV$<m_t<$200~GeV  and 50~GeV$<\mh<$1000~GeV, which are
calculated assuming $(\Delta\frac{1}{\alpha})_{\rm hadrons} =-3.88$
($\delta_\alpha=0$)\protect\cite{piqq_h} for the hadronic vacuum
polarization contribution to $1/\bar{\alpha}(\mmz)$.
}
 
\def\fcaptionoflencfit{%
Fit to the low energy neutral current data in terms of the two universal
charge form factors $\protect\sbar^2(0)$ and $\protect\gzbar^2(0)$.
1-$\sigma$ contours are shown separately for
the $\nu_\mu$--$q$ data\protect\cite{nuq_dat},
the $\nu_\mu$--$e$ data\protect\cite{nue_dat},
the atomic parity violation (APV) data\protect\cite{apv_dat},
and the SLAC $e$--${\rm D}$ 
polarization asymmetry data\protect\cite{ed_dat}:
see eqs.~(\ref{fitofnuqnotgauss}), (\ref{fitofnue}), 
(\ref{fitofapv}) and (\ref{fitofed}), respectively. 
The 1-$\sigma$ contour of the combined fit, eq.~(\ref{fitoflenc}), 
is shown by the thick contour.
The straight dashed line shows the `tree' level prediction
of the minimal SM:
$\bar{\rho} \equiv \protect\gzbar^2(0)/(4\protect\sqrt{2}G_F\mmz) =1$, 
or $\gzbar^2(0) = 4\,\sqrt{2}\,G_F\,\mmz = 0.5486$. 
}
 
\def\fcaptionofed{%
Fit to the SLAC $e$--${\rm D}$ 
polarization asymmetry data\protect\cite{ed_dat}
in terms of the model-independent parameters $2C_{1u}-C_{1d}$ and
$2C_{2u}-C_{2d}$\cite{old_ncfit} of the effective weak Hamiltonian
(\protect\ref{lpv_eq}). Uncertainties due to the sea-quark
contributions (a), the longitudinal to transverse virtual photon 
cross section ratio
$R=\sigma_L/\sigma_T$ (b), and the higher twist effects (c) have been
examined, and the fit (d) is obtained after taking account of all the
uncertainties.
Shown by the solid lines are the tree-level predictions of the SM, 
and the blobs show the predictions at selected $\sin^2\theta_W^{}$ 
values. 
}
 
\def\fcaptionofrunningvsmh{%
The running of the charge form factor $\gzbar^2(q^2)$,
$4\pi/\protect\gzbar^2(\mmz)-4\pi/\protect\gzbar^2(0)$,
as functions of $\mh$ calculated in the SM
for $100\gev<m_t<200\gev$.
The 1-$\sigma$ allowed range from the neutral current 
experiments on the $Z$-pole and at low energies, 
eq.~(\ref{fitofrunning}), 
is also shown for comparison.
}
 
\def\fcaptionofrunningfit{%
The running of the charge form factor $\gzbar^2(q^2)$,
$\sbar^2(q^2)$ and $\ebar^2(q^2)$
as expected from the one-loop contribution of the wino
(fermionic partner of the $W$ in the supersymmetric SM)
to the three neutral current propagators.
(a)~$4\pi/\protect\gzbar^2(\mmz)-4\pi/\protect\gzbar^2(0)\,$;
(b)~$\sbar^2(\mmz)/\bar{\alpha}(\mmz)-\sbar^2(0)/\alpha\,$;
(c)~$\delta_\alpha \equiv 1/\bar{\alpha}(\mmz)-128.72\,$.
The SM contributions are shown for
$m_t=100,\,200\gev$ and $\mh=60,\,1000\gev$.
The singularity at $m_{\rm wino}=\mz/2$ in (a) 
reflects\protect\cite{mz_thr} the deviation of the $Z$ line-shape from 
the standard Breit-Wigner form that has been assumed both 
in the experimental fit and in our theoretical formula.
The 1-$\sigma$ allowed ranges from the neutral current 
experiments on the $Z$-pole and at low energies, 
eq.~(\ref{fitofrunning}), 
are also shown for comparison.
There is no direct measurement of $\delta_\alpha$. 
}
 
\def\fcaptionofncfit{%
2-parameter fit to the combined low energy neutral current data and
the $Z$ parameters.
The latter fit (`LEP$+$SLC') is copied from
Fig.~\protect\ref{figureofzparameterfit} for $\alpha_s(\mz)=0.12$.
The low energy combined fit of Fig.~\protect\ref{figureoflencfit}
has been rescaled to the $\mz$ scale by assuming the SM running of the
two charge form factors, $\protect\sbar^2(q^2)$ and
$\protect\gzbar^2(q^2)$, which depend on $m_t$ and $\mh$.
Uncertainties due to $m_t$ and $\mh$ in the SM predictions for the
running of the form factors are illustrated by drawing the results for
$m_t=100,\,200\gev\,$ and $\mh=60,\,1000\gev\,$ in the same figure. 
The 1-$\sigma$ contour of the combined fit, eq.~(\ref{fitofallnc}), 
is given by the thick contour,
for which the above uncertainties give negligible effects.
}
 
\def\fcaptionofstufit{%
Global fit to the ($S$, $T$, $U$) parameters 
for three $\alpha_s$ values 
and $\delta_\alpha \equiv 1/\bar{\alpha}(\mmz) -128.72 =0$  
and $\delg=0.0055$. 
Four 1-$\sigma$ contours are obtained for each $\alpha_s$, 
by using $m_t=150,\,200\gev$ and $\mh =100,\,1000\gev$
in evaluating the running of the charge form factors: 
see eq.~(\ref{fitofstu}) for a parametrization of the fit 
for $m_t=175\gev$ and $\mh=100\gev$. 
The fourth parameter of the fit, the $\zbb$ vertex form factor 
$\delb(\mmz)$, is allowed to take an arbitrary value, 
free from SM constraints. 
The SM predictions with $\delta_\alpha=0$ and 
$\delg=0.0055$ are also given for $100\gev<m_t<200\gev$ 
and $50\gev<\mh<1000\gev$. 
}

\def\fcaptionofmtmhconstraints{%
Electroweak constraints on ($m_t$, $\mh$) in the minimal SM,
for three selected $\alpha_s$ values at $\delta_\alpha=0$.
Dashed lines show the best $m_t$ values for a given $\mh$,
and the solid contours are for $\chi^2=\chi^2_{\rm min}+1$ and
$\chi^2=\chi^2_{\rm min}+4.61$.
The minimum point of $\chi^2$ is marked by ``$\times$''.
The region $\mh<63\gev$ is excluded by LEP 
experiments\cite{mh_limit}.~
}
 
\def\fcaptionofchisqvsmt{%
Total $\chi^2$ of the SM fit to all the electroweak data
as functions of $m_t$ for $\mh=60$, 300, 1000$\gev$ and
$\alpha_s(\mz)=0.11,\,0.12,\,0.13$.
The uncertainty $\delta_\alpha$ in the hadronic vacuum polarization
contribution to the effective charge $1/\bar{\alpha}(\mmz)$ is shown
for three cases, $\delta_\alpha=-0.1$ (a), $0$ (b), $+0.1$ (c).
The dotted lines are obtained by using the approximate formula
(\protect\ref{total_chisqsm}).
The degree of freedom is 19. 
}
 
\def\fcaptionofchisqvsmh{
Total $\chi^2$ of the SM fit to all the electroweak data
as functions of $\mh$ for $m_t=120, 140, 160, 180, 200~\gev$ and
$\alpha_s(\mz)=0.11,\, 0.12,\, 0.13$.
The hadronic vacuum polarization contribution to the effective charge
$1/\bar{\alpha}(\mmz)$ is fixed by setting $\delta_\alpha=0$.
The dotted lines show our approximation
(\protect\ref{total_chisqsm}).
The degree of freedom is 19.~
}
 
\def\fcaptionofsbarfromasymmetries{%
The universal weak mixing form factor 
$\protect\sbar^2(\mmz)$ as determined from
various asymmetry measurements on the $Z$-pole:
the lepton ($e$, $\mu$, $\tau$)
forward-backward asymmetries\cite{lep93},
the $\tau$ polarization asymmetry\cite{lep93},
the left-right asymmetry\cite{alr94}
and the quark ($b$,$c$) forward-backward asymmetries\cite{lep93}: 
see eqs.~(\ref{fitofzasymmetries}), (\ref{fitofsbarfromnewalr})
and the footnote~\ref{fn:new_and_old_alr}.
Also shown is the deviation `$\chi$' (that is,
$ \chi \equiv (\langle\sbar^2(\mmz)\rangle -0.2302)/
\sigma(\sbar^2(\mmz))$) for each fit individually, where
$\langle\sbar^2(\mmz)\rangle$ and $\sigma(\sbar^2(\mmz))$
denote mean and standard deviation of each fit, respectively.
At the bottom the above $\chi$-values are histogrammed.
}
 
\def\fcaptionofzparameterfitwithnewalr{%
Impact of the left-right asymmetry data\cite{alr94} by the SLD
collaboration.
The band (mean\,(dashed line) and the 1-$\sigma$\,(solid lines))
represents the constraint from the new left-right asymmetry data alone.
The constraints from the fits with and without
(see Figs.~\protect\ref{figureofzparameterfit} and \ref{figureofstufit})
the new asymmetry data are shown by thick and thin lines, respectively,
for $\alpha_s(\mz) = 0.11, 0.12, 0.13$. 
The $\zbb$ vertex form factor $\delb(\mmz)$ is allowed to take 
an arbitrary value, free from SM constraints. \\
(a):
The fits are shown in the
$\protect\sbar^2(\mmz)$--$\protect\gzbar^2(\mmz)$ plane: 
see eq.~(\ref{fitofstuwithnewalr}). 
The SM predictions are obtained by assuming $\delta_\alpha=0$
in the range $100\gev<m_t<260\gev$
and $1\gev<\mh<1000\gev$. \\
(b):
The fits are shown in the $S$--$T$ plane,
where $\delta_\alpha=0$ and $\delg=0.0055$ are assumed,
and $m_t = 174 \gev$ and $\mh=100 \gev$ are used
to calculate the SM running of the charge form factors
between $q^2=0$ and $q^2=\mmz$: 
see eq.~(\ref{fitofstuwithnewalr}) for parametrization of the result. 
The results are insensitive to the actual ($m_t$, $\mh$) values
in the region $m_t>100\gev$ and $\mh>50\gev$ 
(see Figs.~\ref{figureofrunningvsmh} and \ref{figureofrunningfit}). 
The SM predictions are given
in the range $100\gev<m_t<260\gev$ 
and $50\gev<\mh<1000\gev$.
}
 
\def\fcaptionofmtmhconstraintswithnewalr{%
Electroweak constraints on ($m_t$, $\mh$) in the minimal SM,
including the new left-right asymmetry data\cite{alr94},
for three selected $\alpha_s$ values 
$\alpha_s(\mz)=0.11,\,0.12,\,0.13$, and for
(a)~$\delta_\alpha=-0.1$, (b)~$\delta_\alpha=0$,
and (c)~$\delta_\alpha=0.1$.
Dashed lines show the best $m_t$ values for a given $\mh$,
and the solid contours are for $\chi^2=\chi^2_{\rm min}+1$ and
$\chi^2=\chi^2_{\rm min}+4.61$.
The minimum point of $\chi^2$ is marked by ``$\times$''.
The region $\mh<63\gev$ is excluded by LEP 
experiments\cite{mh_limit}.
}
 
\def\fcaptionofchisqvsmtwithnewalr{%
Total $\chi^2$ of the SM fit to all the electroweak data including the
new left-right asymmetry data\cite{alr94} as functions of $m_t$ for
$\mh=$60, 300, 1000~GeV and $\alpha_s(\mz)=$0.11, 0.12, 0.13.
The uncertainty $\delta_\alpha$ in the hadronic vacuum polarization
contribution to the effective charge $1/\bar{\alpha}(\mmz)$ is shown
for three cases, $\delta_\alpha=-0.1$ (a), $0$ (b), $+0.1$ (c).
The dotted lines are obtained by the approximate formula
(\protect\ref{total_chisqsmwithnewalr}).
The degree of freedom is 19. 
}
 
\def\fcaptionofchisqvsmhwithnewalrwiderange{%
Total $\chi^2$ of the SM fit to all the electroweak data including the
new left-right asymmetry data\cite{alr94} as functions of $\mh$ for
$m_t=100 - 200$~GeV, for three selected $\alpha_s$ values
(a)~$\alpha_s(\mz)=0.11$, (b)~$\alpha_s(\mz)=0.11$ and
(c)~$\alpha_s(\mz)=0.11$, at $\delta_\alpha=0$.
The dotted lines show our approximation
(\protect\ref{total_chisqsmwithnewalr})
obtained by fitting the $\chi^2$ values in the region
$63\gev < \mh < 1000\gev$.
The degree of freedom is 19. 
}
 
\def\fcaptionofzparameterfitwithmt{%
2-parameter fits to the $Z$ boson parameters, where in
(a) $\sbar^2(\mmz)$ and $\gzbar^2(\mmz)$ are free parameters,
and in (b) $S$ and $T$ are free parameters.
In both cases the $\zbb$ vertex correction is assumed to be dominated
by the SM contribution, and the $m_t$ value in the vertex correction is
treated as external parameter in the fit.
The 1-$\sigma$ contours are shown for three representative $\alpha_s(\mz)$
values, 0.11 (dashed lines), 0.12 (solid lines), 0.13 (dot-dashed lines).
Also shown are the SM predictions in the range $100\gev<m_t<200\gev$
for $1\gev<\mh<1000\gev$~(a), and for $50\gev<\mh<1000\gev$~(b).
The SM predictions in (a) and the 1-$\sigma$ contours in (b) are obtained
by assuming $\delta_\alpha \equiv 1/\bar{\alpha}(\mmz)-128.72 =0$.
}
 
\def\fcaptionofmtmhconstraintswithnewalrandmt{%
Electroweak constraints on ($m_t$, $\mh$) in the minimal SM,
including the new left-right asymmetry data\cite{alr94}
and the constraint $m_t=174\pm 16$\cite{top_cdf},
for three selected $\alpha_s$ values at $\delta_\alpha=0$.
Dashed lines show the best $m_t$ values for a given $\mh$,
and the solid contours are for $\chi^2=\chi^2_{\rm min}+1$ and
$\chi^2=\chi^2_{\rm min}+4.61$.
The minimum point of $\chi^2$ is marked by ``$\times$''.
The region $\mh<63\gev$ is excluded by LEP 
experiments\cite{mh_limit}.~
}
 
\def\fcaptionofmhconfidencelevelwithnewalr{%
Constraints on the Higgs mass in the SM from all the electroweak data
including the new left-right asymmetry data\cite{alr94}. 
Here the top mass $m_t$ is considered as external parameter
with negligible uncertainty.
Upper (solid
lines) and lower (dashed lines) bound of the Higgs mass at 95\% CL are
shown as functions of $m_t$ for $\alpha_s(\mz)=0.11, 0.12, 0.13$.
The hadronic vacuum polarization contribution to the effective charge
$1/\bar{\alpha}(\mmz)$ is set by
$\delta_\alpha\equiv 1/\bar{\alpha}(\mmz)-128.72=0$.
}
 
\def\fcaptionofchisqvsmhwithnewalrwideeachsector{%
The contributions to $\chi^2$ from each sector of the analysis
in the SM:
(a)~from the $Z$ parameters including the new left-right
asymmetry data\cite{alr94},
(b)~from the low energy neutral current experiments
and (c)~the $\mw$ measurements.
They are calculated as functions of $\mh$ for $m_t=100$--$200\gev$,
at $\alpha_s(\mz)=0.12$ and $\delta_\alpha=0$. 
The degree of freedom is 9 for the $Z$ parameters (a), 
9 for the low energy neutral current experiments (b), 
and 1 for $\mw$ (c).
}
 
\newpage
\section*{\large Figure captions }
\renewcommand{\labelenumi}{Fig.\arabic{enumi}}
\begin{enumerate}

\item \fcaptionofstusm
  \label{figureofstusm}
\item \fcaptionofchargeformfactors
  \label{figureofchargeformfactors}\vspace{-6mm}
\item \fcaptionofgammaz
  \label{figureofgammaz}
\item \fcaptionofsigmah
  \label{figureofsigmah}
\item \fcaptionofrl
  \label{figureofrl}
\item \fcaptionofrb
  \label{figureofrb}
\item \fcaptionofasymmetry
  \label{figureofasymmetry}
\item \fcaptionofzparametersvsmt
  \label{figureofzparametersvsmt}
\item \fcaptionofzparametersvsalphas
  \label{figureofzparametersvsalphas}
\item \fcaptionofnuqparameters
  \label{figureofnuqparameters}
\item \fcaptionofnueparameters
  \label{figureofnueparameters}
\item \fcaptionofedparameters
  \label{figureofedparameters}
\item \fcaptionofapvparameters
  \label{figureofapvparameters}
\item \fcaptionofmw
  \label{figureofmw}
\item \fcaptionofzparameterfit
  \label{figureofzparameterfit}
\item \fcaptionoflencfit
  \label{figureoflencfit}
\item \fcaptionofed
  \label{figureofed}
\item \fcaptionofrunningvsmh
  \label{figureofrunningvsmh}
\item \fcaptionofrunningfit
  \label{figureofrunningfit}
\item \fcaptionofncfit
  \label{figureofncfit}
\item \fcaptionofstufit
  \label{figureofstufit}
\item \fcaptionofmtmhconstraints
  \label{figureofmtmhconstraints}
\item \fcaptionofchisqvsmt
  \label{figureofchisqvsmt}
\item \fcaptionofchisqvsmh
  \label{figureofchisqvsmh}
\item \fcaptionofsbarfromasymmetries
  \label{figureofsbarfromasymmetries}
\item \fcaptionofzparameterfitwithnewalr
  \label{figureofzparameterfitwithnewalr}
\item \fcaptionofmtmhconstraintswithnewalr
  \label{figureofmtmhconstraintswithnewalr}
\item \fcaptionofchisqvsmtwithnewalr
  \label{figureofchisqvsmtwithnewalr}
\item \fcaptionofchisqvsmhwithnewalrwiderange
  \label{figureofchisqvsmhwithnewalr}
\item \fcaptionofzparameterfitwithmt
  \label{figureofzparameterfitwithmt}
\item \fcaptionofmtmhconstraintswithnewalrandmt
  \label{figureofmtmhconstraintswithnewalrandmt}
\item \fcaptionofmhconfidencelevelwithnewalr
  \label{figureofmhconfidencelevelwithnewalr}
\item \fcaptionofchisqvsmhwithnewalrwideeachsector
  \label{figureofchisqvsmhwithnewalrwideeachsector}
\end{enumerate}

\end{document}